\documentclass[twoside]{article}

\usepackage[a4paper]{geometry}

\usepackage{graphicx}
\usepackage{cite,latexsym,amsmath,amssymb,bm,subfigure,slashed,tabularx}
\usepackage{color}

\newcommand{\y}{Y}
\newcommand{\xbj}{{x_{\text{bj}}}}

\newcommand{\singletproj}{{%
\frac{1}{d(R)}%
\parbox{.15cm}{\includegraphics[width=.15cm]{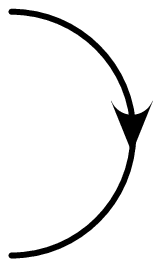}}% 
\ \parbox{.15cm}{\includegraphics[width=.15cm]{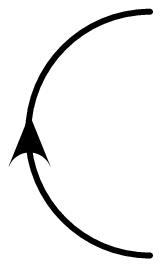}}% 
}}

\newcommand{\tr}{{\rm tr}}               % trace symbol

\newcolumntype{Z}{>{\small}l}

\begin{document}
\ifx\href\undefined\else\hypersetup{linktocpage=true}\fi 
\phantom{a}
\vspace{1cm}
%\title{ \vspace{1cm} 
\noindent{\Large\bf Evolution at small $\xbj$: The Color Glass Condensate}
%\author{
\\
  \vskip 0.5cm
\noindent  {\large Heribert Weigert}%$^{1}$
\\
\\
%$^1$
{\small  Institut f{\"u}r theoretische Physik, Universit{\"a}t Regensburg, 93040 Regensburg, Germany}
%\maketitle
%\begin{abstract}
\vspace{.5cm}
\noindent\begin{center}
\begin{minipage}{.915\textwidth}
  {\small\sf
  When probed at very high energies or small Bjorken $\xbj$, QCD
  degrees of freedom manifest themselves as a medium of dense gluon
  matter called the Color Glass Condensate. Its key property is the
  presence of a density induced correlation length or inverse
  saturation scale $R_s=1/Q_s$.  Energy dependence of observables in
  this regime is calculable through evolution equations, the JIMWLK
  equations, and characterized by scaling behavior in terms of $Q_s$.
  These evolution equations share strong parallels with specific
  counterparts in jet physics. Experimental relevance ranges from
  lepton proton and lepton nucleus collisions to heavy ion collisions
  and cross correlates physics at virtually all modern collider
  experiments.
}
%\end{abstract}
\end{minipage}
\end{center}
\vspace{1cm}
\tableofcontents
\section{Introduction}
\label{sec:introduction}

\subsection{\it QCD in collider experiments at high energies}
\label{sec:it-qcd-collider}

With the advent of modern colliders, from the Tevatron and HERA to
RHIC and planned facilities like LHC and EIC, the high energy
asymptotics of QCD has gained new prominence and importance.  Needless
to say that these facilities have been built with very different
research goals in mind. On the particle physics side one finds most
prominently the search for the Higgs particle as the last missing
ingredient of the standard model and hope for clear
evidence of physics beyond it. Here high energies are needed mainly to
exceed particle production thresholds to facilitate the generation of
telltale signals with sufficiently large cross section.
On the nuclear or heavy ion physics side it is the search for the
quark gluon plasma and its properties that requires high energies to
create the energy densities and temperatures needed to cross the QCD
deconfinement phase transition and create a medium of deconfined
quarks and gluons whose exact nature is still hotly under
debate.

In all these experiments one is talking about enormous center of mass
energies per participant (i.e. nucleon): RHIC operates at center of
mass energies of $200$GeV/nucleon, LHC in its heavy ion mode is aiming
at $2$TeV/nucleon while the pure particle physics experiments will be
conducted via proton proton collisions at $7$TeV. This implies
enormous boost factors between $200$ and several thousands that will
strongly affect the QCD aspects of these experiments. In such an
environment one knows that soft gluon emission is enhanced by
logarithms that arise from $dE/E$ factors in the phase space measure.

In QED the analogous process of photon emission can be treated by
resummation, which gives rise to Sudakov form factors. (These provide
the resolution of the well known infrared ``problem'' in QED.) For the
non-Abelian case the situation is more involved. The (Abelian) Sudakov
type of exponentiation or eikonalization is still present, but
captures only a part of the effects. In QCD the problem of multiple
soft emission is intrinsically nonlinear -- gluons carry color charge
and thus in contradistinction to photons act as sources for further
emission. This provides a growth mechanism for cross sections absent
in QED. Complementarily, after a sufficient number of soft emission
steps, one has to start to consider recombination effects: further
emission into a region that is already populated by a large number of
other color charges will be modified by recombination and absorption,
a mechanism that will invariably slow down further growth in gluon
numbers and lead to saturation.  Both these effects are irrelevant in
QED (suppressed by an additional factor of $\alpha$) and only relevant
in the non-Abelian theory.  There they are usually associated with the
successes and failures of the Balitsky-Fadin-Kuraev-Lipatov (BFKL-)
equation~\cite{Bal-Lip, Lipat, Kuraev:1976ge, Kuraev:1977fs, Lipat-2},
which is the prototype of an evolution equation meant to resum soft
gluon emissions at high energies. This equation carefully takes into
account the multiple emission aspect and in this sense addresses a
specifically non-Abelian issue that is absent in QED, but excludes
recombination effects that arise at large gluon densities and thus is
no longer valid where these become relevant.

The BFKL equation, if taken literally and used beyond this limit of
applicability leads to an untamed, characteristic power-like growth
of cross sections with cm energy, for instance in deep inelastic
scattering (DIS) as conducted in electron proton collisions at HERA.
This in itself would lead to a violation of the unitarity bound on
cross sections.\footnote{Unitarity in QCD requires that cross sections
  asymptotically rise not faster with invariant energy $s$ than $(\ln
  s)^2$. This is usually called Froissart's
  theorem~\cite{Froissart:1961ux}.}

On the other side, at vanishing (or just very small) momentum transfer
the BFKL equation is plagued by a diffusion into the infrared: gluons
are emitted at smaller and smaller momenta and will eventually reach
nonperturbative scales where the assumption of small coupling
underlying the derivation of the equation is no longer valid.

Both of these main problems of the BFKL equation can be cured by
taking into account nonlinear effects in the gluon generation
mechanism at high energies, at least for central collisions, where we
expect gluon densities to become large. This is the domain of a highly
dense gluonic medium and the topic of this review. The main
theoretical tool is an associated evolution (or Wilson type
renormalization group) equation that describes the creation and change
of the medium with increasing energy, the
Jalilian-Marian+Iancu+McLerran+Weigert\-+Leonidov\-+Kovner equation~\cite{
  Jalilian-Marian:1997xn, Jalilian-Marian:1997jx,
  Jalilian-Marian:1997gr, Jalilian-Marian:1997dw,
  Jalilian-Marian:1998cb, Kovner:2000pt, Weigert:2000gi,
  Iancu:2000hn,Ferreiro:2001qy} [the order of names was chosen by Al
Mueller to give rise to the acronym JIMWLK, pronounced ``gym walk''].
The quite distinct combination of features, the generation of large
gluon densities which can arise only in non-Abelian theories, the time
scale differences between soft and hard modes that enter the dynamics
and the onset of recombination and saturation effects have led
McLerran and Venugopalan
%\marginpar{where?} 
to coin the term ``Color Glass Condensate'' (CGC) which subsequently
has been generally adopted as a convenient label for this situation in
particular with regards to the HERA (electron proton), RHIC (heavy
ion) EIC (lepton nucleus) and LHC (both in proton proton and heavy ion
mode) experiments.

It should be kept in mind, however, that the central underlying
concept, the nonlinear effects in soft gluon emission feature
prominently also in other areas. An example that provides
opportunities for fruitful comparison can be found in the study of so
called non-local jet observables. This is a class of observables in
which soft emission from secondary particles --relatively ''hard'' jet
constituents other than the original leading partons-- plays an
important role.  Accordingly jet evolution can be described by
evolution equations that have a, at first sight surprising, strong
structural and conceptual analogy with the JIMWLK equation. For so
called global observables, where the origin of secondary gluon
radiation can be ascribed to the original leading partons, the
equations reduce to liner evolution equations that to a large extent
are dominated by Sudakov type physics -- to a degree that one can even
derive evolution equations without any overt reference to medium
effects. In the generic case, however, the medium is the crucial
ingredient.  This situation will be used throughout this review to
highlight the generic nature of concepts and the mechanisms appearing
in the generation of a dense gluonic medium at high energy, although
phenomenology and applications in both cases are quite different.

Returning to the CGC, one has to highlight its generic feature, the
creation of a density induced correlation length $R_s = 1/Q_s$. The
associated momentum scale $Q_s$ is called the saturation scale. They
simply characterize the onset of the saturation effects induced by the
medium. Beyond $1/Q_s$ color charges start to screen each other and
since the underlying gluon densities grow with energy, $Q_s$ will, by
necessity, follow this trend. $Q_s$ is the characteristic scale of the
medium and responsible for all the ``good'' features that go beyond
the physics of the BFKL equation: Below $Q_s$ the growth of the gluon
density and thus of certain cross sections is checked on the one hand,
and on the other hand the infrared problems are cured: a shrinking
correlation length $R_s$ prevents the appearance of non-perturbative
modes near and below the QCD scale as long as the dense medium is
present, in particular for central collisions.  Generically, one would
expect $Q_s$, as the characteristic scale of the medium, also to be
the relevant scale for the coupling constant in this situation. This
holds true, although it has turned out that the issue is a bit more
subtle than initially expected.

As the list of experiments given at the outset already indicates, the
situation is quite generic. Given the high energy involved in all
these experiments, it is quite obvious that soft gluon emission and
thus JIMWLK like evolution should affect all of them, be it a
relatively simple experiment such as deep inelastic lepton proton
(henceforth referred to by $e p$ or $\gamma^* p$, as done at HERA) or
lepton nucleus scattering ($e A$ or $\gamma^* A$ as planned with the
EIC experiment) or proton proton ($p p$), proton nucleus ($p A$) or
nucleus nucleus ($A A'$) experiments. [$A$ refers to the atomic number
of the nucleus involved.] In all of them one expects CGC physics and
thus the presence of an energy dependent saturation scale $Q_s$ to
affect particle production rates and cross sections. This should allow
to cross correlate many of these experiments in this respect and gain
a clear understanding of the relevance of the CGC in high energy
experiments.  It should be noted, however, that the tools are most
developed for the simple case of deep inelastic scattering of leptons
on protons and nuclei. This example then will also be my starting
point to introduce the physical ideas --with a minimal amount of
formulae-- in the following section.

\subsection{\it The CGC in DIS: high energies, gluon 
densities and the saturation scale}
\label{sec:cgc-dis:-high}

To acquaint the reader with the main features of the CGC, it is
perhaps best to refer to the example in which the physics ingredients
are best understood and the technical tools that were developed to
incorporate them are most directly applicable, deep inelastic
scattering of leptons on (preferably large) nuclei. I will attempt to
keep the technical details in this section to a minimum. All of the
relevant physics ideas collected here will have their theoretical
foundations reviewed in later sections.
%In this general overview, I
%will concentrate on the physics picture and leave formulae aside that
%require more than general textbook knowledge.

\begin{figure}[htbp]
  \centering
  \begin{minipage}{.48\textwidth}
    \includegraphics[width=\textwidth]{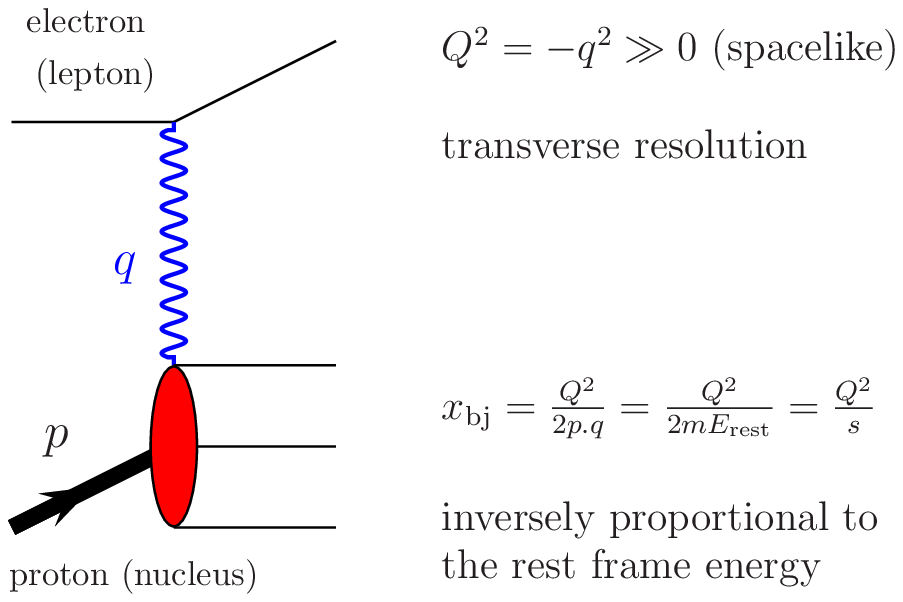}
\vspace{.19\textwidth}
  \caption{\small Kinematics in DIS of leptons off protons or nuclei}
  \label{fig:DISkin}
  \end{minipage}
\hfill
\begin{minipage}{.48\textwidth}
\centering
  \includegraphics[width=.7\textwidth]{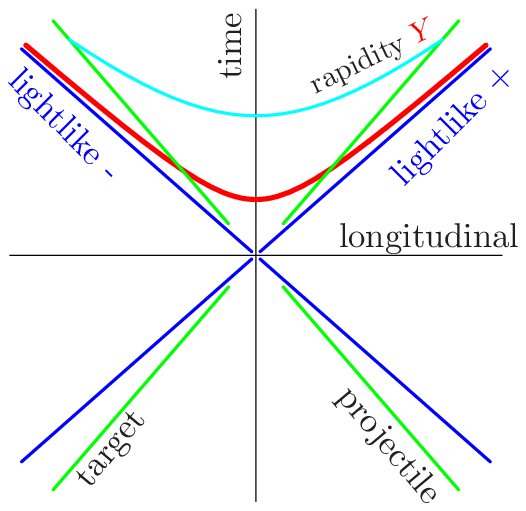}
  \caption{\small 
    Kinematic variables in a Minkowski diagram. Projectile and target
    are separated by a large boost factor $\frac{1}{\xbj}=e^\y$.
    $x^\pm:=\frac{1}{\sqrt{2}}(x^0\pm x^3)$ are the lightlike
    directions in the $x^0$-$x^3$ plane often referred to as ``$+$''
    and ``$-$'' directions.}
\label{fig:lightcone-diag}
\end{minipage}
\end{figure}

\begin{figure}[htbp]
\centering
%\begin{minipage}{8cm}
   \includegraphics[width=7cm]{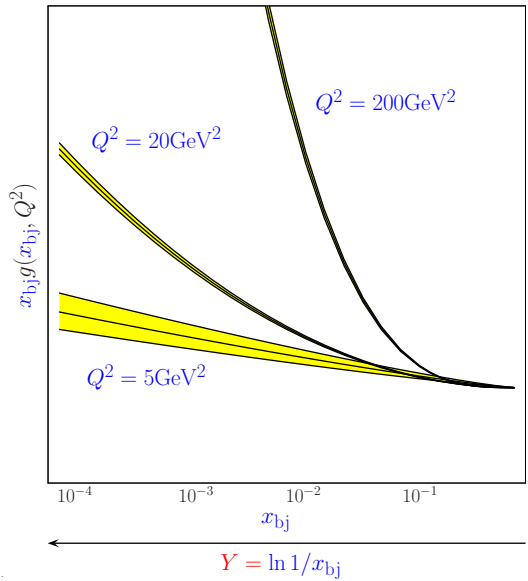}
  \caption{\small 
    Qualitative growth of gluon distributions as from typical DGLAP
    fits of DIS at HERA ($e A$) for different $Q^2$.  }
\label{fig:DGLAPschematics}
%\end{minipage}
\end{figure}

To start with, consider a collision of a virtual photon that imparts a
(large) spacelike momentum $q$ ($Q^2:=-q^2$ is large) on a nucleus of
momentum $p$ as shown in Fig.\ref{fig:DISkin}. Being spacelike, $Q^2$
sets the transverse resolution scale in the problem.  At large
energies the only other relevant invariant is Bjorken-$x$ (or $\xbj$),
defined as $\xbj:= \frac{Q^2}{2 p.q}$ as in Fig.~\ref{fig:DISkin}.
$\xbj$ has the interpretation of a (total) momentum fraction carried
by a struck parton. At small $\xbj$ this has the interpretation of a
light cone momentum fraction $k^+=\xbj p^+$ where $k$ is the momentum
of the probed constituent and $p$ the target momentum.\footnote{This
  is not to be confused with Feynman $x$ or $x_F$, defined as the
  longitudinal momentum fraction of a constituent. The two agree to
  lowest order in the parton model of DIS, but start do differ when
  perturbative emission into the inclusive final state are considered.
  Less inclusive measurements bring $x_F$ into its full right:
  longitudinal momentum fractions of measured particles come into the
  game and relate in specific ways to the longitudinal momentum
  fractions of incoming constituents giving rise to process specific
  definitions of $x_F$.} $\xbj$ directly corresponds to the large
boost factor separating (photon) projectile and target (nucleus) via
$\frac{1}{\xbj}=e^\y$ as shown in
Fig.\ref{fig:lightcone-diag}.\footnote{A more precise definition would
  employ to the worldlines of constituent quarks and gluons in the
  projectile and target wavefunctions which whill appear below.}
At high transverse resolutions the situation is simplest. Many
quantities, amongst them the total cross section, can be addressed in
the framework of an expansion in inverse powers of the transverse
resolution $Q^2$. This is known as twist expansion and Operator
Product Expansion (OPE). Technicalities aside, this expansion has the
character of a density expansion: at high resolution a nucleon in the
target (which may be a proton in the simplest case) appears to be a
dilute system of quarks and gluons. In such a dilute system and in
lowest order of the expansion it is sufficient to express the cross
section entirely in terms of two point functions, the quark and gluon
distributions.  Only higher orders of the expansion would involve
particle correlations -- which are naturally suppressed when densities
a small.  Nevertheless, even the individual terms in this ``density
expansion'' receive quantum corrections which are important.  At large
$Q^2$ the dominant ones carry factors of $(\alpha_s \ln Q^2)^n$ where
a large logarithm ($\ln Q^2$) compensates for the smallness of the
coupling.  These contributions need to be resummed to all orders to
understand the $Q^2$ dependence of quark and gluon distributions at
high resolutions. This can be done diagrammatically --known under the
keyword 'resummation of ladder diagrams'-- or by deriving a
renormalization group equation, a differential equation for the
distribution functions with respect to $\ln Q^2$, the
Dokshitzer-Gribov-Altarelli-Parisi (DGLAP) equations.  By solving
these equations one resums the logarithmically enhanced corrections.
In fact, the differential equations are the only information that can
be extracted perturbatively: their initial conditions, the quark and
gluon distributions at some resolution $Q_0$ {\em do} contain
nonperturbative information about the target.  As a consequence one
needs experimental measurements over a range of $Q^2$ to extract the
gluon distributions at $Q_0$ by a fit consistent with further
evolution. One gains quark and gluon distributions that can be used
also in other experiments on the proton. A quite striking result of
this procedure is a strong growth of the gluon distributions at small
$\xbj$ as sketched qualitatively in Fig.~\ref{fig:DGLAPschematics}.
This rise is driven by soft gluon radiation, (sea) quark distributions
simply follow their rise.

\begin{figure}[htbp]
  \centering
  \includegraphics[width=.485\textwidth]{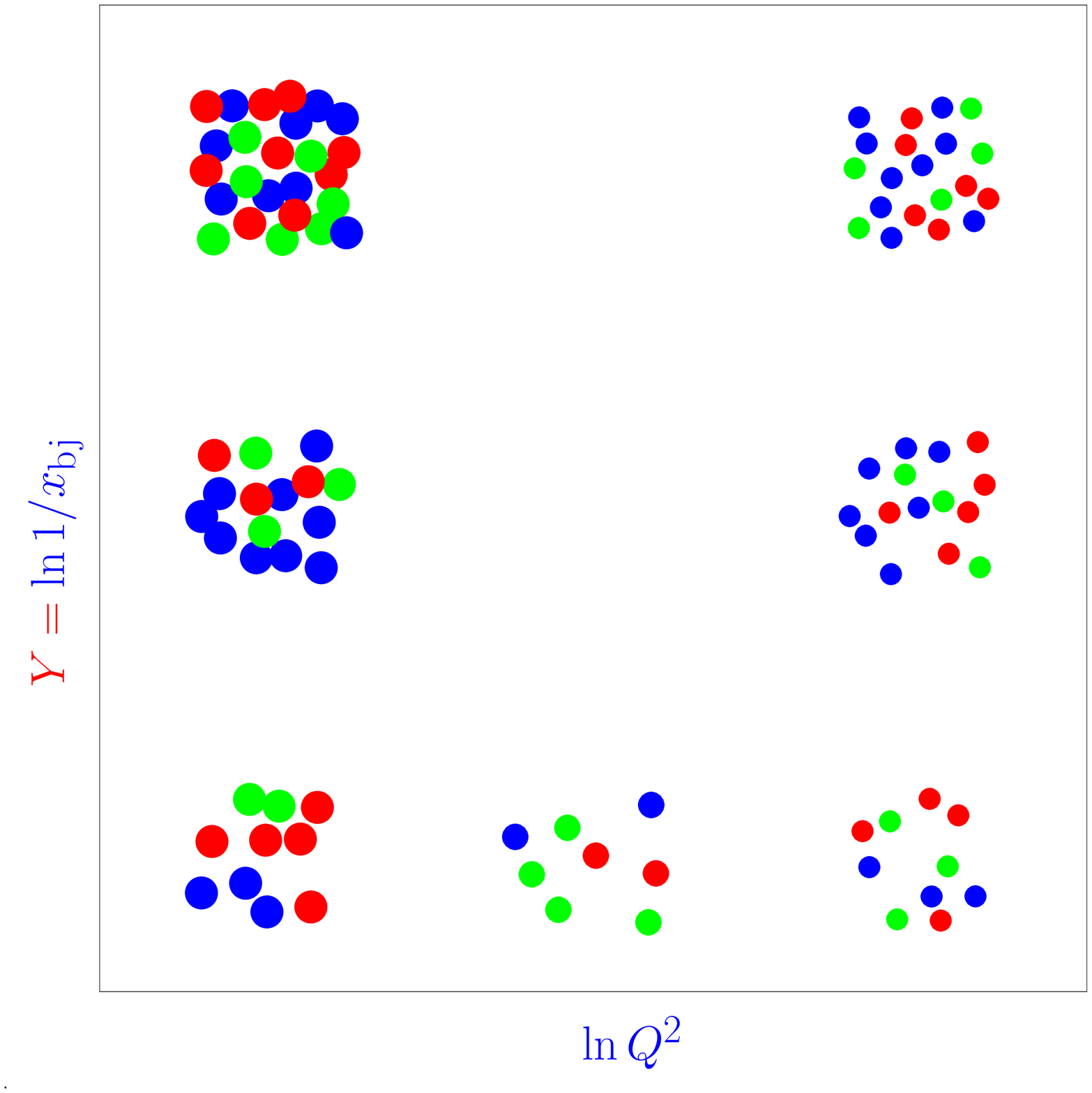}
  \hfill
   \includegraphics[width=.485\textwidth]{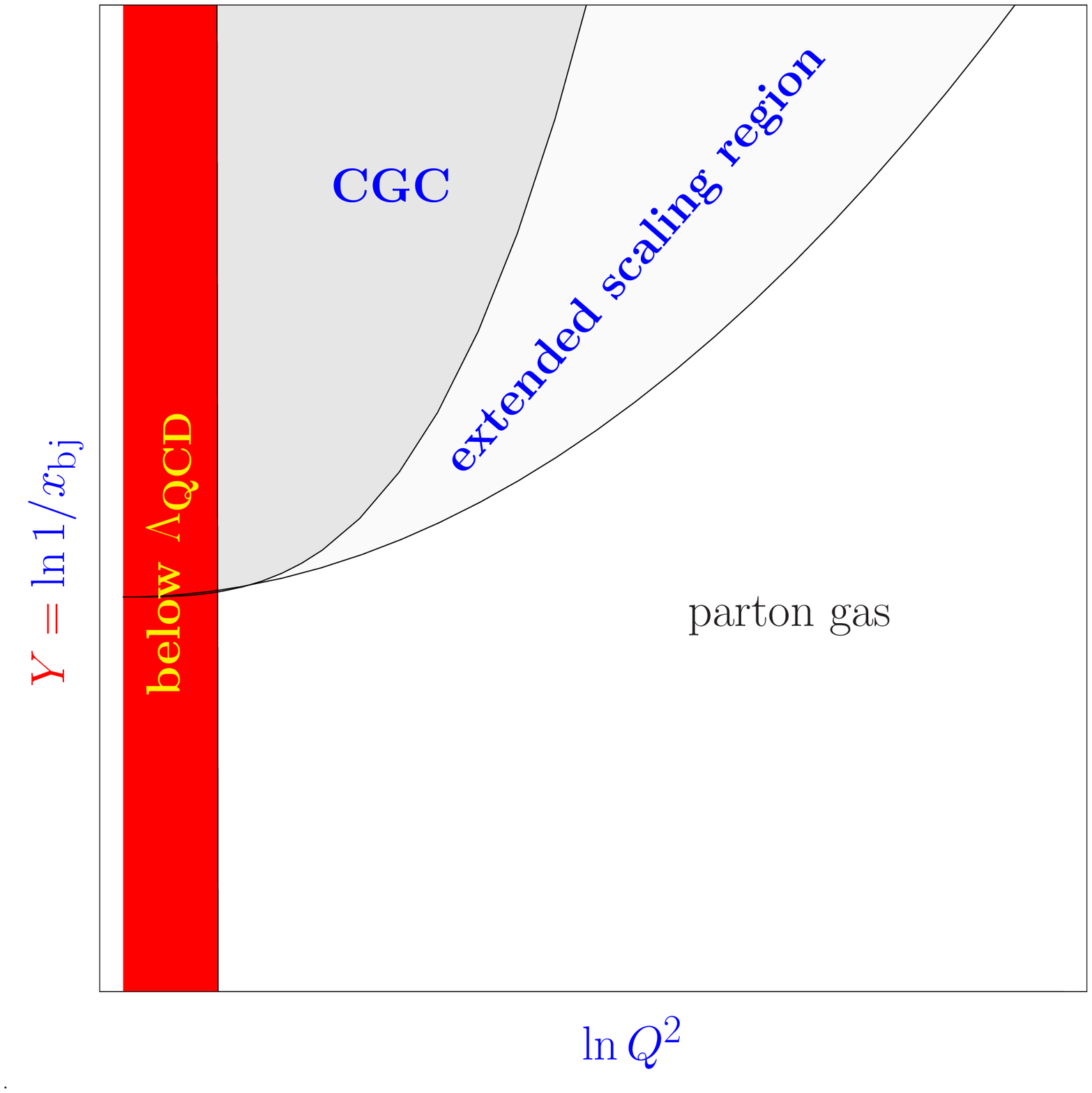}
  \caption{\small Left: 
    Partons in the $\y,\ln Q^2$ plane represented as dots of size
    $1/Q^2$. Numbers grow both with $Q^2$ and $\ln(1/\xbj)$.  Going to
    sufficiently small $\xbj$ at fixed $Q^2$ partons start to overlap.
    Right: corresponding regions of phase space with
    qualitatively different behavior. The lower boudary of the CGC
    region is determined by $Q_s(\xbj)$. The extendeded scaling region will be explaind in Sec.~\ref{sec:scaling-window}.}
  \label{fig:x-Q-plane-density}
\end{figure}

This procedure is self consistent and perturbatively under control as
long as one stays within the large $Q^2$ region.  Starting from a
large enough $Q^2_0$ and going to larger values, one finds that the
objects one counts with the quark and gluon distributions increase in
number but they stay dilute: as pointlike excitations their apparent
sizes follow the transverse resolution scale $1/Q$. At small $\xbj$,
however, the growth of the gluon distribution is particularly
pronounced -- a clear sign that corrections of the form $(\alpha_s
\ln(1/\xbj))^n$) start to become important. Any attempt to track the
$\xbj$-dependence of the cross sections by summing these is immediately
faced with an additional complication: moving towards small $\xbj$ at
{\em fixed} $Q^2$ increases the number of gluons of apparent size
$1/Q$, so that the objects resolved will necessarily start to overlap.
This is depicted in Fig.~\ref{fig:x-Q-plane-density}.
At this point one has long since left the region in which a treatment
at leading order in a $1/Q^2$ expansion (OPE at leading twist) and the
concept of such an expansion itself is meaningful. Instead of single
particle properties like distribution functions one needs to take into
account all (higher twist) effects that are related to the high
density situation in which the gluons in the target overlap. At small
$\xbj$ one is thus faced with a situation in which one clearly needs to
go beyond the standard tools of perturbative QCD usually based on a
twist expansion.

There are different ways to isolate the leading contributions to the
cross section at small $\xbj$ in a background of many or dense gluons
and the history of the field is long~\cite{Mueller:1986wy,
  Mueller:1994rr, Mueller:1994jq, McLerran:1994ni, McLerran:1994ka,
  McLerran:1994vd, Ayala:1995kg, Kovner:1995ja, Ayala:1996hx,
  Kovchegov:1996ty, Balitsky:1996ub, Kovchegov:1997pc,
  Jalilian-Marian:1997xn, Jalilian-Marian:1997jx,
  Jalilian-Marian:1997gr, Jalilian-Marian:1997dw, Balitsky:1997mk,
  Jalilian-Marian:1998cb, Mueller:1999wm, Kovchegov:1999ua,
  Kovner:1999bj, Kovner:2000pt, Weigert:2000gi,
  Iancu:2000hn,Ferreiro:2001qy}.  I will use a physically quite
intuitive picture first utilized in the McLerran-Venugopalan model
which was originally given in the infinite momentum frame of the
nuclear target. 
% Probing this target at small $\xbj$ implies a large
%rapidity difference $\y=\ln(1/\xbj)$ between projectile and target,
%corresponding to a large relative boost factor $e^\y$ of the order of
%several hundred to several thousand in the experiments listed above.
Knowing already that the interaction will be dominated by gluonic
configurations, one tries to isolate the boost enhanced ones among
them. Looking at the gluon field strength of the target for that
purpose, one finds that only the $F^{+ i}$ components are boost
enhanced, all others can be neglected. At the same time one finds a
strong Lorentz contraction in $x^-$ direction, and a time dilation,
correspondingly, in $x^+$ direction.  With such a field strength
tensor, one is left with only one important degree of freedom, which,
by choice of gauge, can be taken to be the + component of a gauge
field. Taking into account time dilation and Lorentz contraction, the
gauge field can then be written as
\begin{equation}
\label{eq:aplus}
\begin{split}
  A =& b+\delta A
\qquad \mbox{with} \qquad
  b^{i,-}=  0,\quad b^+=\beta(\bm{x})\delta(x^-)\ .  
\end{split}
\end{equation}
where $\beta$ is an unspecified function of the transverse components
$\bm{x} := (x^1,x^2)$ of the four vector $x$. This exhibits a leading
contribution $b^+$ which is $x^+$ independent and Lorentz contracted
to a $\delta$-function in $x^-$. The $\delta A$ symbolizes possible
corrections that are kinematically suppressed. They will start to play
a role as quantum corrections in the derivation of evolution
equations.  The boost and Lorentz contraction arguments for the
leading contribution $b$ are $\xbj$ dependent. In the interpretation
of $\xbj$ as a lightcone momentum fraction it determines a resolution
in $x^-$: the $\delta(x^-)$ in \eqref{eq:aplus} is ``localized'' only
with the resolution available with a probe providing the corresponding
momentum fraction. A similar cautionary remark applies to the $x^+$
independence. 

\begin{figure}[htbp]
  \centering
  \subfigure[]{
       \includegraphics[height=2.4cm]{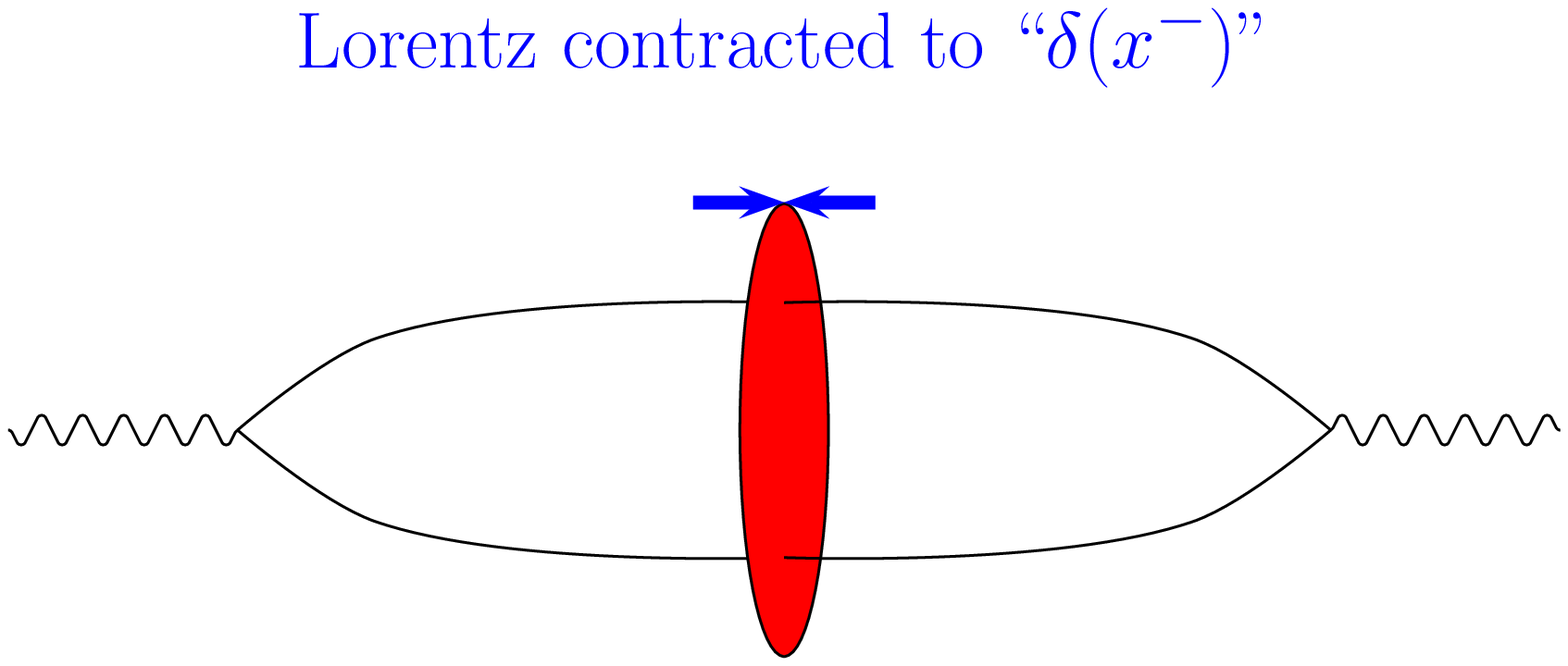} 
       \label{fig:small-x-dis-geom-inf}}
\hfill
  \subfigure[]{
      \includegraphics[height=2.4cm]{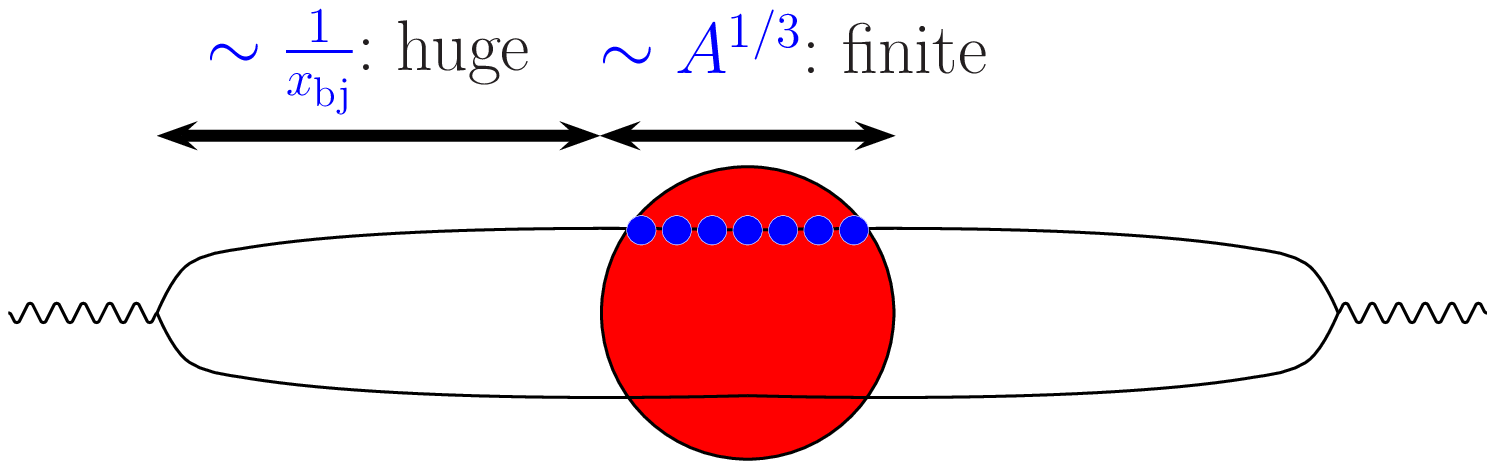}
      \label{fig:small-x-dis-geom-rest}}
  \caption{\small $q\Bar q$ pairs interacting with a nuclear target 
    at small $\xbj$. (a) shows the situation in the targets infinite
    momentum frame with fully Lorentz-contracted target fields, (b)
    shows the situation in the target rest frame in which the
    $\gamma^* q\Bar q$ vertex is far outside the target at a distance
    proportional to the relative boost factor $1/\xbj = e^\y$. }
  \label{fig:small-x-dis-geom}
\end{figure}

Note that mathematically one can always trade a component of a gauge
field for a path-ordered exponential along the
direction that picks up this component:%
%  Note that mathematically one can always trade a
%gauge field that has only a single component for a path ordered
%exponential along the direction that picks up this component:
\begin{equation}
  \label{eq:Udef}
  b^+=i(\partial^+U)U^\dagger \hspace{1cm} U_{\bm{x}}={\sf P}
  \exp \Big\{-i \int\!\!dz^- b^+(z^-,{\bm{x}},0)\Big\}
\ .
\end{equation}
If multiple eikonal interactions are relevant as the high energy
nature of the process would suggest, one expects these path ordered
exponentials to be the natural degrees of freedom, since this is what
they encode: Diagrammatically
\begin{equation}
  \label{eq:U-eikonal-diag}
  U_{\bm{x}}= \sum\limits_{\text{gluons}} 
  \parbox{4cm}{\includegraphics[width=4cm]{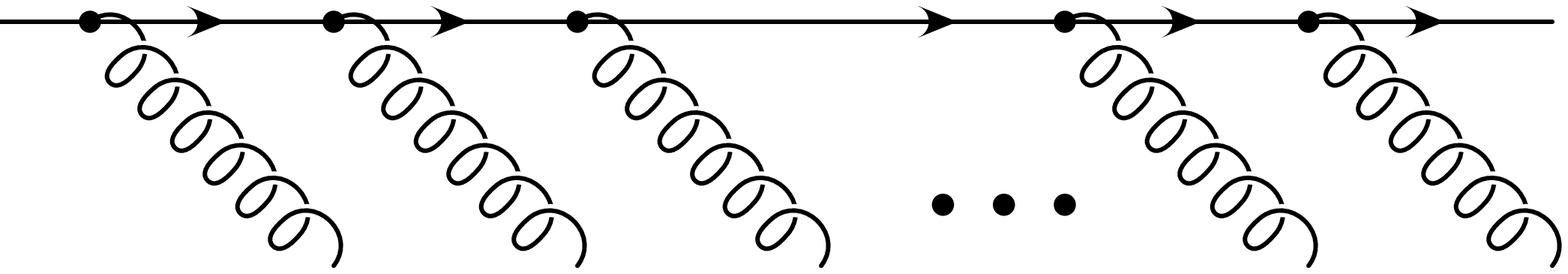}}
\ ,
\end{equation}
where the path ordered integrals along $z^-$ are represented by the
fermion line.  Ignoring the small fluctuations $\delta A$ for a moment,
one has a picture in which the $\gamma^*$ nucleus cross section arises
from a diagram in which the photon splits into a $q\Bar q$ pair which
then interacts with the background field. Due to the $\delta$ like
support of $b^+$ the $q\Bar q$ are not deflected in the transverse
direction during that interaction. This just reflects the largeness of
the longitudinal momentum component of these partons at small $\xbj$.
This is shown in Fig.~\ref{fig:small-x-dis-geom-inf}. To be sure, the
physics content is not frame dependent although it is encoded
differently in, say the rest frame of the target. There, one
encounters neither Lorentz contraction nor time dilation. However the
scale relations are preserved: the photon splits into a $q\Bar q$ pair
far outside the target and its $p^-$ is so large that typical $x^+$
variations of the target are negligible during the interaction. As a
consequence the probe is not deflected in the transverse direction,
picking up any multiple interactions with (gluonic) scattering centers
as it punches though the target. This is shown in
Fig.~\ref{fig:small-x-dis-geom-rest}.

That multiple interactions are of relevance immediately becomes
obvious, once one tries to calculate such diagrams with a background
field method. To this end one evaluates the diagram shown in
Fig.~\ref{fig:small-x-dis-geom} in the background of a field of the
type shown in Eq.~\eqref{eq:aplus} ignoring the small fluctuations.
Such a calculation will prove somewhat involved irrespective of the
method chosen --one of them will be sketched in
Sec.~\ref{sec:eikonalization}-- but the resulting
expression\footnote{An integral over longitudinal momentum fractions
  of the $q$ and $\Bar q$ has been absorbed into the wave function
  part of this expression for clarity -- the full expression is given
  in Eqns.~\eqref{eq:sigma-DIS-explicit} and~\eqref{eq:fermion-wavefunct}}
\begin{equation}
  \label{eq:dipole-cross}
  \sigma_{\mathrm{DIS}}(\xbj,Q^2) = \text{\cal Im}\ 
  \begin{minipage}[c]{3cm}
    \includegraphics[height=.8cm]{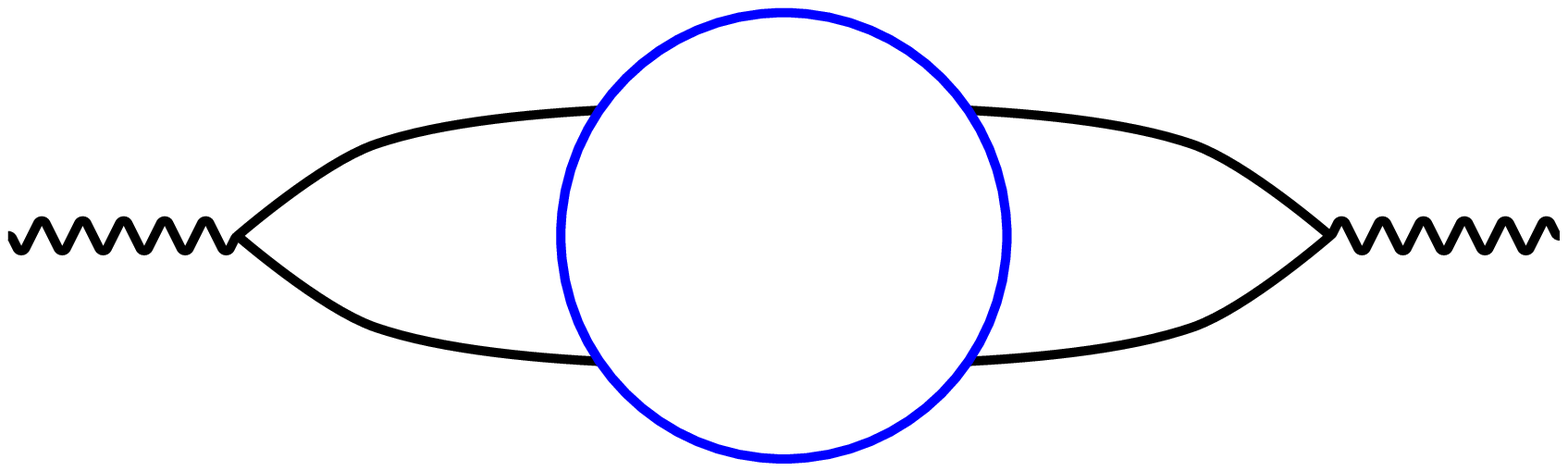}
  \end{minipage}
  =\int\!\!d^2 r 
  \, \vert \psi^2\vert(r^2 Q^2) \hspace{.1cm}
  \int d^2b \ \langle\frac{\tr(1-U_{\bm{x}} U^\dagger_{\bm{y}})}{N_c}\rangle
\end{equation}
has a clearcut interpretation. Beginning with notation,
$\bm{r}=\bm{x}-\bm{y}$ corresponds to the transverse size of the
$q\Bar q$ dipole and $\bm{b}=(\bm{x}+\bm{y})/2$ to its impact
parameter relative to the target.\footnote{$\bm{b}$ defined like this
  is the Fourier conjugate variable to the transverse momentum
  transfer in non-forward matrix elements. See
  also~\eqref{eq:BFKL-Fourier}.} This result shows a clear separation
into a wave function part (the factor $\vert \psi^2\vert(r^2 Q^2)$)
and the dipole cross section part
\begin{equation}
  \label{eq:sigma-dipole}
  \sigma_{\text{dipole}}(\y=\ln(1/\xbj),r^2) := \int d^2b \ 
  \langle\frac{\tr(1-U_{\bm{x}} U^\dagger_{\bm{y}})}{N_c}
  \rangle_{\y=\ln(1/\xbj)}
\ .
\end{equation}
The former has the interpretation as the absolute value squared of the
$q\Bar q$ part of the photon wave function, contains the $\gamma^*
q\Bar q$ vertices and can, alternatively to the background field
method, be calculated entirely within QED.  The wave functions
themselves turn out to be Bessel functions $K_{0,1}$ depending on the
polarization of the virtual photon.  This factor carries all of the
direct $Q^2$-dependence of the cross section.

The dipole cross section part embodies all the interaction with the
background gluon fields in terms of the path ordered exponentials. In
particular, in the absence of a gluon field representing the
interaction with the target, both $U$ factors reduce to unit matrices
and the cross section vanishes. The averaging procedure $\langle \ldots
\rangle_\y$ %
%is understood as an average over the dominant
%configurations at a given $\y$. It 
contains all the information about the QCD action and the target wave
functions that are relevant at small $\xbj$.

Even without exploring any of these in detail, there is one feature
about this interaction which is very characteristic for DIS at small
$\xbj$ which is already visible in the above expression: although
momentum is exchanged between projectile and target, there is no net
color exchange: the $q\Bar q$ pair enters and leaves the interaction
region as a color singlet. The same applies to other intrinsic quantum
numbers such as spin. In a diction dating back to pre-QCD times this
is often referred to as ``pomeron exchange,'' although it is now
acknowledged that the underlying theory of the interaction does not
exhibit a particle cut that would allow an interpretation in terms of
a simple particle exchange.

It is fairly clear from the above discussion, that the isolation of
the leading $b^+$ contribution which defines this average is a
resolution dependent idea: as one lowers $\xbj$, additional modes --up to
now contained in $\delta A$ of Eq.~\eqref{eq:aplus}-- will take on the
features of $b^+$. They will Lorentz contract and their
$x^+$-dependence will freeze. Accordingly the averaging procedure will
have to change. If one writes the average as
\begin{equation}
  \label{eq:average}
  \langle \ldots \rangle_\y = %& 
  \int D[b^+] \ldots W_\y[b^+]
\hspace{.5cm}\text{or, equivalently}\hspace{.5cm}
  \langle \ldots \rangle_\y = %& 
  \int \Hat D[U] \ldots \Hat Z_\y[U]\ ,
\end{equation}
the weights $ W_\y[b^+]$ or $\Hat Z_\y[U]$ will, by necessity, be
$\y=\ln(1/\xbj)$ dependent.

Besides the perspective adopted above, there are equivalent ways to
understand and derive the $\xbj$ dependence of such cross sections in
which the path ordered exponentials do not appear in closed form in
the first calculational step but instead are built up perturbatively.
Among those is the most straightforward procedure in which one allows
the incoming $q\Bar q$ (into which the incoming $\gamma^*$ splits
initially) to branch off additional gluons before it hits the target.
Looking at the corresponding amplitudes this would amount to consider
perturbative corrections to the $\gamma^* \to q\bar{q}$ amplitude
contained in Fig.~\ref{fig:small-x-dis-geom} of the form
  \begin{equation}
    \label{eq:emission}
       \begin{minipage}[c]{1.5cm}
   \begin{center}
     \includegraphics[height=2cm]{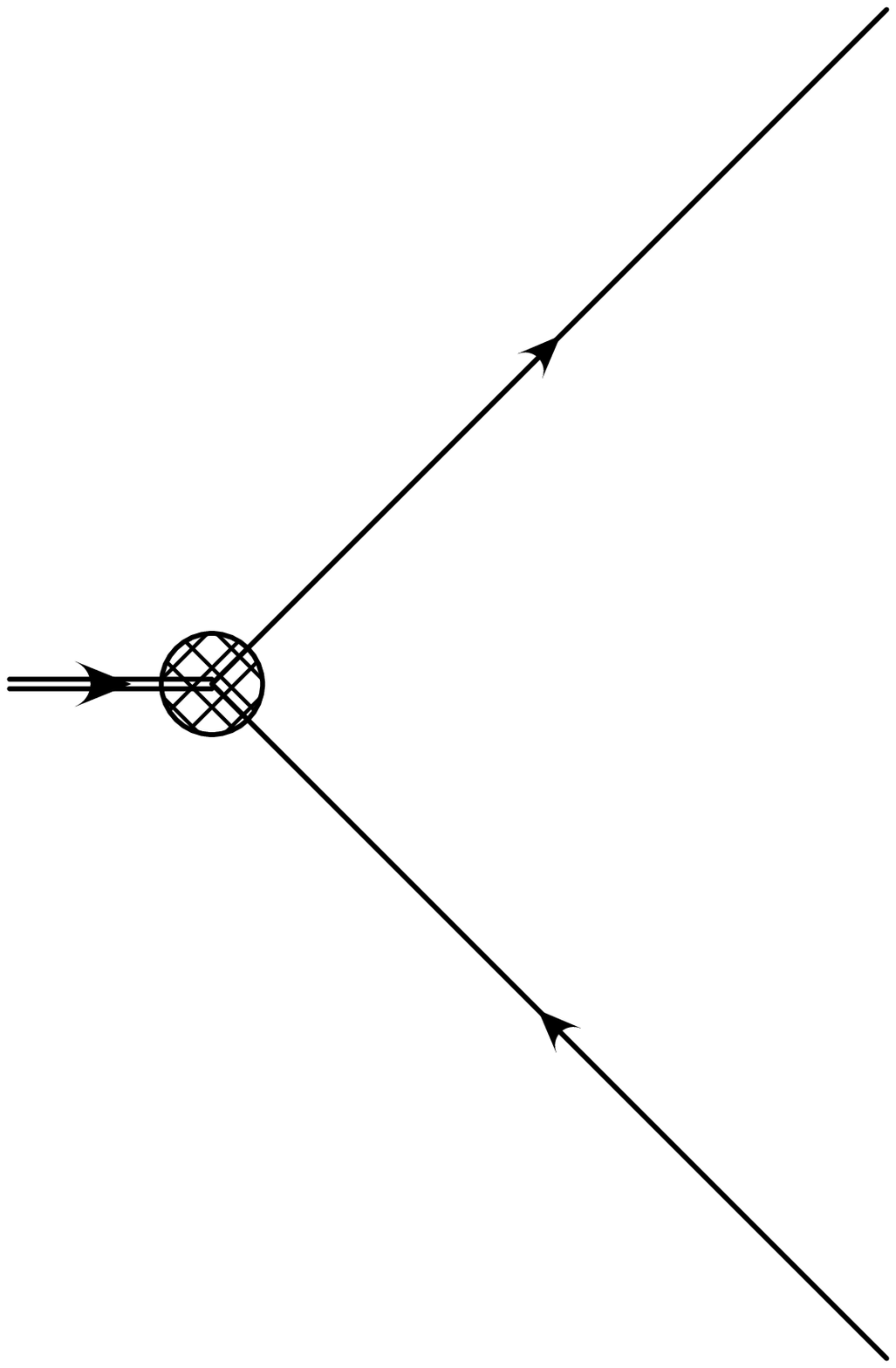}
   \end{center}
 \end{minipage}
+
   \begin{minipage}[c]{2.5cm}
   \begin{center}
     \includegraphics[height=2cm]{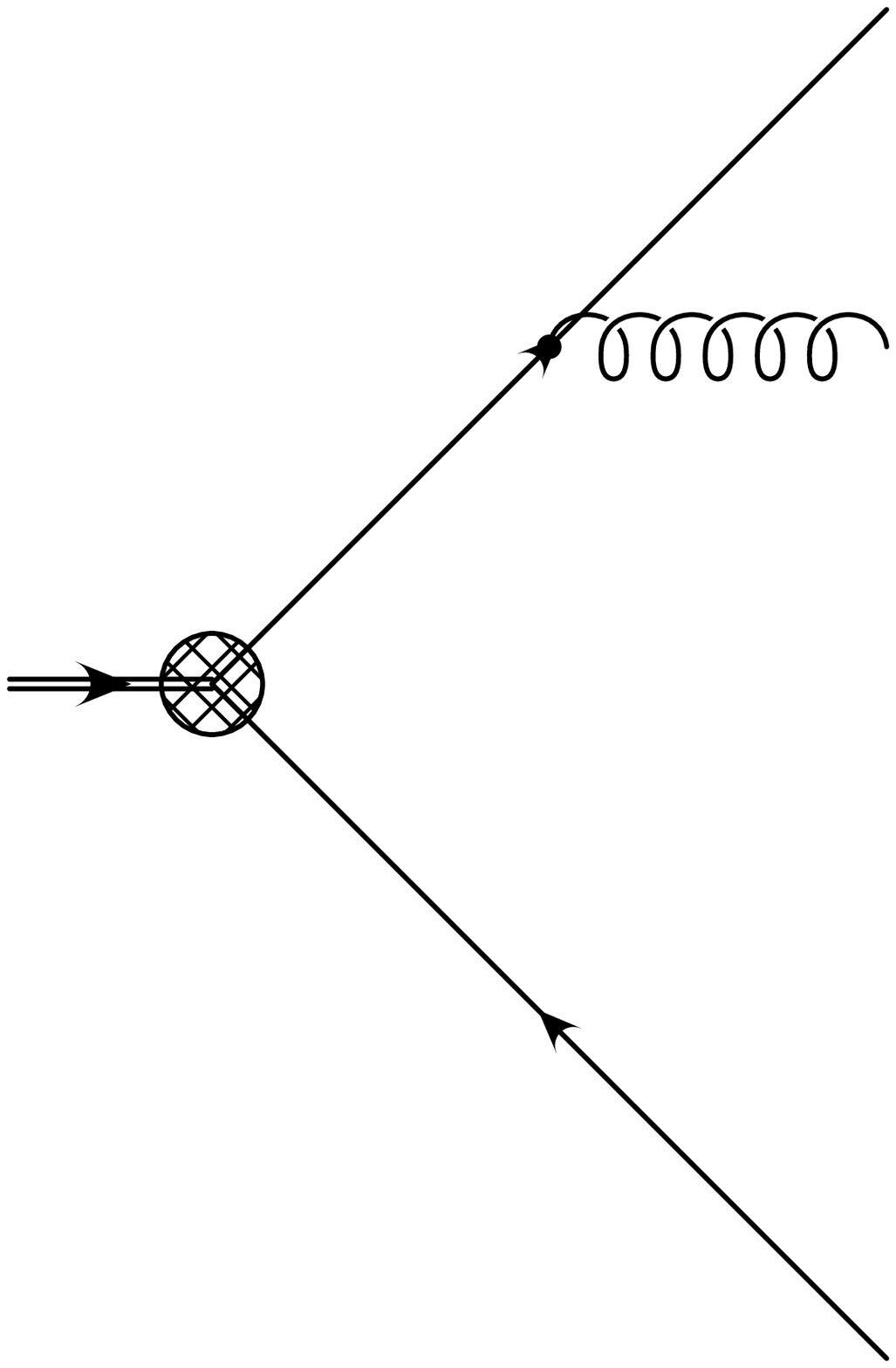}
   \end{center}
 \end{minipage}
+
   \begin{minipage}[c]{2.5cm}
   \begin{center}
     \includegraphics[height=2cm]{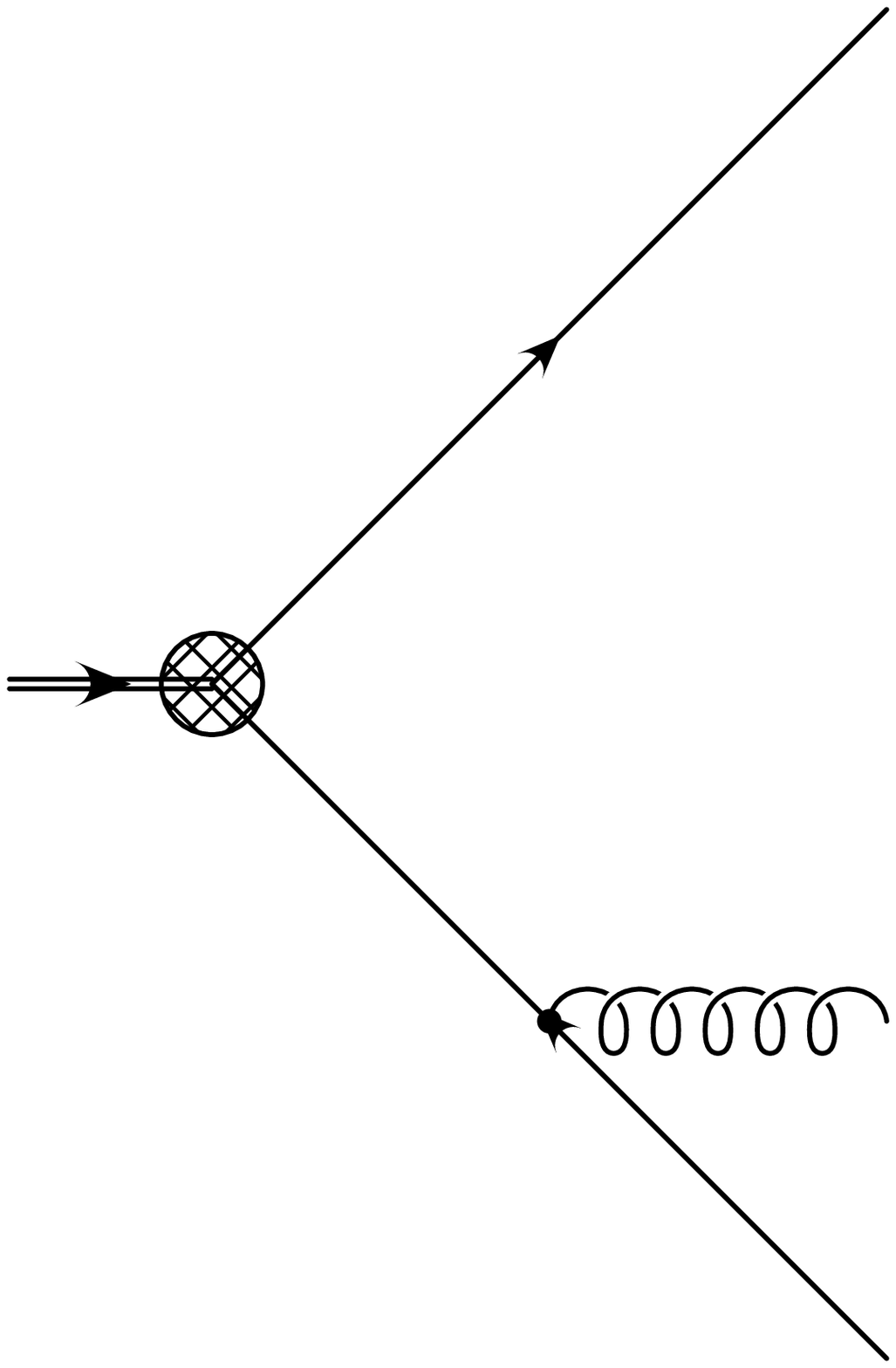}
   \end{center}
 \end{minipage}
+
\begin{minipage}[c]{2.5cm}
   \begin{center}
     \includegraphics[height=2cm]{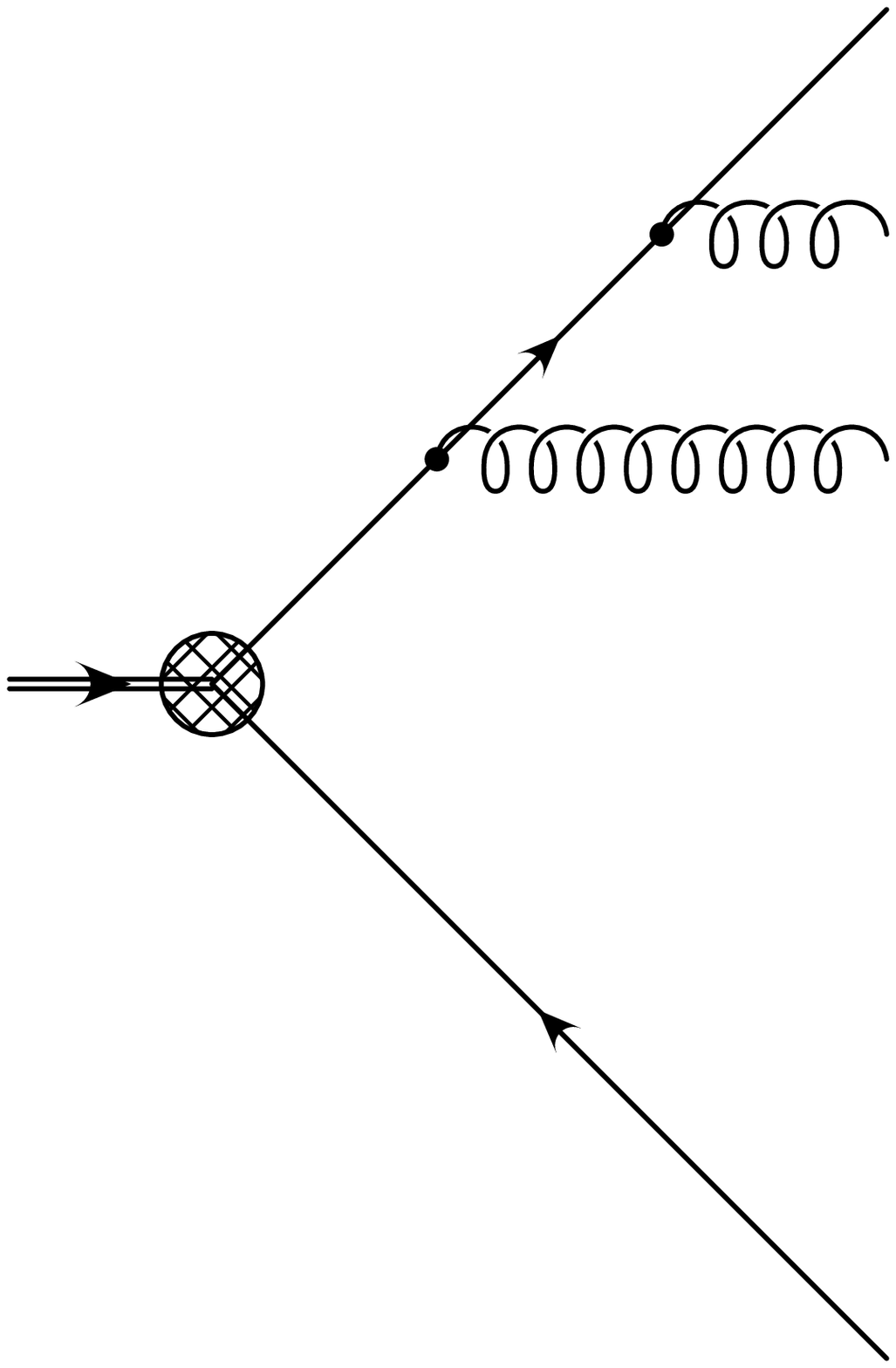}
   \end{center}
 \end{minipage}
+
  \begin{minipage}[c]{2.5cm}
   \begin{center}
     \includegraphics[height=2cm]{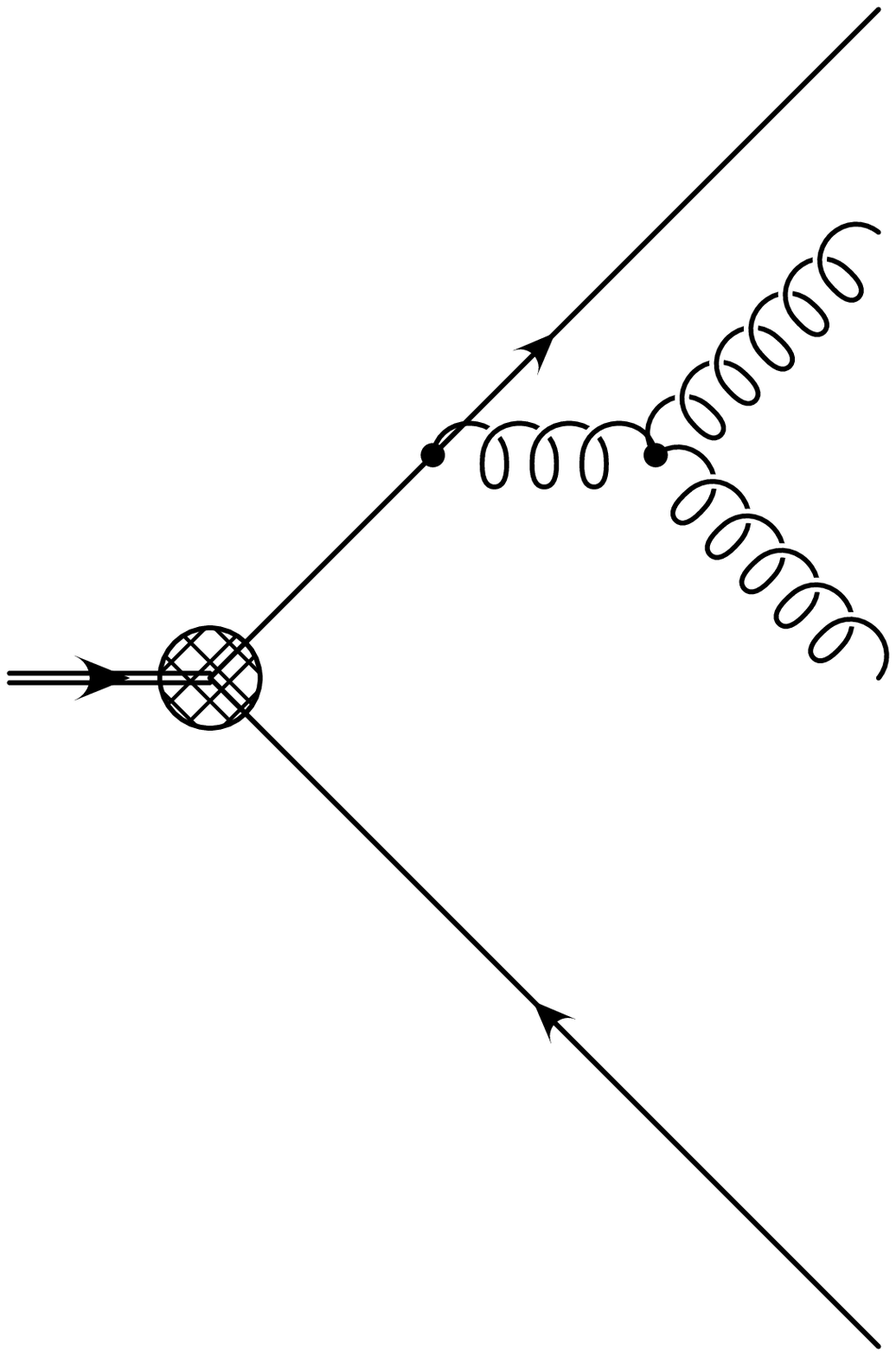}
   \end{center}
 \end{minipage}
+ \ldots
\ \ .
\end{equation}
All but the last of these diagrams would be present already in QED.
It is this last type of contribution that is responsible for the
strong growth and the saturation features of the CGC. That all these
methods lead to the same result will be thoroughly explained in the
main text.

\begin{figure}[htbp]
  \centering
  \includegraphics[width=\textwidth]{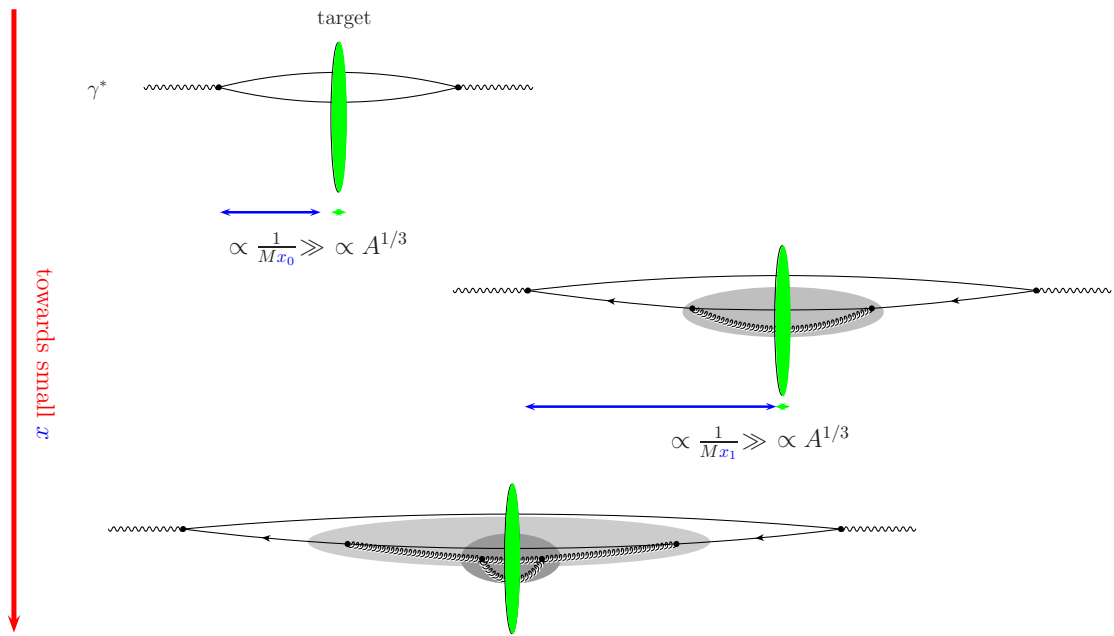}
  \caption{\small RG corrections to the average over background field 
    configurations. The shaded areas contain the contributions
    subsumed into the averaging procedure at successive values of
    $\y=\ln(1/\xbj)$.}
  \label{fig:RG-corrections}
\end{figure}

Fig. \ref{fig:RG-corrections} highlights another feature of the
dominant contributions: as the distance between the $\gamma^*\to
q\bar{q}$ vertex and the target grows with increasing energy like the
corresponding boost factor, the probability of multiple gluon emission
grows accordingly.  This leads to the iterative scheme indicated in
Fig.~\ref{fig:RG-corrections} (with only part of the above diagrams
displayed for brevity).
Note that the leading contributions come, as usual, from ordered
emission,\footnote{Logarithmically enhanced contributions that lead to
  small $x$ evolution equations arise from a region in phase space in
  which the $\xbj$ values in an $n$-gluon amplitude are strictly
  ordered $\xbj_1 > \xbj_1 > \ldots \xbj_n$.  This is in complete
  analogy with transverse momentum ordering in the DGLAP evolution
  case. Ordering in $\xbj$ also implies a hierarchy of longitudinal
  distance scales as indicated in Fig.~\ref{fig:RG-corrections}.} so
that the lines depicted there all have large momentum components in
the photon direction. They themselves will then interact with the
target in a similar way as the original $q\bar{q}$ component: These
punch straight through the target at fixed transverse locations
$\bm{x}$ and $\bm{y}$, giving rise to the $U_{\bm{x}}
U^\dagger_{\bm{y}}$ in the expression for the cross section in
Eq.~\eqref{eq:dipole-cross}. The additional gluons shown in
Eq.~\eqref{eq:emission} and Fig.~\ref{fig:RG-corrections} leave behind
new gluonic Wilson lines $\Tilde U$ (the tilde indicating that they
are in the adjoint representation) at new locations $\bm{z}_i$.  It is
clear from gauge invariance that the second diagram in
Fig.~\ref{fig:RG-corrections} will involve the operator $\Tilde
U_{\bm{z}}^{a b}\ 2\tr( t^a U_{\bm{x}} t^b U^\dagger_{\bm{y}})$ where
$\Tilde U_{\bm{z}}^{a b}$ represents the added gluon. From this
perspective one would expect that calculating additional corrections
would lead to an infinite hierarchy of coupled equations for more and
more complicated correlators of Wilson lines.  This indeed is the
perspective taken by Balitsky in his derivation of this hierarchy of
evolution equations~\cite{Balitsky:1996ub}, the Balitsky hierarchy.
The perspective taken above is different in the sense that these same
diagrams have been interpreted to contribute a change in the averaging
procedure that describes the dipole cross section of
Eq.~\eqref{eq:sigma-dipole}. The $q\bar{q}$ dipole at smaller $\xbj$
is taken to interact with additional gluons and thus the cross section
and averaging procedure changes. In Fig.~\ref{fig:RG-corrections} this
redefinition process is indicated by the shaded areas that include the
target, and subsequently the target and all the gluons softer than the
initial $q\bar{q}$ pair. Both approaches are indeed equivalent as will
become apparent from the mathematical treatment.

As these descriptions indicate, it is possible to calculate the change
of the weights in Eq.~\eqref{eq:average} with $\y$ to a certain
perturbative accuracy -- at the moment complete calculations of these
evolution equations based on independent techniques exist at leading
log accuracy, i.e.\ to accuracy $\alpha_s \ln(1/\xbj)$-- while $\Hat
Z_\y[U]$ at a given $\y_0$ remains incalculable without
nonperturbative input.  This is again the situation one has to face in
any Renormalization Group (RG) setting in which the initial condition
for evolution remains outside the scope of the calculation and must be
determined from experiment.

The renormalization group character of the calculation can be made
explicit by assuming one knew, say $\Hat Z_\y[U]$ at an initial
$\y_0$. This defines the ensemble of initial background fields $b^+$
and one may then integrate over the fluctuations $\delta A$ around
$b^+$ between the old and new cutoffs $\y_0$ and $\y_1$. Taking
the limit $\delta\y=\y_1-\y_0\to 0$ one then gets a
renormalization group equation for $\Hat Z_\y[U]$.

Because one integrates over $\delta A$ in the background of arbitrary
$b$ of the form~\eqref{eq:aplus}, the equation treats the background
field exactly to all orders and captures all nonlinear effects also in
its interaction with the target.  The resulting RG equation is a
functional equation for $\Hat Z_\y[U]$ that is nonlinear in $U$. It
sums corrections to the leading diagram shown in
Eq.~\eqref{eq:dipole-cross} in which additional gluons are radiated
off the initial $q\Bar q$ pair as shown schematically in
Fig.~\ref{fig:RG-corrections}. All multiple eikonal scatterings inside
the target (the shaded areas) are accounted for.

The result of this procedure is the JIMWLK equation alluded to above.
This equation and its limiting cases will be discussed in detail in
Sec.\ref{sec:jimwlk-evol-balitsky}.

\begin{figure}[htbp]
  \centering
  \includegraphics[width=7cm]{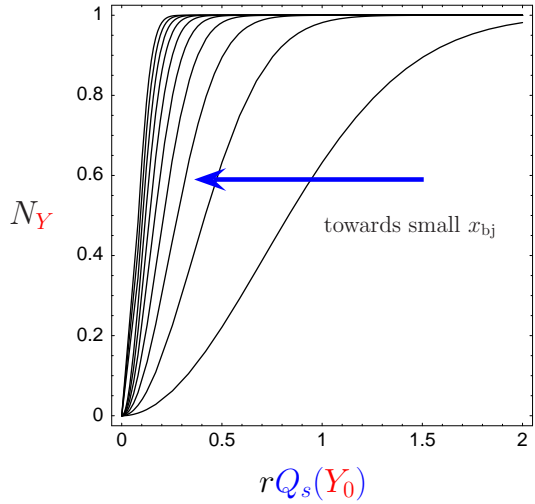}%evolution}
  \caption{\small Generic evolution trend for a one scale dipole 
    correlator $N_{\bm{x y}}$ plotted against $\vert
    \bm{x}-\bm{y}\vert$ as $x\to 0$: the curves move towards the left
    as $\y=\ln(1/\xbj)$ increases, the asymptotic form being a step
    function at the origin. }
  \label{fig:generic-evol}
\end{figure}

Let me close here with a few phenomenological expectations for the key
ingredient of the present phenomenological discussion, the dipole
cross section. Clearly, one expects the underlying expectation value
of the dipole operator,
\begin{equation}
  \label{eq:N-def}
N_{\y\ \bm{x y}}:=\langle\frac{\tr(1-U_{\bm{x}}
U^\dagger_{\bm{y}})}{N_c} \rangle_{\y=\ln(1/\xbj)}
\ ,
\end{equation}
to vanish outside the target, so that the impact parameter integral
essentially samples the transverse target size and thus scales roughly
with $A^{2/3}$.  For central collisions, i.e.\ deep inside the target,
$N_{\y, \bm{x y}}$ is expected to interpolate between zero at small
$\bm{r}=\bm{x}-\bm{y}$ and one at large $\bm{r}$.  Note that $N_{\y,
  \bm{x y}}$ is necessarily limited to lie inside these boundaries as
an immediate consequence of the resummation of the gluon fields into
eikonal factors $U^{(\dagger)}$. Saturation of these limits at small
and large distances is the idea of color transparency for small
dipoles supplemented with saturation for large dipoles and is sketched
schematically in Fig.\ref{fig:generic-evol}.
Focusing on the rightmost curve in Fig.\ref{fig:generic-evol} which
corresponds to $\y_0$ one has a single scale, $Q_s(\y_0)=1/R_s(\y_0)$,
that characterizes the transition between transparent and saturated,
hence its name, saturation scale. Looking back at the definition of
$N_{\y\ \bm{x y}}$ in terms of the relevant fields $U^{(\dagger)}$, it
is clear that the region where it differs from unity is simply the
region where the fields are correlated: $R_s = 1/Q_s$ has the
interpretation of a correlation length.  From this it is
apparent that evolution towards smaller $\xbj$ at fixed $Q^2$ (along
vertical lines in Fig.\ref{fig:x-Q-plane-density}), which will add
more gluons, will necessarily lead to shorter correlation length and
thus larger saturation scales: The curve will move toward the left as
indicated, with smaller and smaller $R_s(\y)$ characterizing the
changeover between the two regimes. Since small $\xbj$ evolution
equations respect the generic saturation features shown in
Fig.~\ref{fig:generic-evol}, $Q_s$ is also the scale that
characterizes the onset of color recombination effects that slow down
and eventually stop the growth of the dipole cross section for large
dipole sizes.

\begin{figure}[htb]
  \centering
        \includegraphics[width=3.5cm]{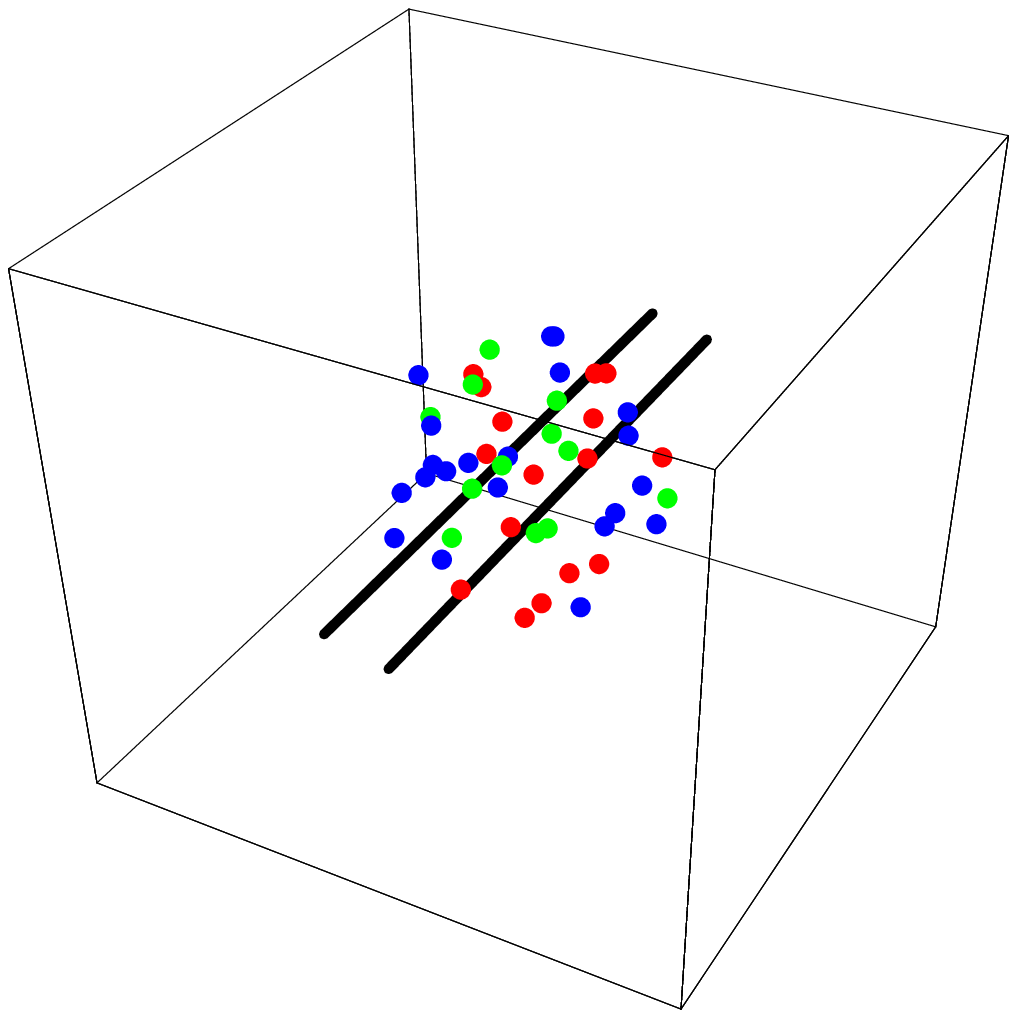}
\hspace{5cm}
          \includegraphics[width=3.5cm]{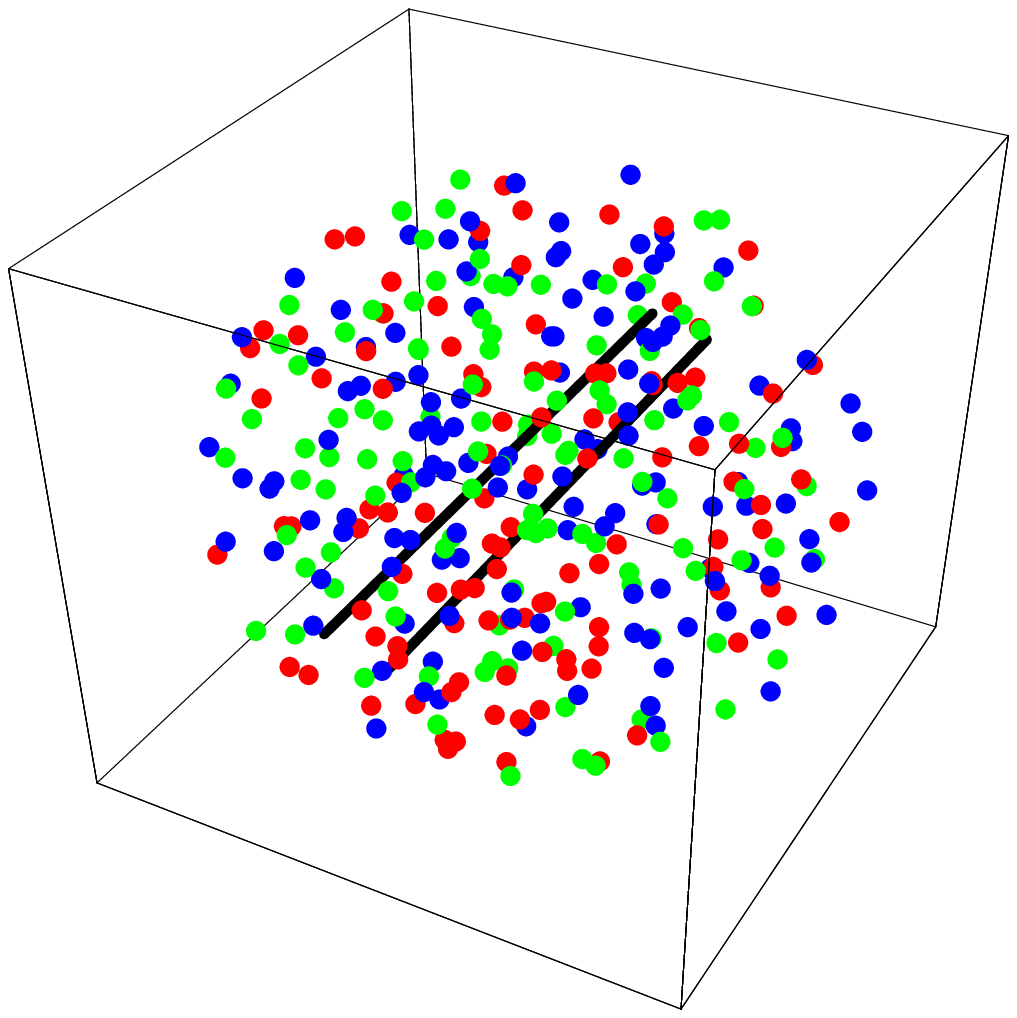}
  
  \caption{\small The origin of growth of $Q_s$ with $A$ in dilute targets 
    with decorrelated scattering centers in longitudinal direction for
    a small and a large nuclear target. The lines indicate the $q\Bar
    q$ pair, the dots the uncorrelated scattering centers.}
  \label{fig:naive-Qs-A-scaling}
\end{figure}

From here one can immediately understand that going from small to
large hadronic targets (i.e. by increasing the target's atomic number
$A$) one also increases the apparent gluon density in this kinematic
domain: The starting point again is that the energy is assumed to be
high enough that the projectile would punch straight through the
target so that multiple scattering contributions sum into the eikonal
factors entering $N_{\y\ \bm{x y}}$. In large nuclei, one scatters
off gluons from many independent nucleons along the $q\Bar q$
trajectory as sketched in Fig.~\ref{fig:naive-Qs-A-scaling}.
Following the Lorentz contraction argument from above, the whole depth
of the nucleus contributes to the charge density seen at a point in
the transverse plane. Since, due to confinement, gluons from different
nucleons --encountered at different longitudinal positions-- are
necessarily decorrelated, this naturally leads to a rescaling of the
transverse correlation length $R_s^2$ by the inverse nuclear radius.
Since the nuclear volume scales with the atomic number $A$, this
implies scaling of $Q_s^2$ with $A^{1/3}$:
\begin{equation}
  \label{eq:naive-Qs-A-scaling}
Q_s^A({\y_0})^2 \sim Q_s^p({\y_0})^2  \ {A^{1/3}}
\ .
\end{equation}
This behavior will reappear in a model that was developed to describe
qualitative features of saturation effects at small $\xbj$, the
McLerran-Venugopalan (or MV) model, and explains the interest in
studying DIS at small $\xbj$ off large hadronic targets.  The argument
leading to Eq.~\eqref{eq:naive-Qs-A-scaling} is simplistic and a more
careful treatment will lead to a modification of the $A$ dependence
--although an enhancement typically persists. The most obvious
source of such modifications are correlations induced by evolution
towards smaller $\xbj$. This is why the scaling relation is written
for some initial (not too small) $\y_0$ at which such effects can be
expected to be small.

\begin{figure}[htb]
  \centering \includegraphics[height=4.4cm]{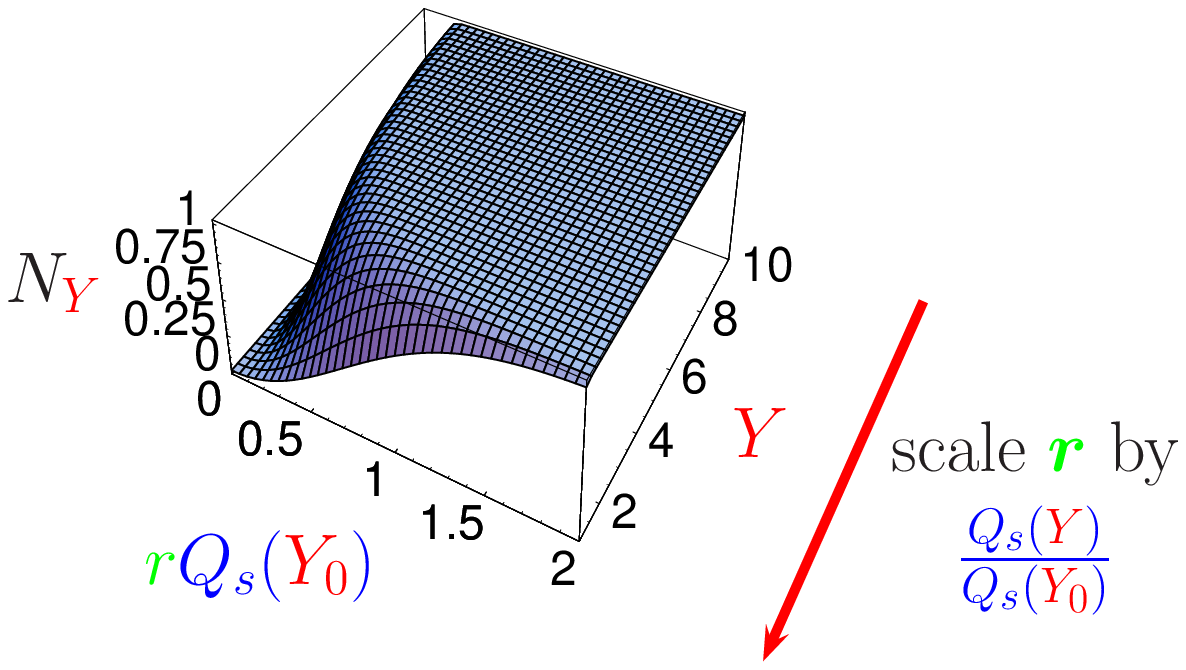}
  \hspace{1.5cm} 
  \includegraphics[height=4.4cm]{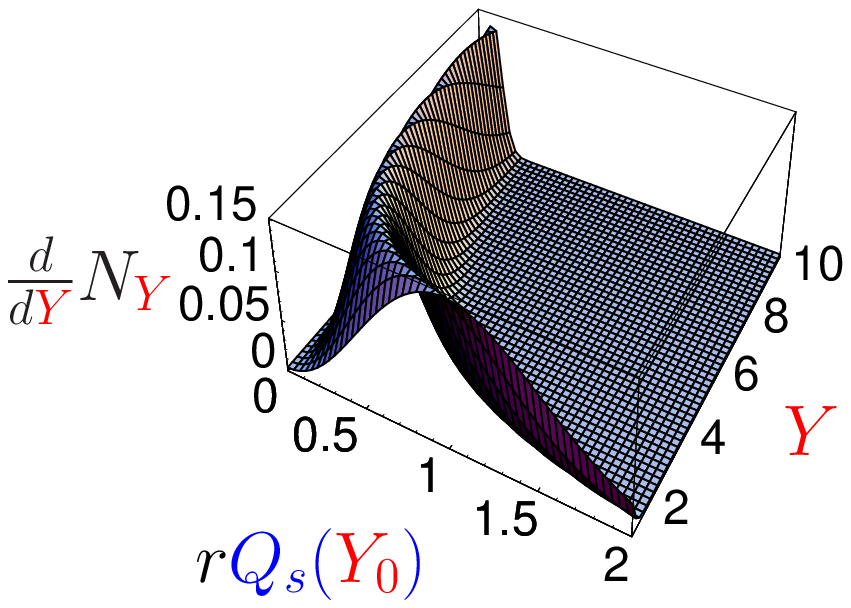}
  \caption{\small
    Dipole function (left) and its rapidity derivative (right) for
    central collisions plotted against dipole size. The latter shows
    that the change induced by evolution happens at larger and larger
    momentum scales. This implies IR safety.}  \label{fig:IR-safety}
\end{figure}

\begin{figure}[htb]
  \centering \includegraphics[height=4.5cm]{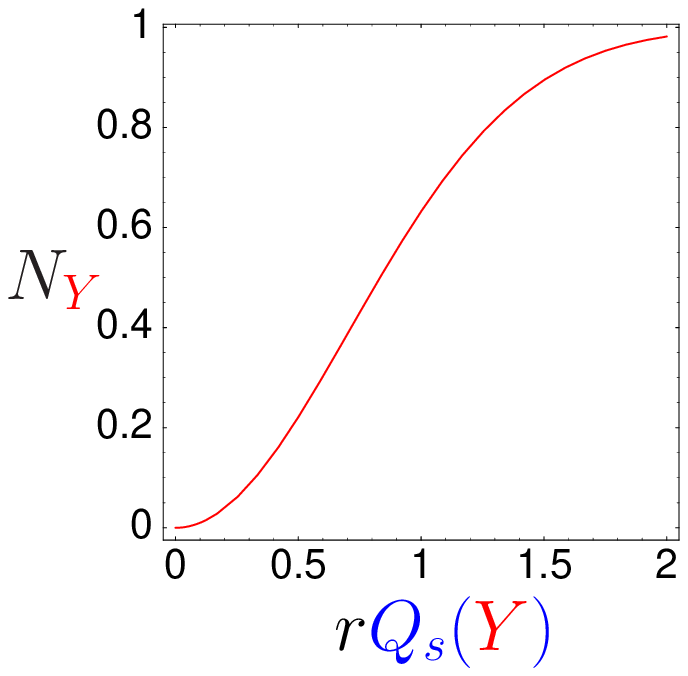}
  \hspace{2cm} \includegraphics[height=4.5cm]{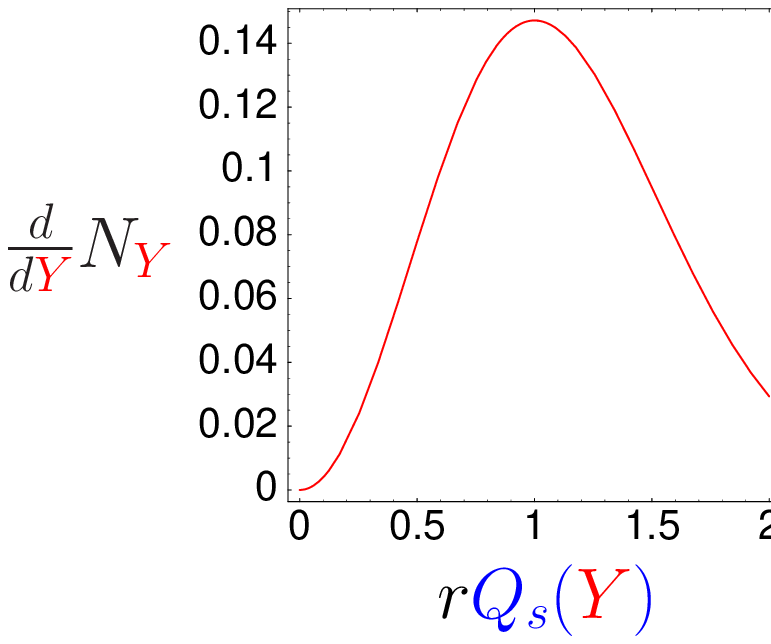}
  \caption{\small
    $Q_s$ scaled version of Fig.\ref{fig:IR-safety} assuming that
    geometric scaling holds.}
  \label{fig:IR-safety-scaled}
\end{figure}

Returning to the $\xbj$ dependence sketched in
Fig.~\ref{fig:generic-evol}, it is crucial to note that such behavior
has far reaching consequences as it leads to perturbative consistence
of the RG approach described above. If one takes the $\y$ dependence
of $N$ sketched in Fig.~\ref{fig:generic-evol} as given, any RG
equation that leads to this behavior will predict the largest changes
at momentum scales of the order of $Q_s(\y)$ -- this is where the
infinitesimal change between neighboring curves and thus the r.h.s. of
the corresponding evolution equation peaks. Contributions at momenta
below $Q_s$ are strongly suppressed. As $Q_s$ grows with $\y$,
evolution moves away from $\Lambda_\text{QCD}$ and thus remains
perturbative. This is illustrated in Fig.~\ref{fig:IR-safety}, which
shows the curves of Fig.~\ref{fig:generic-evol} in a 3d plot on the
left and the scales active in the evolution equation i.e. $\partial_\y
N_\y$ on the right.
Fig.~\ref{fig:IR-safety} anticipates a further characteristic feature
of evolution towards small $\xbj$ that, at this point, can not be
predicted, but was observed in DIS data by Golec-Biernat and
W{\"u}sthoff (G-B+W)~\cite{Golec-Biernat:1998js, Golec-Biernat:1999qd,
  Stasto:2000er}, namely scaling with $Q_s(\y)$, a phenomenon called
geometric scaling~\cite{Stasto:2000er} by Stasto, Golec-Biernat, and
Kwiecinski. They have found that for $\xbj$ values below $10^{-2}$ all
HERA data show beautiful scaling, if in Eq.\eqref{eq:sigma-dipole} one
assumes the scale in the $\bm{r}$ dependence carries at the same time
all the $\y=\ln(1/\xbj)$-dependence according to
\label{eq:GB+W-scaling-ansatz}
\begin{equation}
  \label{eq:BK-scaling-0}
  N_{\y\ \bm{x y}} = N( (\bm{x}-\bm{y})^2 Q_s(\y)^2 )
\ .
\end{equation}
Such scaling takes all the curves represented in
Fig.~\ref{fig:IR-safety} and maps them onto a single curve as shown in
Fig.~\ref{fig:IR-safety-scaled}.
The same carries through to the DIS cross section via
Eq.~\eqref{eq:dipole-cross} and leads to the successful fit of the
HERA data just alluded to.

There are many more physics ideas associated with the label CGC, such
as the existence of an extended scaling region at resolutions $Q$
above $Q_s$ indicated already in Fig.~\ref{fig:x-Q-plane-density},
which is quite important for the conceptual consistency of the scaling
fit to HERA data, the relevance of the CGC to the initial conditions
of heavy ion collisions, consequences for particle multiplicities and
the Cronin effect, as well as an intimate relationship with other QCD
phenomena that involve soft gluon emission such as jet physics.  Among
the core observations is the fact that the JIMWLK evolution equation
intrinsically leads to the scaling behavior discussed in
Figs.~\ref{fig:IR-safety} and~\ref{fig:IR-safety-scaled}.

Many phenomenological applications are based on this scaling behavior
and I strongly recommend to have a first glance at
Sec.~\ref{sec:pheno} to get an idea of the broad range of affected
phenomena before delving into the main text. This should provide
additional a strong motivation to understand the underlying
theoretical tools and evolution equations.

\subsection{\it How to read this review}
\label{sec:how-read-this}

Having given a brief overview over the core physics ideas in the
previous subsection, it is left to the body of the paper to
substantiate these claims and flesh out the context from which they
arise.

This review contains two interwoven threads of arguments, one of them
centered around the derivation of evolution equations that govern soft
gluon emission from QCD at small coupling but large gluonic densities,
the other aimed at exploring their consequences and interpretation
using models and numerical work as well as phenomenological
applications. They can of course not be fully separated. Clearly the
theoretical grounding is necessary for any application, but also the
other direction is vital: for instance the numerical work
in~\cite{Rummukainen:2003ns} has had consequences on the understanding
of the necessity of running coupling corrections, not only in a
quantitative, but also a conceptual sense. Similarly, the discovery of
geometric scaling in HERA data has prompted a look for scaling
features in the evolution equations. This in mind, the paper develops
both threads in stages. It presents a different perspective on the
derivation and interpretation of the underlying evolution equations
than other existing reviews~\cite{Iancu:2002xk,Iancu:2003xm}, not the
least by highlighting the parallels with nonlinear effects in jet
physics in Sec.~\ref{sec:jimwlk-soft-gluon}.  I have also been aiming
at exposing the close connection between the McLerran Venugopalan
model with the BK equation in Sec.~\ref{sec:jimwlk-to-bk}. These
formulations have not been shown elsewhere, but are hidden under the
surface of many discussions of the Glauber-Mueller type nature of the
multiple scattering phenomena omnipresent in small $\xbj$ physics. I
have attempted to present an up to date description of the role of
BFKL physics as the driving force of small $\xbj$ evolution and have
included results of numerical simulations of the full JIMWLK equations
as well as a thorough discussion of running coupling in the BK case.

Sec.~\ref{sec:eikonalization} is devoted to the first important type
of nonlinearity encountered: eikonalization of soft gluon fields into
path ordered exponentials often also called eikonal
factors\footnote{This term has found various applications: besides
  this usage, also exponentiation of low order contributions in
  correlators like~\eqref{eq:N-def} is often labelled as
  eikonalization.} or Wilson lines. Sec.\ref{sec:prop-cross-sect} aims
at explaining their appearance in cross sections and propagators and
Sec.\ref{sec:it-mclerr-venug} discusses an immediate consequence,
boundedness of the underlying correlators, in terms of the MV model
and its generalizations. The link to gluon distributions, density
effects and $A$ scaling also appear in this context.

Sec.~\ref{sec:jimwlk-evol-balitsky} then gives an overview over the
formulation of the JIMWLK equation and the associated physics ideas. It
starts off firmly in the formal thread with the full derivation in
Sec.~\ref{sec:jimwlk-evol-balitsky} -- I consider this quite
informative, but the reminder can be followed without having read this
in a first go. Sec.~\ref{sec:evol-equat-expl} explains the JIMWLK
Hamiltonian, its properties and the nature of the equation as a
Fokker-Planck equation. Sec.~\ref{sec:it-geometric-scaling-eqn-idea}
formulates the idea of scaling in the context of these evolution
equations and Sec.~\ref{sec:jimwlk-to-bk} discusses the relation of
JIMWLK, BK and BFKL (in both forward and nonforward cases) equations
from the perspective of $N_c$ and density expansions.  How the
underlying growth pattern is imprinted by BFKL physics is discussed in
Sec.~\ref{sec:growth-bfkl}.  This gives rise to the idea of a scaling
window as emphasized in Sec.~\ref{sec:scaling-window} and already
shown in Fig.~\ref{fig:x-Q-plane-density} as the extended scaling
region.

Sec.~\ref{sec:from-fokker-planck} performs a translation of the JIMWLK
equation to a Langevin type formalism: Sec.\ref{sec:illustrating-idea}
gives a toy example to illustrate the idea while
Sec.\ref{sec:numer-results-from} gives the result which is the basis
for numerical simulations. Sec.\ref{sec:JIMWLK-shower-operators}
introduces shower operators which will, in
Sec.~\ref{sec:jimwlk-soft-gluon} find an interpretation as operators
that generate the soft gluon cloud in the photon wavefunction and
counts among the conceptual developments.

Sec.~\ref{sec:jimwlk-soft-gluon} leads up to this interpretation and
helps to understand that JIMWLK evolution shows a structure which is
generic for soft gluon emission physics. This is illustrated with an
example from the context of non-global jet observables for which
Sec.~\ref{sec:analogy} and \ref{sec:from-jimwlk-bk} give evolution
equations with clear structural analogies to BK and JIMWLK. Next is a
derivation of these equations which may be skipped in a first reading.
In this derivation one first constructs the leading part of generic
soft n-gluon amplitudes (corresponding to the photon wave functions)
in Sec.~\ref{sec:ampl-strongly-order}. As a second step one utilizes
the iterative nature of the construction process to extract a
corresponding evolution equation in the next subsections.
Sec.~\ref{sec:twotypes} highlights the two types of nonlinearities
encountered in the derivation --eikonalization and nonlinear
evolution-- and explores $N_c$ or correlator factorization once more
in the jet situation.

Sec.~\ref{sec:jets-dense-media} briefly discusses the lessons to be
learned for small $\xbj$ evolution from the preceding section and
provides a first glance at what is to be expected if jets are created
inside a dense medium.

Sec.~\ref{sec:numer-results} and~\ref{sec:runn-coupl-effects} discuss
numerical results from solving JIMWLK and BK evolution equations.
Discussed are scaling,~\ref{sec:numerical-JIMWLK}, correlator
factorization violations,~\ref{sec:size-fact-viol}, and the problems
of fixed coupling simulations in the UV,~\ref{sec:bk-jimwlk-compared}.
With running coupling UV phase space is under control and allows
quantitative studies of $Q_s$ and the scaling
features,~\ref{sec:it-running-coupling-phase-space}. This subsection
contains the core results on scaling behavior.
Sec.~\ref{sec:evol-a-depend} returns to $A$ scaling already introduced
in the MV model and discusses the changes imposed by evolution.

Sec.~\ref{sec:pheno} tries to give a flavor of the phenomenology
associated with the previous theoretical results. Highlighted are
geometric scaling, initial conditions of heavy ion collisions,
consequences for the Cronin effect and a few more.

Experimental efforts in connection with RHIC, LHC and EIC increase the
pressure to come up with more observable consequences and better tools
to make predictions -- a few ideas are mentioned in the conclusions,
Sec.~\ref{sec:conclusions}.

% spellmark 1

\section{Eikonalization at high energies} 
\label{sec:eikonalization}

\subsection{\it Propagators and cross sections}
\label{sec:prop-cross-sect}

There are two distinct types of nonlinearities present in the
formulation of cross sections and their $\xbj$ dependence as presented
in Sec.\ref{sec:cgc-dis:-high}, the first is the resummation of the
gluon field into path-ordered exponentials $U^{(\dagger)}_{\bm{x}}$
collinear to the projectile, the second is the nonlinearity of the
evolution in correlators of this type of field. The first aspect is
present already in the Abelian theory, the second appears only in the
non-Abelian case but can not be separated from the first. Without the
first step, the identification of the relevant variables, all the
following steps have no proper foundation and formulae like
Eq.\eqref{eq:dipole-cross} remain mysterious. Let me, therefore, start
with identifying the variables and a brief derivation of this formula
for the DIS cross section.

The starting point is the kinematics. A collision at large energies
singles out a longitudinal direction, say $x^3$, as the collision
axis.  In such a frame the large energy is characterized by a big
rapidity separation between target and projectile.
% which leads to Fig.~\ref{fig:lightcone-diag}. 
To judge how a projectile, say the $q\Bar q$ dipole emerging at lowest
order in the photon wave function of Eq.~\eqref{eq:dipole-cross}, sees
the target, it is best to consider the color Field generated by the
target. Remembering that light cone components $x^\pm
:=\frac{1}{\sqrt{2}}(x^0\pm x^3)$ are eigendirections of a boost in
$3$ direction ($x^\pm\to e^{\pm \y} x^\pm$ with $\gamma = e^\y =
\frac{1}{\sqrt{1-v^2/c^2}}$) it becomes clear that the components of
the field strength tensor are enhanced by a factor $e^\y$ in the plus
components and suppressed by a factor $e^{-\y}$ in the minus
components. At the same time Lorentz contraction ``shrinks'' the
support of the color field to a (near) delta function in $x^-$ and
time dilatation renders it (nearly) independent on $x^+$. The leading
components are
\begin{equation}
  \label{eq:leading-F}
F^{i +}(x) = \delta(x^-)\partial^i\beta(\bm{x})  \ .
\end{equation}
On the r.h.s. I have already made use of the gauge freedom to express
$F^{i +}$ via a single scalar function which can be associated with
the $b^+$ of Eq.~\eqref{eq:aplus} and~\eqref{eq:Udef}.  

To calculate cross sections like~\eqref{eq:dipole-cross}, one needs to
consider current correlators of the form $\langle J^\mu(u) J^\nu(v)
\rangle$ in the background field representing the target and subtract
noninteracting counterpart (to go from an S-matrix to a T-matrix
contribution).  To lowest order, the cross
section~\eqref{eq:dipole-cross} then takes the form
\begin{equation}
  \label{eq:cross-sect-diag}
  \sigma_{\mathrm{DIS}}(\xbj,Q^2) = \text{Im}\ 
  \begin{minipage}[c]{3cm}
    \includegraphics[height=.8cm]{dipole}
  \end{minipage}
  =\text{Im}\left\langle\parbox{3cm}{\includegraphics[width=3cm]{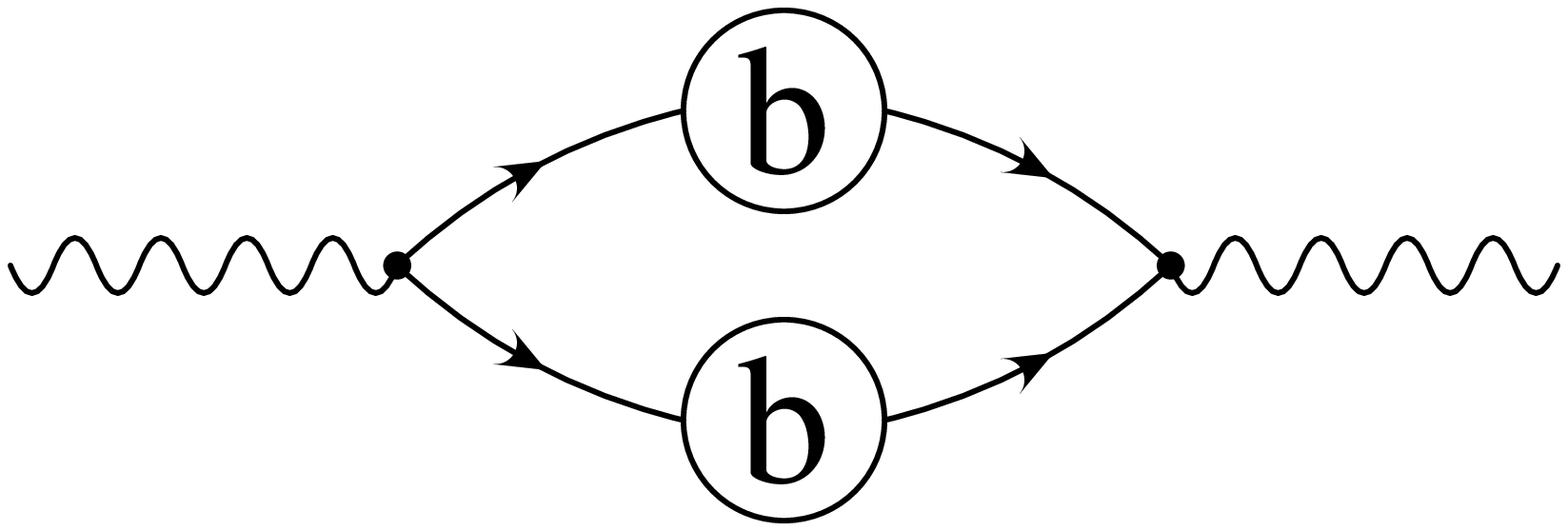}}
  -  \parbox{3cm}{\includegraphics[width=3cm]{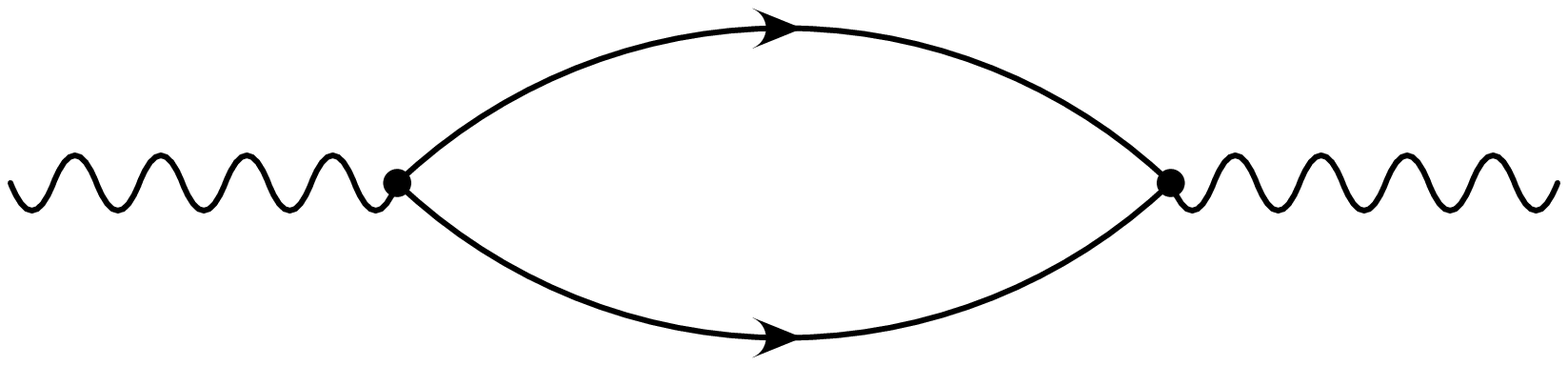}}\right\rangle
\end{equation}
where the interaction with the target is now explicitly trough the
fermion propagators in a background field
$\parbox{1.6cm}{\includegraphics[width=1.6cm]{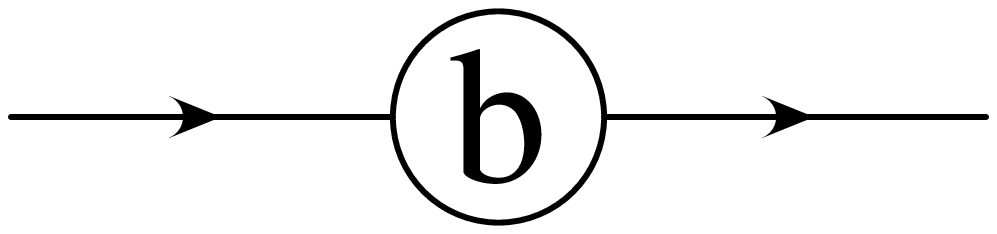}}$ and the
average $\langle \ldots \rangle$ is over background fields $b$ as
mandated by the interaction with the target. The simplification lies
in the fact that the types of field allowed to enter the interaction
is restricted to the type in~\eqref{eq:leading-F}.  With it one can
find results for cross sections and also include perturbative
corrections that will modify~\eqref{eq:cross-sect-diag} once quantum
fluctuations around $b$ become large.

The latter will be discussed in detail in later sections, here I am
interested in the physics aspect of the calculation leading
to~\eqref{eq:dipole-cross}. The core element here is the background
field propagator which encodes all the interaction with the target. It
is sufficient to study this propagator to understand the origin of the
eikonal factors in the dipole cross section as well as the general
form of the result. The key physics insight will be that the
interaction is solely mediated through an exchange of momentum.

Besides these physics results, gauge invariance is a somewhat delicate
issue: While parametric enhancement and Lorentz contraction of
components in the Field strength tensor can be argued in a gauge
independent manner, the extraction of path-ordered exponentials that
lead to formulae like Eq.~\eqref{eq:dipole-cross} appears to be gauge
dependent: What happens in a gauge in which one forces the ``+''
component of the gauge field to zero? This will be briefly touched
upon towards the end of this section.

To explore the situation, consider the propagator of a scalar field in
the background of some gauge field.\footnote{This is in essence the
  simplified version of an argument used first
  in~\cite{Balitsky:1996ub} from which the spin degrees of freedom are
  exorcized for clarity.} This quantity has also been used as the core
ingredient to the gluon propagator in~\cite{Weigert:2000gi} in the
derivation of the final form of the JIMWLK equation. Using first a
Schwinger parameter representation and then standard (nonrelativistic)
Feynman-pathintegral techniques in which the Schwinger parameter takes
the role of ``time,'' one writes its propagator as
\begin{equation}
  \label{eq:scalarprop}
  \frac{-i}{D^2[b]}(x,y) = \int_0^{\infty} ds e^{- s i D^2[b] } 
  =  \int^{\infty}_0 ds \int_y^x [dz] 
  e^{-\int^s_0 d\kappa \frac{(\dot z(\kappa))^2}{4} }
 \ {\sf P} \exp\Big\{- i g \int_y^x dz^\mu b_\mu(z)\Big\}
\end{equation}
where the path-integral is over trajectories $z$, parametrized by
$\kappa$, that connect $x$ (at $\kappa=0$) to $y$ (at $\kappa=s$).
The expression at $b=0$ is the free propagator.  Based on the
kinematical arguments above, the background field is of the form given
in~\eqref{eq:aplus}: $b^\mu(z)=g^\mu_{\ +}\delta(z^-)\beta(\bm{z})$.
This implies that only the ``$-$'' component of the trajectories are
probed and only a dependence on the transverse coordinates $\bm{z}$
remains. Due to the $\delta(z^-)$ shape one finds free propagation
unless $x$ and $y$ are on different sides of the $z^-=0$ hyperplane,
in which case a nontrivial contribution from the path-ordered
exponential emerges with free propagation before and after.  Since the
field is concentrated at $z^-=0$, one may deform the eikonal factors
outside the $z^-=0$ hyperplane into straight line eikonals according
to
\begin{equation}
  \label{eq:straightline-factors}
  {\sf P} \exp\Big\{- i g \int_y^x dz^\mu b_\mu(z)\Big\}
  \to 
  {\sf
  P}\exp\Big\{-i g\int\limits_{y^-}^{x^-}\!dz^-\,
b^+(0,{\bm{z}},z^-)\Big\}
\end{equation}
and one has to distinguish 4 cases depending on which side of $0$ one
places $x^-$ and $y^-$. For $x^->0>y^-$ it is possible to extend the
path to cover the whole line from, $\bm{z}-\infty n$ to $\bm{z}+\infty
n$ ($n$ a lightlike vector in ``$-$'' direction). This is what has
been denoted $U_{\bm{z}}$ throughout:
\begin{equation}
  \label{eq:Uz-def}
  U_{\bm{z}}:={\sf
  P}\exp\Big\{-i g\int\limits_{\bm{z}-\infty n}^{\bm{z}+\infty n}
\!dz^-\, b^+(0,{\bm{z}},z^-)\Big\}
\ .
\end{equation}
For $x^-<0<y^-$ one obtains $U^\dagger_{\bm{z}}$ and for same side
propagation --with $x^-,y^->0$ or $x^-,y^-<0$-- one simply gets the
unit matrix and thus the free propagator. Following the lines
of~\cite{Balitsky:1996ub} one eventually arrives at an explicit
expression for the scalar propagator of the form
\begin{align}
  \label{eq:KGpropexpl}
\left[\frac{-i}{D^2[b]}\right]_{a b}(x,y) 
%=  G_{ab}(x,y)}
%\\ & = & i \int\!\! \frac{d^4p}{(2\pi)^4} \sum_c
%  \frac{\phi_a^{(p,c)}(x) \left(\phi^{(p,c)}\right)^\dagger_a(y)}{
%     2p^-\left(p^+ -\frac{p_\perp^2+m^2-i\epsilon}{2p^-}\right)}
%\nonumber \\ & = &
%i(-i) 
%\int\!\! \frac{d^2p_\perp dp^-}{2p^- (2\pi)^3} \sum_c
%  \left[\theta(x^--y^-)\theta(p^-)
%    - \theta(y^--x^-)\theta(-p^-)\right]
%  \left[
%    \phi_a^{(p,c)}(x) \left(\phi^{(p,c)}\right)^\dagger_a(y)
%  \right]_{p^+ = \frac{p_\perp^2+m^2}{2p^-}}
%\nonumber \\ & = &
= &
\int\!\! \frac{ dp^-}{2p^- (2\pi)^3}
  \left[\theta(x^--y^-)\theta(p^-)
    - \theta(y^--x^-)\theta(-p^-)\right]
    \int\!\! {d^2{\bm{p}}'  d^2{\bm{q}}' }
    \nonumber \\ &%&\hspace{1cm} 
\times
    \left[
    %\varphi_a^{(p',b')}(x)
      e^{-i p'.x}
      \int\! \frac{ d^2{\bm{z}}}{(2\pi)^2}\
      e^{-i({\bm{p}}'-{\bm{q}}'){\bm{z}}}
      \ {\sf P}\exp\Big\{-i g\int\limits_{y^-}^{x^-}\!dz^-\, b^+(0,{\bm{z}},z^-)\Big\}
 %     \tilde U^{-1}(x^-,y^-,{\bm{z}}) _{a b}
    %\left(\varphi^{(q',a')}\right)^\dagger_a(y)
      e^{i q'.y}
  \right]%_{%
%\begin{minipage}{3cm}\scriptsize
%${p'}^+ %= p^+\vert_{p^+=\frac{p_\perp^2+m^2}{ 2p^-}}
%       -\frac{p_\perp^2-{p'}^2_\perp}{ 2p^-}
%       = \frac{{p'}^2_\perp+m^2}{ 2p^-}$\\
%${q'}^+= \frac{{q'}^2_\perp+m^2}{ 2p^-}$\\
%${p'}^-={q'}^-=p^-$
%  \end{minipage}}
%\nonumber
\end{align}
where some of the components of these momenta are interrelated
according
to ${p'}^+ %= p^+\vert_{p^+=\frac{p_\perp^2%+m^2}{ 2p^-}}
-\frac{{\bm{p}}^2-{{\bm{p}}'}^2}{ 2p^-} = \frac{{{\bm{p}}'}^2%+m^2
}{
  2p^-}$, ${q'}^+= \frac{{{\bm{q}}'}^2
%+m^2
}{ 2p^-}$ and ${p'}^-={q'}^-=p^-$. The massive version simply has
on-shell ``$+$'' momenta shifted by $m^2/2p^-$.  An alternative
calculation of this propagator using a spectral method can be found
for example in~\cite{Ayala:1995kg}.

This background field propagator can now be used in the diagrams
of~\eqref{eq:cross-sect-diag}:
\begin{align}
 {\text{Im}}\left( \label{eq:cross-sect-prop}
  \int d^4x e^{-iq.x}  \int d^4y e^{-iq.y}  q^\mu 
  \left[\frac{-i}{D^2[b]}\right]_{a b}(x,y)  q^\nu  
  \left[\frac{-i}{D^2[b]}\right]_{a b}(y,x)  
  - (\text{same at } b=0)\right)
\end{align}
where the momentum flowing through the diagram is $q$, the vertices at
$x$ and $y$ carry factors of the incoming (outgoing) momentum $q^\mu$
($q^\nu$) [For scalar ``quarks'' the current is $J^\mu(u)
=\phi^\dagger(u)\stackrel{\leftrightarrow}{\partial}\!\!{}^\mu\phi(u)$
instead of the $J^\mu(u) = \Bar\psi(u)\gamma^\mu\psi(u)$ for
fermions].  The remaining calculation is straightforward but tedious.
I will only report the key features needed to understand the result:
\begin{itemize}
\item The minus component of the momenta is conserved. It is easy to
  convince oneself by inspection of the theta functions that the
  corresponding loop momentum will be constrained to lie between $0$
  and $q^-$. Parametrizing it as a dimensionless fraction of $q^-$ one
  sees a fraction $\alpha$ flowing trough one of the lines while a
  fraction $1-\alpha$ enters the other. This momentum fraction
  integral remains present to the very end.
\item Since all propagation is free unless $x$ and $y$ lie on
  different sides of the $z^-=0$ hyperplane, all ``same side''
  contributions cancel between the two terms
  of~\eqref{eq:cross-sect-prop}.
\item In the remaining contributions one encounters eikonal factors of
  the form~\eqref{eq:Uz-def} in the combination $\tr
  U^\dagger_{\bm{z}_1} U_{\bm{z}_2}/N_c$. The second term at $b=0$
  will simply subtract a $1$ since all other ingredients are equal.
  The trace originates from the contraction of color indices at the
  $\gamma^* q\Bar q$ vertex. This is the origin of the dipole operator
  $\tr(1-U_{\bm{z}_1} U^\dagger_{\bm{z}_2})/N_c$ and allows the key
  observation that no color is exchanged with the target.
\item All integrals besides those over the transverse coordinates
  ${\bm{z}_1}$, ${\bm{z}_2}$ can be performed. The momentum exchange
  with the target is parametrized by this coordinate dependence. 
\item The remaining $x^-$ and $y^-$ integrals have the form
  $\int_0^\infty d x^- \int_{-\infty}^0 d y^-$ or $\int_{-\infty}^0 d
  x^- \int_0^\infty d y^-$. These cover regions outside the $z^-=0$
  hyperplane where the $q\Bar q$ pair propagates freely.  Each $x^-$
  and $y^-$ integral can be identified as an integral representation
  of a modified Bessel or McDonald function.  They describe the
  amplitudes for the $\gamma^*\to q\Bar q$ transitions before and
  after the interaction with the target -- the ``virtual photon
  wavefunctions.'' [See~\cite{Hebecker:1998kv} for a direct
  calculation of these amplitudes.]
\end{itemize}
The final result, reads
\begin{eqnarray}
\label{eq:sigma-DIS-explicit}
\sigma_{\text{DIS}}(\y,Q^2)& = & \int d^2{\bm r}\int\limits^1_0\!d\alpha\
|\psi(\alpha,{\bm r}^2,Q^2)|^2
\ 
\int d^2b \ 
  \langle\frac{\tr(1-U_{\bm{z}_1} U^\dagger_{\bm{z}_2})}{N_c} \rangle
\ .
\end{eqnarray}
For longitudinal polarizations (these are
chosen for closer similarity to the fermion results with physical
transverse polarizations) 
and vanishing masses one finds
\begin{equation}
  \label{eq:scalar-long-wavefunct}
  |\psi_{L}(\alpha,{\bm r}^2,Q^2)|^2 = 
  \frac{3\alpha_{\text{em}}}{4\pi^2}
  \alpha(1-\alpha)Q^2 K_0^2(\sqrt{\alpha(1-\alpha)Q^2})
\ .
\end{equation}

The physically relevant case of spin $1/2$ quarks can be derived
following the same lines and leads to a very similar result -- it is
essentially only the Bessel functions that are modified in the end
(see Eq.~\eqref{eq:fermion-wavefunct}). However, starting the
calculation with the fermion propagator in the small $x$ background
field adds technical complexity to intermediate stages of the
calculation~\cite{Balitsky:1996ub}. Since they do not influence the
core physics content I will only briefly sketch what has to be done in
addition the steps outlined above.

One first has to derive the fermion propagator starting from the
equivalent to the l.h.s. of~\eqref{eq:scalarprop}
\begin{eqnarray}
  \label{eq:ferm-expl}
  S[A](x,y) & = &[i \slashed D+m]_x 
  \big( -D^2+g\sigma^{\mu\nu}F_{\mu\nu}-m^2\big)^{-1}(x,y)
\end{eqnarray}
($\sigma^{\mu\nu}\propto[\gamma^\mu,\gamma^\nu]$ generates spinor
rotations) and ends up with the expression
\begin{align}
  \label{eq:ferm-fully-expl}
  S[A](x,y)  = &  -i[i \slashed D+m]_x \int\frac{dp^-}{(2\pi)^32p^-}
  \big[\theta(x^--y^-)\theta(p^-)-\theta(y^--x^-)\theta(-p^-)\big]
\int\! d^2{\bm{p}'} d^2{\bm{q}'}
\nonumber \\ &
  e^{-ip'.x}\int\frac{d^2\!z}{(2\pi)^2}e^{-i({\bm{p}'}-{\bm{q}'}){\bm{z}}} 
   \ {\sf P}\exp\Big\{-i g\int\limits_{y^-}^{x^-}\!ds
  \big(\frac{dw^-}{ds}A^++\frac{\sigma F}{2p^-}\big)(0,{\bm{z}},w^-(s))\Big\}
   e^{iq'.y}
%\nonumber \\  = &
%-i \int\frac{dp^-}{(2\pi)^32p^-}
%  \big[\theta(x^--y^-)\theta(p^-)-\theta(y^--x^-)\theta(-p^-)\big]
%\int\! d^2{\bm{p}}' d^2{\bm{q}'}
%\nonumber \\ &&
%\Big[ \slashed p'+m %-g\gamma^- A^+_x
%\Big]
%\nonumber \\ &&
%  e^{-ip'.x}\int\frac{d^2\!z}{(2\pi)^2}e^{-i({\bm{p}'}-{\bm{q}'}){\bm{z}}} P\exp\bi
%g[i g\int\limits_{y^-}^{x^-}\!ds
%  \big(\frac{dw^-}{ds}b^++\frac{\sigma F}{2p^-}\big)(0,{\bm{z}},w^-(s))\big]
%   e^{iq'.y}\nonumber \\ &&
% +\gamma^-\delta(x^--y^-)\int\frac{dp^-d^2p_\perp }{(2\pi)^32p^-} e^{-ip.(x-y)}
\end{align}
with exactly the same restrictions on momenta as in the scalar case,
Eq.~\eqref{eq:KGpropexpl}.  In fact the $\sigma F$ contribution
simplifies considerably, if one takes into account that $(\sigma F)^2
= 0$ due to $\sigma^{i -}\sigma^{j -} = 0$ (for arbitrary transverse
indices $i$ and $j$).  The remaining contribution does not enter the
DIS cross section at all.  The final result is again of the
form~\eqref{eq:sigma-DIS-explicit} although polynomials in the
numerator lead to different types of Bessel functions. The
wavefunction contribution now is given by
\begin{equation}
  \label{eq:fermion-wavefunct}
\vert \psi_{T,L}\,(\alpha,\bm{r},Q^2) \vert^2  \,=\,  
\frac{6\, \alpha_{\text{em}}}{4\,\pi^2 } \, \sum_{f}\,e_f^2 \, \begin{cases} 
[\,\alpha^2+(1-\alpha)^2\,]\:\overline{Q}^2\,K_1^2\,(\overline{Q}\, r) 
\,+\, m_f^2\: K_0^2\,(\overline{Q}\, r)\,
\cr\cr
4\,Q^2\,\alpha^2\,(1-\alpha)^2\,K_0^2\,(\overline{Q}\, r)\;.
\end{cases}
\end{equation}
where subscripts $T$ and $L$ refer to transverse and longitudinal
polarizations of the photon and $\overline{Q}^2 \:=\:
\alpha\,(1-\alpha)\,Q^2\,+\,m_f^2$.

There exist many alternative derivations in the
literature.~\cite{Buchmuller:1995mr, Buchmuller:1996xw,
  Hebecker:1999ej} for example build up the wavefunctions
perturbatively and then calculate the expression for the cross
section. In these approaches eikonalization --the appearance of the
path ordered exponentials-- has to be obtained perturbatively from
diagrams like the ones in Eq.~\eqref{eq:emission}. The key observation
there is that in the kinematic situation at small $\xbj$, in which,
say, a quark, with high energy of momentum $p$ that emits a soft gluon
of momentum $k$, the product of the propagator of the emitting
particle and the emission vertex, simplifies drastically and takes the
form of what is often called an eikonal current
\begin{equation}
  \label{eq:eikonal-current}
  J^\mu_{p k} := \frac{p^\mu}{p.k +i\epsilon}
\end{equation}
to which the soft gluon couples as $J^\mu_{p k} b_\mu(k)$.  With $p$
pointing in ``$-$''-direction this object involves a $k^+$ and a $b^+$
only, and after Fourier transformation of the expression, one begins
to recognize that the denominator induces the $\theta$-function
structure needed to build up the path-ordering in the expressions
above.  Iterating the procedure carefully leads to
Eq.~\eqref{eq:U-eikonal-diag}. This correspondence becomes important
again when one starts to write down functional expressions that
implement soft gluon clouds in both small $\xbj$ and jet physics, the so
called shower operators first encountered here in connection with a
Langevin rewrite of the JIMWLK equation.

All of these approaches share a common technical difficulty: as soon
as one tries to choose a gauge in which the ``+'' component (the
component entering the path ordered exponentials $U_{\bm{z}}$)
vanishes, the above arguments appear to fail pathologically. Since
such a choice has been used in the original formulation of the
McLerran-Venugopalan model and in early versions of the JIMWLK
equation, I need to explain how this confusion is resolved. Such a
calculation shifts the leading contributions into transverse
components of the gauge field $\Bar b_i(x)$ where the bar serves to
characterize a gauge in which $\bar b^+=0$. One then proceeds to show
that there is only a single degree of freedom contained in the two
components of $\Bar b_i$ by showing that they can always be written in
terms of a group valued field $U_{x^-,\bm{x}}$ via
\begin{align}
  \label{eq:aplusgauge}
  \Bar b_i(x) = -\frac{1}{i g} U_{x^-,\bm{x}}^\dagger
   \ \partial_i U_{x^-,\bm{x}} 
\ .
\end{align}
This expression is of the form $\theta(x^-)\Bar\beta_i(\bm{x})$, it
has support over ``half'' of space time instead of only on a
hyperplane at $x^-=0$. Only by comparing to the calculation presented
above does one realize that $U_{x^-,\bm{x}}$ is nothing but the gauge
transform that relates the two (barred and unbarred) gauges
\begin{equation}
  \label{eq:gaugetransf}
   U_{x^-,\bm{x}} = {\sf P}\exp\Big\{-i g\int\limits_{-\infty}^{x^-}\!dz^-\,
   b^+(0,{\bm{x}},z^-)\Big\} 
\ ,
\end{equation}
although within the barred formulation, such an interpretation is not
available.  As was to be expected, one merely ends up reshuffling the
same degree of freedom by changing the gauge. Nevertheless, the
calculation of the propagators via path-integral
representations~\eqref{eq:scalarprop} becomes quite nontrivial due to
the altered support of the background field and a method based on
spectral representations and wave functions in the presence of these
background fields becomes more efficient.

%which are related to the above by
%\begin{align}
%  \label{eq:aplusgauge}
%  \Tilde \Bar b_i(x) = -\frac{1}{i g} 
%  {\sf  P}\exp\Big\{i g\int\limits_{-\infty}^{x^-}\!dz^-\, 
%   b^+(0,{\bm{x}},z^-)\Big\}
%   \ \partial_i \
%   {\sf P}\exp\Big\{-i g\int\limits_{-\infty}^{x^-}\!dz^-\,
%   b^+(0,{\bm{x}},z^-)\Big\}
%\ .
%\end{align}
%where the bar on the gauge field on the left hand side indicates that
%it belongs to a different gauge than the unbarred field on the right
%hand side. In this gauge indeed $\Bar b^+=0$, but

%{\color{red} The result of such a calculation is the same as before, with the same
%eikonal factors $U$ carrying the physical degrees of freedom, although
%the relationship to the gauge field is more intricate. Additional
%gauge factors at $x$ and $y$ clearly drop out in the diagram for the
%cross section: They cancel at the $\gamma^* \to q\Bar q$ vertices.}

\subsection{\it The McLerran-Venugopalan model and 
its generalizations}
\label{sec:it-mclerr-venug}

The McLerran-Venugopalan (or MV) model was formulated as a means to
describe cross sections at small but fixed $\xbj$, like the DIS cross
section in Eq.~\eqref{eq:dipole-cross}, by parametrizing the averaging
procedure on the r.h.s. with a suitable ansatz. In essence, one
assumes that the dominant configurations $b$ of Eq.~\eqref{eq:aplus}
are governed by a Gaussian weight. This was argued to be reliable in
the case of large nuclei and formulated in terms of color sources in
the large hadronic target. The large field $b$ of the introduction can
be thought of as generated from these sources via the Yang-Mills
equations.  Just as the background field in the gauge chosen in the
introduction, the corresponding color current is of the form
\begin{equation}
  \label{eq:colorcharge}
  J^\mu(x) = \delta^{\mu +} \delta(x^-) \rho({\bm{x}})
\end{equation}
where $\rho({\bm{x}})=\rho^a({\bm{x}})t^a$ is a matrix in color space
that describes the color charge density in the target.  As explained
in the introduction, at $Q^2$ large enough to be in the perturbative
domain, the probe can transversally resolve colored structures inside
individual constituent nucleons.  In longitudinal direction the
situation is different.  Just as with the color field $b$ created from
it, the $ \delta(x^-) $ structure of the current is to be taken with
respect to a longitudinal resolution imposed by the value of $\xbj$ in
the experiment.  In going beyond that resolution, or by viewing the
situation in the target rest frame, one should think of $
\rho({\bm{x}}) $ as the integral in $x^-$ of color charges at a given
transverse position ${\bm{x}}$: $\rho({\bm{x}}) :=\int dx^- \rho(x^-,
{\bm{x}})$. A probe with $\xbj$ so small that it can not resolve
longitudinal internal structures
% over the whole longitudinal extent of the hadron ($\propto A^{1/3}$) 
will couple to this integral directly instead of the individual color
charges inside the constituent nucleons. Since the latter are color
neutral on their own, the target sees a incoherent superposition of
color charges. Let me denote the incoherent sum of color charges in a
tube extending the full length of the nucleus with transverse area
$1/Q^2$ (according to transverse resolution) ${\cal Q}^a(x)$. This may
be written in terms of $\rho$ as
\begin{equation}
  \label{eq:color-charge-res}
  {\cal Q}^a(x) := \int d^2{\bm{y}}\ \phi_{Q^2}(\bm{x}-\bm{y})
  \int dx^- \rho^a(x^-, {\bm{y}})
\end{equation}
where $\phi_{Q^2}$ represents a coarse graining function adapted to
the transverse resolution scale $1/Q^2$ (by its normalization $\int
d^2{\bm{y}}\ \phi_{Q^2}(\bm{y})\sim 1/Q^2$) whose precise nature will
not be important. If we ignore geometrical complications and assume
uniform longitudinal thickness (``cylindrical nuclei'') ${\cal Q}^a$
should obey (locally, in the coarse grained sense)
\begin{align}
  \label{eq:charge-corr}
  \langle  {\cal Q}^a  \rangle = 0 
  \hspace{2cm}  
  \langle  {\cal Q}^a  {\cal Q}^b  \rangle = 
  \delta^{a b} \frac{1}{Q^2}\frac{g^2 A}{\pi R_A^2} 
\ .
\end{align}
The average here can be thought of as a configuration average. The
first of these equalities states color neutrality and the second gives
the ``typical charge'' squared.  With $R_A = R_{\text{proton}}
A^{1/3}$ both the individual charge of Eq.\eqref{eq:color-charge-res}
and the correlator in Eq.\eqref{eq:charge-corr} scale with $A^{1/3}$,
so that asymptotically $[ {\cal Q}^a, {\cal Q}^b] = i f^{a b c} {\cal
  Q}^c \ll {\cal Q}^2$ and the charges can be treated as commuting
objects. This is the original McLerran-Venugopalan argument that would
lead to correlators of color charges $\langle {\cal Q}^{a_1} \ldots
{\cal Q}^{a_1}\rangle$ that are fully determined by the two point
function and hence by a Gaussian functional weight. This weight is
characterized by a width $\mu_A^2$, which should be thought of as
$\xbj$ and $Q^2$ dependent.  It is common use to not explicitly
discuss the coarse graining in transverse space and write this
distribution in terms of $\rho({\bm{x}})$. At this point one should
caution that although the $x^-$ integral
in~\eqref{eq:color-charge-res} will scale with $A^{1/3}$, the actual
proportionality constant can only be found in a more careful
treatment. As indicated, geometry alone leads to an additional factor
that is generically smaller than $1$. At a given impact parameter
$\bm{b}$, this could be taken into account by writing $\pi R^2
S_A({\bm{b}})$ instead of $A^{1/3}$~\cite{Levin:2003nc}. Here
$S_A({\bm{b}})$ is the Wood-Saxon formfactor and $R$ is the gluon
radius of the nucleon, the size of the gluon distribution in the
transverse plane.  This would still imply scaling of $Q_s^2$ with
$A^{1/3}$ at large enough nuclei but would lead to a noticeable
reduction for small nuclei. The naive approximation of a cylindrical
nucleus on the other hand would amount to $Q_s^A(\xbj) = Q_s^p(\xbj)^2
A^{1/3}$.

A slightly different treatment of color charges has been adopted in
the calculation of gluon distributions
in~\cite{Jalilian-Marian:1997xn} and in the derivation of the BK
equation by Kovchegov~\cite{Kovchegov:1999ua}. Here one would resolve
the nucleons in the nucleus and treat the color charges at different
$x^-$ as belonging to different nucleons. Again one can ignore all
higher order correlators of $\rho$ and describe the
$\rho$-distribution by a distribution local in transverse and
longitudinal space
\begin{align}
  \label{eq:rhodist}
  W_{\text{MV}}[\rho] := {\cal N} \exp\Big\{
  -\frac{1}{2}\int dx^- d^2{\bm{x}}\ 
  \frac{\rho^a(x^-,\bm{x})\rho^a(x^-,\bm{x})}{\mu^2(x^-)}
  \Big\}
\ .
\end{align}
The link between the two descriptions is provided by the requirement
that the widths in the distributions be related by $\mu_A^2=\int
dx^-\mu^2(x^-)$ wherever this integral is large enough for the
original arguments leading to a Gaussian weight to apply.  The first
formulation directly uses a large surface density of gluons (as
parametrized by $\mu_A^2$), the latter formulation explicitly builds
it from a locally small ``per nucleon'' volume-density $\mu^2(x^-)$.
Instead of the original classicality argument it is the decorrelation
in $x^-$ which implies that charge commutators never play a role in
the evaluation of correlators like the dipole function and thus that
the Gaussian weight~\eqref{eq:rhodist} is adequate. The latter
description is actually capable of interpolating between two opposing
limits as one increases $\mu^2(x^-)$, say as a consequence of soft
gluon radiation and coarse graining in $x^-$ as one progresses towards
smaller $\xbj$.  Both versions can be found under the name
McLerran-Venugopalan model.  A reliable treatment of the region in
between the extremes where individual $\mu^2(x^-)$ grow large requires
additional information that lies beyond this level of modeling.
Information of this type is expected to arise from small $\xbj$
evolution equations such as the JIMWLK equation. It is because of the
link with one of the original derivations of the BK equation that I
will further discuss~\eqref{eq:rhodist} and its generalizations.

Note that due to its local nature and the gauge transformation
properties of the current $J$,~\eqref{eq:rhodist} is a gauge invariant
choice.  The Gaussian leads to a local correlator for
$\rho(x^-,{\bm{x}})$ both in $x^-$ and ${\bm{x}}$ where $\mu^2(x^-)$
turns out to be related to the gluon density.

The above uses the case of large nuclei in which individual nucleons
are used as decorrelated sources of a cumulatively large number of
gluons, but the resulting description in terms of a Gaussian
weight may also apply to small nuclei or even a proton in a situation
where evolution towards small $\xbj$ builds up sizeable gluon fields,
provided these fields are not strongly correlated over a large $\xbj$
range. [$A$ dependence of this alternative source of gluons would of
course be different, see also Sec.~\ref{sec:evol-a-depend}.]  In this
sense one may substitute small $\xbj$ for large $A$ in the above
reasoning.

It is instructive to calculate dipole functions and gluon
distributions in the McLerran-Venu\-go\-palan model which allows one to
formulate a number of (possibly simplistic) phenomenological
expectations that should be closer to reality the larger $A$ and the
smaller $\xbj$ become.  The simplest objects of interest are the
dipole cross section at fixed $\xbj$, which contains the operator $\tr
U_{\bm{x}} U^\dagger_{\bm{y}}$ and, somewhat more complicated, the
gluon distribution, which contains the operator $\tr( U \partial_i
U^\dagger)_{\bm{x}} ( U \partial_i U^\dagger)_{\bm{y}} $. Since most
correlators of interest --as well as the evolution equations
considered below-- can be expressed via path ordered exponentials
which depend on $b^+$ instead of $\rho$ with $-\partial_i\partial_i
b^+=\rho$ (from the $+$ component of the Yang-Mills equation), it is
natural to replace $ \rho$ by $b^+$ and write the weight of the
McLerran-Venugopalan model as
\begin{align}
  \label{eq:bdist}
  W_{\text{MV}}[b^+] := {\cal N} \exp\Big\{
  -\frac{1}{2}\int dx^- d^2{\bm{x}}\
  \frac{(\partial_i\partial_i b^{+ a})(x^-,\bm{x})
    (\partial_j\partial_j b^{+ a})(x^-,\bm{x})}{\mu^2(x^-,Q^2)}\Big\}
\ .
\end{align}
It is clear that the $b$-correlator remains local in $x^-$ but becomes
nonlocal in the transverse direction. Still, gauge invariance is
guaranteed by its equivalence to the original~\eqref{eq:rhodist}.

I will explain below that the local nature in $x^-$ --which leads to
the absence of commutator contributions in simple correlators-- is, if
not equivalent to, then at least compatible with the large $N_c$ limit
and is an important structural feature of the model.  The specific
nature of transverse nonlocality will be changed through gluon
emission as induced by small $x$ evolution be it in the JIMWLK or BK
evolution equations.  I will anticipate this and allow distributions
with correlators of the form
\begin{equation}
  \label{eq:b-corr}
 \langle b^{+ a}_{x^-,\bm{x}} b^{+ a}_{y^-,\bm{y}} \rangle  
 = \delta(x^--y^-)G_{x^-; \bm{x y}}
\end{equation}
with a more general transverse coordinate dependence. Such a
generalization has also been argued for by Mahlon and
Lam~\cite{Lam:1999wu} in order to implement overall color neutrality
by imposing conditions on $G_{x^-; \bm{x y}}$ that were incompatible
with the original MV form.\footnote{In this generalization gauge
  invariance still holds provided the $\xbj$-dependence is derived
  using one of these evolution equations -- they resum the gauge
  invariant, leading $\xbj$-dependence and the initial condition is
  gauge invariant. An illustration that some apparent gauge dependence
  may be spurious at small $\xbj$ is provided by the dipole cross
  section itself: This is gauge invariant in the sense that the
  observable predominantly probes configurations $b$ as
  in~\eqref{eq:aplus}. Hence, it does not matter along which curves
  the Wilson lines are closed at infinity, any curve provides the same
  factor $1$ which is not displayed in the formulae -- this
  information is not probed (suppressed) at small $\xbj$ and the
  result is gauge invariant ``to leading order at small $\xbj$.'' The
  same reasoning applies to the evolution equations themselves. }

For clarity of interpretation it is useful to change notation still a
bit further and replace $x^-$ in the above by $\y$ itself. This is
possible, since evolution towards smaller $\xbj$ (larger $\y$) is
related to a change of resolution in $x^-$ as discussed above. A more
detailed discussion of this is given
in~\cite{Iancu:2000hn,Ferreiro:2001qy}, but an early version can
already be found in~\cite{Jalilian-Marian:1997xn}.  To summarize, one
ends up interpreting the $U$ factors as path ordered exponentials
\begin{equation}
  \label{eq:pathord}
  U_{\bm{x}} = {\sf P}\exp\Big\{ -i g\int\!\!dz^- b^+_{z^-,\bm{x}} \Big\} 
  =: {\sf P}\exp\Big\{ -i \int\!\!d\y\ \alpha_{\y,\bm{x}}\Big\} 
\end{equation}
where $\alpha$ can be taken to be defined via this change of
variables. The above implies a Gaussian weight for $\alpha$ of the
form
\begin{align}
  \label{eq:Gaussian-A}
  W_\y[\alpha]  := \exp \Big(-\frac{1}{2}\int\limits^\y\!\! d\y'\! 
  \int\!\!d^2\!u\, d^2\!v\ 
  \alpha^a_{\y',\bm{u}} G^{-1}_{\y',\bm{u v}}  \alpha^a_{\y',\bm{v}}\Big)
\intertext{with} 
\langle \alpha_{\y\bm{x}} \alpha_{\y'\bm{y}} \rangle 
=   \delta(\y'-\y'') G_{\y'',\bm{u v}} \theta(\y-\y')
\end{align}
Simple correlators are calculable, for instance the (singlet) two
point functions with fields in the fundamental and adjoint
representations come out to be
\begin{subequations}
  \label{eq:simple-corr}
\begin{align}
\label{eq:simple-corr-1}
  \langle \tr( U_{\bm{x}} U^\dagger_{\bm{y}}) \rangle = & 
%  N_c e^{-C_{\text{f}}  \int\limits^\y\!\! d\y' (G_{\y',\bm{x y}}
%    -\frac{G_{\y',\bm{x x}}+G_{\y',\bm{y y}}}{2})} 
%  = 
N_c\ e^{-C_{\text{f}}  {\cal G}_{\y,\bm{x y}}}
\\ 
   \langle \Tilde\tr( \Tilde U_{\bm{x}} \Tilde U^\dagger_{\bm{y}}) \rangle 
   = & 
%2 N_c C_{\text{f}} e^{-N_c  \int\limits^\y \!\!d\y' (G_{\y',\bm{x y}}
%    -\frac{G_{\y',\bm{x x}}+G_{\y',\bm{y y}}}{2})} 
%  = 
2 N_c C_{\text{f}}\ e^{-N_c  {\cal G}_{\y,\bm{x y}}}
\end{align}
\end{subequations}
with 
\begin{equation}
  \label{eq:tilde-G}
  {\cal G}_{\y,\bm{x y}}:=\int\limits^\y \!\!d\y' \left( G_{\y',\bm{x y}}
    -\frac{G_{\y',\bm{x x}}+G_{\y',\bm{y y}}}{2}\right)
\end{equation}
a symmetric function of $\bm{x}$ and $\bm{y}$ that vanishes in the
local limit. The color factors are generic: the prefactor is simply
the dimension of the representation, the factor in the exponent is the
first Casimir.

Since such Glauber-Mueller type exponentiations are very
characteristic for multiple scattering events in the CGC I will, once,
outline how to arrive at this result. This will also allow me to
highlight the role played by color flow in this type of event. I will
start with a diagrammatic interpretation of the lowest order
contribution to, say~\eqref{eq:simple-corr-1}. One finds
\begin{align}
  \label{eq:G-CalG-diag}
 - N_c C_{\text{f}}\int\limits^\y \!\!d\y'\left( G_{\y',\bm{x y}}
    -\frac{G_{\y',\bm{x x}}+G_{\y',\bm{y y}}}{2}\right) =
%\hspace{.1cm}
\parbox{.5cm}{\includegraphics[width=.2cm]{Gtrleft}\vspace{.85cm}}
\hspace{-.25cm}
\left( \hspace{.2cm}
\parbox{1.3cm}{\includegraphics[width=1.3cm]{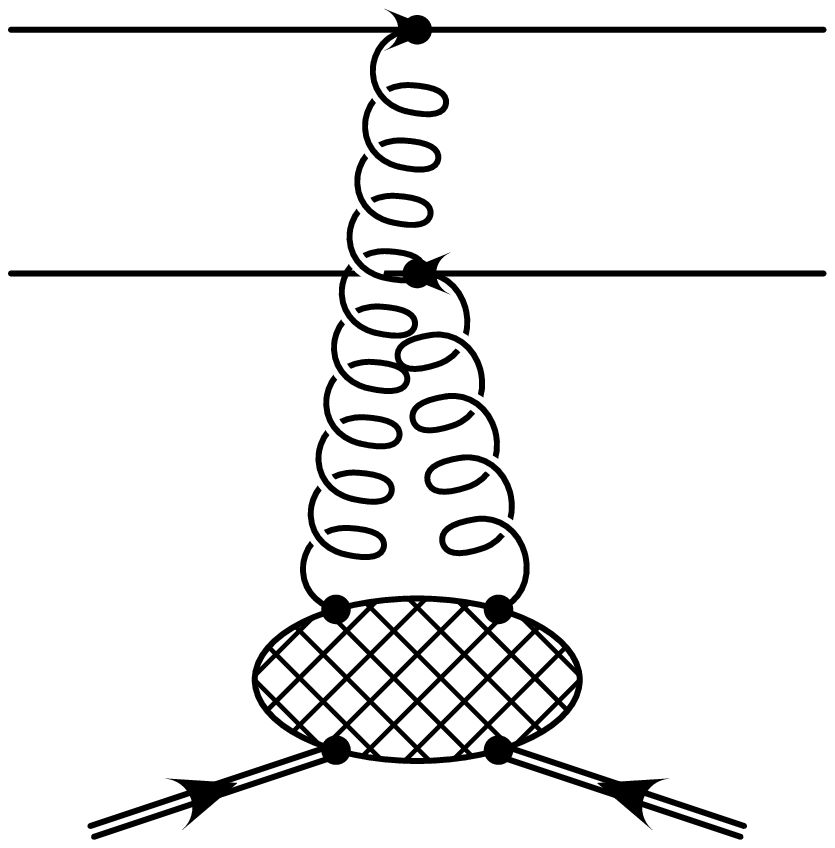}}
+%\frac{1}{2}
\left(\hspace{.2cm}
\parbox{1.3cm}{\includegraphics[width=1.3cm]{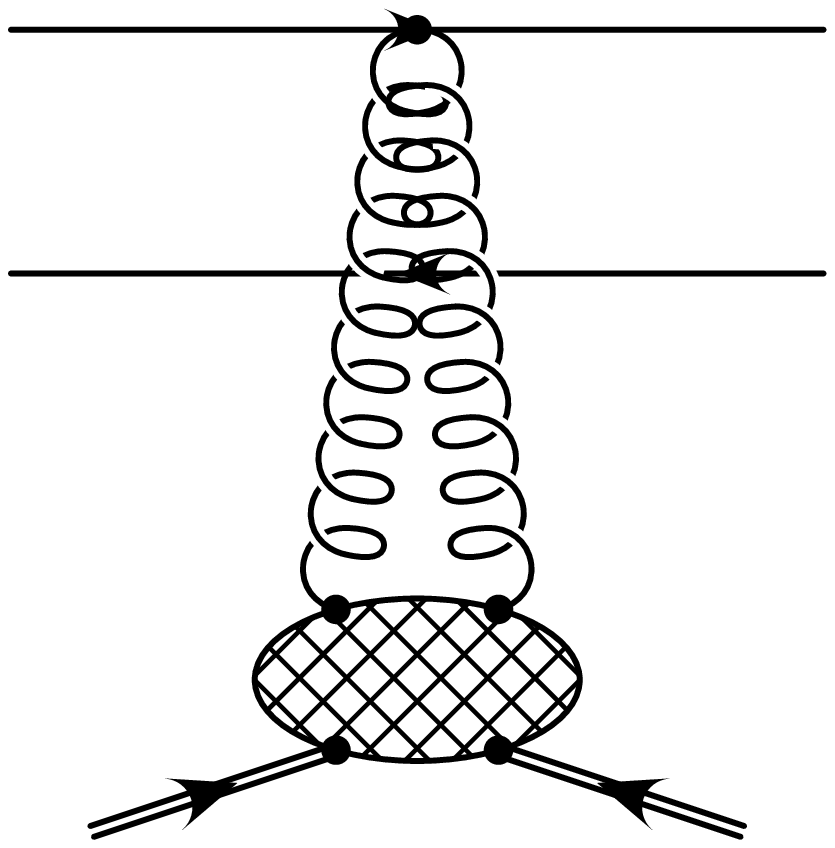}}
+\parbox{1.3cm}{\includegraphics[width=1.3cm]{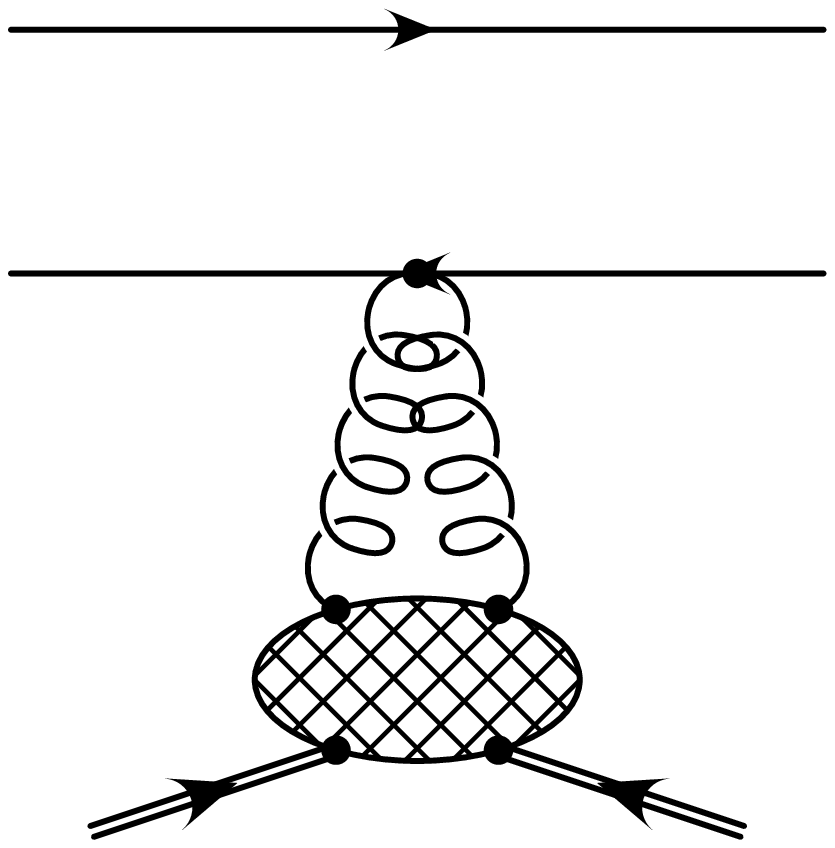}}
\hspace{.2cm}\right)
\right)
%\hspace{.1cm}
\parbox{.5cm}{\includegraphics[width=.2cm]{Gtrright}\vspace{.85cm}}
 .
\end{align}
The horizontal lines correspond to the Wilson lines expanded to the
order indicated by the number of gluon insertions. This implies the
relative factor of $-\tfrac{1}{2}$ of the last two diagrams compared
to the first as shown on the l.h.s..  The hatched blob stands for $G$
and the lower legs indicate that these correlators represent
interaction with the target. The gluons hook into the Wilson lines at
$\y'$, which is then integrated over up to $\y$, the value
characterizing the functional weight~\eqref{eq:Gaussian-A}. The
external closing lines
``\parbox{.2cm}{\includegraphics[width=.2cm]{Gtrleft}}'' and
``\parbox{.2cm}{\includegraphics[width=.2cm]{Gtrright}}'' stand for
Kronecker deltas on the external color indices and implement the color
trace. The path ordered exponentials furnish two $\y$ integrals per
factor of $G$ and since the $\y$ structure of the weight is {\em
  local} only one of them remains.  It turns out that the
contributions to all orders can be rearranged into a locally
subtracted version of the two point function called ${\cal G}$ in the
above. Explicitly displaying the $\y$ integrals, the relevant
precursor to ${\cal G}$, which still contains nontrivial color
structure is
\begin{equation}
  \label{eq:tilde-G-diag}
\int\limits^\y \!\!d\y'
%\hspace{.2cm}\parbox{.5cm}{\includegraphics[width=.5cm]{Gtrleft}}
\hspace{.2cm}
\parbox{1.3cm}{\centering $\y'$\\ \includegraphics[width=1.3cm]{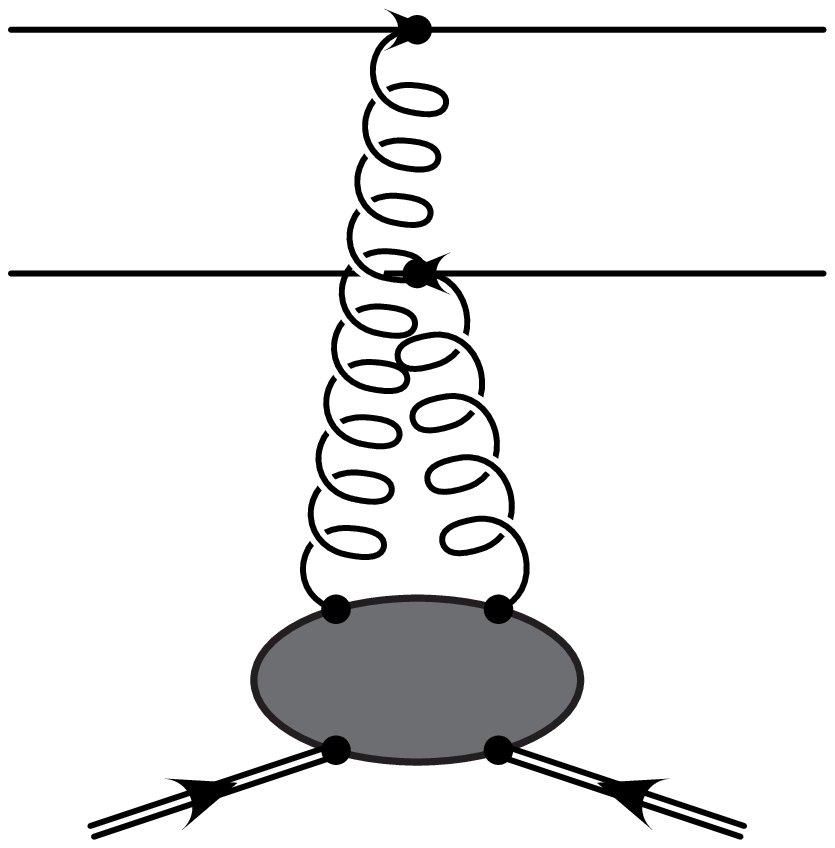}}
%\hspace{.2cm}\parbox{.5cm}{\includegraphics[width=.5cm]{Gtrright}}
\hspace{.2cm}
:=
  \int\limits^\y \!\!d\y' 
%\hspace{.2cm}\parbox{.5cm}{\includegraphics[width=.5cm]{Gtrleft}}
\hspace{.2cm}\left(
\hspace{.2cm}\parbox{1.3cm}{\centering $\y'$\\ \includegraphics[width=1.3cm]{Gboth}}
+%\frac{1}{2}
\left(\hspace{.2cm}
\parbox{1.3cm}{\centering $\y'$\\ \includegraphics[width=1.3cm]{Gupper}}
+\parbox{1.3cm}{\centering $\y'$\\ \includegraphics[width=1.3cm]{Glower}}
\hspace{.2cm}\right)
\right)
%\big( G_{\y',\bm{x y}}
%    -\frac{G_{\y',\bm{x x}}+G_{\y',\bm{y y}}}{2}\big)
%\hspace{.2cm}\parbox{.5cm}{\includegraphics[width=.5cm]{Gtrright}}
%\hspace{.2cm}
\ .
\end{equation}
On the r.h.s. the color structure of the first digram is $t^a\otimes
t^a$ while that of the other two is simply $C_R \cdot 1\otimes 1 $
with $C_R$ the first Casimir of the representation in question. To
arrive at a generic expression for higher orders in $G$, locality in
$\y$ is essential. It implies that the n-th order contribution is a
simple iteration of this structure:
\begin{equation}
  \label{eq:n-th-order-S}
  \int\limits^\y d\y^n \int\limits^{\y^n} d\y^{n-1} \ldots \int\limits^{\y^1} d\y^{0} \ \ 
  \parbox{1.3cm}{\centering $\y_n$\\ \includegraphics[width=1.3cm]{CalG-neu}}
  \cdots 
  \parbox{1.3cm}{\centering $\y_0$\\ \includegraphics[width=1.3cm]{CalG-neu}}
\end{equation}
Note that the color exchange with the target is zero in each
individual factor.  Consequently an insertion
of~\eqref{eq:tilde-G-diag} will never mix {\em inequivalent}
multiplets in the multiplet decomposition of Wilson lines considered.
For the correlators in~\eqref{eq:simple-corr} this means that it is
sufficient to calculate the singlet channel contribution of an
individual insertion and iterate that to all orders.\footnote{A closer
  examination of the multiplet structures would reveal the following:
  For the $q\Bar q$ case an insertion of~\eqref{eq:tilde-G-diag} can
  be written as a diagonal $2\times 2$ matrix in a space consisting of
  a singlet and an octet.  For the $g g$ case the matrix operates on
  $8\otimes 8 = 1 \oplus 8\oplus 8\oplus 10\oplus \Bar{10}\oplus 27 $
  (in SU(3)) and will indeed mix the two equivalent adjoint
  representations therein.}  Since there is only one singlet, the
multiplet of interest completely decouples from all others.  Writing
the singlet projection operator generically as $\singletproj$ for a
2-Wilson-line example in a representation $R$ with dimension $d(R)$,
one finds that the color structures of all three terms
in~\eqref{eq:n-th-order-S} in the singlet projection are reduced to
$C_R\ \singletproj$
% $\frac{\tr( t^a_R
%  t^a_R)}{d(R)}\singletproj$ with $C_R=\frac{\tr( t^a_R t^a_R)}{d(R)}$
so that one identifies
\begin{equation}
  \label{eq:calG-singletproj}
  %\int^\y d\y' 
  \singletproj\hspace{.5cm}
  \parbox{1.3cm}{\centering $\y'$\\ \includegraphics[width=1.3cm]{CalG-neu}}
  \hspace{.5cm} \singletproj = %{\cal G}_{\y; {\bm x y}}
  -\left( G_{\y',\bm{x y}}
    -\frac{G_{\y',\bm{x x}}+G_{\y',\bm{y y}}}{2}\right)
  \ C_R  \hspace{.5cm}\singletproj 
\end{equation}
as the relevant building block for iteration in the singlet channel
of~\eqref{eq:n-th-order-S}.  Extending the integration range of all
integrals up to $\y$ then restores powers of ${\cal G}$ and furnishes
the $1/n!$ factors needed to arrive at the simple exponential form
shown in~\eqref{eq:simple-corr}. The external color trace acts on the
overall singlet projector and provides the $d(R)$ in front of the
exponential factor.

The fact that there is no color transferred to the target means that
in terms of color, an individual insertion is planar. Locality in $\y$
ensures that this remains the case for multiple insertions: all the
diagrams entering the above correlators are {\em planar} in their
color structure and in this sense correspond to a large $N_c$
limit~\cite{'tHooft:1974hx}.

Quite generically, one expects correlators of Wilson lines to show
some exponentiating properties in that their logarithm is in some
sense generically simpler than the object itself, just as
in~\eqref{eq:simple-corr} -- but in general without any reference to
Gaussian weights or the $N_c$ limit. Such examples are provided by
general calculations of Sudakov type form factors, Drell-Yan
production~\cite{Beneke:1995pq} or certain jet observables. The latter
case will be discussed in Sec.~\ref{sec:jimwlk-soft-gluon} and the
simplification there is in terms of logarithmically enhanced
contributions as stated in~\eqref{eq:jet-exponentiation}. In general
the naturalness of such a parametrization gives no clue as to whether
it should arise from simple averaging procedures as employed here.
Their use in this case is mainly motivated by the MV model and an
application to the evolution equations with its resulting
simplifications and interpretation to be discussed in
Sec.~\ref{sec:jimwlk-to-bk}. The comparison with the Sudakov example
should provide enough of a hint as to how to think about extending
such a picture at least in principle.

Returning to the planar diagrams encountered with MV type models one
can start to prepare for better phenomenological understanding. To
this end consider first the gluon distribution which was originally
calculated in~\cite{Jalilian-Marian:1997xn} in the McLerran
Venugopalan model in a somewhat more pedestrian manner. 

Since we deal with a gauge theory the definition of the gluon
distribution is not simply formulated as a naive number operator
construction. It involves a bilinear in field strength operators
instead of a simple bilinear of fields. In $A^+=0$ gauge, the
expression can be rendered as
\begin{equation}
\label{eq:GDF}
\xbj G(\xbj,Q^2)=\frac{1}{\pi}\int {d^2k \over (2 \pi)^2}\,\Theta(Q^2-
{\bm{k}}^2)\bigl\langle F^{i+}_a(\xbj,\bm{k})
F^{i+}_a(\xbj,-\bm{k})\bigr\rangle
\ .
\end{equation}
This is written in a form that exhibits its relationship with the 
unintegrated gluon density
\begin{equation}
\label{eq:phidef}
\varphi_\y({\bm{k}})\,:=\,\frac{4\pi^3}{N_c^2-1}\,
\frac{1}{\pi R^2}\,\frac{d^3 N}{d\y d^2{\bm{k}}}\,=\,\frac{1}{\pi R^2}\,
\frac{
\langle F^{i+}_a(\xbj,\bm{k})F^{i+}_a(\xbj,-\bm{k}) \rangle}{N_c^2-1}
\ .
\end{equation}
In a general gauge one has to Fourier transform the expressions back
to coordinate space and insert a Wilson line in $x^-$ direction.  In
the same $A^+=0$ gauge, the Fourier transform of the $F F$ correlator
of the unintegrated Gluon density involves the operator $\tr( U
\partial_i U^\dagger)_{\bm{x}} ( U\partial_i U^\dagger)_{\bm{y}} $
(c.f.  Eq.\eqref{eq:aplusgauge}) which is perfectly calculable in this
class of MV like models. One may even consider
\begin{align}
  \label{eq:gluedist}
  \langle 
[U \partial_i
U^\dagger ]^a_{\bm{x}} [ U \partial_j U^\dagger]^b_{\bm{y}} 
 \rangle = &
  -\delta^{ab}
  \frac{(\partial_i^{\bm{x}}\partial_j^{\bm{y}} {\cal G}_{\y; {\bm{x y}}})}%
  {N_c\ {\cal G}_{\y; {\bm{x y}}}}
  \bigg\{
  1-e^{%\Big[
  -N_c {\cal G}_{\y; {\bm{x y}}}}%\Big]
  \bigg\}
\ .
\end{align}
It is clearly ${\cal G}$ and its gradients that determine the size of
the gluon distribution. Many phenomenological applications exist that
fall into this class of parametrization and it is worth to give a
first flavor of the issues discussed phenomenologically by listing
some of them.  The original MV expression
of~\cite{Jalilian-Marian:1997xn} emerges
%\begin{eqnarray}
%     G_{ij}^{ab}(y,{\bm{x}};y^\prime,{\bm{x}}^\prime) & = &  
% -\delta^{ab}\left( \nabla_i \nabla_j^\prime 
%\gamma({\bm{x}}-{\bm{x}}^\prime) \right)
% {1 \over { N_c\left[ 
%  \gamma ({\bm{x}}-{\bm{y}})-\gamma(0)\right]}} 
%\nonumber \\ && 
%  \left(1 - \exp\left\{g^4N_c \chi(y,Q^2) 
%[\gamma({\bm{x}}-{\bm{y}})-\gamma(0)]\right\} \right)
%\end{eqnarray}
by setting ${\cal G}_{\y; {\bm{x y}}} = g^4\chi(\y,Q^2)
[\gamma({\bm{x}}-{\bm{y}})-\gamma(0)]$, where $ \gamma ({\bm{x}}) =
\frac{1}{ {\bm{\partial}}^4} ({\bm{x}}) = {1 \over {8\pi}} {\bm{x}}^2
\ln ({\bm{x}}^2\Lambda^2_\mathrm{QCD}) + \gamma(0) $ and
$\chi(\y,Q^2):=\int^\y d\y' \mu^2(\y',Q^2)$ is the total color charge
sampled inside the target at a given transverse resolution $Q^2$. The
gluon distribution proper emerges in either case after traces over
color and Lorentz-indices are taken. The MV result reads (see
\cite{Jalilian-Marian:1997xn})
\begin{equation*}
\frac{4(N_c^2-1)}{N_c {\bm{x}}^2}
     \left[1-\left({\bm{x}}^2\Lambda_\mathrm{QCD}^2\right)^{
     \frac{g^4N_c}
      {8\pi}\chi(\y,Q^2){\bm{x}}^2}\right]
\label{eq:MV-gluedist-coord}
\end{equation*}
and leads to a unintegrated gluon distribution of the form
\begin{equation}
  \label{eq:MV-gluedist-mom}
  \varphi_\y^{\text{MV}}({\bm{k}}) =\frac{N^2_c-1}{ 4\pi^4 \alpha_s N_c}
  \int {d^2{\bm x} \over {\bm x}^2}
  \left(1-e^{-{\bm x}^2Q_s^2(\y)/4}\right)\,
  e^{i\, {\bm{k}} \cdot {\bm x}}\, .
\end{equation}
In this expression $\chi(\y,Q^2)$, which parametrizes the color
charge in longitudinal direction, has been reinterpreted in terms of a
Golec-Biernat+W{\"u}sthoff type saturation scale $Q_s$.  Equally often
one finds $Q_s$ in this expression identified with the gluon density
of individual nucleons in the nuclear target~\cite{Kharzeev:2000ph,
  Kharzeev:2001yq, Kharzeev:2001gp, Baier:2003hr}:
\begin{equation} Q_s^2(\y,{\bm
  x}^2,b) ={4\pi^2\alpha_s N_c\over N_c^2-1}\, \xbj G(\xbj,1/{\bm x}^2)
{\rho_{\text{part}}(b)\over 2}\, .
  \label{eq:Qs-Gnuc}
\end{equation}
This allows for additional impact parameter ($\bm{b}$-) dependence and
introduces $G(\xbj, {\bm{k}}^2)$, the {\it nucleon} gluon distribution
(with the identification ${\bm{k}}^2 = 1/{\bm x}^2$) and a nuclear
profile function $\rho$ that counts participants at fixed impact
parameter.\footnote{See~\cite{Kowalski:2003hm, Bondarenko:2003ym,
    Gotsman:2004ra} for studies of impact parameter dependence which
  remain largely outside the scope of this review.}  Often the gluon
distribution in the nucleon is taken to be of the simple perturbative
form compatible with the logarithmic structure of $\gamma$ in the MV
model. Another model for ${\cal G}$ to be listed in this context is
the celebrated Golec-Biernat W\"usthoff model used to fit the HERA
data.  Accordingly this was used not to describe gluon densities but
dipole correlators via
\begin{equation}
  \label{eq:GBW-dip}
 \langle\frac{ \tr(1-U_{\bm{x}}U_{\bm{y}}^\dagger)}{N_c}\rangle_\y 
 = 1-e^{-{\bm x}^2Q_s^2(\y)/4}
\end{equation}
in the same spirit as~\eqref{eq:MV-gluedist-mom}, but with a
parametrization of the $\y$ dependence added to cope with $\xbj$
dependence in the data.  All these interpretations share the same
$A^{1/3}$ scaling present in the MV model. Here is a collection of
generic features:
\begin{itemize}
\item Dipole correlators of the form
  $\tr(1-U_{\bm{x}}U_{\bm{y}}^\dagger)/N_c$ will naturally be bounded
  by $1$ at large distances and will show the color transparency +
  saturation asymptotics as soon as ${\cal G}$ grows with distance.
\item Any growth of ${\cal G}$ with $\y$ will result in the
  qualitative behavior sketched in Fig.~\ref{fig:generic-evol}.
\item As long as the assumption of uncorrelated scattering centers
  holds, one would expect that ${\cal G}$ scales like $A^{1/3}$. This
  clearly will enhance the importance of the nonlinearities for large
  nuclei. Going to large nuclei in this sense has a similar effect as
  going to small $\xbj$. Quantum corrections that drive the change
  towards small $\xbj$, however, will induce correlations and this will
  not occur when going to larger $A$. The consequent change in the
  naive $A^{1/3}$ scaling will be discussed in
  Sec.~\ref{sec:evol-a-depend}.
\end{itemize}
The first two of these are consistency requirements, the last
property, scaling of the exponent has been used to reinterpret the
expression in many different ways as already indicated. This aspect
connects to the realm of model building.

While any reference to small $\xbj$ and $A$ scaling are specific to
the problem at hand, the exponentiation of leading order contributions
in expressions~\eqref{eq:simple-corr} and~\eqref{eq:gluedist} is not.
Already in QED, in the calculation of soft photon bremsstrahlung,
occurs exponentiation of this type.  The interpretation of the object
in the exponent there is that of a probability to not emit soft
photons below a certain experimental resolution. Other calculations
involving soft gauge bosons will acquire a similar form, in particular
in situations in which the second nonlinearity mentioned at the
beginning of the previous section, the creation of additional ``hard''
particles, is excluded.  An example are contributions to jet
observables from soft gluons going into the ``empty'' region outside
the hard jets (see Sec.~\ref{sec:jimwlk-soft-gluon}).

The first decisive step beyond this stage is the derivation of an
evolution equation that determines the $\xbj$ or $\y$ dependence of
${\cal G}$ and its generalizations. This is the topic of the next
section.

% spellmark 2

\section{JIMWLK  evolution and the Balitsky hierarchy}
\label{sec:jimwlk-evol-balitsky}

After having identified the relevant variables for a description of
QCD scattering at high energies --which constitutes the first
resummation, the eikonalization of gluons into Wilson lines $U$-- one
now needs to calculate how general correlators of such fields change
with $\xbj$. This leads to the second resummation necessary, this time
in the guise of an RG equation.  Sec.~\ref{sec:syst-deriv} gives a
brief overview on how this calculation can be most efficiently
organized and can be skipped by those only interested in a discussion
of the results, which will be given in the subsections following it.

\subsection{\it A systematic derivation }
\label{sec:syst-deriv}

From the discussion above it is clear that one needs to understand
perturbative, logarithmically enhanced corrections to correlators of
Wilson lines of the general form
%\begin{equation}
%  \label{eq:typcorr}
$  \big\langle U^{(\dagger)}_{\bm{x}_1}\otimes\ldots\otimes
  U^{(\dagger)}_{\bm{x}_n}\big\rangle_b $
%\end{equation}
with $U^{(\dagger)}$ in the fundamental representation.\footnote{This
  is completely general, as one may write any higher representation as
  a (local) product of $U^{(\dagger)}$. For example adjoint links
  emerge as a combination of two fundamental ones by virtue of $\Tilde
  U^{a b}=2\tr[t^a U t^b U^\dagger]$.}  
%The $\langle\ldots\rangle_b$
%average is taken with the QCD action in the presence of a complicated
%target wave function and thus the corrections alluded to are the best
%one can hope to calculate.

One way to do this efficiently is to introduce a generating
functional for such correlators and perform the calculation directly
for this general case. I therefore introduce
\begin{equation}
  \label{eq:barcalZdef}
  \Bar{\cal Z}[J^\dagger,J] := 
  \langle e^{{\cal S}_{\mathrm{ext}}^{q\bar q}[b,J^\dagger,J]} 
  \rangle_b
\end{equation}
where
\begin{equation}
  \label{eq:minuscurrent}
  {\cal S}_{\mathrm{ext}}^{q\bar q}[A,J^\dagger,J] =  
  \int\!\! d^2\bm{x} \Big\{
  \tr \big(
      (J^{\dagger}_{\bm{x}})^t U_{\bm{x}}[A^+] \big)
   +
   \tr \big(
      J_{\bm{x}}^t
     U^\dagger_{\bm{x}}[A^+] 
 \big)\Big\}
\end{equation}
is an external source term,
%\footnote{Note that I purposely write $\tr
%  A^t B$ to unclutter notation in what follows, {\em not} $\tr A B$,
%  having in mind that $\frac{\delta}{\delta A_{i j}}\tr A^t
%  B=\frac{\delta}{\delta A_{i j}} A_{k l} B_{k l} = B_{i j}$ while the
%  other variant would give $B_{j i}$. Also, $x^+$ has been suppressed,
%  assuming all $U$'s to be located at $x^+=0$ as discussed above.}
correlators %(Eq.~(\ref{eq:typcorr})) 
are extracted via $J^{(\dagger)}$
derivatives in the usual way
%\begin{subequations}
  \begin{align}
\label{eq:makeUs}
     \frac{\delta}{\delta J}
  e^{{\cal S}_{\mathrm{ext}}^{q\bar q}[U,U^\dagger,J^\dagger,J]} = %& 
  \ 
  U^\dagger e^{{\cal S}_{\mathrm{ext}}^{q\bar q}[U,U^\dagger,J^\dagger,J]}
\ ,
&
\hspace{1cm} %\\
  \frac{\delta}{\delta J^\dagger}
  e^{{\cal S}_{\mathrm{ext}}^{q\bar q}[U,U^\dagger,J^\dagger,J]} = %& 
  \
  U e^{{\cal S}_{\mathrm{ext}}^{q\bar q}[U,U^\dagger,J^\dagger,J]}
\\ \intertext{or explicitly}
 \label{eq:typcorr}
\frac{\delta}{\delta J_1}\cdots 
\frac{\delta}{\delta J_n}
\frac{\delta}{\delta J^\dagger_1}\cdots 
\frac{\delta}{\delta J^\dagger_m}
\Bar{\cal Z}[J^\dagger,J] =& \ \big\langle 
U_1^\dagger \otimes \ldots \otimes U_n^\dagger
\otimes 
U_1 \otimes \ldots \otimes U_m \big\rangle_b
\ .
  \end{align}
%\end{subequations}
  
  So far the physics of small $\xbj$ is encoded in the type of
  correlation functions considered --exclusively correlators of link
  operators $U^{(\dagger)}$-- all the rest is mathematical convenience
  to help summarize the result.  To extract the logarithmic
  corrections, one now expands the gluon field $A$ around $b$ (whose
  correlators are assumed to be known) and keeps fluctuations to the
  order $\alpha_s\ln(1/\xbj)$. That is to say that one will expand
  around $b$ to one loop accuracy and select the terms carrying a
  $\ln(1/\xbj)$ factor.  This way one will be able to infer the change
  of correlation functions as one lowers $\xbj$. This is the second
  statement about physics or rather what one can learn about it
  through such an approach.

Now turn back to the actual calculation. At one loop one needs at most
second order in fluctuations:\footnote{ In writing
  Eq.~(\ref{eq:Bsecorder}) one has anticipated that $\big\langle
  \delta A^+_u\big\rangle_{\delta A}[b]=0$ as in the free case.  This is a
  consequence of the structure of the propagator in this background
  field and has been used repeatedly~\cite{Balitsky:1996ub,Kovner:2000pt}.}
\begin{align}
\label{eq:Bsecorder}
\begin{split}
  \langle &  e^{{\cal S}_{\mathrm{ext}}^{q\bar q}[b+\delta
    A,J^\dagger,J]}  \rangle_{b,\delta A}
=\\  & =   \langle \left(1+\delta
    A_x \frac{\delta}{\delta b_x} + \frac 1 2 \delta A_x
    \frac{\delta}{\delta b_x} \delta A_y \frac{\delta}{\delta b_y} +
    {\cal O}(\delta A^3)\right) e^{{\cal S}_{\mathrm{ext}}^{q\bar
      q}[b,J^\dagger,J]} \rangle_{b,\delta A} 
\\  & =   \langle e^{{\cal
      S}_{\mathrm{ext}}^{q\bar q}[b,J^\dagger,J]} \rangle_b +\frac 1
  2\langle \delta A_x \frac{\delta}{\delta b_x} \delta A_y
  \frac{\delta}{\delta b_y} e^{{\cal S}_{\mathrm{ext}}^{q\bar
      q}[b,J^\dagger,J]} \rangle_{b,\delta A} +\ldots 
\\  & =  
   \langle e^{{\cal S}_{\mathrm{ext}}^{q\bar q}[b,J^\dagger,J]} 
   \rangle_b 
\\ & \ \
  +\frac 1 2\langle\langle \delta A_x \delta A_y \rangle_{\delta A}[b]
%\\ & \ \ \times
  \left(2\left(\frac{\delta}{\delta b_x}{\cal S}_{\mathrm{ext}}^{q\bar
      q}[b,J^\dagger,J]\right)\left( \frac{\delta}{\delta b_y} {\cal
      S}_{\mathrm{ext}}^{q\bar q}[b,J^\dagger,J]\right) +\frac{\delta}{\delta
      b_x} \frac{\delta}{\delta b_y} {\cal S}_{\mathrm{ext}}^{q\bar
      q}[b,J^\dagger,J]\right) 
\\ & \ \ \ \ \times e^{{\cal S}_{\mathrm{ext}}^{q\bar
      q}[b,J^\dagger,J]} \rangle_b +\ldots
\ \ .
\end{split}
\end{align}
The second term is the calculable perturbative correction to
$\bar{\cal Z}[J^\dagger,J]$, the generating functional for all
generalized distribution functions of the target, an object which can
not be calculated as such with present tools.

To understand the individual terms in Eq.~(\ref{eq:Bsecorder}) in
detail requires the calculation of the one loop corrections to path
ordered exponentials $U[b]^{(\dagger)}$ in the presence of a
background field $b$ of the form Eq.~(\ref{eq:aplus}).  Before diving
into this it may be helpful to specialize once more to the DIS example
from above to illustrate the physics content of the terms in
Eq.~(\ref{eq:Bsecorder}). Here one needs to look at\footnote{After
  integration over impact parameter (simply $\bm{x} +
  \bm{y}$, if one puts the target at the origin in transverse
  space) this is the dipole cross section that features prominently in
  calculations of $\gamma^* p$ or $\gamma^* {\cal A}$ cross sections
  at small $\xbj$ \cite{Hebecker:1999ej, Hebecker:1998kv,
    Kovchegov:1999yj}.}
\begin{equation}
  \label{eq:dipolecross}
  \begin{split}
      \tr\Big(\bm{1}-\frac{\delta}{\delta
        J_{\bm{x}}^\dagger}
      \frac{\delta}{\delta J_{\bm{y}}}\Big)
\big\langle & e^{{\cal
        S}_{\mathrm{ext}}^{q\bar q}[b+\delta A,J^\dagger,J]}
    \big\rangle_{b,\delta A}\Big\vert_{J\equiv 0} =
  \big\langle\tr(\bm{1}-U_{\bm{x}}[b]\,
  U^\dagger_{\bm{y}}[b])\big\rangle_b
  %\nonumber 
%  \\ & \hspace{4cm} 
+\mbox{quantum corrections}\ .
  \end{split}
\end{equation}
The quantum corrections are induced by fluctuations $\delta A$
incorporating physics below the $\xbj$ values for which the original
$b$ were a good approximation. One would like to use these quantum
corrections to redefine the $b$ average to adapt to the new, lower
values of $\xbj$. If this is possible the averaging procedure
$\langle\ldots\rangle_b$ becomes $\xbj$ dependent. Of course one needs
to perform this step in general for the whole generating functional,
not only the specific correlator \eqref{eq:dipolecross}. This means
one has to carry out this step for the second term in
Eq.~(\ref{eq:Bsecorder}). This sets the task: {\em If one is able to
  calculate these quantum corrections before taking the $b$ average,
  that is, for all relevant $b$, one can deduce how the
  $U^{(\dagger)}$ correlation functions or the weight $Z[U,U^\dagger]$
  that defines $\langle\ldots\rangle_b$ evolves with $\xbj$.}

Clearly there are two generic types of corrections corresponding to
the two terms in
\begin{align}
\label{eq:chisigmafirst}
\frac 1 2\langle \delta A_x \delta A_y \rangle_{\delta A}[b]
&
\left(2\left(\frac{\delta}{\delta b_x}{\cal S}_{\mathrm{ext}}^{q\bar
    q}[b,J^\dagger,J]\right)\left( \frac{\delta}{\delta b_y} {\cal
    S}_{\mathrm{ext}}^{q\bar q}[b,J^\dagger,J]\right) +\frac{\delta}{\delta
    b_x} \frac{\delta}{\delta b_y} {\cal S}_{\mathrm{ext}}^{q\bar
    q}[b,J^\dagger,J]\right)
   % \nonumber \\ & \times 
e^{{\cal S}_{\mathrm{ext}}^{q\bar
    q}[b,J^\dagger,J]}
\ .
\end{align}
The first term represents contributions where the gluon propagator
connects two quarks ($U$s) or antiquarks ($U^\dagger$s) as well as a
quark to an antiquark. In addition there are pure self energy
corrections dressing one quark or antiquark line instead of connecting
two of them, represented by the second term.

Particular correlators are again selected taking any number of $J$
derivatives and setting $J$ to zero. One point functions get
contributions from the second term only, while both terms contribute
to anything with more than two $J^{(\dagger)}$ derivatives. Clearly it
is sufficient to calculate
\begin{align}
  \label{eq:chifirst}
  \left.
    \frac{1}{2}
    \frac{\delta}{\delta J^{(\dagger)}_x}\frac{\delta}{\delta
      J^{(\dagger)}_y}\right\vert_{J^{(\dagger)}\equiv 0}
  %\circ 
  \frac 1 2\langle \delta A_u
  \delta A_v \rangle_{\delta A}[b]\ 2\ \left(\frac{\delta}{\delta
      b_u}{\cal S}_{\mathrm{ext}}^{q\bar q}[b,J^\dagger,J]\right)
  \left(  \frac{\delta}{\delta b_v} {\cal S}_{\mathrm{ext}}^{q\bar
      q}[b,J^\dagger,J] \right)
\end{align}
and
\begin{align}
  \label{eq:sigmafirst}
  \left.\frac{\delta}{\delta J^{(\dagger)}_x}
  \right\vert_{J^{(\dagger)}\equiv 0}
  %\circ 
  \frac 1 2 \langle \delta A_u \delta A_v \rangle_{\delta A}[b]
  \left(\frac{\delta}{\delta b_u} \frac{\delta}{\delta b_v} {\cal
      S}_{\mathrm{ext}}^{q\bar q}[b,J^\dagger,J]\right)
\end{align}
to completely reconstruct Eq.~(\ref{eq:chisigmafirst}) as these terms
are precisely second respectively first order in $J^{(\dagger)}$.
Eq.~(\ref{eq:chisigmafirst}) then defines the change in all other
correlators and one has reached the goal of finding the fluctuation
induced corrections.

The task is clear and the terms appearing in the actual calculation
are best visualized diagrammatically. For the gluon exchange diagrams
corresponding to Eq.~(\ref{eq:chifirst}) one defines
\begin{align}
  \label{eq:chidiags}
  \begin{split}
    %\alpha_s
    \ln(1/\xbj)\cdot \Bar\chi^{q q}_{\bm{x}\bm{y}} 
    :=
  \begin{minipage}[m]{2.6cm} 
    \includegraphics[height=2.6cm]{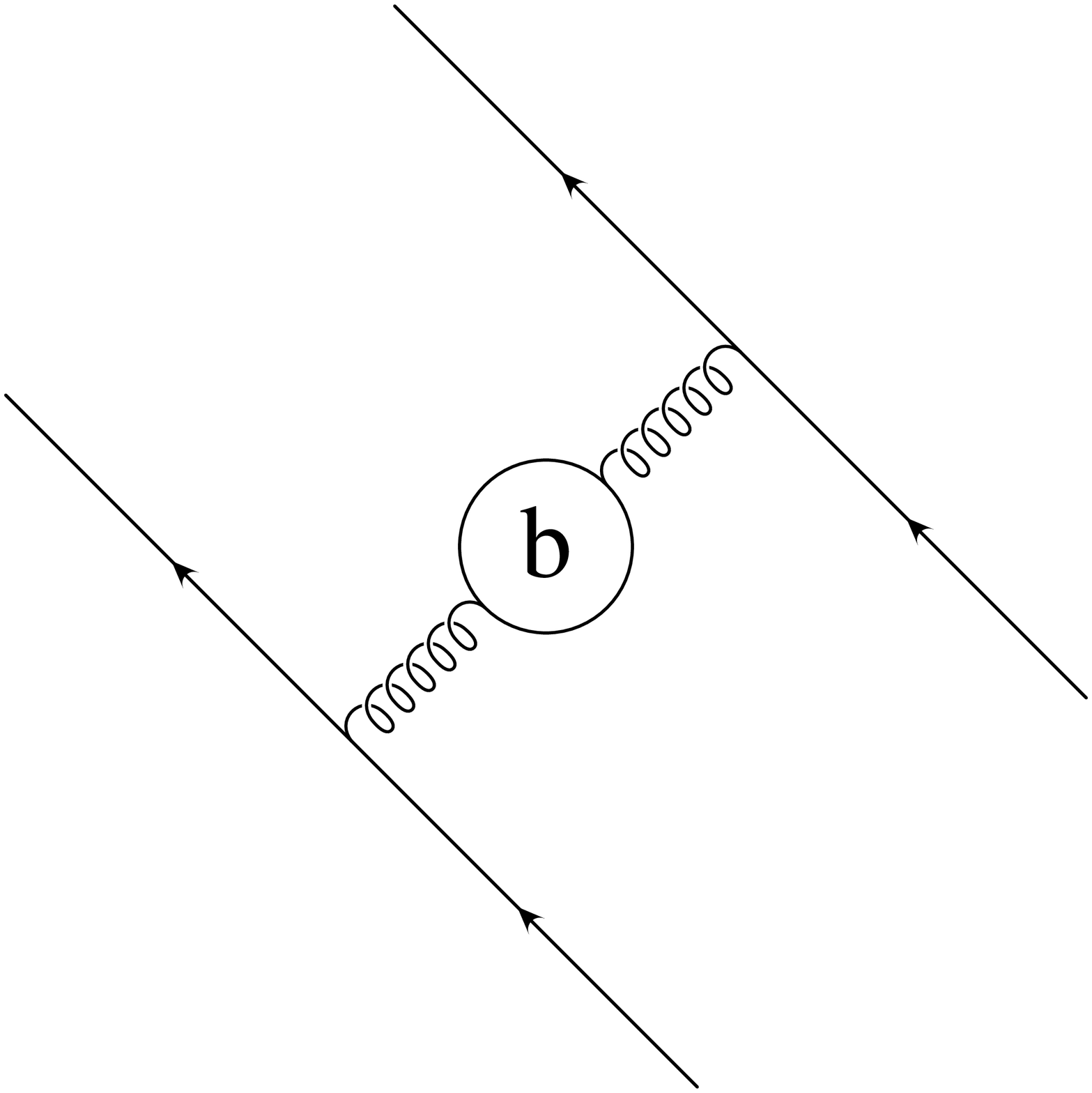}
  \end{minipage}
  ; & \hspace{.6cm}% \alpha_s
  \ln(1/\xbj)\cdot \Bar\chi^{q\Bar
    q}_{\bm{x}\bm{y}} 
  :=
  \begin{minipage}[m]{2.6cm} 
    \includegraphics[height=2.6cm]{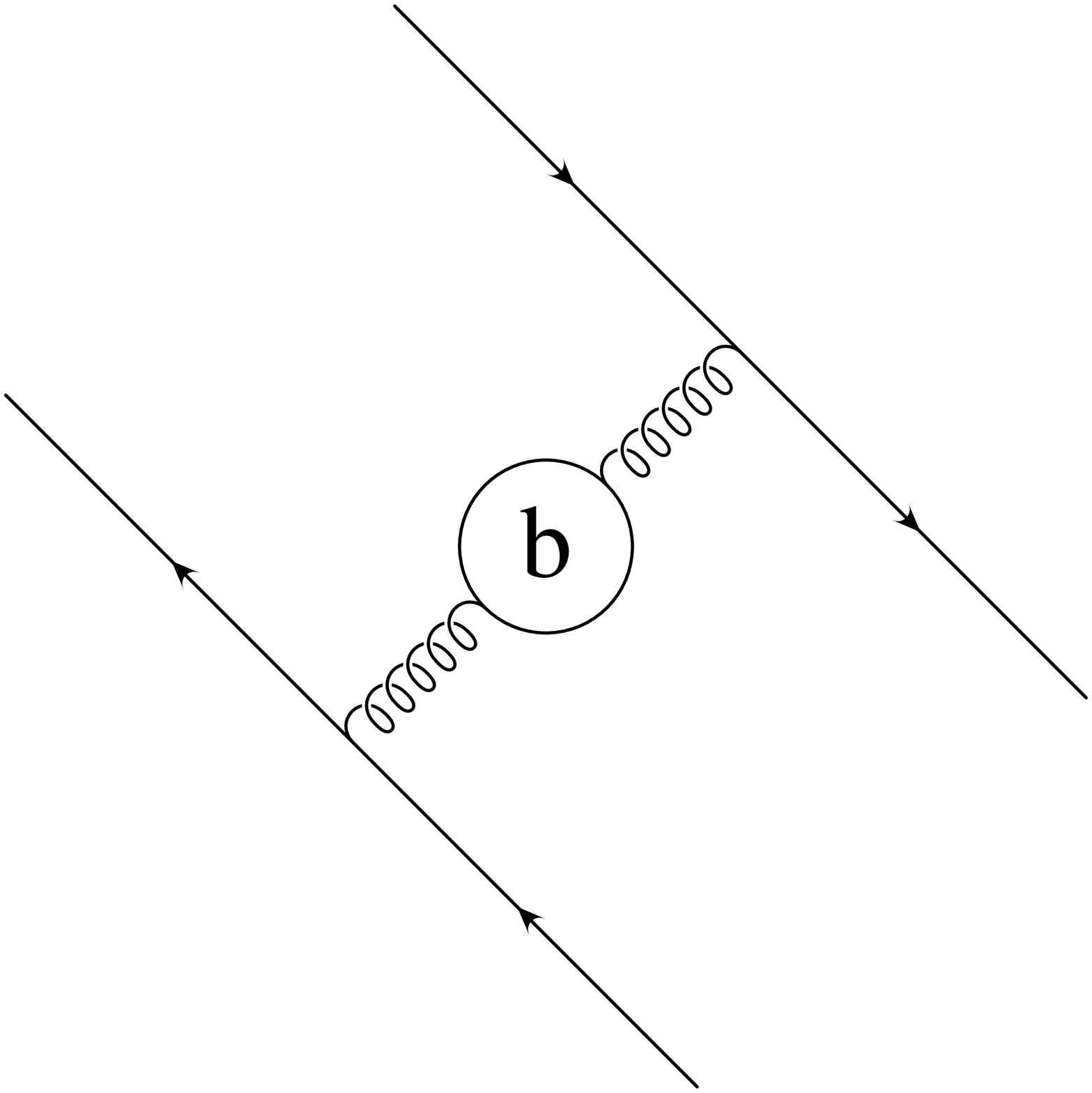}
  \end{minipage}
%  \\
%  %\alpha_s
%  \ln(1/\xbj)\cdot 
%  \Bar\chi^{\Bar q q}_{\bm{x}\bm{y}} 
%  :=
%  \begin{minipage}[m]{3cm} 
%    %\epsfysize=3cm \epsfbox{chiqbq}
%    \includegraphics[height=3cm]{chiqbq}
%  \end{minipage};
%  & \hspace{.6cm} %\alpha_s
%  \ln(1/\xbj)\cdot \Bar\chi^{\Bar q\Bar
%    q}_{\bm{x}\bm{y}} 
%  :=
%  \begin{minipage}[m]{3cm} 
%    %\epsfysize=3cm \epsfbox{chiqbqb}
%    \includegraphics[height=3cm]{chiqbqb}
%  \end{minipage}
%  \ .
  \end{split}
\end{align}
and analogously for $\Bar\chi^{\Bar q q}_{\bm{x}\bm{y}}$ and $\Bar\chi^{\Bar q\Bar q}_{\bm{x}\bm{y}}$. 
As already seen in~\cite{Balitsky:1996ub,Kovner:2000pt}, these split naturally
into $x^-$ ordered contributions when one combines the structure of
the vertices (the $b$ derivatives of $S_\mathrm{ext}$ in
Eqns.~(\ref{eq:chifirst}) and~(\ref{eq:sigmafirst})) and the gluon
propagator in the background field. Take $\Bar\chi^{q\Bar q}$ as an
example:\footnote{Representations for $\Bar\chi^{\Bar q
    q}_{\bm{x}\bm{y}}$
  $\Bar\chi^{qq}_{\bm{x}\bm{y}}$ and $\Bar\chi^{\Bar
    q\Bar q}_{\bm{x}\bm{y}}$ result from reversing the
  quark lines accordingly.}
\begin{align}
  \label{eq:chidiagstimeordered}
  \begin{minipage}[m]{2.5cm} 
    \includegraphics[height=2.5cm]{chiqqb}
  \end{minipage}
  & =
  \begin{minipage}[m]{2.5cm} 
    \includegraphics[height=2.5cm]{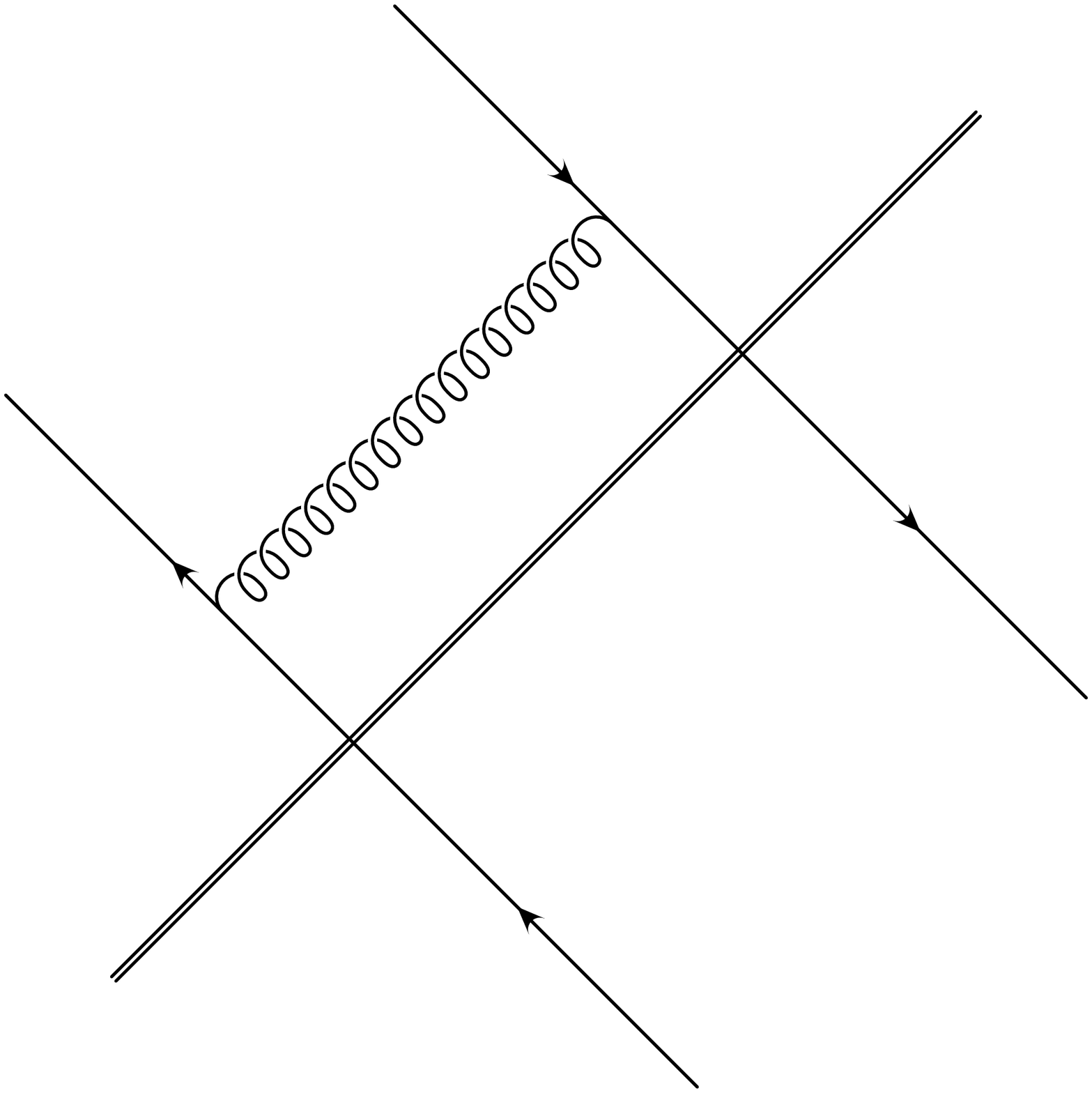}
  \end{minipage} 
  + \begin{minipage}[m]{2.5cm} 
    \includegraphics[height=2.5cm]{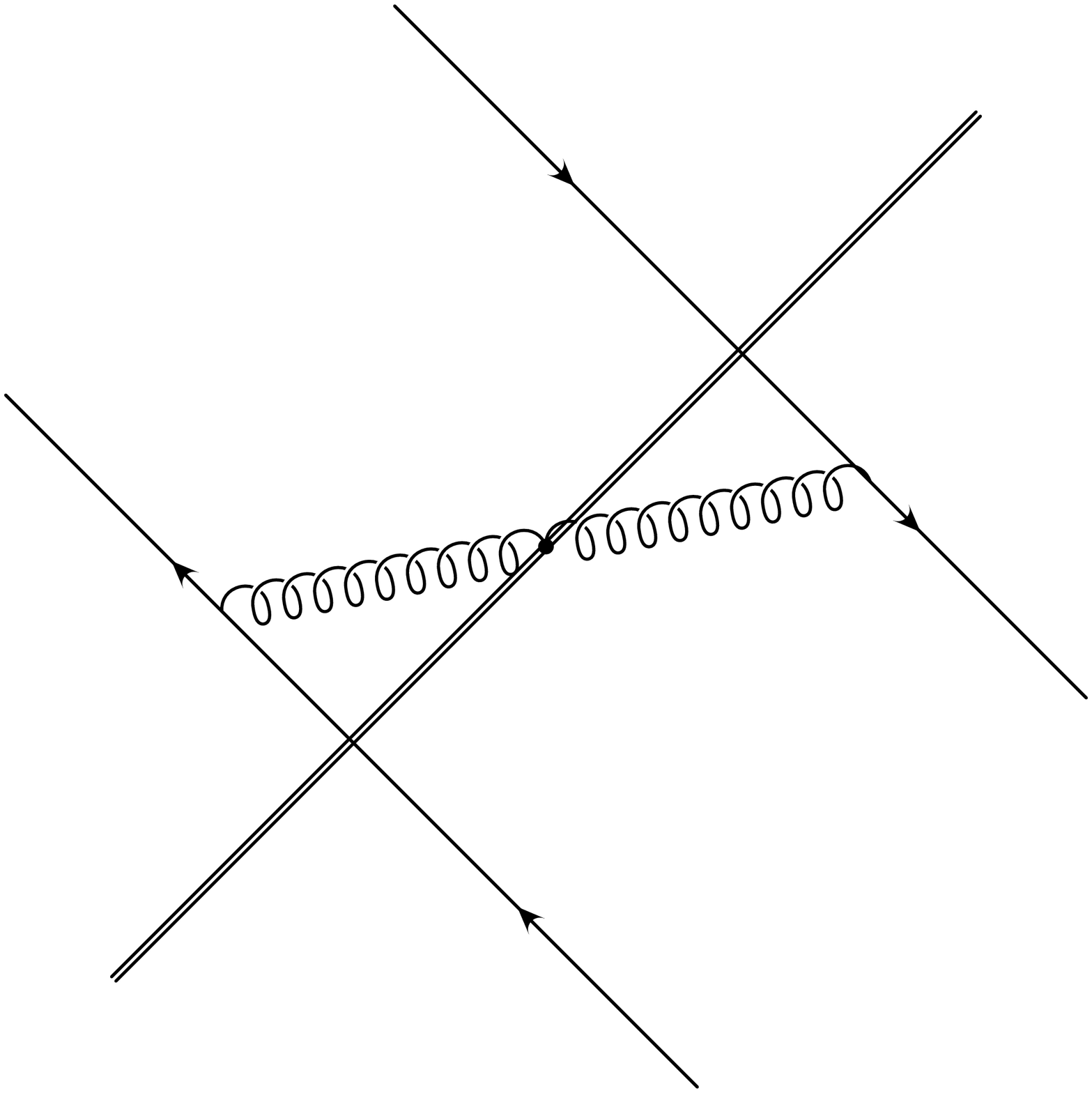}
  \end{minipage} 
  + \begin{minipage}[m]{2.5cm} 
  \includegraphics[height=2.5cm]{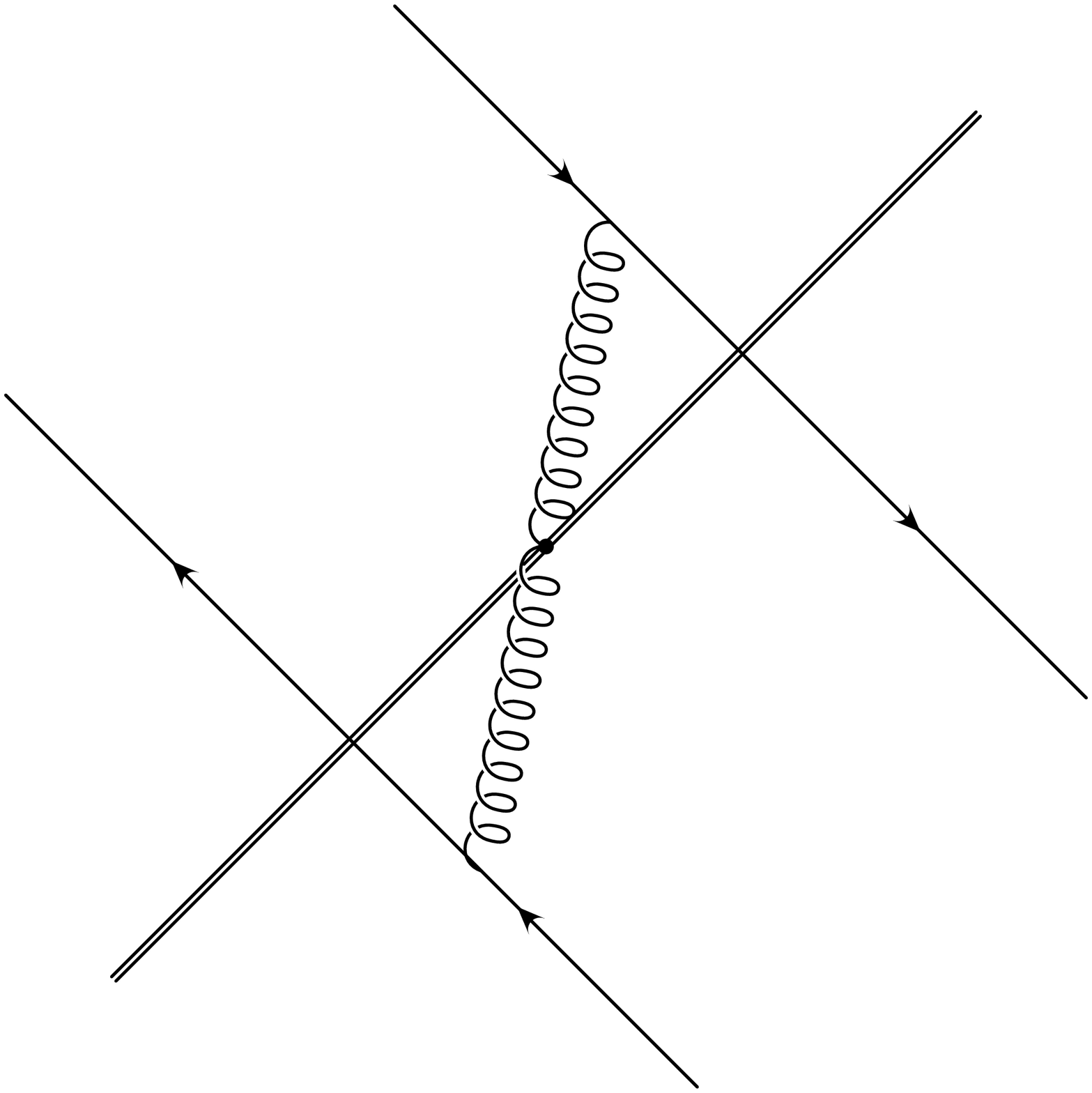}
  \end{minipage} 
  +\begin{minipage}[m]{2.5cm} 
    \includegraphics[height=2.5cm]{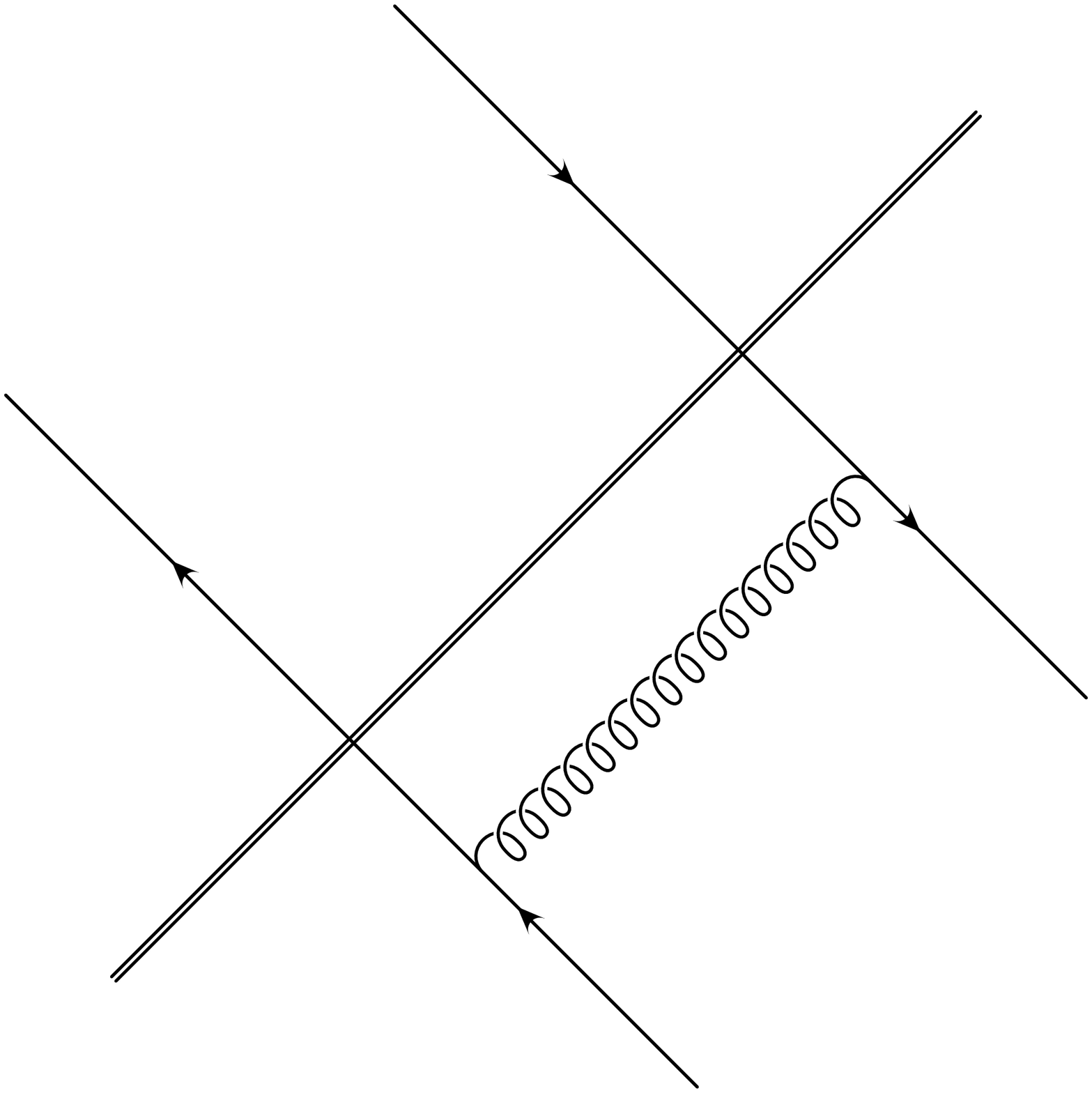}
  \end{minipage}
\ .    
\end{align}
The quark self energy correction in Eq.~(\ref{eq:sigmafirst}) and its
$x^-$ ordered parts are given by
%\begin{subequations}
% 
\begin{align}
  \label{eq:sigmadiags}
 %\alpha_a 
  \ln(1/\xbj)\ \Bar\sigma^q_{\bm{x}} &=
    \begin{minipage}[m]{2cm} 
      \includegraphics[height=2cm]{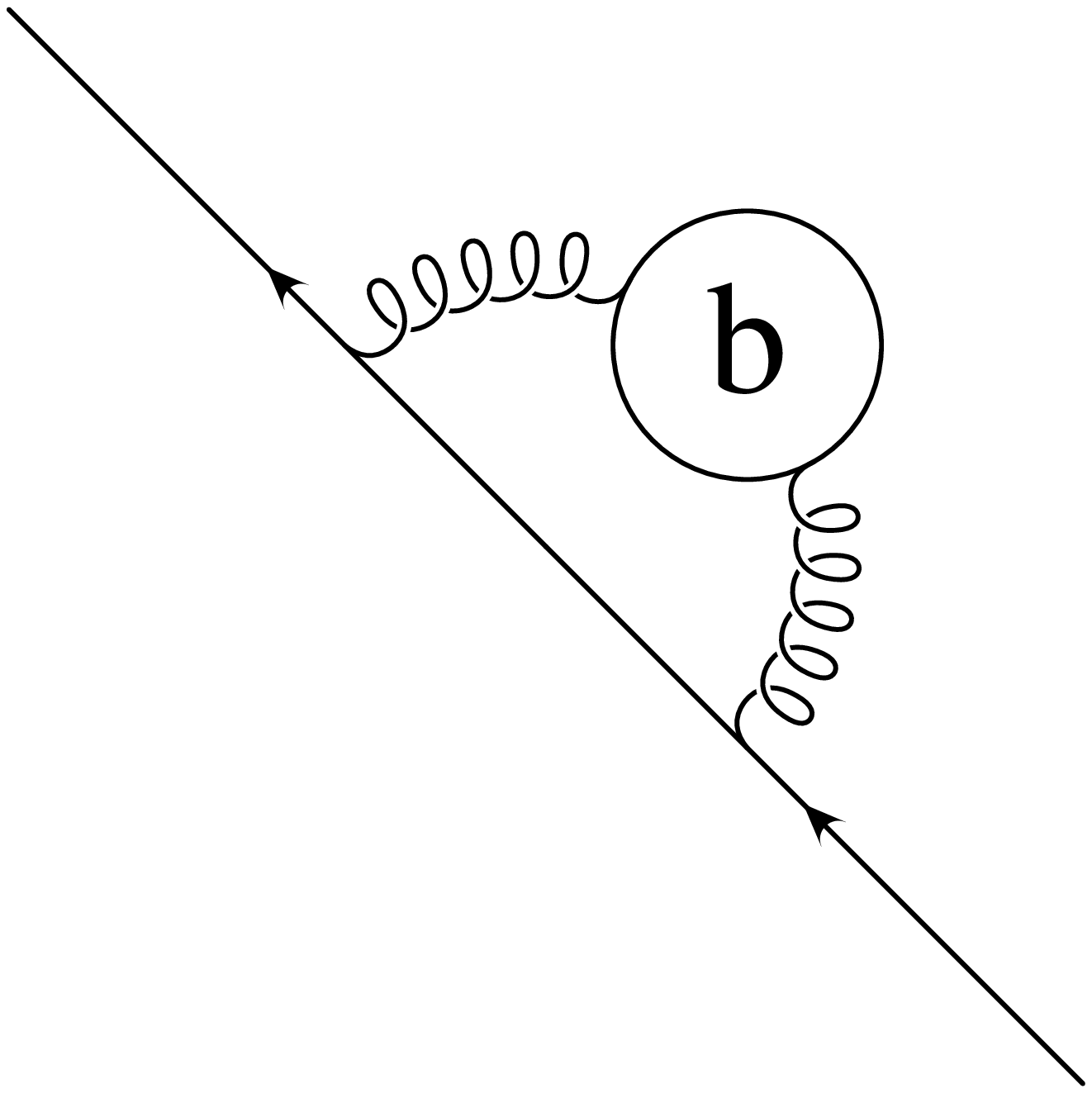}
    \end{minipage}  
    =
    \begin{minipage}[m]{2.1cm} 
      \includegraphics[height=2.1cm]{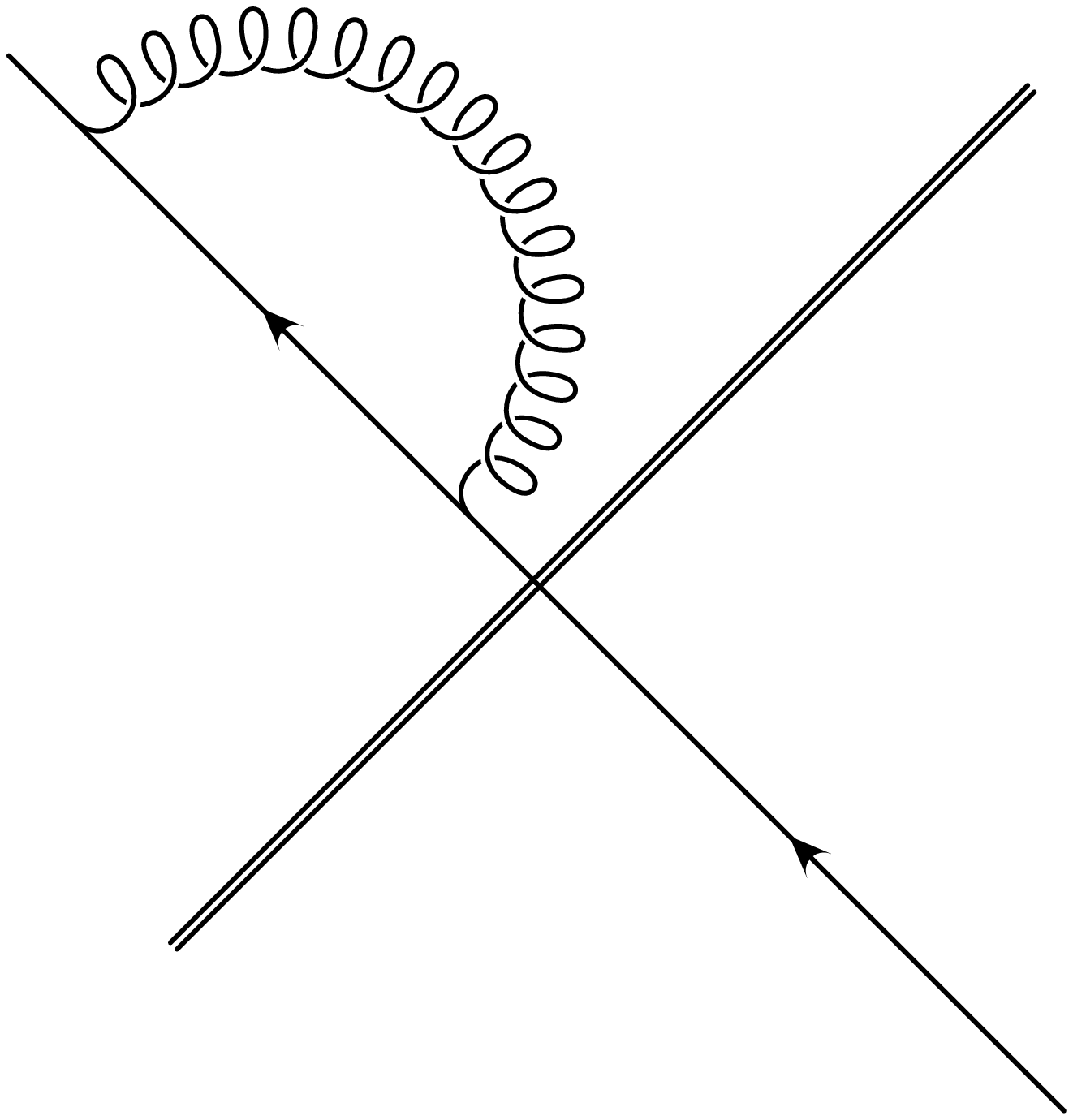}
    \end{minipage}  
    + \begin{minipage}[m]{2.1cm} 
      \includegraphics[height=2.1cm]{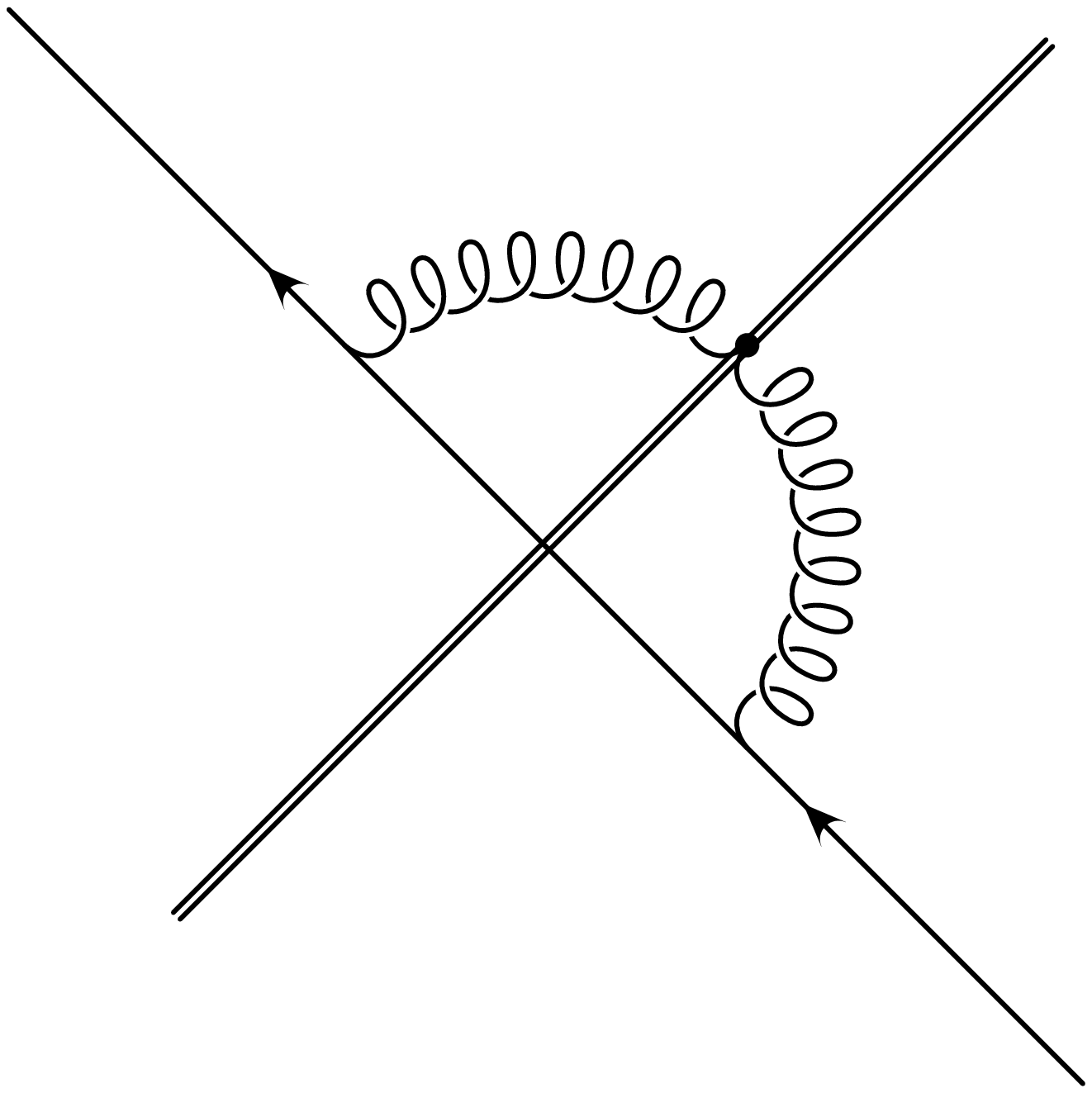}
    \end{minipage}  
    + \begin{minipage}[m]{2,1cm} 
      \includegraphics[height=2cm]{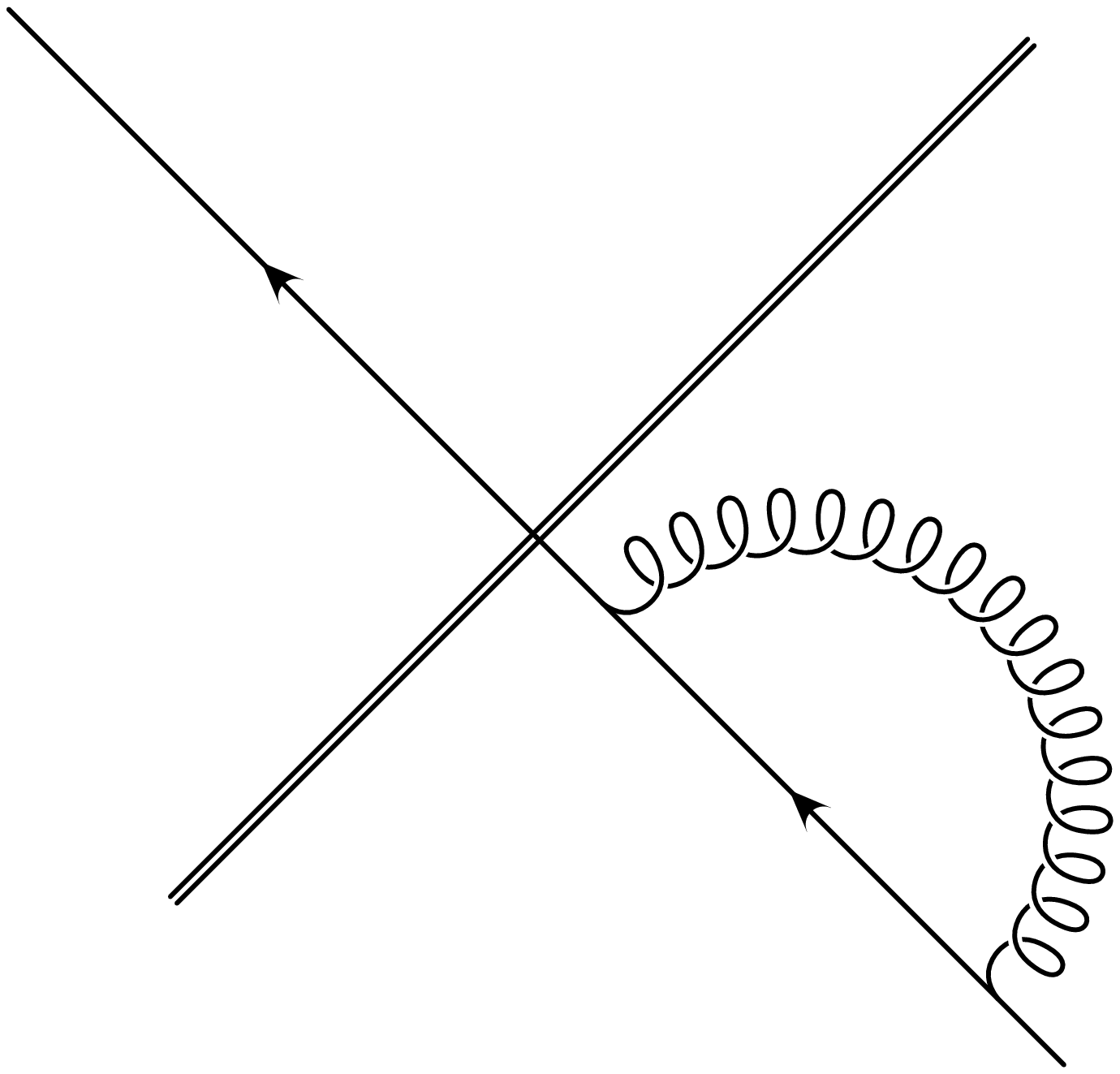}
    \end{minipage}  
%    \\
%  %  \alpha_s
%    \ln(1/\xbj)\ \Bar\sigma^{\Bar q}_{\bm{x}} &=
%    \begin{minipage}[m]{2cm} 
%      %\epsfysize=2cm \epsfbox{sigmaqb}
%      \includegraphics[height=2cm]{sigmaqb}
%    \end{minipage}
%    =
%    \begin{minipage}[m]{2.1cm} 
%      %\epsfysize=2cm \epsfbox{sigmaqb.tup}
%      \includegraphics[height=2cm]{sigmaqb-tup}
%    \end{minipage}
%    +
%    \begin{minipage}[m]{2.1cm} 
%      %\epsfysize=2cm \epsfbox{sigmaqb.tint}
%      \includegraphics[height=2cm]{sigmaqb-tint}
%    \end{minipage}
%    +
%    \begin{minipage}[m]{2.1cm} 
%      %\epsfysize=2cm \epsfbox{sigmaqb.tdown}
%      \includegraphics[height=2cm]{sigmaqb-tdown}
%    \end{minipage}
    \hspace{.5cm} ,
\end{align}
%\end{subequations}
the antiquark diagrams are completely analogous.  Again,
the various components of $\Bar\chi$ correspond to gluon exchange
diagrams while the $\Bar\sigma$s are self energy
corrections. %\footnote{These are referred to as ``real'' and
%  ``virtual'' respectively in \cite{Balitsky:1996ub}.}  
As for the time ordered ones, there are those with and without
interaction with the target (with the gluon line crossing the $x^+$
axis or not respectively as shown in
Eqns.~(\ref{eq:chidiagstimeordered}) and (\ref{eq:sigmadiags})). These
latter versions determine the $U^{(\dagger)}$ dependence of $\bar\chi$
and $\bar\sigma$ in detail.  They have been calculated several times
using different methods. The common feature of all of them is that
diagrams without target interaction leave the number of
$U^{(\dagger)}$s invariant, while those with target interaction insert
an additional adjoint $U^{(\dagger)}$ at the point of interaction.
This implies a nonlinearity in the evolution: Any correlator with a
fixed number of fundamental representation $U^{(\dagger)}$s will
couple to other correlators with two additional factors of fundamental
representation $U^{(\dagger)}$s in each step in $\xbj$.

The diagrams have been calculated in~\cite{Balitsky:1996ub}
or~\cite{Kovner:2000pt}, here I only quote the results for $\bar\chi$
and $\bar\sigma$.
% (see also App.~\ref{sec:balkernel}). 
For convenience, define the integral kernel
\begin{equation}
  \label{eq:confkernel}
  {\cal K}_{\bm{x z y}} :=
  \frac{(\bm{x}-\bm{z}).(\bm{z}-\bm{y})}{%
  (\bm{x}-\bm{z})^{2}(\bm{y}-\bm{z})^{2}} 
  = 
  {\cal K}_{\bm{y z x}}
\ .
\end{equation}
Explicit expressions for the components of
$\bar\sigma$ are given by
\begin{subequations}
  \label{eq:barsigmadef}
  \begin{align}
    %\alpha_s
    [\Bar\sigma^q_{\bm{x}}]_{i j} := &
-
    \frac{\alpha_s}{2\pi^2} \int\!\! d^2z\  
{\cal K}_{\bm{x z x}}
%K(\bm{x},\bm{z},\bm{x})
    \big([U_{\bm{z}}]_{i j}
\tr(U_{\bm{x}}U^\dagger_{\bm{z}})
-N_{c}[U_{\bm{x}}]_{i j}
    \big) 
    \\
    %\alpha_s
[\Bar\sigma^{\Bar q}_{\bm{x}}]_{i j} := &
-
    \frac{\alpha_s}{2\pi^2} \int\!\! d^2z\ 
{\cal K}_{\bm{x z x}}
%K(\bm{x},\bm{z},\bm{x})
    \big([U^\dagger_{\bm{z}}]_{i j}
\tr(U^\dagger_{\bm{x}}U_{\bm{z}})
      -N_{c}[U^\dagger_{\bm{x}}]_{i j}
    \big)
\ .
  \end{align}
\end{subequations}
while those of $\bar\chi$  read
{\small
\begin{subequations}
  \label{eq:barchidef}
\begin{align}
  %\alpha_s 
  & 
[\Bar\chi^{q\Bar q}_{\bm{x}\bm{y}}]_{i j\,k l} := 
  %-
  \frac{\alpha_s}{2\pi^2} \int\!\! d^2z\ 
{\cal K}_{\bm{x z y}}
%\\ \nonumber & \times
\Big(
[U_{\bm{z}}U^\dagger_{\bm{y}}]_{i l}
[U^\dagger_{\bm{z}}U_{\bm{x}}]_{k j}+
[U_{\bm{x}}U^\dagger_{\bm{z}}]_{i l}
[U^\dagger_{\bm{y}}U_{\bm{z}}]_{k j}
-[U_{\bm{x}}U^\dagger_{\bm{y}}]_{i l}\delta_{k j}
-\delta_{i l}[U^\dagger_{\bm{y}}
U_{\bm{x}}]_{k j}\Big)
\\  
%\alpha_s 
& 
[\Bar\chi^{\Bar q q}_{\bm{x}\bm{y}}]_{i j\,k l} := 
%\alpha_s 
[\Bar\chi^{q\Bar q}_{\bm{y}\bm{x}}]_{k l\,i j} 
\nonumber \\ &  \hspace{1cm} =
%-
\frac{\alpha_s}{2\pi^2} \int\!\! d^2z\ {\cal K}_{\bm{y z x}}
%\\ \nonumber & \times
\Big(
[U_{\bm{z}}U^\dagger_{\bm{x}}]_{k j}
[U^\dagger_{\bm{z}}U_{\bm{y}}]_{i l}+
[U_{\bm{y}}U^\dagger_{\bm{z}}]_{k j}
[U^\dagger_{\bm{x}}U_{\bm{z}}]_{i l}
-[U_{\bm{y}}U^\dagger_{\bm{x}}]_{k j}\delta_{i l}
-\delta_{k j}[U^\dagger_{\bm{x}}U_{\bm{y}}]_{i l}\Big)
\\
%  \alpha_s 
& [\Bar\chi^{q q}_{\bm{x y}}]_{i j\,k l} := 
-
  \frac{\alpha_s}{2\pi^2} \int\!\! d^2z\ 
{\cal K}_{\bm{x z y}}
%\\ \nonumber & \times
\Big(
[U_{\bm{z}}]_{i l}[U_{\bm{y}}
U^\dagger_{\bm{z}}U_{\bm{x}}]_{k j}
+[U_{\bm{x}}U^\dagger_{\bm{z}}
U_{\bm{y}}]_{i l}[U_{\bm{z}}]_{k j}
-[U_{\bm{x}}]_{i l}[U_{\bm{y}}]_{k j}
-[U_{\bm{y}}]_{i l}[U_{\bm{x}}]_{k j}
\Big)
\\
%  \alpha_s 
& 
[\Bar\chi^{\Bar q\Bar q}_{\bm{x}\bm{y}}]_{i j\,k l} 
:= 
-
  \frac{\alpha_s}{2\pi^2} \int\!\! d^2z\ 
{\cal K}_{\bm{x z y}}
%\\ \nonumber & \times
\Big(
[U^\dagger_{\bm{z}}]_{i l}
[U^\dagger_{\bm{y}}U_{\bm{z}}
U^\dagger_{\bm{x}}]_{k j}
+[U^\dagger_{\bm{x}}U_{\bm{z}}
U^\dagger_{\bm{y}}]_{i l}
[U^\dagger_{\bm{z}}]_{k j}
-[U^\dagger_{\bm{x}}]_{i l}
[U^\dagger_{\bm{y}}]_{k j}
-[U^\dagger_{\bm{y}}]_{i l}
[U^\dagger_{\bm{x}}]_{k j}
\Big)
\end{align}
\end{subequations}
}

At this point all the calculational effort pays off:
Using the above definitions for $\Bar\chi$ and $\Bar\sigma$, and
remembering that ${\cal S}_{\mathrm{ext}}^{q\bar q}[A,J^\dagger,J]$
depends on $A$ via $U^{(\dagger)}$ only, recasts
Eq.~(\ref{eq:Bsecorder}) as
{\small
\begin{align}
\label{eq:viachisigma}
\begin{split}
  \langle & e^{{\cal S}_{\mathrm{ext}}^{q\bar q}[b+\delta
    A,J^\dagger,J]}  \rangle_{b,\delta A}
%\\ & = 
-   \langle e^{{\cal S}_{\mathrm{ext}}^{q\bar q}[b,J^\dagger,J]} 
   \rangle_b 
\\ & = 
  \frac 1 2\langle\langle \delta A_x \delta A_y \rangle_{\delta A}[b]
  \left(2\left(\frac{\delta}{\delta b_x}{\cal S}_{\mathrm{ext}}^{q\bar
      q}[b,J^\dagger,J]\right)\left( \frac{\delta}{\delta b_y} {\cal
      S}_{\mathrm{ext}}^{q\bar q}[b,J^\dagger,J]\right)
  +\frac{\delta}{\delta
      b_x} \frac{\delta}{\delta b_y} {\cal S}_{\mathrm{ext}}^{q\bar
      q}[b,J^\dagger,J]\right) 
%\\ & \ \ \times 
  e^{{\cal S}_{\mathrm{ext}}^{q\bar
      q}[b,J^\dagger,J]} \rangle_b 
\\ & \ \ \ \ +\ldots
\\ & =
%\alpha_s
\ln(\frac{\xbj_0}{\xbj})\, \big\langle \left\{\frac{1}{2}
\tr\left[
\begin{pmatrix}
  \Bar\chi^{q q} & \Bar\chi^{q \Bar q}
\\
  \Bar\chi^{\Bar q q} &\Bar\chi^{\Bar q \Bar q} 
\end{pmatrix}[U,U^\dagger]{\cdot}
\left(
\begin{pmatrix}
  J^\dagger\\ J
\end{pmatrix}\otimes (J^\dagger,J)
\right)
\right]
+(\Bar\sigma^q,\Bar\sigma^{\bar q})[U,U^\dagger]
\begin{pmatrix}
  J^\dagger\\ J
\end{pmatrix}
\right\}
e^{{\cal S}_{\mathrm{ext}}^{q\bar q}[b,J^\dagger,J]}
\  \big\rangle_b
\ .
\end{split}
\end{align}
}The logarithm comes from an integration over $\delta A$ with momenta
in a finite interval of $\xbj$ values. Accordingly the equation has the
form of a finite difference equation for $\bar{\cal Z}[J^\dagger,J]$
with respect to $\ln 1/\xbj$.

This gives rise to the final RG equation in differential form, which I
write using condensed notation with $\bm{J}=(J^\dagger,J)$
and $\frac{\delta}{\delta\bm{J}}=(\frac{\delta}{\delta
  J^\dagger},\frac{\delta}{\delta J})$ as
\begin{equation}
  \label{eq:barcalZevol}
  \frac{\partial}{\partial \ln 1/\xbj} \Bar{\cal Z}[\bm{J}] = 
  %\alpha_s 
  \left\{\frac{1}{2}
    \bm{J}_u \bm{J}_v
\Bar\chi_{u v} 
[\frac{\delta}{\delta\bm{J}}]
+\bm{J}_u
\Bar\sigma_u[\frac{\delta}{\delta\bm{J}}]
\right\}
\Bar{\cal Z}[\bm{J}]
\end{equation}
where $u$ and $v$ now also stand for $q$ and $\bar q$ in addition to
color and transverse coordinates.
%\begin{equation}
%  \label{eq:barcalZevollong}
%  \frac{\partial}{\partial \ln 1/\xbj} \Bar{\cal Z}[J^\dagger,J] = 
%  \alpha_s \left\{\frac{1}{2}
%\tr\left[
%\begin{pmatrix}
%  J^\dagger\\ J
%\end{pmatrix}\otimes (J^\dagger,J)
%{\cdot}
%\begin{pmatrix}
%  \Bar\chi^{q q} & \Bar\chi^{q \Bar q}
%\\
%  \Bar\chi^{\Bar q q} &\Bar\chi^{\Bar q \Bar q} 
%\end{pmatrix}[\frac{\delta}{\delta J^\dagger},\frac{\delta}{\delta J}]
%\right]
%+(J^\dagger, J)
%\begin{pmatrix}
%\Bar\sigma^q \\ \Bar\sigma^{\bar q} 
%\end{pmatrix}[\frac{\delta}{\delta J^\dagger},\frac{\delta}{\delta J}]
%\right\}
%\Bar{\cal Z}[J^\dagger,J]
%\end{equation}
%For comparison with previous notation see App. \ref{sec:notation}.
Eq.~(\ref{eq:barcalZevol}) represents a full set of coupled evolution
equations for all correlators of $U^{(\dagger)}$ also known as the
Balitsky hierarchy. It is obtained from the above by taking
$J^{(\dagger)}$ derivatives evaluated at $J^{(\dagger)}=0$ as
demonstrated in Eq.~(\ref{eq:dipolecross}) for the dipole cross
section.\footnote{The convention for index contraction again is
  defined in order to simplify notation. Products are defined as in
  $J^\dagger\Bar\sigma^q:= [J^\dagger]_{i j} [\Bar\sigma^q]_{i j}$
  with integration over transverse coordinates implied.}

Although Eq.\eqref{eq:barcalZevol} contains all the information we are
looking for it is still not written in an optimal way. In fact it
requires a functional Fourier transform from $\bm{J}$ to $\bm{U}$ to
expose its statistical nature as a Fokker-Planck equation. Let me do
this, first taking $U$ and $U^\dagger$ as independent variables.

If one recalls the definition of $\Bar{\cal Z}[\bm{J}] := \langle
e^{{\cal S}_{\mathrm{ext}}^{q\bar q}[b,\bm{J}]} \rangle_b$ and
the initial idea of interpreting $\langle\ldots\rangle_b$ as a
statistical average with associated statistical weight $\Bar Z$
according to
\begin{equation}
  \label{eq:statweight}
  \langle\ldots\rangle_b = \langle\ldots\rangle_{\bm{U}}
  =\int\!D[\bm{U}] ( \Bar Z[\bm{U}] \ldots )\ , 
\end{equation}
one may try to rewrite Eq.~(\ref{eq:barcalZevol}) as an equation for
$\Bar Z$. Indeed --remembering that ${\cal S}_{\mathrm{ext}}^{q\bar
  q}[b,\bm{J}]={\cal S}_{\mathrm{ext}}^{q\bar
  q}[\bm{U},\bm{J}]$ is actually given as a functional
of $U^{(\dagger)}$--
\begin{equation}
  \label{eq:barcalZevol_1}
  \begin{split}
    \frac{\partial}{\partial \ln 1/\xbj}  \int\!D[\bm{U}] 
    \Bar Z[\bm{U}] 
    & 
    e^{{\cal S}_{\mathrm{ext}}^{q\bar
        q}[\bm{U},\bm{J}]} = %\alpha_s 
    \int\!D[\bm{U}] 
    \Bar Z[\bm{U}]
    %\\ & \times 
    \left\{\frac{1}{2} 
      \Bar\chi_{u v}[\bm{U}] 
      \bm{J}_u \bm{J}_v
      +\Bar\sigma_u [\bm{U}]
      \bm{J}_u
    \right\} e^{{\cal S}_{\mathrm{ext}}^{q\bar
        q}[\bm{U},\bm{J}]}\ .
  \end{split}
\end{equation}
%
%\begin{equation}
%  \label{eq:barcalZevol_1_long}
%  \begin{split}
%    \frac{\partial}{\partial \ln 1/\xbj} & \int\!D[U,U^\dagger] \Bar
%    Z[U,U^\dagger] e^{{\cal S}_{\mathrm{ext}}^{q\bar
%        q}[U,U^\dagger,J^\dagger,J]} = \alpha_s \int\!D[U,U^\dagger]
%    \Bar Z[U,U^\dagger] 
%\\ & 
%\times \left\{\frac{1}{2} \tr\left(
%\begin{pmatrix}
%  \Bar\chi^{q q} & \Bar\chi^{q \Bar q}
%  \\
%  \Bar\chi^{\Bar q q} &\Bar\chi^{\Bar q \Bar q}
%\end{pmatrix}[U,U^\dagger] 
%{\cdot}
%\begin{pmatrix}
%  J^\dagger\\ J
%\end{pmatrix}\otimes (J^\dagger,J)
%\right) +(\Bar\sigma_q ,\Bar\sigma_{\bar q})[U,U^\dagger]
%\begin{pmatrix}
%  J^\dagger\\ J
%\end{pmatrix}
%\right\} e^{{\cal S}_{\mathrm{ext}}^{q\bar
%    q}[U,U^\dagger,J^\dagger,J]}\ .
%  \end{split}
%\end{equation}
Using the fact that if one treats $U$ and $U^\dagger$ as independent
variables one finds, conjugate to Eqns.~(\ref{eq:makeUs}), that
%\begin{subequations}
\begin{align}
\label{eq:makeJs}
%\begin{align}
  \frac{\delta}{\delta U}\, e^{{\cal S}_{\mathrm{ext}}^{q\bar
      q}[\bm{U},\bm{J}]} = & \ J^\dagger\, e^{{\cal
      S}_{\mathrm{ext}}^{q\bar q}[\bm{U},\bm{J}]}\ ,
 \hspace{1cm}% \\
  \frac{\delta}{\delta U^\dagger}\, e^{{\cal
      S}_{\mathrm{ext}}^{q\bar q}[\bm{U},\bm{J}]} = %& 
  \ J\,
  e^{{\cal S}_{\mathrm{ext}}^{q\bar q}[\bm{U},\bm{J}]}\ .
\end{align}  
%\end{subequations}
Hence, Eq.~(\ref{eq:barcalZevol_1}) can be rewritten as an equation
for $\bar Z$ shown in~\eqref{eq:barZevol}. Note that by treating $U$
and $U^\dagger$ as independent integration variables [this is
necessary for~(\ref{eq:makeJs}) to hold and to perform the translation
step] one violates the $SU(N_c)$ condition $U U^\dagger =\bm{1}$. The
variables $U^{(\dagger)}$ introduced here are more general
(invertible) $N_c\times N_c$ matrices at this point. This will be
rectified shortly by using properties of $\bar\chi$ and $\bar\sigma$
and hence the QCD derived nature of evolution for $\bar Z$. Leaving
this for later, the evolution equation for $\bar Z$ is given by
\begin{equation}
  \label{eq:barZevol}
  \begin{split}
    \frac{\partial}{\partial \ln 1/\xbj} & \Bar Z[\bm{U}] 
                                %\\ = &\
    =
   %\alpha_s 
    \left\{\frac{1}{2} 
            \frac{\delta}{\delta \bm{U}_u}
            \frac{\delta}{\delta \bm{U}_v}
          \Bar\chi_{u v}[\bm{U}]
      -\frac{\delta}{\delta \bm{U}_u}
        \Bar\sigma_u[\bm{U}]
    \right\} \Bar Z[\bm{U}]
\ ,
  \end{split}
\end{equation}
where the $U^{(\dagger)}$ derivatives act on both the kernels
$\bar\chi,\bar\sigma$ and $\bar Z$. This equation exhibits exactly the
same structures as the JKLW equation, an earlier version formulated in
terms of the charge densities $\rho$ ~\cite{ Jalilian-Marian:1997xn,
  Jalilian-Marian:1997jx, Jalilian-Marian:1997gr,
  Jalilian-Marian:1997dw, Jalilian-Marian:1998cb, Kovner:2000pt},
although the variables and details of the kernels are, of course,
different. Both have the form of a Fokker-Planck (FP) equation, simply
due to the fact that the original equation as represented in
Eq.~(\ref{eq:barcalZevol_1}) had only terms quadratic and linear in
$\bm{J}$.

As with any other FP equation, the fact that the right hand side is a
total derivative implies that the normalization of $\bar Z$ --its
$U,U^\dagger$ integral-- is conserved under $\xbj$ evolution and, if one
chooses this norm to be $1$, allows to interpret $\bar Z$ as a
probability distribution at any stage of the evolution.

Clearly this formulation is completely equivalent to the original
version involving the generating functional $\bar{\cal Z}$.  Instead
of obtaining equations for correlators by functional differentiation,
one now extracts them, again as in any FP formulation, by multiplying
with the desired mononomial in $U^{(\dagger)}$ followed by integrating
the result over $U,U^\dagger$.

Before exploring how to reconcile this Fourier transformation step
with the $SU(N_c)$ group nature of the physical $U$ fields I use the
observation --based on a lengthy but straightforward calculation using
the explicit form of $\bar\chi$ and $\bar\sigma$-- that
\begin{align}
  \label{eq:forceterm}
  \frac{1}{2} 
       \bigg(
            \frac{\delta}{\delta \bm{U}_v}
          \Bar\chi_{u v}[\bm{U}]\bigg)-
        \Bar\sigma_u[\bm{U}] =0 
\end{align}
(for both the $q$ and $\bar q$ components) to rewrite the
Fokker-Planck equation in a form that somewhat resembles Brownian
motion on a curved space
\begin{equation}
  \label{eq:BrownianFP}
    \frac{\partial}{\partial \ln 1/\xbj}  \Bar Z[\bm{U}]  =  
   %\alpha_s
    \frac{1}{2}  \frac{\delta}{\delta \bm{U}_u} 
          \Bar\chi_{u v}[\bm{U}]
            \frac{\delta}{\delta \bm{U}_v}
            \Bar Z[\bm{U}]
            \ .
\end{equation}
For consistency one now needs to be able to impose the group
constraint on this equation.  What one needs in fact is, that for
physical distributions, which are necessarily of the form
\begin{equation}
  \label{eq:zphys0}
  \bar Z_{\mathrm{phys}}[U,U^\dagger] = \delta(U U^\dagger-\bm{1})
  \delta(\det U-1)\Hat Z[U] \ ,
\end{equation}
the constraint factor $ \delta(U U^\dagger-\bm{1}) \delta(\det
U-1)$ is left invariant by the evolution operator, i.e. that one may
commute it though the delta functions.

In order to get a feeling for what is involved, it is useful to
collect some ideas of what to expect, if this were actually possible.
The main step in rewriting the evolution within the group consists of
turning $\frac{\delta}{\delta U}$ and $\frac{\delta}{\delta
  U^\dagger}$ from independent to dependent derivatives via
$UU^\dagger=1$. What is to replace them?

Pondering this question a little (c.f. App. C of \cite{Weigert:2000gi} for a
pedestrian approach), one inevitably ends up considering the natural
derivatives along $SU(N_c)$, its left and right invariant vector
fields. The former would be given by\footnote{It is easy to see that
  this is the correct form as both variants amongst themselves and
  with each other satisfy the $SU(N_c)$ commutation relations at each
  point.}
\begin{equation}
  \label{eq:Lieder}
  i\nabla^a_U:= [U t^a]_{i j} \frac{\delta}{\delta U}_{i j} 
   =[-t^a U^{-1}]_{i j}
    \frac{\delta}{\delta U^{-1}}_{i j}\ ,
\end{equation}
so that as a first step in this direction one would like to reexpress
things via $[U t^A]_{i j} \frac{\delta}{\delta U}_{i j}$ and $[-t^A
U^\dagger]_{i j} \frac{\delta}{\delta U^\dagger}_{i j}$ instead of $
\frac{\delta}{\delta U}_{i j}$ and $\frac{\delta}{\delta U^\dagger}_{i
  j}$.  Indeed, if one now just tries out what this transformation
will do, one finds remarkable results for the transformed components
of $\bar\chi$:\footnote{To achieve this technically, one needs to be
  able to represent all of ${\delta}/{\delta U^{(\dagger)}_{i j}}$s
  components. Accordingly one has to replace the index
  $a\in\{1,\ldots, N_c^2-1\}$ by ${\sf A}\in\{0,\ldots, N_c^2-1\}$ and
  supply $t^0=1/\sqrt{2 N_c}\ \bm{1}$, such that $\tr(t^{\sf
    A} t^{\sf B})=\delta^{{\sf A} {\sf B}}/2$ for all ${\sf A},{\sf
    B}$. Then the transformation is easily accomplished.  (see App. C
  of \cite{Weigert:2000gi} for details).} 
%As already advertised, the results are quite striking:
\begin{itemize}
\item in $SU(N_c)$ all four of the transformed components are equal. 
\item in $SU(N_c)$ all of them vanish if either ${\sf A}$ or ${\sf B}$ is $0$.
  Only octet components survive. All matrices involved now are
  explicitly in the adjoint representation. 
\item The unique common form even factorizes naturally into the
  ``square'' of a much simpler factor. The result, after pulling out a
  minus sign for later convenience, will be denoted by $\hat\chi^{a
    b}_{{\bm{x}}{\bm{y}}}[U]$ and reads
\begin{equation}
  \label{eq:chisun}
  \begin{split}
%  \alpha_s 
    \hat\chi^{a b}_{\bm{x}\bm{y}}&[U]:=  
%\\ = &   -\frac{\alpha_s}{2\pi^2} \int\!\! d^2z\ 
%{\cal K}_{\bm{x z y}}
%\Big(( \Tilde U^{-1}_{\bm{z}} \Tilde U_{\bm{y}})^{a b}
%+( \Tilde U^{-1}_{\bm{x}} \Tilde U_{\bm{z}} )^{a b}
%-( \Tilde U^{-1}_{\bm{x}} \Tilde U_{\bm{y}} )^{a b}
%-\Tilde{\mathbf{1}}^{a b}
%\Big) 
%\\ = &   
-
\frac{\alpha_s}{%2
\pi^2} \int\!\! d^2z
\bigg(\frac{({\bm{x}}-{\bm{z}})_i}{%
({\bm{x}}-{\bm{z}})^2} 
\big[ \Tilde{\mathbf{1}}-\Tilde U^{-1}_{\bm{x}} 
\Tilde U_{\bm{z}}\big]^{a c}\bigg)
\bigg(\frac{({\bm{z}}-{\bm{y}})_i}{%
({\bm{z}}-{\bm{y}})^2} 
\big[\Tilde{\mathbf{1}}- \Tilde U^{-1}_{\bm{z}} 
\Tilde U_{\bm{y}} \big]^{c b}\bigg)
\ .
  \end{split}
\end{equation}
\end{itemize}
The diagonalizing factors of $\hat\chi$ deserve a name for further
reference
\begin{equation}
  \label{eq:edef}
  [{\cal E}^{a b}_{{\bm{x}} {\bm{y}}}]_i:= 
  \sqrt{\frac{\alpha_s}{%2
      \pi^2}}
  \bigg(\frac{({\bm{x}}-{\bm{y}})_i}{%
    ({\bm{x}}-{\bm{y}})^2} 
  \big[ \Tilde{\mathbf{1}}
  -\Tilde U^{-1}_{\bm{x}} 
  \Tilde U_{\bm{y}}\big]^{a b}\bigg)
  \ .
\end{equation}
%If the FP equation can be written entirely within $SU(N_c)$ --and this
%will be demonstrated below-- one may also expect that a description in
%terms of a Langevin process with a completely decorrelated Gaussian
%random noise becomes possible due to the factorized form
%of~(\ref{eq:chisun}) via~(\ref{eq:edef}).

The results of the above transformation of $\bar\chi$ are so
suggestive that it seems obvious what the form of the FP operator on
physical configurations has to be:
$%\begin{equation}
%  \label{eq:physFP}
 % \alpha_s
-  \frac{1}{2}i\nabla^a_{U_{\bm{x}}} 
  \hat\chi^{a b}_{\bm{x} \bm{y}} 
  i\nabla^b_{U_{\bm{y}}}
%  \ .
%\end{equation}
$.
%as one has seen that $\bar\chi$ reduces to $\hat\chi$ through a
%unitary transformation followed by a restriction to $SU(N_c)$. 
The chain of argument to prove this is the following. One sets out to
show that
\begin{equation}
  \label{eq:physcomm}
  \frac{\delta}{\delta\bm{U}_u} \Bar\chi_{u v} 
  \frac{\delta}{\delta\bm{U}_v}
  \,\,
  \delta(UU^\dagger-1)\delta(\det U-1)=
  \delta(UU^\dagger-1)\delta(\det U-1)
  \,\,
  \nabla^a_{U_{\bm{x}}} 
  \hat\chi^{a b}_{\bm{x} \bm{y}}
  \nabla^b_{U_{\bm{y}}}
  \ .
\end{equation}
This amounts to verifying that
\begin{equation}
  \label{eq:condition}
  \frac{\delta}{\delta\bm{U}_u} \Bar\chi_{u v} 
  \frac{\delta}{\delta\bm{U}_v} 
  \,
  U^{(\dagger)}_{\bm{w}_1}\otimes\ldots\otimes 
  U^{(\dagger)}_{\bm{w}_n}\Big\vert_{UU^\dagger=1} =
  \nabla^a_{U_{\bm{x}}} 
  \hat\chi^{a b}_{\bm{x} \bm{y}} 
  \nabla^b_{U_{\bm{y}}}
  \,
  U^{(-1)}_{\bm{w}_1}\otimes\ldots \otimes
  U^{(-1)}_{\bm{w}_n}
\end{equation}
for arbitrary mononomials $U^{(\dagger)}_{\bm{w}_1} \otimes
\ldots \otimes U^{(\dagger)}_{\bm{w}_n}$, a task that once
more is best performed via a generating functional.  This strategy has
the added benefit that it makes it readily apparent how the two
operators lead to the same hierarchy of equations for the correlators,
which provided the starting point for all these deliberations.  The
details of this exercise are given in App.~D of~\cite{Weigert:2000gi}.

This leads to what is now known as the JIMWLK equation, the RG/FP
equation on the physical configuration space, where
\begin{equation}
  \label{eq:zphys}
  \bar Z_{\mathrm{phys}}[U,U^\dagger] = \delta(U U^\dagger-\bm{1})
  \delta(\det U-1)\hat Z[U] \ .
\end{equation}
The constraint factor may now be absorbed into the measure to form a
functional Haar measure according to
\begin{equation}
  \label{eq:measure}
  \Hat D[U]:= D[\bm{U}] \delta(U U^\dagger-\bm{1})
  \delta(\det U-1)
\ .
\end{equation}
This constitutes the natural measure to use for averaging correlators
of the form $\big\langle U^{(-1)}_{\bm{w}_1}\otimes\ldots
\otimes U^{(-1)}_{\bm{w}_n} \big\rangle_U$ with the
probability distribution $\hat Z[U]$.  The RG/FP equation simplifies
to
\begin{equation}
  \label{eq:finalFP}
%  \begin{split}
    \frac{\partial}{\partial\ln 1/\xbj} \,
    % & 
   \hat Z[U](\xbj)
   %\\  = &
   = 
   %\alpha_s
   \,- \frac{1}{2}
   i\nabla^a_{U_{\bm{x}}}
   \hat\chi^{a b}_{\bm{x}\bm{y}}
   i\nabla^b_{U_{\bm{y}}}\,\hat Z[U](\xbj)
\ .
%  \end{split}
\end{equation}
This formulation is fully equivalent to the initial one, but with a
large amount of redundancy removed.

A crucial observation is that the evolution operator on the right hand
side of Eq.\eqref{eq:finalFP}
\begin{equation}
  \label{eq:HFPdef}
  H_{\text{JIMWLK}}:= \frac{%\alpha_s
  1}{2} 
  i\nabla^a_{U_{\bm{x}}} 
  \hat\chi^{a b}_{\bm{x} \bm{y}} 
  i\nabla^b_{U_{\bm{y}}}
\end{equation}
is actually self adjoint and may be called a Fokker-Planck
Hamiltonian. It has been shown to be positive semidefinite and to
posses a (trivial) fixed point at infinite energy with vanishing
correlation length~\cite{Weigert:2000gi}.

\subsection{\it The JIMWLK equation and 
  the Balitsky hierarchy}
\label{sec:evol-equat-expl}

I will begin this section with a short summary of the results of the
previous subsection and then turn to their interpretation.  The JIMWLK
equation describes the evolution of correlators of Wilson lines
collinear to the projectile direction which are defined through
\begin{equation}
  \label{eq:average-def}
  \langle \ldots \rangle_\y = \int \Hat D[U] \ldots \Hat Z_\y[U]
\end{equation}
where $\Hat D[U]$ denotes a functional Haar measure, in keeping with
the group valued nature of the field variables $U$. The evolution
equation can be compactly written as
\begin{equation}
  \label{eq:JIMWLK}
  \partial_\y\Hat Z_\y[U] = -H_{\text{JIMWLK}}[U] \Hat Z_\y[U]
\ .
\end{equation}
This form  highlights its statistical nature as a (functional) diffusion
equation. The Fokker-Planck Hamiltonian is defined as
\begin{subequations}
  \label{eq:JIMWLK-Hamiltonian}
\begin{align}
  \label{eq:JIMWLK-Hamiltonian-chi}
 H_{\text{JIMWLK}} := & \frac{1}{2}  \ i\nabla^a_{\bm{x}} 
 \chi^{a b}_{\bm{x y}}i\nabla^b_{\bm{y}}
\\
\label{eq:chidef}
\chi^{a b}_{\bm{x y}} := & 
 -\frac{\alpha_s}{\pi^2}\ \int\!\! d^2\!z\ {\cal K}_{\bm{x z y}} 
  \big[ (1-U^\dagger_x U_z) (1-U^\dagger_z U_y)\big]^{a b} 
\\ 
\label{eq:JIMWLK-K}
  {\cal K}_{\bm{x z y}} = & 
  \frac{(\bm{x}-\bm{z})\cdot(\bm{z}-\bm{y})}{(\bm{x}-\bm{z})^2
    (\bm{z}-\bm{y})^2}
\end{align}
\end{subequations}
and has been shown to be positive semidefinite.  In
Eqns.~\eqref{eq:JIMWLK-Hamiltonian}, both a summation and an
integration convention is applied to repeated indices and coordinates.

$i\nabla^a_{\bm{x}}$ is a functional derivative operator defined as
\begin{align}
  \label{eq:Lie-simpledef}
  i\nabla^a_{\bm{x}} := -[U_{\bm{x}} t^a]
  \frac{\delta}{\delta U_{\bm{x}, i j}}
  \ .
\intertext{$\frac{\delta}{\delta U_{\bm{x}, i j}}$ is the ordinary 
  functional (or variational) derivative w.r.t. the components of 
  the $U$ field:}
\frac{\delta}{\delta U_{\bm{x}, i j}} U_{\bm{y}, k l} 
 = \delta_{i k} \delta_{j l} \delta^{(2)}_{\bm{x y}}
\end{align}
where $ \delta^{(2)}_{\bm{x y}} := \delta^{(2)}(\bm{x}-\bm{y}) $ for
compactness.

This operator in fact corresponds to the left invariant vector field
on the group manifold --for a physicist, the most familiar
interpretation might be that as a vielbein or moving frame. For the
present purpose, only a few operational facts are needed, which are
briefly summarized as follows:
\begin{itemize}
\item
Operationally the definition~\eqref{eq:Lie-simpledef} leads to
\begin{subequations}
  \label{Lie-der}
\begin{align}
  i\nabla^a_{\bm{x}} U_{\bm{y}} := & 
  -U_{\bm{x}} t^a \delta^{(2)}_{\bm{x y}}
  \ ,
  \hspace{1cm}
   i\nabla^a_{\bm{x}} U^\dagger_{\bm{y}} := 
    t^a U^\dagger_{\bm{x}} \delta^{(2)}_{\bm{x y}} \ .
\intertext{There is, of course, a corresponding definition for the
right invariant vector fields $i\Bar\nabla^a_{\bm{x}}$:}
   i\Bar\nabla^a_{\bm{x}} U_{\bm{y}} := & 
   t^a U_{\bm{x}} \delta^{(2)}_{\bm{x y}}
   \ ,
  \hspace{1cm}
   i\Bar\nabla^a_{\bm{x}} U^\dagger_{\bm{y}} := 
   - U^\dagger_{\bm{x}} t^a \delta^{(2)}_{\bm{x y}}
   \ .
\end{align}
\end{subequations}
\item Their main properties are the commutation relations (which I display
leaving the functional nature aside for a second)
\begin{subequations}
\begin{align}
  \label{eq:comm-rules}
  [i\nabla^a,i\nabla^b] =  i f^{a b c} i\nabla^c 
  \hspace{1cm}
   [i\Bar \nabla^a,i\Bar \nabla^b] =  i f^{a b c} i\Bar \nabla^c  
   \hspace{1cm}
   [i\Bar \nabla^a,i \nabla^b] = & 0 
\ .
\end{align}
$\nabla$ and $\Bar\nabla$  are interrelated by 
\begin{equation}
  \label{eq:nablarel}
  i\nabla^a_{\bm{x}} = 
  -[\Tilde U^\dagger_{\bm{x}}]^{a b} i\Bar\nabla^b_{\bm{x}};
  \hspace{1cm}
  i\Bar\nabla^a_{\bm{x}} =
  -[\Tilde U^\dagger]^{a b}_{\bm{x}} i\nabla^b_{\bm{x}}
\end{equation}
\end{subequations}
and ``representation conscious:'' With the above definitions for the
action on $U$ and $U^\dagger$ in the $q$ and $\Bar q$ representation,
it automatically follows from representation theory that acting
on $U$ or $U^\dagger$ in an arbitrary representation produces
analogous formulae with the generators appearing on the r.h.s. in that
representation.
\item The mathematical concepts of right and left invariant vector
  fields are used to generate right and left translations on the group
  manifold. For the present purpose this is best expressed by writing
%\footnote{Note that
%  nevertheless, both correspond to what is called a left action of the
%  group, as $f(e^{-i\omega} e^{-i\y}, U) = f(e^{-i\omega}
%  ,f(e^{-i\y}, U))$ in both cases.}
%\begin{align*}
%  \label{eq:translations}
$  e^{-i\omega^a (i\nabla^a)} U = U e^{i\omega^a t^a} $
%  \hspace{1cm} 
and
$  e^{-i\omega^a (i\Bar\nabla^a)} U =e^{-i\omega^a t^a} U $.
%\ .
%\end{align*}
Functional forms of this of course involve an integral in the exponent
as in
\begin{align}
  \label{eq:translations-func}
%$
  e^{-i\int_x\omega^a_x (i\nabla^a_x)} U_y = U_y e^{i\omega^a_y t^a}  
%$.
  \hspace{1cm} \text{or} \hspace{1cm}
  e^{-i\int_x\omega^a_x (i\Bar\nabla^a_x)} U_y =e^{-i\omega^a_y t^a} U_y 
\ .
\end{align}
This property has been used in~\cite{Weigert:2003mm} to write an
operator that creates soft gluon emission in jets. Below, it will be
explained how this can be used to summarize efficiently the soft gluon
content of the projectile wave function entering for example in DIS.
%}
\end{itemize}
Using these tools, $H_{\text{JIMWLK}}$ is elegantly written as
\begin{align}
  \label{eq:JIMWLK-Hamiltonian-2}
 H_{\text{JIMWLK}} = &
   -\frac{1}{2} \frac{\alpha_s}{\pi^2}\ {\cal K}_{\bm{x z y}}\ 
  \big[ i\nabla^a_x i\nabla^a_y+i\Bar\nabla^a_x i\Bar\nabla^a_y
  +\Tilde U_z^{a b}(i\Bar\nabla^a_xi\nabla^b_y
  +i\nabla^a_x i\Bar\nabla^b_y) \big]
\end{align}
[integration convention for $x, z, y$].  While the factorized form of
Eq.~\eqref{eq:JIMWLK-Hamiltonian} is most useful in a derivation of a
Langevin description of the evolution that allows a numerical
implementation, this second form is more economical in the derivation
of evolution equations for given correlators that follow as a
consequence from Eq.~\eqref{eq:JIMWLK}.

This form also clearly distinguishes two types of terms: those that
add a new gluon to the diagrams at $\bm{z}$ and those that don't.
Diagrammatically this separation corresponds to splitting the time
ordered diagrams of Sec.~\ref{sec:syst-deriv} according to whether the
new gluon crosses the $x^-=0$ line and thus interacts with the target
or is reabsorbed before or after without interaction and thus acts as
a virtual correction to the projectile wave function.

\begin{figure}[htb]
  \centering
  \includegraphics[width=\textwidth]{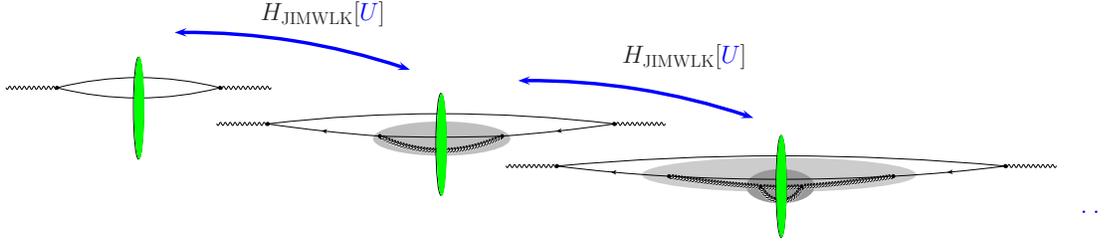}
  \caption{\small 
    The JIMWLK operators adds gluons into the photon wavefunction, 
    virtual corrections are not shown}
  \label{fig:addgluons}
\end{figure}

This allows to make contact with Eq.~\eqref{eq:emission} used to
illustrate the importance of soft gluon emission in the virtual photon
wavefunction. The contributions shown there directly correspond to real
diagrams (adding new factors of $U$ into the expressions), the virtual
contributions have been omitted.  It is easy to see now how iterated
application of the JIMWLK operator adds new gluons, if not on an
amplitude level but at least for the DIS cross section as a whole, as
shown in Fig.~\ref{fig:addgluons}.
To do so, one needs to write the evolution equation for a general
correlator $O[U]$. To explore the first step shown in
Fig.\ref{fig:addgluons}, one might choose $O[U]$ to be
\begin{subequations}
 \label{eq:S-def}
\begin{equation}
 \Hat S_{\bm{x y}} = \tr(U^\dagger_{\bm{x}} U_{\bm{y}})/N_c 
\end{equation}
which appears in the top left ``bare'' diagram. For convenience, I
also define a shorthand for its $\y$ dependent expectation value
\begin{equation}
 S_{\y; \bm{x y}} := \langle \Hat S_{\bm{x y}} \rangle_\y \ .
\end{equation}  
\end{subequations}
To obtain the evolution equation, one simply multiplies both sides
of~\eqref{eq:JIMWLK} with $O[U]$ and averages
with~\eqref{eq:average-def}. The result then is an evolution equation
for $\langle O[U] \rangle_\y$
\begin{equation}
  \label{eq:OU-evo}
  \partial_\y \langle O[U] \rangle_\y
  =
  - \langle H_{\text{JIMWLK}} O[U] \rangle_\y
  \ .
\end{equation}
Because of the nonlinear nature of $H_{\text{JIMWLK}}$ the r.h.s. will in
no case be expressible by $\langle O[U] \rangle_\y$ alone, but
instead involves other correlators as well. The evolution equation of
these new quantities will then also be needed, and the argument
repeats itself, ultimately leading to an infinite coupled hierarchy of
evolution equations, the Balitsky hierarchy.  For $\Hat S_{\bm{x y}}$,
one immediately arrives at
{\small\begin{align}
  \label{eq:twopoint-JIMWLK}
  \partial_\y \big\langle \Hat S_{\bm{x y}}\big\rangle_\y = & 
  \frac{\alpha_s}{2\pi^2}
  \Big\langle\Big[{\cal K}_{\bm{u} \bm{z} \bm{v}}\Big(
  i\nabla^a_{{\bm{u}}}i\nabla^a_{{\bm{v}}}
 +
 i\Bar\nabla^a_{{\bm{u}}}i\Bar\nabla^a_{{\bm{v}}}\Big)
   + {\cal K}_{\bm{u} \bm{z} \bm{v}}\Big(\big[\Tilde U_{\bm{z}}\big]^{a
   b}(i\Bar\nabla^a_{{\bm{u}}}i\nabla^b_{{\bm{v}}}
 +i\Bar\nabla^a_{{\bm{v}}}i\nabla^b_{{\bm{u}}})\Big)\Big]
\frac{\tr(U_{\bm{x}}U^\dagger_{\bm{y}})}{N_c} \Big\rangle_\y 
\nonumber \\  = &
   \frac{\alpha_s}{2\pi^2}
  \Big\langle\Big[
  \Tilde {\cal K}_{\bm{x} \bm{z} \bm{y}}
 % \big(
 % 2 {\cal K}^{(1)}_{\bm{x} \bm{z} \bm{y}} -
 % {\cal K}^{(1)}_{\bm{x} \bm{z} \bm{x}} -{\cal K}^{(1)}_{\bm{y}
 %   \bm{z} \bm{y}}\big) 
\Big(
-2 %\frac{N_c^2-1}{2 N_c}
C_{\text{f}} \frac{\tr(U_{\bm{x}}U^\dagger_{\bm{y}})}{N_c}\Big)
%\nonumber \\ & \hspace{.3cm}
+ \Tilde {\cal K}_{\bm{x} \bm{z} \bm{y}} \
%\big(
%  2 {\cal K}^{(2)}_{\bm{x} \bm{z} \bm{y}} -
%  {\cal K}^{(2)}_{\bm{x} \bm{z} \bm{x}} -{\cal K}^{(2)}_{\bm{y}
%    \bm{z} \bm{y}}\big)    
  \frac{\big[\Tilde U_{\bm{z}}\big]^{a b} 2\ \tr(t^a U_{\bm{x}}t^b U^\dagger_{\bm{y}})}{N_c}
   \Big] \Big\rangle_\y
\end{align}
}where
\begin{equation}
  \label{eq:BK-kernel-from-JIMWLK}
  \Tilde {\cal K}_{\bm{x} \bm{z} \bm{y}} = \big(
  2 {\cal K}%^{(i)}
  _{\bm{x} \bm{z} \bm{y}} -
  {\cal K}%^{(i)}
  _{\bm{x} \bm{z} \bm{x}} -{\cal K}%^{(i)}
  _{\bm{y}
    \bm{z} \bm{y}}\big)
\ . 
\end{equation}
Observe on the r.h.s. the operator $\big[\Tilde U_{\bm{z}}\big]^{a b} 2\ 
\tr(t^a U_{\bm{x}}t^b U^\dagger_{\bm{y}})$ already anticipated in the
corresponding discussion of Sec.~\ref{sec:cgc-dis:-high}. This term
corresponds to the real contribution shown in the middle diagram of
Fig.~\ref{fig:addgluons}, while the other term is a virtual
contribution not explicitly shown. I will return to explore this
viewpoint in more detail in Sec.~\ref{sec:jimwlk-soft-gluon}

To compare with the factorizing limit and in particular the BK
equation, one applies the Fierz identity
\begin{equation}
  \label{eq:Fierz}
  \big[\Tilde U_{\bm{z}}\big]^{a b}
  \tr(t^a U_{\bm{x}}t^b U^\dagger_{\bm{y}}) =
  \frac{1}{2}\Big(\tr( U_{\bm{x}}U^\dagger_{\bm{z}})
           \tr( U_{\bm{z}}U^\dagger_{\bm{y}})
           -\frac{1}{N_c}\tr( U_{\bm{x}}U^\dagger_{\bm{y}})
  \Big)
\end{equation}
to rewrite Eq.~\eqref{eq:twopoint-JIMWLK} as
\begin{equation}
    \label{eq:pre-BK-S}
    \partial_\y \langle \Hat S_{\bm{x y}}\rangle_\y 
    = 
    \frac{\alpha_s N_c}{2\pi^2} \int d^2z \ \Tilde{\cal K}_{\bm{x}
    \bm{z} \bm{y}}\,\langle \Hat S_{\bm{x z}} \Hat S_{\bm{z y}} 
  -  \Hat S_{\bm{x y}} 
  \rangle_\y 
  \ .
\end{equation}
Alternatively in terms of
\begin{equation}
  \label{eq:dipole-expect}
  N_{\y; \bm{x y}} = \langle \Hat N_{\bm{x y}} \rangle_\y 
\ ,
\hspace{2cm}
 \Hat N_{\bm{x y}} = \tr(1-U^\dagger_{\bm{x}} U_{\bm{y}})/N_c 
\end{equation}
this equation reads
\begin{equation}
  \label{eq:dipoleOP}
  \partial_\y \langle {\Hat N}_{\bm{x y}}\rangle_\y  =
  \frac{\alpha_sN_c}{2\pi^2}\int\!\!d^2 z
  \ \Tilde{\cal K}_{\bm{x}
    \bm{z} \bm{y}}\,
  %\frac{(\bm{x}-\bm{y})^2}{%
  %  (\bm{x}-\bm{z})^2(\bm{z}-\bm{y})^2} 
  \Big\{  
  \big\langle {\Hat N}_{\bm{x z}}+{\Hat N}_{\bm{z y}}-
  {\Hat N}_{\bm{x y}} 
  \big\rangle_\y 
  - \big\langle  {\Hat N}_{\bm{x z}}
  {\Hat N}_{\bm{z y}}\big\rangle_\y 
  \Big\}  
\end{equation}

These immediately reduce to a closed equation for $S_{\bm{x y}}$ or
$N_{\bm{x y}}$ respectively, if one assumes factorization in the
nonlinearity on the r.h.s. by replacing
\begin{equation}
  \label{eq:factorization}
  \langle \Hat S_{\bm{x z}} \Hat S_{\bm{z y}}  
  \rangle_\y \to S_{\y; \bm{x z}}\>  S_{\y; \bm{z y}}  
\hspace{1cm}\text{or equivalently}\hspace{1cm}
 \langle \Hat N_{\bm{x z}} \Hat N_{\bm{z y}}  \rangle_\y \to N_{\y; \bm{x z}}\>  N_{\y; \bm{z y}}   \ .  
\end{equation}
The result is known as the BK equation and typically written in terms
of $N_{\y; \bm{x y}}$:
\begin{equation}
  \label{eq:BK}
  \partial_\y N_{\y; \bm{x y}}  =
  \frac{\alpha_sN_c}{2\pi^2}\int\!\!d^2 z
  \ \Tilde{\cal K}_{\bm{x}
    \bm{z} \bm{y}}\,
  %\frac{(\bm{x}-\bm{y})^2}{%
  %  (\bm{x}-\bm{z})^2(\bm{z}-\bm{y})^2} 
  \Big\{  
   N_{\y; \bm{x z}}+N_{\y; \bm{z y}}-
  N_{\y; \bm{x y}} 
  - N_{\y; \bm{x z}} N_{\y; \bm{z y}} 
  \Big\}  
\ .
\end{equation}
To come back to the Balitsky hierarchy, there are several equivalent
options on how to write the next equation in the hierarchy built on
top of $ N_{\y; \bm{x z}}$ if factorization does not hold in
Eq~\eqref{eq:dipoleOP}. One may consider the evolution equation of
$\langle \Hat N_{\bm{x z}} \Hat N_{\bm{z y}} \rangle_\y$ or $\langle
\Hat S_{\bm{x z}} \Hat S_{\bm{z y}} \rangle_\y$ as well as $\langle
\big[U_{\bm{z}}\big]^{a b} 2\ \tr(t^a U_{\bm{x}}t^b
U^\dagger_{\bm{y}})\rangle_\y$ or any other combination on the
r.h.s. of this ``pre-BK equation.'' In any case, $H_{\text{JIMWLK}}$ as
  given in~\eqref{eq:JIMWLK-Hamiltonian-2} allows to arrive at these
  equations most efficiently.

\subsection{\it General properties of JIMWLK and the fixed 
  point at infinite energy}
\label{sec:gener-prop-jimwlk-fixed}

There are a number of quite general properties which have been shown
to hold already on the JIMWLK level~\cite{Weigert:2000gi}. The most
generic of these concerns the interpretation of the JIMWLK equation in
a probabilistic sense. Due to the fact that the r.h.s. can be viewed
as a total derivative (c.f. Eq.~\eqref{eq:finalFP}), or equivalently
by looking at Eq.~\eqref{eq:OU-evo} with $O[U]=1$, one finds that
``probabilities'' (the normalization of $\Hat Z_\y[U]$) are conserved:
\begin{equation}
  \label{eq:norm-cons}
  \partial_\y \int\Hat D[U] Z_\y[U] = 0
\ .
\end{equation}
In~\cite{Weigert:2000gi} it has been shown that the Fokker-Planck
Hamiltonian in the JIMWLK equation is positive semidefinite. This
leads to stable asymptotic behavior: a finite evolution step is given
by
\begin{equation}
  \label{eq:finite-evol}
  \Hat Z_\y[U] = e^{-H_{\mathrm{JIMWLK}} (\y-\y_0)} \Hat Z_{\y_0}[U]
\ ,
\end{equation}
which, given positivity, means that evolution slowly erases high lying
modes.  The lowest lying mode, the solution with zero eigenvalues, on
the other hand, is simply $\Hat Z_\y[U]=1$, as the Fokker-Planck
Hamiltonian Eq.~\eqref{eq:JIMWLK-Hamiltonian-2} acts with at least one
derivative. This solution acts therefore as the fixed point at
$\y=\infty$ and carries the (conserved) normalization of all these
``states.'' This state then acts as a fixed point of evolution at
infinite energy. In this situation ($\Hat Z_\y[U]=1$) all there is to
define correlators is the Haar measure, which acts locally in each
point in transverse space. A first implication is that the correlation
length in this situation is strictly zero. A second one is the fact
that, via Eq.~\eqref{eq:finite-evol} the system moves toward this
solution -- in RG parlance this solution acts as an attractive fixed
point.  This is fully in line with the physics expectations laid out
in the introduction: going to higher and higher energy adds more and
more gluons of the same size, increases density and necessarily leads
to a shrinking correlation length. This element of the physics picture
then is already visible on the abstract level of positivity of the
spectrum and existence of an attractive fixed point in this evolution
equation.

\subsection{\it Geometric scaling in BK and JIMWLK -- the idea}
\label{sec:it-geometric-scaling-eqn-idea}

At this point one is in a position to make first contact with the
scaling idea of Golec-Biernat and W{\"u}sthoff that provides such an
efficient parametrization of the small $\xbj$ HERA data. The argument is
due to Iancu, Itakura, and McLerran~\cite{Iancu:2002tr}. They assume
(in the context of the fixed coupling BK equation just derived) that
$N_{\y\ \bm{x y}}$ depends only on the combination
$(\bm{x}-\bm{y})^2 Q_s(\y)^2=:\bm{r}^2 Q_s(\y)^2 $.  Then
$\y$-derivatives can be traded for $\bm{r}^2$ derivatives and the
l.h.s.\ of the BK equation is easily rewritten as
\begin{equation}
    \label{eq:BK-lhs-scaling}
\partial_\y N({\bm{r}^2 Q_s(\y)^2}) =
  \bm{r}^2 \partial_{\bm{r}^2} N({\bm{r}^2 Q_s(\y)^2}) \partial_\y
  \ln Q_s(\y)^2 =: \bm{r}^2 \partial_{\bm{r}^2} N({\bm{r}^2
    Q_s(\y)^2}) 2 \lambda(\y) \ ,    
\end{equation}
where the last equality defines a new quantity, the evolution rate
$\lambda(\y)=\partial_\y\ln Q_s(\y)$.  Taking into account the
spatial boundary conditions imposed by ``color
transparency+saturation,'' i.e. the fact that $N(\bm{r}^2=0)=0$ and
$N(\bm{r}^2=\infty)=1$, allows to take the zeroth moment of the BK
equation --to integrate with $\frac{d^2r}{\bm{r}^2}$--~in order to
isolate $\lambda$ on the l.h.s.. The r.h.s then provides an integral
expression for it in form of its zeroth moment:
\begin{equation}
    \label{eq:lambda-integral-def}
    2 \pi \lambda(\y) = \frac{\alpha_s N_c}{2\pi^2} \int 
    \frac{d^2r d^2z }{\bm{u}^2\bm{v^2}}\big( N(\bm{u}^2Q_s^2) 
      + N(\bm{v}^2Q_s^2) - N(\bm{r}^2Q_s^2) 
      - N(\bm{u}^2Q_s^2) N(\bm{v}^2Q_s^2)
      \big)
\end{equation}
where I have set $\bm{u}=\bm{x}-\bm{z}$ and $\bm{v}=\bm{z}-\bm{y}$ for
compactness. Due to scale invariance of the integral, the r.h.s.\ is
indeed independent of $\y$: the scale common to all $N$ can be chosen
at will without changing the integral, $\lambda$ is constant.  Using
the definition of $\lambda$ in Eq.~\eqref{eq:BK-lhs-scaling}, this
$\y$ independence of $\lambda$ immediately implies that $Q_s$ grows like
a power in $\xbj$:
\begin{equation}
  \label{eq:Qspower}
  Q_s(\y)= e^{\lambda(\y-\y_0)} Q_s(\y_0) 
  = \Big(\frac{\xbj_0}{\xbj}\Big)^\lambda  Q_s(\xbj_0)
\ .
\end{equation}
This simple power like growth induced by constant $\lambda$ only holds
at fixed coupling. Running coupling effects render $\lambda$ $\y$
dependent and thus modify the $\y$ dependence of $Q_s$ with the
general trend to slow down evolution. These effects will be discussed
in detail later on. For the moment, let me return to the implications
of scaling features in small $\xbj$ evolution equations as such: if
such scaling is to occur, the shape of the solution to the BK equation
is strongly constrained -- one has to simultaneously satisfy the
scaling condition and the evolution equation. These combined
requirements should be sufficient to completely fix the shape of the
scaling solution in $\bm{r}$, a conclusion that has been confirmed by
numerical work again and again.

The above argument only applies {\em if} the scaling situation is ever
reached in evolution. Recently, Munier and
Peschanski~\cite{Munier:2003vc,Munier:2003sj,Munier:2004xu} have
argued analytically that this feature should be attractive and emerge
quite generally as soon as the ``color transparency+saturation''
condition is met on the initial condition. This was done using an
approximation that maps the BK equation onto equations of the type of
the Fisher and Kolmogorov-Petrovsky-Piscounov (KPP)
equation~\cite{FISHER,KPP} which are known to posses ``traveling
wave'' solutions which in this context appears as scaling. While
numerical work on the BK equation had demonstrated this earlier on for
certain initial conditions~\cite{Braun:2000wr, Kimber:2001nm,
  Armesto:2001fa, Levin:2001et, Lublinsky:2001bc, Lublinsky:2001yi,
  Gotsman:2002yy, Golec-Biernat:2001if, Albacete:2003iq} this is a
nice theoretical confirmation despite the need for an additional
approximation.

This has very important implications: {\em If} such a regime is
reached, the situation is {\em qualitatively} different from the
situation encountered, say, in the case of DGLAP evolution. There,
features of the initial conditions and the noncalculable
nonperturbative information contained therein will be slowly washed
out but never completely forgotten. In the scaling regime everything
is determined by the (perturbatively calculated) evolution equation --
the nonperturbative input of the initial condition is completely
forgotten in the sense that it is sufficient to state that scaling is
reached and that the saturation scale has a given value to fully
determine the system at this and all further $\y$. The initial
condition will be visible only {\em before} one enters the scaling
regime and will determine the $Q_s$ value at which scaling initially
emerges, but has no further influence on the evolution {\em beyond}
that point.  That this behavior is characteristic for this type of
evolution equation, also in the more general JIMWLK case, has been
explored extensively in numerical
simulations~\cite{Rummukainen:2003ns} and will be further discussed in
Sec.~\ref{sec:numer-results}.

The discussion above applies to the BK equation, but it can be equally
applied to the JIMWLK equation: With the scaling dimension of the $U$
fields at $1$, all the above scaling arguments go through, if one
assumes {\em all} correlators of $U$ fields to carry $\y$-dependence
only via $Q_s(\y)$. In principle this is not impossible and might
emerge from evolution in a natural way, but it is easy to envision
that it is at least violated in a particular physics situation via the
initial condition to evolution. This should then be seen in
experimental data over a certain range of $\xbj$ values.

The experimental implications are also striking. As an example
consider the DIS cross section of Eq.\eqref{eq:dipole-cross}. If the
dipole cross section depends on $\y$ via $Q_s$ only, one should expect
the DIS cross section to scale with $Q/Q_s(\y)$ as discussed in
Sec.~\ref{sec:geometric-scaling-exp}. This is the phenomenon
originally called geometric scaling~\cite{Stasto:2000er}.

\subsection{\it From JIMWLK and BK to BFKL, large $N_c$  and density
expansions}
\label{sec:jimwlk-to-bk}

Another important result is the statement that the BFKL equation
emerges as the small density limit of the JIMWLK (and as a corollary,
also of the BK) equation. This was first demonstrated
in~\cite{Jalilian-Marian:1997jx}. The argument involves an expansion
of the dipole correlators $\langle N_{\bm{x y}} \rangle_\y $ in powers
of the gluon field and thus in terms of gluon densities. As a
consequence the nonlinear terms in Eq.~\eqref{eq:dipoleOP} become
negligible, and it turns out that Fourier-transforming the remainder
leads directly to the BFKL equation.  To state precisely what the
expansion parameter for this low density limit may be, it is
sufficient to consider the ansatz~\eqref{eq:Gaussian-A} and to
rephrase the BK equation in terms of ${\cal G}$. Since ${\cal G}$ is a
two gluon correlator, it is then easy to see that the low density
expansion can be made precise as an expansion in powers of ${\cal G}$.
The name density expansion is justified in the sense that ${\cal G}$
also fully determines the gluon density as seen from
Eq.~\eqref{eq:gluedist}, but one has to remember that large ${\cal G}$
does not literally translate into large gluon densities.

The evolution I want to consider is in fact the evolution equation for
the two point function before factorization into the BK form proper,
i.e. Eq.~\eqref{eq:pre-BK-S}, which I write explicitly as
\begin{equation}
  \label{eq:dipole-to-higher}
  \partial_\y \langle \tr( U_{\bm{x}} U^\dagger_{\bm{y}}) \rangle = 
  \frac{\alpha_s}{2\pi^2} \int\!\!d^2z\  \Tilde {\cal K}_{\bm{x z y}} \Big(
  -2C_{\text{f}}\ \langle \tr(U_{\bm{x}}U^\dagger_{\bm{y}}) \rangle +
  \langle \big[\Tilde U_{\bm{z}}\big]^{a b}
  2\tr(t^a U_{\bm{x}}t^b U^\dagger_{\bm{y}}) \rangle 
   \Big)
\ .
\end{equation}
With the two point averages already known from
Eq.~\eqref{eq:simple-corr-1} all that is left to calculate is the
three point function. This can be calculated along the same lines as
the various two point functions. One finds
\begin{equation}
  \label{eq:simple-3-point}
  \langle [\Tilde U_{\bm{z}}]^{a b}\ 2 \tr( t^a U_{\bm{x}} t^b U^\dagger_{\bm{y}})\rangle 
  = 2 N_c C_{\text{f}} e^{-\frac{1}{2} \big(
      N_c({\cal G}_{\y,{\bm{x z}}} +{\cal G}_{\y,{\bm{y z}}})
      - \frac{{\cal G}_{\y,{\bm{x y}}}}{N_c}\big)}
\ .
\end{equation}
Instead of an explicit construction let it suffice here that this
result satisfies all necessary limiting relations. Using, where needed,
the Fierz identity~\eqref{eq:Fierz}
%\begin{equation}
%  \label{eq:coupling-fierz}
%    \big[\Tilde U_{\bm{z}}\big]^{a b}\
%  2 \tr(t^a U_{\bm{x}}t^b U^\dagger_{\bm{y}}) =
% \tr( U_{\bm{x}}U^\dagger_{\bm{z}})
%           \tr( U_{\bm{z}}U^\dagger_{\bm{y}})
%           -\frac{1}{N_c}\tr( U_{\bm{x}}U^\dagger_{\bm{y}})
%\end{equation}
one gets coincidence limits for the fields that have to be respected
also for their correlators:
\begin{subequations}
  \label{eq:concidence}
\begin{align}
  \lim\limits_{y\to z} \big[\Tilde U_{\bm{z}}\big]^{a b}\
  2 \tr(t^a U_{\bm{x}}t^b U^\dagger_{\bm{y}}) = & 
  2 C_{\text{f}} \ \tr( U_{\bm{x}}U^\dagger_{\bm{z}})
\\
  \lim\limits_{y\to x} \big[\Tilde U_{\bm{z}}\big]^{a b}\
  2 \tr(t^a U_{\bm{x}}t^b U^\dagger_{\bm{y}}) = & 
  \Tilde\tr( \Tilde U_{\bm{x}} \Tilde U^\dagger_{\bm{z}})
\\
  \lim\limits_{x,y\to z} \big[\Tilde U_{\bm{z}}\big]^{a b}\
  2 \tr(t^a U_{\bm{x}}t^b U^\dagger_{\bm{y}}) = & 2 N_c C_{\text{f}}
\ .
\end{align}  
\end{subequations}
Obviously Eqns.~\eqref{eq:simple-corr} and~\eqref{eq:simple-3-point}
are compatible with these requirements.

Incidentally this truncates the nonlinear equation for the correlators
of $U$ to a single, highly nonlinear equation for $ {\cal G}$.  Using
the above, Eq.~\eqref{eq:dipole-to-higher} reads
\begin{equation*}
  %\label{eq:tilde-G-evo}
   \partial_\y  e^{-C_{\text{f}}{\cal G}_{\y,\bm{x y}}} 
   = \frac{\alpha_s}{2\pi^2} \int\!\!d^2z\  \Tilde {\cal K}_{\bm{x z y}} \Big(
   -2C_{\text{f}} e^{-C_{\text{f}}  {\cal G}_{\y,\bm{x y}}}
   +2  C_{\text{f}} e^{-\frac{1}{2} \big(
      N_c({\cal G}_{\y,{\bm{x z}}} +{\cal G}_{\y,{\bm{y z}}})
      - \frac{{\cal G}_{\y,{\bm{x y}}}}{N_c}\big)}
   \Big)
\end{equation*}
or, equivalently,
\begin{equation}
  \label{eq:tilde-G-evo-short}
 \partial_\y {\cal G}_{\y,\bm{x y}}  =  \frac{\alpha_s}{\pi^2} \int\!\!d^2z\  
 \Tilde {\cal K}_{\bm{x z y}} \Big(
 1-  e^{-\frac{ N_c}{2} \big(
 {\cal G}_{\y,{\bm{x z}}} +{\cal G}_{\y,{\bm{y z}}}
 - {\cal G}_{\y,{\bm{x y}}}\big)}
\Big)
\ .
\end{equation}

Since one involves only planar diagrams in the construction of the
correlators due to locality of the weight in $\y$, one might have
reached this result also via a different route: One rewrites the
bracketed expression on the r.h.s of \eqref{eq:dipole-to-higher}
once more using the Fierz identity~\eqref{eq:Fierz}
%\begin{align*}
%-2 C_{\text{f}}\tr(U_{\bm{x}}U^\dagger_{\bm{y}})
%+2
%  \big[\Tilde U_{\bm{z}}\big]^{a b}
%  \tr(t^a U_{\bm{x}}t^b U^\dagger_{\bm{y}})
%& =
%\tr( U_{\bm{x}}U^\dagger_{\bm{z}})
%           \tr( U_{\bm{z}}U^\dagger_{\bm{y}})
%           -N_c\tr( U_{\bm{x}}U^\dagger_{\bm{y}})
%\end{align*} 
to map it onto the form given already in~\eqref{eq:dipoleOP}. Now one
may factorize the nonlinearity in $\Hat N$ in the large $N_c$ spirit to
obtain the BK equation proper.  Then one only needs to introduce an
exponential parametrization of the form~\eqref{eq:simple-corr-1}, {\em
  without any reference to the Gaussian weight of
  Eq.~\eqref{eq:Gaussian-A} at all} and approximate $C_{\text{f}}\to
N_c/2$ to arrive again at~\eqref{eq:tilde-G-evo-short}. This
highlights the link of the BK equation with the large $N_c$ expansion
quite clearly and demonstrates that~\eqref{eq:tilde-G-evo-short} is
fully equivalent to the more common forms of the BK equations shown
earlier.

It is important to note that at this point one has strictly speaking
left the realm of the original MV model with its assumption of
uncorrelated scattering centers in longitudinal direction. Aside from
the initial condition, ${\cal G}$, as it emerges as a solution
to~\eqref{eq:tilde-G-evo-short}, will no longer scale with $A^{1/3}$.
This will be discussed further in Sec.~\ref{sec:evol-a-depend}.

For small $ {\cal G}$~\eqref{eq:tilde-G-evo-short} immediately reduces
to the coordinate space BFKL equation in the form
\begin{equation}
  \label{eq:tilde-G-BFKL}
 \partial_\y {\cal G}_{\y,\bm{x y}}  =  \frac{\alpha_s N_c}{2\pi^2} \int\!\!d^2z\  
 \Tilde {\cal K}_{\bm{x z y}}  \big(
 {\cal G}_{\y,{\bm{x z}}} +{\cal G}_{\y,{\bm{y z}}}
 - {\cal G}_{\y,{\bm{x y}}}\big)
\ .
\end{equation}
That this equation indeed relates to the better kown momentum
space version requires a little work, but the correspondence is
expected: it has been demonstrated for the forward limit of the JIMWLK
and BFKL equations already in~\cite{Jalilian-Marian:1997jx}. Recently,
Bartels, Lipatov and Vacca~\cite{Bartels:2004ef} have shown explicitly
that the lowest order part of~\eqref{eq:tilde-G-evo-short} corresponds
to the full non-forward BFKL equation (at leading order in
$\ln(1/\xbj)$).  Since this equation is usually written for a gluon
4-point function, this requires an explanation.

In this low density limit, all that Eq.~\eqref{eq:tilde-G-evo-short}
does is to iterate the BFKL variant of what is called reggeized two
gluon exchange between projectile and target. This allows to draw both
sides more symmetrically than in Eqns.~\eqref{eq:G-CalG-diag}
and~\eqref{eq:tilde-G-diag} and to expose the two coordinates
(labelled $\bm{u}$ and $\bm{v}$ below) at which the exchanged pair of
gluons couples to the target. This corresponds to the step from a the
solution of a Bethe Salpeter type equation to the corresponding Greens
function. This latter object will be denoted ${\cal G}_{\y;\ \bm{x
    y},\bm{u v}}$, with Fourier-transform
$\delta^{(2)}({\bm{q}}-{\bm{q}}')\Phi_{\y}({\bm{k}},{\bm{k}}';{\bm{q}}')$.
Diagrammatically, this is written as
\begin{align}
  \label{eq:calG-4-coords}
   {\cal G}_{\y;\ \bm{x y},\bm{u v}} = 
   \parbox{3cm}{ \includegraphics[height=2.5cm]{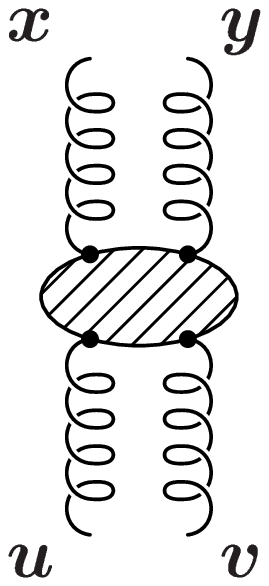}}
\hspace{1cm}
\Phi_{\y}({\bm{k}},{\bm{k}}';{\bm{q}}')= 
\parbox{3cm}{ \includegraphics[height=2.5cm]{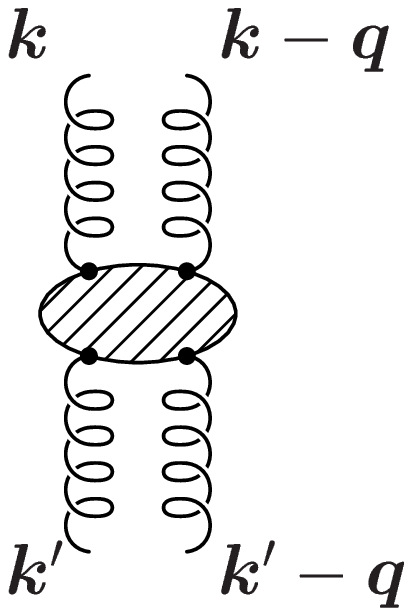}}
\ .
\end{align}
[The definition in terms of underlying gluon correlators now contains
local subtractions of the type shown in~\eqref{eq:tilde-G-diag} both
for top and bottom coordinates.] Explicitly, with this assignment of
variables
\begin{align}
  \label{eq:BFKL-Fourier}
  \delta^{(2)}({\bm{q}}-{\bm{q}}')\Phi_{\y}({\bm{k}},{\bm{k}}';{\bm{q}}')
=\int d^2 x\,  d^2 y\, d^2 u \, d^2 v\
%\phantom{xxxxxxxxxxxxxxxxxxxxxxxxxxxxx}
%\nonumber \\
%\cdot  
{\cal G}_{\y;\ \bm{x y},\bm{u v}} \
e^{i{\bm{k}}{\bm{x}}+i({\bm{q}}-{\bm{k}}){\bm{y}}-i{\bm{k}}'{\bm{u}}
  -i({\bm{q}}'-{\bm{k}}'){\bm{v}}}
\end{align}
and to make contact with the standard momentum space formulation
reduces to a simple exercise in Fourier transformation.

The connection between the  result of these manipulations and the
conformal formulation in coordinate space is well documented in the
literature\footnote{See for example~\cite{Lotter:1996vk}, for a
  pedagogical presentation.} so that taking this route might be easier
to the non-expert than the direct calculation
of~\cite{Bartels:2004ef}. Only by using conformal symmetry tools is it
possible to write complete solutions to the nonforward BFKL equation
-- but this is outside the scope of this review.

While the authors in~\cite{Bartels:2004ef} arrive at the BFKL limit in
the usual way, by simply dropping the nonlinearity in
Eq.~\eqref{eq:dipoleOP}, and hence refer to $N$ as the object of their
evolution equation\footnote{Note that both $N$ and ${\cal G}$ vanish
  in the local limit, a fact that is called the Moebius representation
  in~\cite{Bartels:2004ef}. } their calculations make it clear that
they discuss a two gluon object coupling to the photon wavefunction or
``impact factor.'' This object has to be identified with ${\cal G}$ as
done above.

The form of the nonlinearity in Eq.~\eqref{eq:tilde-G-evo-short} is
intriguing in its exponential nature. This directly corresponds to the
eikonal nature of the multiple scattering induced by the presence of
complete Wilson lines in the definition of correlators $S_{\y; {\bm{x
      y}}}$ or $N_{\y; {\bm{x y}}}$.  This becomes important as soon
as one attempts to go beyond leading order in density, as becomes
necessary if one tries to step beyond the large $N_c$ limit. For
example~\cite{Bartels:2004ef} aims at providing a $1/N_c$ correction
to the BK equation. They find a contribution which contains the square
of $\frac{ N_c}{2} \big( {\cal G}_{\y,{\bm{x z}}} +{\cal G}_{\y,{\bm{y
      z}}} - {\cal G}_{\y,{\bm{x y}}}\big)$, the factor in the
exponent of the ($N_c$-leading)
expression~\eqref{eq:tilde-G-evo-short}. If the eikonalization feature
of JIMWLK and BK equations is taken seriously, this can not be done in
a way consistent with the density expansion without also including
higher order contributions from the nonlinearity on the r.h.s.
of~\eqref{eq:tilde-G-evo-short}. At least for large nuclei this
becomes an important consideration.

\subsection{\it Growth in BFKL and BK, the 
Mueller-Triantafyllopoulos description}
\label{sec:growth-bfkl}

To identify BFKL as the low density limit of JIMWLK and BK implies
that linear BFKL evolution is present in the more general framework
and should manifest itself in those areas of phase space in which
densities are not yet large. BFKL is known for  fast and unlimited
growth of its solutions, the ${\cal G}$ or $\Phi$ above. The role of
the nonlinear effects is clearly to tame this growth and to ensure
that the objects of evolution, $S_{\y;{\bm{x y}}}$ or $N_{\y; {\bm{x
      y}}}$ for example, with their interpretation as correlators of
eikonal factors $U_{\bm{x}}^{(\dagger)}$ stay bounded between $0$ and
$1$.\footnote{In the case of the BK equation, which can be written
  without any reference to this interpretation via eikonal factors,
  this boundedness has to be a property of the evolution equation
  itself. \cite{Banfi:2002hw} has given an argument in the context of
  the Banfi-Marchesini-Smye (BMS) equation, why this is the case.
  Their argument directly translates to the BK case due to a structural
  analogy of the BMS and BK equations that will be discussed in more
  detail in Sec.~\ref{sec:jimwlk-soft-gluon}.}

Mueller and Triantafyllopoulos~\cite{Mueller:2002zm,
  Triantafyllopoulos:2002nz} have provided a very elucidating
perspective on this matter that employs BFKL as the driving force
behind all growth at small $\xbj$ which turns out to be surprisingly
successful. The ideas entering this model naturally incorporate the
notion of a scaling window~\cite{Iancu:2002tr} -- a region in phase
space at distances shorter than $1/Q_s$, i.e. away from the high
density limit, at which the presence of $Q_s$ exerts its influence.

As indicated, one cornerstone of these ideas is the strong growth and
diffusive character of BFKL evolution which can be most easily
understood in the forward case (i.e. the above equation at $q=0$). At
$q=0$ the above equations can be diagonalized by power functions
\begin{equation}
\label{eq:eigenzero}
e^{(\nu,n)}({\bm{k}}) = 
2\pi \sqrt{2} \,
({\bm{k}}^2)^{-\frac{3}{2}
-i\nu} e^{-i n \phi}\phantom{xx};\;\nu \in {\mathbb{R}} , n\in {\mathbb{Z}}
\end{equation}
in momentum- as well as in coordinate space. The eigenvalues are given
by
\begin{equation}
\chi(\nu,n):=\frac{\alpha_s N_c}{\pi}\left[2\psi(1)-\psi(\frac{1+|n|}{2}+i\nu)
                               -\psi(\frac{1+|n|}{2}-i\nu)\right]
\ ;\hspace{2cm}
\chi(\nu):=\chi(\nu,0)
\label{eq:chi-BFKL}
\end{equation}
with $\psi$ defined as the logarithmic derivative of the
$\Gamma$-function ($\psi(x)=\Gamma'(x)/\Gamma(x)$) and $\chi(\nu)$ as
a shorthand for $\chi(\nu,0)$. The solution then reads
\begin{equation}
\Phi_{\y}({\bm{k}},{\bm{k}}')=
 \sum_{n=-\infty}^{+\infty}
\int_{-\infty}^{+\infty} \frac{d \nu}{2 \pi}
\  e^{(\nu,n)}({\bm{k}}) \   
e^{\y\chi(\nu,n)}\ e^{(\nu,n)\,\ast}({\bm{k}}')
\ .
\label{eq:solzero}
\end{equation}

Of particular interest is the large $\y$ asymptotics of this
equation. There the contributions will be dominated by the modes with
$n=0$, shown here in the coordinate space version to underline the
analogy~\cite{Balitsky:1997mk}:
\begin{equation}
  \label{eq:N-BFKL-sol}
  {\cal G}_{\y; \bm{x y}, \bm{u v}} = \int \frac{d\nu}{2\pi^2} %d^2r_0
  \
  ((\bm{x}-\bm{y})^2)^{-\frac{1}{2}+i\nu} 
  e^{\y\chi(\nu) %+i \nu \ln(\bm{x}-\bm{y})^2/\bm{r}_0^2
  } 
 ((\bm{u}-\bm{v})^2)^{-\frac{1}{2}-i\nu} 
 % \ {\cal G}_{\y_0,\bm{r}_0}
= \int \frac{d\Tilde\nu}{2\pi^2 i} e^{\y \chi(\Tilde\nu) - (1-\Tilde\nu)(\rho-\rho')}
\end{equation}
where $\rho:=\ln(\bm{x}-\bm{y})^2 \mu^2$ and
$\rho':=\ln(\bm{u}-\bm{v})^2 \mu^2$. The integration contour in the
$\nu$ (or Mellin-) integral is to be taken parallel to the imaginary
axis, to the right of all singularities; in the last version I have
substituted $\Tilde \nu := i\nu$ with an appropriate change in
definition of $\chi$.

Further insight emerges by taking the saddle point approximation to in
Eq.~\eqref{eq:solzero} or~\eqref{eq:N-BFKL-sol}.  \footnote{The result
  does not depend on $\mu$.} This selects $\nu_0=0$ as the extremal
value in the $\nu$ integration. The result displays a factorizing
structure,
\begin{align}
  \label{eq:BFKL-saddle-fact}
  e^{\y\chi({\Tilde\nu}_0)}
%e^{(\rho-\rho')(1-{\Tilde\nu}_0)} 
  \int \frac{d{\Tilde\nu}}{2\pi^2 i} e^{\y\frac{1}{2}\chi''({\Tilde\nu}_0)({\Tilde\nu}-{\Tilde\nu}_0)^2
    +({\Tilde\nu}-{\Tilde\nu}_0)(\rho-\rho')}
= e^{\y\chi({\Tilde\nu}_0)}
%\left(\frac{(\bm{x}-\bm{y})^2}{(\bm{u}-\bm{v})^2}\right)^{1-{\Tilde\nu}_0} 
\frac{1}{\sqrt{\chi''({\Tilde\nu}_0)\y}} 
e^{-\frac{(\rho-\rho')^2}{\chi''({\Tilde\nu}_0)\y} } 
\end{align}
where the  first factor, $e^{\y\chi({\Tilde\nu}_0)} =
\left(\frac{\xbj_0}{\xbj}\right)^{\chi({\Tilde\nu}_0)}$, describes the
strong, power like growth in $1/\xbj$ alluded to already in the
introduction. The second factor, which originates from the Gaussian
integral in ${\Tilde\nu}$, describes diffusion in the transverse plane
of dipole sizes:
\begin{equation}
  \label{eq:psidiffdef}
\psi_\y(\rho-\rho') :=\frac{1}{\sqrt{\chi''({\Tilde\nu}_0)\y}}\ 
e^{-\frac{(\rho-\rho')^2}{2 \chi''({\Tilde\nu}_0)\y} }   
\end{equation}
is immediately recognizable as the solution to the diffusion equation
\begin{align}
  \label{eq:diff-transv}
  \partial_\y \psi_\y(\rho) = \frac{\chi''({\Tilde\nu}_0)}{2}
  \partial_\rho^2\psi_\y(\rho)
\end{align}
with initial condition $ \psi_0(\rho)= \delta(\rho)$.  The same
picture holds in momentum space as well. This is the well known IR
problem of the BFKL equation at one loop accuracy: even an initially
short range configuration spreads to nonperturbatively large
distances. Since the main focus here is in nonlinear effects I will
refrain from a discussion of two loop corrections to BFKL.

This diffusion and growth now has to be contrasted with what happens
in JIMWLK and BK evolution. The nonlinearity tames the growth and, at the
same time, prevents diffusion into the infrared. At small ${\cal G}$,
though, evolution is correctly described by BFKL dynamics.

Mueller and Triantafyllopoulos therefore have suggested to simulate
the presence of the nonlinear term by an absorptive barrier in the
above diffusion process that is to be self consistently adjusted such
that the solution to the so modified BFKL equation never grows large.
The result can be interpreted as a change of the dominant
contributions to the integral in, say~\eqref{eq:N-BFKL-sol}, in such a
way that one now follows the saddle point but allows diffusion only
from small dipoles at size ranges where ${\cal G}$ is not yet large.
Creation of new objects in the already saturated domain is eliminated
via absorption across the boundary.

In order to explore the first part of this argument and to understand
how to choose the absorptive boundary, one can look for directions in
$\y-Q$ phase space in which BFKL evolution remains bounded by
%This may be (approximately) achieved by 
simultaneously looking for a saddle point and $\y$ dependent starting
sizes $\rho'(\y)$, for which the exponent in Eq.~\eqref{eq:N-BFKL-sol}
vanishes and thus ${\cal G}$ remains approximately
constant~\cite{Mueller:2002zm, Iancu:2002tr}. The conditions to
satisfy in this case are
\begin{subequations}
    \label{eq:muellercond}
  \begin{align}
\y \chi'(\Tilde\nu_c) + (\rho-\rho'(\y)) = 0 & 
\hspace{1cm} \text{saddle point} \\
\label{eq:almost-const-G}
\y \chi(\Tilde\nu_c) + (1-\Tilde\nu_c)(\rho-\rho'(\y)) = 0 & 
\hspace{1cm} \text{(almost) const. } {\cal G} 
  \end{align}
\end{subequations}
with solutions
\begin{subequations}
    \label{eq:TM-sols}
  \begin{align}
\frac{\chi'(\Tilde\nu_c)}{\chi(\Tilde\nu_c)} = & \frac{1}{1-\Tilde\nu_c} \\
\rho-\rho'(\y) = & \frac{\chi(\Tilde\nu_c)}{ 1-\Tilde\nu_c } \y 
\ . % \nonumber
\intertext{(The saddle point value $\Tilde\nu_c$ remains implicit, but 
  changes slightly from the BFKL value $\Tilde\nu_0$.) Translating the 
  latter back into distances $\rho =:\ln(r^2\mu^2)$ and 
  $\rho'(\y)=:\ln R_0^2(\y)\mu^2$ one finds}
R_0^2(\y) = & r^2 e^{- \frac{\chi(\Tilde\nu_c)}{ 1-\Tilde\nu_c } \y }
\ .
  \end{align}
\end{subequations}
Using this as a saddle point for the integral in ${\cal G}$ one
finds
%\footnote{A similar scaling form with a leading
%  $(Q_s(\y)/Q)^{1-\Tilde\nu_0}$ 
%  had been given earlier in\cite{Iancu:2002tr}.}
\begin{align}
  \label{eq:MT-saddle}
  {\cal G}_{\y; r, R_0(\y)} \approx & 
\left(\frac{r^2}{R_0(\y)}\right)^{(1-{\Tilde\nu}_c)} 
  \int \frac{d{\Tilde\nu}}{2\pi^2 i} e^{\y\frac{1}{2}
    \chi''({\Tilde\nu}_c)({\Tilde\nu}-{\Tilde\nu}_c)^2
    +({\Tilde\nu}-{\Tilde\nu}_c)\ln\left(\frac{r^2}{R_0(\y)}\right)}
\nonumber \\ = & 
\left(\frac{r^2}{R_0(\y)^2}\right)^{(1-{\Tilde\nu}_c)} 
%\left(\frac{(\bm{x}-\bm{y})^2}{(\bm{u}-\bm{v})^2}\right)^{1-{\Tilde\nu}_0} 
\frac{e^{-\frac{(\rho-\rho')^2}{\chi''({\Tilde\nu}_c)\y} }}{\sqrt{\chi''({\Tilde\nu}_c)\y}} 
\ .
\end{align}
A list of observations apply~\cite{Mueller:2002zm}:
\begin{itemize}
\item Within the diffusion radius the dominant factor is
  $\left(r^2/R_0(\y)\right)^{(1-{\Tilde\nu}_c)} $. The
  solution has scaling form, just as argued above for the solutions of
  the JIMWLK and BK equations after some initial settling down. The
  scale variable is $R_0$ which has exponential $\y$ dependence,
  again as required in the BK case. Moreover the coefficient is known
  from Eq.~\eqref{eq:TM-sols}.
\item Apart from a mild correction through the square root factor,
  $R_0(\y)$ describes lines of constant ${\cal G}$. ${\cal G}$ evolved
  in this manner will not grow exponentially. This is because emission
  from large objects has been prohibited in sharp contrast to the
  diffusion of the BFKL equation.
\end{itemize}
Of course the argument can be equally phrased in momentum space which
essentially replaces $r\to 1/Q$ and $R_0\to 1/Q_0$. 

One may remove the nonscaling square root factor by replacing
condition~\eqref{eq:almost-const-G}, with the requirement of truly
constant $\cal G$ and work with the resulting definition of $Q_0 \to
Q_s$ without introducing absorptive boundaries. However, evolution
speed $\lambda =\partial_\y \ln Q_s$, will necessarily change if one
really implements the effect of the nonlinearity in form of the
absorptive boundary. Mueller and
Triantafyllopoulos~\cite{Mueller:2002zm, Triantafyllopoulos:2002nz}
continue to formalize their argument by also including running
coupling effects. In their treatment, the scale for the coupling is
set by the size of the emitting object, the parent ``dipole.'' These
refinements do not change the generic conclusions given above but they
ensure scaling and provide quantitative results for $\lambda$ and
$Q_s$. Due to the approximations made, their results are valid
asymptotically at large $\y$ and contain an unknown constant (called
$\Tilde Y$ below) which is related to the (uncontrolled) small $\y$
details.  I write their result for the evolution as
in~\cite{Rummukainen:2003ns}:
\begin{equation}
  \label{eq:dionlambda}
  \lambda := \partial_\y \ln Q_s(\y) 
  = \frac{.90}{\sqrt{\y+\Tilde Y}} - \frac{0.47}{(\y+\Tilde Y)^{5/6}} 
  + \text{higher inverse powers}
\ .
\end{equation}
This can be integrated to give $Q_s$ at the price of an additional
integration constant $\Tilde c$:\footnote{Alternatively, one could
  have directly used the expressions for $Q_s$
  in~\cite{Mueller:2002zm, Triantafyllopoulos:2002nz} which contain a
  free parameter with the same role.}
\begin{equation}
  \label{eq:dionfit}
  \frac{Q_s(\y)}{\Lambda_{\text{QCD}}} = \Tilde c \cdot 
  \exp[ 2\cdot 0.9 (\y+\Tilde Y)^{\frac{1}{2}} 
  -6\cdot 0.47 (\y+\Tilde Y)^{\frac{1}{6}}]
\ .
\end{equation}
In particular the coefficient of the second term is affected by the
presence of the absorptive boundary. Its value will turn out to be
important in the comparison with numerical simulations shown in
Fig.~\ref{fig:assesing-sqrt-running}.

Clearly running coupling effects do slow down evolution quite
considerably, the leading contribution goes from an exponent linear in
$\y$ to a square root in the exponent but they do not destroy
scaling. The functional form of the diffusive factor in
Eq.~\eqref{eq:MT-saddle} is modified through the correct determination
of lines of constant ${\cal G}$ but functional form of the leading
$\left(Q^2/Q_s(\y)^2\right)^{-(1-{\Tilde\nu}_c)}$ is unaltered
(although ${\Tilde\nu}_c$ changes slightly from the solution
to~\eqref{eq:TM-sols})~\cite{Mueller:2002zm}:
\begin{align}
  \label{eq:MT-scaling-running}
{\cal G}_{\y; Q,Q_s(\y)} 
= C \left(\frac{Q^2}{Q_s(\y)^2}\right)^{-(1-{\Tilde\nu}_c)} 
\left[\ln\frac{Q^2}{Q_s(\y)^2} + \frac{1}{1-{\Tilde\nu}_c}\right]
%\frac{1}{\sqrt{\chi''({\Tilde\nu}_c)\y}} 
%e^{-\frac{(\rho-\rho')^2}{\chi''({\Tilde\nu}_c)\y} } 
\ .
\end{align}
$\Tilde\nu_c$ (or $\Tilde\nu_0$ in pure BFKL) is usually called the
anomalous dimension governing the leading large $k$ behavior of the
unintegrated gluon distribution through~\eqref{eq:gluedist}. This
result is expected to hold in what is called the scaling window which
will be discussed next.

\subsection{\it Scaling above $Q_s$, the existence of 
a scaling window}
\label{sec:scaling-window}

The scaling form of Eq.\eqref{eq:MT-scaling-running} is valid at small
${\cal G}$ which immediately implies distances shorter than $1/Q_s$.
Consequently, one has to ask how far away from $Q_s$ this can be
trusted. This, in fact, is important not only for the approximate
solution discussed above but also as a limitation on the range of
validity of the evolution equations, be it JIMWLK or BK. That this
range is limited as a matter of principle should be clear from the
derivation of these equations, which selects contributions of the form
$\alpha_s \ln(1/\xbj)$ and rejects contributions enhanced by other
logarithms, such as $\alpha_s \ln(Q^2/\mu^2)$ as subleading. If one
looks at the phase space plot of Fig.~\ref{fig:x-Q-plane-density}, one
would naively expect $\ln(1/\xbj)$ corrections to be most important
above the diagonal, while $\ln(Q^2/\mu^2)$ corrections should be most
important below the diagonal, with a region in between in which double
logarithmic corrections of the form $\alpha_s
\ln(1/\xbj)\ln(Q^2/\mu^2)$ are most important. But where can one
safely neglect these other logarithms?

Even before Mueller and Triantafyllopoulos constructed their
scaling solutions,~\cite{Iancu:2002tr} have in fact argued that the
range of validity of the evolution equations and also of scaling
expressions of the type~\eqref{eq:MT-scaling-running} is surprisingly
large. It is expected to extend over a large scaling window defined by
the condition
\begin{equation}
  \label{eq:scaling-window-ineq}
  Q_s^2 \leq Q^2 \leq Q_s^4/\Lambda_{\text{QCD}}^2
\end{equation}
or, more stringently
\begin{equation}
  \label{eq:scaling-window-ineq-stringent}
  1 \lesssim \ln(Q^2/Q_s^2) \ll \ln(Q_s^2/\Lambda_{\text{QCD}}^2)
\ .
\end{equation}
The original argument is based on the diffusion radius of BFKL if one
follows evolution along lines of (almost) constant $\cal G$ as in
Eq.\eqref{eq:MT-saddle}. This sets a natural range for the validity of
scaling formulae like~\eqref{eq:MT-scaling-running}.

Besides the theoretical considerations above, the existence of a
scaling window is also important phenomenologically. The key example
is the G-B+W idea to fit the HERA data with a scaling ansatz. In the
HERA range $Q_s$ comes out to lie in between $1-2$GeV$^2$, while the
momenta entering the fit range extend up to $Q^2$ of several hundred
GeV$^2$.  Thus it is  natural to expect these data to exhibit
scaling features only if a scaling window exists that reaches far
beyond $Q_s$. In a sense, the success of the G-B+W fit directly poses
the question for the existence of such a window whose theoretical
origin was only understood much later in~\cite{Iancu:2002tr}.

% spellmark 3

\section{Fokker-Planck and Langevin%: numerical methods  and ``shower operators''
}
\label{sec:from-fokker-planck}

At this point all of the regions marked in the phase space plot of
Fig.~\ref{fig:x-Q-plane-density} have been fully explained in terms of
a QCD evolution equation although for some of the steps certain
approximations have been used. To go beyond this stage numerical work
is needed. To allow for an efficient numerical treatment of the JIMWLK
equation a reformulation is required. This takes the form of a
translation of the Fokker-Planck equation to a Langevin description.
This is done in Secs.~\ref{sec:illustrating-idea}
and~\ref{sec:numer-results-from}.  Interestingly enough, in this
process one finds theoretical tools that also help in interpreting the
gluon radiation process and directly lead to the diagrammatic
interpretation of Eq.~\eqref{eq:emission}.  This step is performed in
Sec.~\ref{sec:JIMWLK-shower-operators} while the interpretation will
become clear only in Sec.~\ref{sec:jimwlk-soft-gluon} by way of a
surprisingly close analogy with the theory of QCD jets.

\subsection{\it From Fokker-Planck to Langevin: Illustrating the idea}
\label{sec:illustrating-idea}

Although this topic is covered in many textbooks, the presentation is
often somewhat old-fashioned and the intimate connection with
path-integrals is not always mentioned. For this reason I demonstrate
the relationship of Fokker-Planck and Langevin formulations with the
aid of a simple toy example, a particle diffusion problem, that, for
ease of comparison, shares some of the features of our problem. The
toy model equation is
\begin{equation}
  \label{eq:toyFP}
  \partial_\y P_\y(\bm{x}) = - H_{\text{FP}}  \ P_\y(\bm{x})
\end{equation}
with a Fokker-Planck Hamiltonian
\begin{equation}
  \label{eq:toyFPHam}
  H_{\text{FP}} = \frac{1}{2} i\partial_{\bm{x}}^\mu\, 
  \chi_{\mu\nu}(\bm{x})\, i\partial_{\bm{x}}^\nu \ .
\end{equation}
The particle coordinates $\bm{x}$ (in $D$ dimensional space)
correspond to the field variables of the original problem and the
probability distribution $P_\y$ to $\Hat Z_\y$. It serves to define
correlation function of some operator $O(\bm{x})$ via $\langle
O(\bm{x}) \rangle_\y = \int d^Dx\ O(\bm{x}) P_\y(\bm{x})$.

Eq.~(\ref{eq:toyFP}) admits a path-integral solution which is, as
usual, built up from infinitesimal steps in $\y$: this is a solution
$P(\y',\bm{x}';\y,\bm{x})$ with initial condition
\begin{equation}
  \label{eq:toyinitcond}
  P(\y,\bm{x}';\y,\bm{x}) = \delta^{(D)}(\bm{x}'-\bm{x})
\end{equation}
that is composed of many, infinitesimally small steps at times
$\y_k:=\y+k\epsilon$. In this solution,  the product rule
{\small
\begin{equation}
  \label{eq:pathint-sol0}
   P(\y',\bm{x}';\y,\bm{x}) =\int d^Dx_{n-1} \ldots d^Dx_1\ 
   P(\y',\bm{x}';\y_{n-1},\bm{x}) P(\y_{n-1},\bm{x}_{n-1};\y_{n-2},\bm{x}_{n-2})\ldots P(\y_{1},\bm{x}_{1};\y,\bm{x})
\end{equation} 
expresses the finite $\y'-\y$ solution $P(\y',\bm{x}';\y,\bm{x})$ in
terms of infinitesimal steps $P(\y_i,\bm{x}_i;\y_{i-1},\bm{x}_{i
  -1})$. %Solutions with 
Arbitrary initial conditions $
P_{\y_0}(\bm{x})$ are then recovered via $P_\y(\bm{x})= \int d^Dy
P(\y,\bm{x};\y_0,\bm{y}) P_{\y_0}(\bm{y})$.}

The derivation of the infinitesimal step from $\y_{i-1}$ to
$\y_i=\y_{i-1}+\epsilon$ is textbook material. It reads (the index
$i$ on the coordinate also refers to the time step, not to the vector
component)
\begin{equation}
  \label{eq:toyP}
  P(\y_i,\bm{x}_i;\y_{i-1},\bm{x}_{i -1})= 
  N \int\!\!d^Dp_i \ e^{-\epsilon \big( \frac{1}{2}
    \bm{p}_i^\alpha \bm{p}_i^\beta \chi^{\alpha \beta}(\bm{x}_{i-1}) 
    + i \bm{p}_i\cdot 
    \big(\frac{\bm{x}_i-\bm{x}_{i-1}}{\epsilon}-\sigma(\bm{x}_{i-1} )\big)
    \big)}+{\cal O}(\epsilon^2)
\end{equation}
where $\sigma_\mu(\bm{x}) := \frac{1}{2} i\partial_{\bm{x}}^\alpha
\chi_{\alpha\mu}(\bm{x})$ and $N$ is a coordinate independent
normalization factor.

From here, the step to the Langevin formulation is trivial: as this
expression is quadratic in the momenta one can trivially rewrite it
with the help of an auxiliary variable $\bm{\xi_i}$ [$i$ again the
step label] as
\begin{equation}
  \label{eq:toyPlang0}
   P(\y_i,\bm{x}_i;\y_{i-1},\bm{x}_{i-1})= 
   N \int\!\! d^D\xi_i\ \sqrt{\det(\chi(\bm{x}_i))}\
   e^{-\frac{1}{2}\bm{\xi}_i\chi^{-1}(\bm{x}_i)\bm{\xi}_i}\ 
   \delta^{D}\big(\bm{x}_i-[\bm{x}_{i-1}
   +\epsilon(\bm{\xi}_i
   +\sigma(\bm{x}_{i-1})]\big)\ .
\end{equation}
The $\delta^{D}(\ldots)$ arises from the momentum integration and
determines $\bm{x}_i$ in terms of $\bm{x}_{i-1}$ and the correlated
noise $\bm{\xi}_i$.\footnote{Re-expressing the delta function via a
  momentum integral and performing the Gaussian integral over
  $\bm{\xi}_i$ immediately recovers Eq.~\eqref{eq:toyP}} The equation
for $\bm{x}_i$,
\begin{equation}
  \label{eq:toyLang1}
  \bm{x}_i=\bm{x}_{i-1}+\epsilon\big(\bm{\xi}_i+\sigma(\bm{x}_{i-1})\big)\ ,
\end{equation}
is called the Langevin equation. It contains both a deterministic term
[the $\sigma(\bm{x}_{i-1})$ term] and a stochastic term [the
$\bm{\xi}$ term]. To fully define the problem without any reference to
its path-integral nature, one then needs to state the Gaussian nature
of the (correlated) noise separately. This is the formulation often
found in textbooks.  The path-integral version would appear to be much
more elucidating and up to date.\footnote{One of the main sources of
  confusion with stochastic differential equations, the choice of
  discretization [the details of the $x_i$ and $x_{i-1}$ dependence]
  and its impact on the form of the equation itself, finds its natural
  explanation there. It corresponds to the choice of discretization in
  the path integral solution Eq.~(\ref{eq:toyP}). The choice here is
  called the Stratonovich form. One might have as well have given the
  Ito form by choosing a different discretization for
  Eq.~(\ref{eq:toyP}) or yet another version with the same physics
  content.}

If, as in the JIMWLK case, $\chi$ factorizes as
\begin{equation}
  \label{eq:toyfac}
  \chi_{\mu\nu}(\bm{x})={\cal E}_{\mu a}(\bm{x}){\cal E}_{\nu a}(\bm{x})\ ,
\end{equation}
a simple redefinition of the noise variable leads to a version of the
Langevin formulation with a Gaussian {\em white} noise:
\begin{equation}
  \label{eq:toyPlang1}
   P(\y_i,\bm{x}_i;\y_{i-1},\bm{x}_{i-1})= N \int\!\! d^n\xi_i\ 
   e^{-\frac{1}{2}\bm{\xi}_i^2}\ 
    \delta^{D}\big(\bm{x}_i-[\bm{x}_{i-1}+\epsilon({\cal E}(\bm{x}_{i-1})\bm{\xi}_i+\sigma(\bm{x}_{i-1})]\big) \ .
\end{equation}
The correlation is now absorbed into the Langevin equation through the
appearance of ${\cal E}$\footnote{The dimensions $D$ and $n$ of
  configuration space and noise now need not be equal as is
  illustrated by the JIMWLK example. Again, write the
  $\delta$-function as a momentum integral and carry out the Gaussian
  integral in $\bm{\xi}$ to recover Eq.~\eqref{eq:toyP} from
  Eq.~\eqref{eq:toyPlang1}.} which reads
\begin{equation}
  \label{eq:toyLang2}
  \bm{x}_i=\bm{x}_{i-1}+\epsilon\big({\cal E}(\bm{x}_{i-1})\bm{\xi}_i
  +\sigma(\bm{x}_{i-1}) \big)
\ . 
\end{equation}
Wherever this form is available, it will be the most efficient
version to use in a numerical simulation, as the associated white noise
can be generated more efficiently than the correlated noise of the
more general case.

What is left is to sketch a numerical procedure to implement the
solution given in~\eqref{eq:pathint-sol0}, \eqref{eq:toyP}
using the expressions Eq.~\eqref{eq:toyPlang0} or
Eq.~\eqref{eq:toyPlang1}. As a first step one replaces the average with
the probability distribution $P_\y(\bm{x})$ by an ensemble average
\begin{equation}
  \label{eq:toy-ensemble-average}
  \langle O(\bm{x}) \rangle_\y = \int\! d^Dx\  O(\bm{x})\  P_\y(\bm{x}) 
  \approx \frac{1}{N}
  \sum\limits_{U\in {\sf E}[ P_\y]}  O(\bm{x}) 
\end{equation}
where $N$ is the size of the (large) ensemble which is taken according
to the probability distribution $P_\y(\bm{x})$. This one can easily
do for the initial condition at $\y_0$. The Langevin equation then
allows to propagate each ensemble member  in $\y$ by
$\epsilon$, thereby providing  a new ensemble at
$\y_1=\y+\epsilon$. Iteration then creates a chain of ensembles
for a discrete set of $\y_i$ that provide an approximation to a set
of $P_{\y_i}(\bm{x})$ in that they allow one to measure any
correlator $\langle O(\bm{x}) \rangle_\y$ according to
Eq.~\eqref{eq:toy-ensemble-average}.

The above clearly shows that a ``Langevin system'' is a way to write a
path-integral solution to a differential equation that is particularly
easy to implement numerically. Many, often confusing issues, such as
the time step discretization issues, find their natural resolution in
this path integral setting.

\subsection{\it A Langevin formulation for JIMWLK evolution}
\label{sec:numer-results-from}

Translating Fokker-Planck equations to Langevin equations for
numerical work is not only useful in the case of a particle system as
portrayed in the previous subsection. Also in the case of the JIMWLK
equation this is the only promising angle of attack. Besides the
evolution ``time''-discretization needed in any example, it is also
necessary to discretize transverse space to render the number degrees
of freedom finite. Similar methods have been discussed in lattice
gauge theory under the general heading stochastic quantization.

In the present context, this approach has already been suggested
in~\cite{Weigert:2000gi} and a derivation has been given
in~\cite{Blaizot:2002xy} to which I refer the reader for details. To
affect the translation in analogy to the example above, one may
follow~\cite{Weigert:2000gi} and write the JIMWLK equation as
\begin{align}
  \label{RG-new-standard}
  \partial_\y \Hat Z_\y[U] 
%& = \frac{\alpha_s}{2}
%  i\nabla^a_{\bm{x}}\chi^{a b}_{\bm{x y}} i\nabla^b_{\bm{y}} 
%  \Hat Z[U]
% \nonumber \\ 
& =
 -%\alpha_s
 \nabla^a_{\bm{x}}\Big[\frac{1}{2}\nabla^b_{\bm{y}} \Hat\chi^{a b}_{\bm{x y}} 
-\big(\frac{1}{2}\nabla^b_{\bm{y}} 
\Hat\chi^{a b}_{\bm{x y}} \big)\Big]\Hat Z_\y[U]
\end{align}
and define 
\begin{equation}
  \label{eq:newsigdef}
  \Hat\sigma^a_{\bm{x}}:= \frac{1}{2}\nabla^b_{\bm{y}} 
\Hat\chi^{a b}_{\bm{x y}}= - i \big(\frac{1}{2}\frac{\alpha_s}{\pi^2}\int\!\!d^2z 
\frac{1}{({\bm{x}}-{\bm{z}})^2} \Tilde \tr( \Tilde t^a \Tilde U_{\bm{x}}^\dagger  
\Tilde U_{\bm{z}})\Big)
\end{equation}
(the ``$\sim$'' indicating adjoint matrices and traces). 

One then abandons the description in terms of weight functionals $\Hat
Z_\y$ in favor of one in terms of ensembles of fields which are
governed by the corresponding Langevin equations.

Explicitly, to calculate any observable
$O[U]$ of the fields $U$ one  writes
\begin{equation}
  \label{eq:ensemble-average}
  \langle O[U] \rangle_\y = \int\!\Hat D[U] O[U] \Hat Z_\y[U] 
  \approx 
  \frac{1}{N} \sum\limits_{U\in {\sf E}[\Hat Z_\y]}  O[U]
\end{equation}
where, separately at each $\y$, the sum is over an ensemble ${\sf
  E}[\Hat Z_\y]$ of $N$ configurations $U$ whose members were
created randomly according to the distribution $\Hat Z_\y$. Clearly,
for $N\rightarrow\infty$, the ensemble and $\Hat Z_\y$ contain the
same information.

The Langevin equation then schematically\footnote{This is a continuous
  time version of the equation that strictly speaking is not unique.
  The path-integral derivation shown in~\cite{Blaizot:2002xy} and
  maybe more transparently in the appendix
  of~\cite{Rummukainen:2003ns} makes it clear that we are to take a
  ``retarded'' prescription here in which the derivative on the l.h.s\ 
  is taken as a finite difference and the fields on the r.h.s.\ are
  determined at the previous time step.}  reads
\begin{equation}
  \label{eq:Langevin}
  \partial_\y\, [U_{\bm{x}}]_{i j} 
%  = [U_{\bm{x}} i t^a]_{i j} 
%  \Big[\omega^a_{\bm{x}}+\alpha_s\Hat\sigma^a_{\bm{x}}\Big]
  = [U_{\bm{x}} i t^a]_{i j} \Big[\int\!\! d^2y\, 
  [{\cal E}_{\bm{x} \bm{y}}^{a b}]_k 
  [\xi^b_{\bm{y}}]_k+\Hat\sigma^a_{\bm{x}}\Big]
\end{equation}
where
\begin{equation}
  {\cal E}^{ab}_{{\bm{x}}{\bm{y}}} = \left(\frac{\alpha_s}{\pi^2}\right)^{1/2} 
  \frac{({\bm{x}}-{\bm{y}})_k}{({\bm{x}}-{\bm{y}})^2} 
    [ 1 - \Tilde U^\dagger_{\bm{x}} \Tilde U_{\bm{y}} ]^{ab} 
\end{equation}
is the ``square root'' of $\chi$, 
$\chi^{ab}_{{\bm{x}}{\bm{y}}} = {\cal E}^{ac}_{{\bm{x}}{\bm{z}}} {\cal E}^{cb}_{{\bm{z}}{\bm{y}}}$,
and $\xi$ are independent Gaussian random variables with
correlators determined according to
\begin{equation}
  \label{eq:etacorr}
%  \langle\ldots \rangle_\omega = \det{\hat\chi}^{1/2} 
%  \int\!D[\omega]\, (\ldots)\, 
%  e^{-\frac{1}{2}\omega\hat\chi^{-1} \omega}
%\hspace{.5cm} \mbox{and} \hspace{.5cm}
  \langle\ldots \rangle_\xi = \int\!D[\xi]\, (\ldots)\,
  e^{-\frac{1}{2}\xi \xi}
\ .
\end{equation}
%Note the factor of $i$ which is essential to render this an
This clearly represents an equation for an infinitesimal change of an
element of $SU(N_c)$. In fact, the components of
$\omega^a_{\bm{x}}:=\Big[\int\!\!  d^2y\, [{\cal E}_{\bm{x} \bm{y}}^{a
  b}]_k [\xi^b_{\bm{y}}]_k+\Hat\sigma^a_{\bm{x}}\Big]$ can be directly
interpreted as the ``angles'' parametrizing a local gauge
transformation in transverse space.

It is of particular importance to note that the possibility to
formulate the stochastic term via a completely decorrelated Gaussian
noise $\xi$, that is to say with $\langle \xi^{a,i}_{\bm{x}}
 \xi^{b,j}_{\bm{y}} \rangle = \delta^{a b} \delta^{i j} 
\delta^{(2)}_{\bm{x}\bm{y}}$, reduces the numerical
cost for a simulation like this considerably.

All that is left to do conceptually is the explicit discretization and
the creation of an initial condition in form of an ensemble of $U$
fields at $\y_0$.

A reasonable choice for the initial condition is an ensemble that
leads to a shape for $\langle U^\dagger_{\bm{x}} U_{\bm{y}}\rangle$
that is compatible with the generic physics requirements as shown in
Fig.~\ref{fig:generic-evol}, for instance with
\begin{equation}
   \langle U^\dagger_{\bm{x}} U_{\bm{y}}\rangle \propto  
        \exp\left(\frac{-({\bm{x}}-{\bm{y}})^2}{4R^2}\right) .
\end{equation}
The scale $R$ here is close to the initial saturation scale $R_s$
(depending on the precise definition of $R_s$).

A possible choice for the discretizations~\cite{Rummukainen:2003ns} is
as follows: one discretizes the Langevin equation in
Eq.~\eqref{eq:Langevin} using a regular square lattice of $N^2$ sites
(volume $(Na)^2$, where $a$ is the lattice spacing), with periodic
boundary conditions. In dimensionless units ($a = 1$) and defining a
rescaled, discrete, evolution time
\begin{equation}
  \label{eq:resc-evol-tau}
  s := \frac{\alpha_s}{\pi^2}\y
\end{equation}
the Langevin equation becomes 
\begin{equation} 
U_{{\bm{x}}; s + \delta s} =
U_{{\bm{x}};s} \exp[i t^a \omega^a_{{\bm{x}};s}] 
\end{equation}
where
\begin{align}
  \omega^a_{{\bm{x}},s}  = &  \sqrt{\delta s} 
        \sum_{{\bm{y}}} K^i({\bm{x}}-{\bm{y}}) 
        [ 1 - \tilde U^\dagger_{{\bm{x}};s}\tilde U_{{\bm{y}};s}]^{ab} \xi^{i,b}_{\bm{y}} 
     %   \nonumber\\  & 
        -   \delta s \sum_{{\bm{y}}} S({\bm{x}}-{\bm{y}}) \frac12 \Tilde \tr 
        [i \Tilde t^a \tilde U^\dagger_{{\bm{x}},s} \tilde U_{{\bm{y}},s}]. 
  \label{eq:omega}  
\end{align}
Here $K^i({\bm{r}}) = \frac{{\bm{r}}^i}{{\bm{r}}^2}$, $S({\bm{r}}) =
\frac1{{\bm{r}}^2}$ and $\xi$ is a Gaussian noise with $ \langle
\xi^{i,a}_{{\bm{x}}}\xi^{j,b}_{{\bm{y}}} \rangle =
\delta^{i j}\delta^{a b}\delta_{{\bm{x}},{\bm{y}}}$. $K_i({\bm{r}})$
and $S({\bm{r}})$ are taken to be $ =0$ if ${\bm{r}} =0$. This is
justified by the fact that the apparent singularities at these points
are completely canceled in the JIMWLK Hamiltonian these Langevin
equations are equivalent to.

As a consequence of this additional discretization, one needs to
carefully check that the results are free of any artifacts in the
sense that the numerical results can be reliably extrapolated to
infinite volume, that the spatial continuum limit can be extracted and
that ``time'' discretization errors are negligible. Most of these are
rather technical issues and I refer to~\cite{Rummukainen:2003ns} for
details. A careful study of the field theoretical continuum limit,
however, has led to a clearer understanding of the nature and
importance of running coupling corrections for the selfconsistency of
the method. This will be reviewed carefully in
Sec.~\ref{sec:numer-results}.

\subsection{\it Shower operators: the operators that 
generate soft gluon clouds}
\label{sec:JIMWLK-shower-operators}

The translation to Langevin form was motivated above with numerical
implementations. There is in addition a conceptual side effect of this
reformulation, whose importance is not obvious at this point of the
discussion. I will nevertheless start to prepare for the discussion in
Sec.~\ref{sec:jets-dense-media} by writing an apparently formal
solution to the Langevin equation that contains a version of an
operator dubbed a ``shower operator'' in~\cite{Weigert:2003mm}. This
operator can be interpreted to build up the ordered soft emission part
of the projectile wavefunction, the soft gluon cloud, although this
is not fully apparent from the present context. This interpretation
will become obvious by comparison with the treatment of soft emission
in jets in Sec.~\ref{sec:jimwlk-soft-gluon}. Let me now come to the
formal solution itself. To this end I rewrite the r.h.s. of
Eq.~\eqref{eq:Langevin} in terms of a ``single emission operator''
(also this terminology will become self explaining in the discussion
of amplitudes in Sec.~\ref{sec:jimwlk-soft-gluon}):
\begin{equation}
  \label{eq:Langevin-operator}
  \partial_\y\, U_{\y, \bm{x}}
  =  i \int\!\! d^2u\, d^2v\, 
  \Big[\Big( [{\cal E}_{\bm{u} \bm{v}}^{a b}]_k 
  [\xi^b_{\bm{v}}]_k+\Hat\sigma^a_{\bm{u}}\Big)i\nabla^a_{\bm{u}}\Big]_\y
  U_{\y, \bm{x}}
\ .
\end{equation}
This can now be integrated using $\y$-ordering:
\begin{equation}
  \label{eq:Langevin-formalsol}
   U_{\y, \bm{x}} = {\sf P}_\y \exp\Big\{i \int_{\y_0}^\y d\Tilde\y  \int\!\! d^2u\, d^2v\,  \Big[
  \Big( [{\cal E}_{\bm{u} \bm{v}}^{a b}]_k 
  [\xi^b_{\bm{v}}]_k+\Hat\sigma^a_{\bm{u}}\Big)i\nabla^a_{\bm{u}}\Big]_{\Tilde\y} \Big\} U_{\y_0, \bm{x}}
\ .
\end{equation}
The operator
\begin{equation}
  \label{eq:small-x-shower}
  {\sf P}_\y \exp\Big\{i \int_{\y_0}^\y d\Tilde\y  \int\!\! d^2u\, d^2v\,  \Big[
  \Big( [{\cal E}_{\bm{u} \bm{v}}^{a b}]_k 
  [\xi^b_{\bm{v}}]_k+\Hat\sigma^a_{\bm{u}}\Big)i\nabla^a_{\bm{u}}\Big]_{\Tilde\y} \Big\}
\end{equation}
is the shower operator responsible for soft gluon emission. Please
note the analogy with Eq.~\eqref{eq:translations-func}: the only
special feature is the complicated nature of what plays the role of
the angle variable $\omega$. This now contains $U$ at earlier $\y$ and
in a nonlocal manner. This is a nice illustration of how the JIMWLK
equation implements the resummation of the second nonlinearity: the
change in $U$ turns nonlinear in $U$ itself. One might be tempted to
call that ``second eikonalization:'' eikonal factors iteratively enter
the exponents themselves. That these operators indeed have strong
analogies with path ordered exponentials is obvious from the
appearance of $\y$ ordering. This leads to useful properties as will
be seen in identities like Eq.~\eqref{eq:Xivar}.

The relevance of this type of operator lies in the fact that it not
only appears in the Langevin context, but also in the construction of
amplitudes with arbitrary numbers of soft gluons. n-th order
truncations simply correspond to n-gluon amplitudes.  An example at
small $x$ is provided by the virtual photon wave function entering the
DIS cross sections in Eq.~\eqref{eq:dipole-cross}, now taken to all
orders in soft gluon emission.  In this sense the operator
in~\eqref{eq:small-x-shower} creates the ``Cold Gluon Cloud'' [Alex
Kovner's favorite rendering of CGC] and clearly summarizes the same
physics as the JIMWLK equation. I will now present an alternative
derivation of nonlinear evolution equations of the JIMWLK type based
on soft gluon amplitudes and gluon shower operators. To show the
generality of the concepts I will do this for a new physics example, a
specific type of jet observable.

% spellmark 4

\section[JIMWLK and soft gluon emission: analogues in jet physics]{JIMWLK and soft gluon emission: comparing with and learning 
from jet physics}
\label{sec:jimwlk-soft-gluon}

At first sight it would appear to be fairly far fetched to expect
saturation type physics of any kind in a jet measurement, in particular
if the jets are created in ``empty space,'' for example by an $e^+
e^-$ collision. There one is typically looking at observables that
have long been described purely perturbatively as multiple emission
processes without any reference to density at all. Although it is
conceivable that in certain corners of phase space --for example by
looking at in some sense ``strongly focussed'' jet events-- that these
particles start to overlap in a way comparable to the small $\xbj$
situation, in the sense that one would have to take density effects
into account, most would have relegated such phenomena into the realm
of extremely rare events with little experimental relevance.

This has changed with the identification of so called non-global jet
observables by Dasgupta and Salam~\cite{Dasgupta:2001sh,
  Dasgupta:2002bw} already for $e^+ e^-$ collisions. Furthermore, it
turns out that for these same observables the importance of the
nonlinear effects is increased if the jets originate in or travel
though a dense medium created by other means -- for example in a heavy
ion collision. It is to be expected that many jet observables
accessible in the heavy ion context will acquire non-global
characteristics and become sensitive to medium and density effects.
This type of consideration will become particularly important in the
heavy ion program at the LHC. I will turn to a brief discussion of
this situation in Sec.~\ref{sec:jets-dense-media}, but will present
the underlying theory in the simpler $e^+ e^-$ case. Already in this
situation one finds evolution equations with similar structure and
interpretation of the nonlinear effects as in the small $\xbj$ case
(c.f.  Eq.~\eqref{eq:BMS-2} and the BK equation~\eqref{eq:BK-S}).

\begin{figure}[htb]
  \centering
\includegraphics[height=4cm]{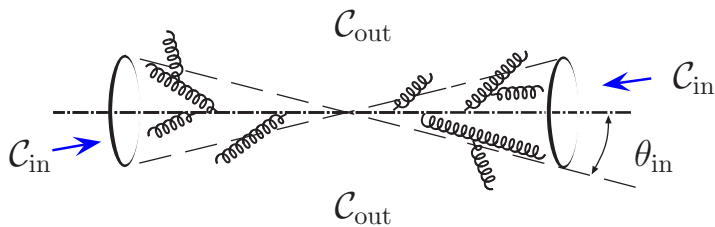}%besen2-neu}
  \caption{\small%
    Two jet event in $e^+ e^-$ with prescribed geometry and soft
    emission into the region away from the jets. ${\cal
      C}_{\text{in}}$ and ${\cal C}_{\text{out}}$ denote the regions
    inside and outside the jets which for intuitiveness are
    characterized by an opening angle $\theta_{\text{in}}$.}
  \label{fig:e+e-jet}
\end{figure}

To formulate an archetypical observable of this type, consider a $e^+
e^-$ collision at total invariant energy $Q^2$ and look at a 2 jet
event (considering 3 and more jet events as a perturbative correction
for the time being) as depicted in Fig.~\ref{fig:e+e-jet}.
Typical soft- and collinear-safe jet observables are for example
broadening $B$ or thrust $T$, both defined as sums over kinematic
variables of measured hadrons. To be explicit,
$B=\sum_h\frac{|(p_t)_h|}{Q}$, $1-T=\sum_h\frac{|(p_t)_h|
  e^{-|\eta_h|}}{Q}$ where $(p_t)_h$ and $\eta_h$ are (three dim.)
hadron transverse momentum and rapidity w.r.t. the jet axis. Setting
$V = B$ or $1-T$ the corresponding jet shape distributions are then
defined as
\begin{equation}
  \label{eq:jet-shape-global}
  \Sigma(Q,V) = \sum\limits_n \int \frac{d\sigma_n(Q)}{\sigma_T(Q)}
  \theta(V-\sum\limits_h v_h)
\end{equation}
where $d\sigma_n(Q)$ is the $n$-hadron distribution, $\sigma_T(Q)$ the
total cross section and $v_h$ is a single hadrons contribution to the
observable. The most important
perturbative contributions to such variables are enhanced by soft and
collinear phase space logarithms, a single emission carries at most
both of them.

Thrust and broadening belong to a large class of observables in which
these enhanced contributions exponentiate~\cite{Catani:1993ua} to give
a systematic perturbative expansion for the exponent of the
form\footnote{This result {\em is} nontrivial: For example it states
  the absence of contributions of the form $\alpha_s^n \ln^{m} V$ with
  $m=n+2\ldots 2n$ in the exponent. Only one of the ``exponentiated''
  gluons carries both logarithms.}
\begin{equation}
  \label{eq:jet-exponentiation}
  \ln\Sigma(Q,V) = \sum\limits_{n=1}^\infty \alpha_s(Q^2)^n
  \left(A_n \ln^{n+1}V + B_n\ln^n V+\ldots\right)
\ .
\end{equation}
The $A_n$ series is referred to as double logarithmic and $B_n$ as
single logarithmic, the remaining contributions are not logarithmically
enhanced. The variables considered up to this point are referred to as
global: all final state hadrons are counted in the sums $\sum_h
v_h$, no matter where they end up in relation to the physical jet.
Non-global observables are more restrictive, they measure for instance
contribution only from hadrons that end up in the region away from the
physical jet. To prepare for such a measurement one distinguishes the
region containing the jets (${\cal C}_{\text{in}}$) from the region
safely away from the jets (${\cal C}_{\text{out}}$) as indicated in
Fig.~\ref{fig:e+e-jet}. To provide a prototypical example, consider
the ``inter jet energy flow'' defined by
\begin{equation}
  \label{eq:Sig-ee}
  \Sigma_{e^+e^-}(Q,E_{\text{out}})=
  \sum_n\int\frac{d\sigma_n(Q)}{\sigma_T(Q)}\cdot
  \Theta\!\left(E_{\text{out}}-\sum_{h\,\in\,C_{\text{out}}}\omega_h\right)
  \ .
\end{equation}
Here $v_h=\omega_h$ is the energy of the measured hadron and the sum
is restricted to ${\cal C}_{\text{out}}$. No event with more than a
total energy $E_{\text{out}}$ outside the jets is counted. In a
perturbative calculation, the phase space in energy extends up to $E
\sim Q/2$ which will give rise to large logarithms in
$(Q/2)/E_{\text{out}}\sim E/E_{\text{out}}$, but there will be no
collinear contribution since the measured particles are emitted at
large angles (from the jet as a source). This implies that in the
representation~\eqref{eq:jet-exponentiation} only the single
logarithmic terms contribute. While this is an important difference
with the earlier global examples, it is not immediately evident, why
this observable should bear any similarities with the dipole cross
sections of the CGC. The careful analysis of Dasgupta and
Salam~\cite{Dasgupta:2001sh, Dasgupta:2002bw} has uncovered that in
distinction to the global case these single logarithmic contributions
are sensitive to the internal structure of the jet. To the required
accuracy the measured soft particles can not be treated as if emitted
by the initial hard $q\Bar q$ pair. Through emission from secondary
``hard'' particles, the observable becomes sensitive to the internal
structure of the jet. To retain this information one must keep track
of the {\em nonlinear} multiple emission structure inside the jets.
Nevertheless, the connection with BK and JIMWLK evolution was not
apparent until a beautiful paper by Banfi, Marchesini, and Smye
(BMS)~\cite{Banfi:2002hw} reformulated the numerical approach
of~\cite{Dasgupta:2001sh, Dasgupta:2002bw} (in the large $N_c$ limit)
for exactly the observable described above in the form of an evolution
equation.
%\marginpar{{\color{red} what is the equation in:
%    $Q/E_{\text{out}}$ or $E/E_{\text{out}}$ }}
%With $E_{\text{out}}$ much smaller than the hard scale of the jet, QCD
%radiation into this region is small and these observables offer an
%ideal opportunity to study nonperturbative effects. The central result
%of~\cite{Banfi:2002hw} is an evolution equation that resums, for large
%$Q/E_{\text{out}}$ and {\em in the large $N_c$ limit}, all leading
%terms arising from large angle soft emission. I will refer to this
%equation as the BMS equation. 
%Subsequent analysis of the consequences
%implied by this equation presented in the same publication confirm
%features found using numerical methods by Dasgupta and
%Salam~\cite{Dasgupta:2001sh, Dasgupta:2002bw}. 
The equation itself is nonlinear in the part describing contributions
arising {\em inside} the jet regions and linear outside.  The
contributions inside, quite surprisingly for most experts, have a
striking analogy with the BK equation.  For small angle emission in a
certain frame, the analogy becomes even more striking: the
contributions of BMS inside the jet regions become identical to the BK
equation. That such an analogy is restricted to the in-region, is
clear: in the derivation of the BK equation one assumes, in fact
requires, the presence of non-linear effects {\em everywhere} and any
geometric exclusiveness of the type imposed by jets was simply never
considered. What appeared to be more surprising is the relation of jet
physics to BFKL dynamics, which dominates the BK equation in the small
density limit. This link has been established
in~\cite{Marchesini:2003nh} and indeed comes into play in the limit of
small angle emission in a judiciously chosen frame.

Subsequently in~\cite{Weigert:2003mm}, I have shown that the analogy
goes further: there exists a functional evolution equation that
generalizes the BMS equation to finite $N_c$ which parallels the
JIMWLK equation.  This should be not too surprising, as BK emerges
from JIMWLK by means of a factorization assumption at least strongly
linked with the large $N_c$ limit. There are useful lessons to be
learned for small $\xbj$ physics from both the structure and the
derivation of this new jet evolution equation. The structure teaches
us about possible consequences of exclusive constraints on
observables. It shows, for instance, how non-emission requirements
lead to Sudakov like resummations in the restricted areas --there only
the first of the resummations into eikonal factors is at work--,
whereas the unrestricted areas remain subject to nonlinear evolution
and experience both types of nonlinearities. The derivation was built
on earlier results on amplitudes~\cite{Bassetto:1983ik,Fiorani:1988by}
and thus gives direct access to them. This should provide greater
technical flexibility for any generalizations that might turn out to
be desirable also on the small $\xbj$ side.  Not least has the
presentation in the introduction been heavily influenced by the
results derived in this context.

\subsection{\it The analogy of BMS and BK equations}
\label{sec:analogy}

To present the analogy, let me give a short account of the BMS
equation before I compare with the BK equation. I will closely follow
the exposition in~\cite{Banfi:2002hw}.  The observable considered by
BMS has been given already in~\eqref{eq:Sig-ee}. The process is
dominated by iterative soft emission from eikonalized gluons that are
not deflected from their original trajectory. The natural kinematic
variables are thus the directions of the emitters ($a,b,k,p,\bar p$ in
what follows) and the corresponding energies. Phase space integrals
will be written in terms of solid angles and energies.

A Mellin transformation represents the cross section in terms of
transition probabilities $G_{ab}(E,E_{\text{out}})$
\begin{equation}
  \label{eq:Sig-ee1}
  \Sigma_{e^+e^-}(E,E_{\text{out}})\>=\>\int
\frac{d\nu\,e^{\nu E_\text{out}}}{2\pi i \nu}
  \>G_{p\bar p}(E,\nu^{-1})
\simeq G_{p\bar p}(E,E_{\text{out}})\>,
\end{equation}
where $p$ and $\Bar p$ are the directions of the original $q$ and
$\Bar q$ respectively and a saddle point approximation has been taken.

$G_{ab}(E,E_{\text{out}})$, or $G_{a b}(E)$ for short where the
explicit $E_{\text{out}}$ dependence is not needed, satisfies the BMS
equation
\begin{equation}
  \label{eq:BMS-1}
\begin{split}
  E\partial_E\>G_{ab}(E)=\int \frac{d^2 \Omega_k}{4\pi}\,\Bar\alpha_s
  w_{ab}(k)\left[u(k)\,G_{ak}(E)\cdot G_{kb}(E)-G_{ab}(E)\right]
\ ,
\end{split}
\end{equation}
%-----------------------------------
where $\Bar\alpha_s:=\alpha_s N_c/\pi$ 
%-----------------------------------
and
\begin{equation}
  \label{eq:wab}
  w_{ab}(k)=\frac{(p_ap_b)}{(p_ak)(kp_b)}=
  \frac{1-\cos\theta_{ab}}{(1-\cos\theta_{ak})(1-\cos\theta_{kb})}
\end{equation}
is the soft emission kernel. Conventions are adapted to lightlike
momenta and the energies, $\omega_p, \omega_q$ have been factored out
in $(p q)$ compared to the 4-vector product: $p.q=\omega_p \omega_q (p
q) = \omega_p \omega_q(1-\cos\theta_{p q})$.

The jet geometry is encoded in the definition of $u(k)$, which
emerges from a Mellin factorization of the energy constraint in
Eq.~\eqref{eq:Sig-ee}:
\begin{equation}
  \label{eq:Theta-fac}
  \Theta\Big(\!E_{\text{out}}\!-\!
  \sum_{i\in{\cal C}_{\text{out}}}\omega_i\!\Big)=
  \int\frac{d\nu\,e^{\nu E_{\text{out}}}}{2\pi i \nu}\, \prod_i u(k_i)\>,
  \quad   u(k)=\Theta_{\text{in}}(k)+e^{-\nu\omega}\Theta_{\text{out}}(k)\>,
\end{equation}
with the $\Theta_{\text{in}}$ and $\Theta_{\text{out}}$ having support
inside and outside the jet regions respectively. 
%----------------

The strategy employed by BMS in the derivation of Eq.~\eqref{eq:BMS-1}
is surprisingly straightforward. The evolution equation was derived
from the knowledge of the structure of the real emission part alone.
Virtual corrections were reintroduced using the requirement of
real-virtual cancellation in the infrared.

The real emission part of $G$ needed for this argument can be written
as
\begin{equation}
  \label{eq:GR-ee}
  G_{p\bar p}^{({\text{real}})}(E,E_{\text{out}})=1\!+\!
\sum_{n=1}^{\infty}\!\int\prod_{i=1}^n
\left\{\Bar\alpha_s \frac{d\omega_i}{\omega_i}\frac{d^2 \Omega_i}{4\pi}u(k_i)\,
\Theta(E\!-\!\omega_i)\!\right\} W_n(p k_{1} \ldots k_{n} \bar p)
\end{equation}
where the phase space of soft gluons is cut by $E$. $W_n$ is the large
$N_c$ factorized version of the transition probability of a hard
$q\Bar q$ color singlet into $(q\Bar q)_{\text{hard}}
g^n_{\text{soft}}$,
\begin{equation}
  \label{eq:Wn}
  W_n(p k_{1} \ldots k_{n} \bar p)\>=\>\frac{(p\bar p)}
  {(pk_1)(k_1k_2)\ldots(k_n\bar p)}\>
%,\qquad
%  (qq')\equiv 1-\cos\theta_{qq'}
\ .
\end{equation}
The line of argument then starts by taking a logarithmic derivative
$E\partial_E$ of Eq.~\eqref{eq:GR-ee} which uniquely fixes the
quadratic term in Eq.~\eqref{eq:BMS-1}. This term describes the
creation of a real gluon in the final state with its energy at the
phase space boundary $E$. This term is affected by the jet-geometry as
signalled by the factor $u(k)$. The virtual corrections are then taken
into account by subtraction of a linear term (no real gluon in the
final state, virtual corrections only renormalize the existing part
components of the wave function) with a coefficient that ensures
infrared finiteness through real virtual cancellation. Virtual
corrections occur globally, both inside and outside the jet region,
thus the factor $u(k)$ is absent. This yields the second term in
Eq.~\eqref{eq:BMS-1}.

That this indeed ensures infrared finiteness is best understood after
rewriting~\eqref{eq:BMS-1} in the form
%or equivalently
\begin{equation}
  \label{eq:BMS-2}
  \begin{split}
  E\partial_E\>G_{ab}(E)
  &=-E\partial_E\>R_{ab}^{(0)}(E)\cdot G_{ab}(E)\\
  &+\int \frac{d^2 \Omega_k}{4\pi}\,\Bar\alpha_s
  w_{ab}(k)\,u(k)\,\left[\,G_{ak}(E)\cdot G_{kb}(E)-G_{ab}(E)\,\right]
\ .
  \end{split}
\end{equation}
where $R_{ab}^{(0)}(E)$ is called the
single log Sudakov radiator for bremsstrahlung emission
\begin{equation}
  \label{eq:Rad}
  R_{ab}^{(0)}(E)\>=\>\int_0^E\frac{d\omega}{\omega}
  \int\frac{d^2 \Omega_k}{4\pi}\,\Bar\alpha_s\,w_{ab}(k)\,
  [1-u(k)]\>=\>\Delta\cdot r_{ab}\>,
\end{equation}
%splitting off the Sudakov term as in Eq.~\eqref{eq:BMS-2}. 
BMS show that to good accuracy one may take $u(k)$ to restrict the
phase space integrals to the outside region for the Sudakov radiator
term and to the inside region for the remainder of
Eq.~\eqref{eq:BMS-2}.

With integration in the Sudakov term  restricted to the out region,
one recognizes that there is no danger of encountering any ill effects
from the poles of the kernel in this term. For the other term one
needs to prove (see~\cite{Banfi:2002hw}) that $G_{a a}(E) = 1$. Then
the quadratic and the linear terms cancel where the kernels diverge,
rendering the expression finite.

For more details on the ingredients as well as the physics of this
equation see~\cite{Banfi:2002hw, Dasgupta:2001sh, Dasgupta:2002bw,
  Bassetto:1983ik, Fiorani:1988by} and references therein.

It is now straightforward to compare the BMS equation,
Eq.~\eqref{eq:BMS-1}, to the BK equation in its incarnation in terms
of $S_{\y; \bm{x y}}$:
 \begin{equation}
    \label{eq:BK-S}
    \partial_\y S_{\y; \bm{x y}} 
    = 
    \frac{\alpha_s N_c}{2\pi^2} \int d^2z \ \Tilde{\cal K}_{\bm{x}
    \bm{z} \bm{y}}\,
  ( S_{\y; \bm{x z}} S_{\y; \bm{z y}} 
  -  S_{\y; \bm{x y}} )
\ .
  \end{equation}
Ignoring for the moment the Sudakov radiator term in
Eq.~\eqref{eq:BMS-2}, or setting $u(k) \to 1$ for the time being, the
similarity of Eqns.~\eqref{eq:BMS-2} and~\eqref{eq:BK-S} is striking.
A one to one relationship of structures emerges if one maps directions
onto transverse coordinates, the kernels onto each other and tries to
view $G_{a b}(E)$ as the average of some $\tr(U^\dagger_a U_b)/N_c$
(where $a$ and $b$ represent the directions of the leading hard
particles):
\begin{equation}
  \label{eq:interpretation}
  G_{a b}(E)
  \ \overset{{\text{\normalsize?}}}{\longleftrightarrow}\
  \langle \tr(U^\dagger_a U_b)/N_c \rangle_E 
  \ .
\end{equation}
Note that this is fully compatible with the requirement that $G_{a
  a}(E)=1$, needed to ensure real virtual cancellation in the second
term of Eq.~\eqref{eq:BMS-2}.  While this appears to be quite
intuitive and fully in line with the physical interpretation, it is
not immediately clear how precisely to perform that latter part of the
translation.  One would ask, for instance, how in detail to arrive at
a consistent definition of these $U$s and the averaging process
suggested by Eq.~\eqref{eq:interpretation} in the light of the tree
like branching process underlying the physics of the BMS equation.
Additional questions would be how to reconcile their role in
probabilities with that in amplitudes and many more in the same vein.

That the interpretation proposed in Eq.~\eqref{eq:interpretation}
should be possible suggests itself even more strongly, if one uses the
observation made by BMS that in the small-angle-emission limit in a
carefully chosen frame where the measure turns flat and the kernels
agree completely:
\begin{equation}
  \label{eq:w-small-angle}
  w_{a b}(k)  \to \Tilde {\cal K}_{{\Hat a} {\Hat k} {\Hat b}} 
\ .
\end{equation}
This is the region where Marchesini and
Mueller~\cite{Marchesini:2003nh} have established the link between jet
and BFKL-dynamics.

%The benefit of demonstrating such a correspondence should be obvious
%to any reader familiar with the color glass condensate (CGC): If there
%is such an interpretation of the objects entering the BMS equation,
%then there is hope one might generalize it along similar lines as the
%BK equation, which can be viewed as the factorized limit of a more
%general functional equation, the JIMWLK equation which I will turn to
%next, in order to show how it reduces to BK under certain assumptions.
%This will be the analogy used to suggest a generalization of the BMS
%equation.

%If such a correspondence can be established one would expect the BMS
%equation to posses a generalization that shows the parallels with the
%JIMWLK equation.

\subsection{\it From JIMWLK to BK: generalizing BMS by analogy}
\label{sec:from-jimwlk-bk}

To understand what the corresponding nonfactorized equation might look like, recall
the chain of argument leading from the JIMWLK equation
Eq.~\eqref{eq:JIMWLK} via Eq.~\eqref{eq:twopoint-JIMWLK} to the
pre-factorized from of the BK equation~\eqref{eq:pre-BK-S}.  From this
short calculation it is pretty obvious that the linear and nonlinear
terms arise in essence from the first and second term in
Eq.\eqref{eq:JIMWLK-Hamiltonian-2}:
\begin{align*}
%  \label{eq:JIMWLK-Hamiltonian-2-1}
 H_{\text{JIMWLK}} = &
   -\frac{1}{2} \frac{\alpha_s}{\pi^2}\ {\cal K}_{\bm{x z y}}\ 
  \big[ i\nabla^a_{\bm{x}} i\nabla^a_{\bm{y}}
  +i\Bar\nabla^a_{\bm{x}} i\Bar\nabla^a_{\bm{y}}
  +\Tilde U_{\bm{z}}^{a b}(i\Bar\nabla^a_{\bm{x}} i\nabla^b_{\bm{y}}
  +i\nabla^a_{\bm{x}} i\Bar\nabla^b_{\bm{y}}) \big]
\ .
\end{align*}
Note that this is in line with the interpretation of these terms as
virtual and real contributions respectively. To arrive at different
coefficients in linear and nonlinear terms as in Eq.\eqref{eq:BMS-1}
one simply has to allow for different Kernels respectively.

To write down the desired Fokker-Planck Hamiltonian in the jet case,
one needs
\begin{enumerate}
\item to use the correspondences of variables and kernels noted in the
  above. In this step it proves useful, if I slightly redefine the
  functional aspect of the invariant vector fields to produce $\delta$
  functions adapted to the structure imposed by the solid angle
  measure. To this end I will write
  \begin{align}
    \label{eq:nabla-def}
    i\nabla^a_p U_p = t^a U_p \Bar\delta(p-q)
  \end{align}
  where $\Bar\delta(p-q)$ contains Jacobian factors such that
  \begin{align}
    \label{eq:bar-delta}
    \int\frac{d\Omega_k}{4\pi} \Bar\delta(p-q) f(q) = f(p)
    \ .
  \end{align}
  This will simplify expressions considerably.  
\item The different coefficient in real and virtual parts are
  implemented by allowing for different integration kernels ${\cal
    K}^{(i)}$ $i=1,2$ for the two terms. Clearly both analogues of
  ${\cal K}^{(i)}$ will be proportional to $w_{p q}(k)$ with the
  proportionality factor $f^{(i)}(k)$ carrying the jet geometry via
  some $u(k)$ dependence.\footnote{Note that the calculations are
    somewhat simplified by the fact that $w_{p p}(k) = 0$. The
    relation between kernels on the functional side to those on the
    BMS side, as given by Eq.~\eqref{eq:BK-kernel-from-JIMWLK} for
    JIMWLK/BK pair, becomes a matter of simple proportionality.}
\end{enumerate}
The result of this exercise is a Fokker-Planck Hamiltonian that has been
dubbed $H_{\text{ng}}$ (for ``non-global'') in\cite{Weigert:2003mm},
here I refer to it generically by $H_{\text{jet}}$.  I will present it
in two forms, to parallel the two versions of the BMS
equation,~\eqref{eq:BMS-1} and~\eqref{eq:BMS-2}. To this end I will
introduce an additional function $\Tilde f^{(2)}(k)$ that will serve
as the coefficient of the generalization of the Sudakov radiator term.
One has the following forms of the Hamiltonian:
\begin{subequations}  
\label{eq:H_ng-all}
\begin{align}
  \label{eq:H_ng-0}
  H_{\text{jet}} :=  &-\frac{\alpha_s}{2\pi}\ w_{u v}(k) \Big[
   f^{(1)}(k)\big( i\nabla^a_{{\bm{u}}}i\nabla^a_{{\bm{v}}}
 +i\Bar\nabla^a_{{\bm{u}}}i\Bar\nabla^a_{{\bm{v}}}\big)
 +f^{(2)}(k)\big[U_{\bm{k}}\big]^{a
   b}\big(i\Bar\nabla^a_{{\bm{u}}}i\nabla^b_{{\bm{v}}}
 +i\Bar\nabla^a_{{\bm{v}}}i\nabla^b_{{\bm{u}}}\big)\Big]
%\nonumber \\
%= & i\nabla^a_{u}\chi^{a b}_{u v}i\nabla^b_{v}
%\end{align}
%\begin{align}
\intertext{in analogy with~\eqref{eq:BMS-1}, and}
  \label{eq:H_ng}
  H_{\text{jet}} :=  &-\frac{\alpha_s}{2\pi}\ w_{u v}(k) \Big[ 
  \Tilde f^{(1)}(k)\Big(
  i\nabla^a_{{\bm{u}}}i\nabla^a_{{\bm{v}}}
 +i\Bar\nabla^a_{{\bm{u}}}i\Bar\nabla^a_{{\bm{v}}}\Big)
   \nonumber \\ &
 +  %\Big[ 
   f^{(2)}(k)\Big( i\nabla^a_{{\bm{u}}}i\nabla^a_{{\bm{v}}}
 +i\Bar\nabla^a_{{\bm{u}}}i\Bar\nabla^a_{{\bm{v}}}+\big[U_{\bm{k}}\big]^{a
   b}(i\Bar\nabla^a_{{\bm{u}}}i\nabla^b_{{\bm{v}}}
 +i\Bar\nabla^a_{{\bm{v}}}i\nabla^b_{{\bm{u}}})\Big)\Big]
%\nonumber \\
%= & i\nabla^a_{u}\chi^{a b}_{u v}i\nabla^b_{v}
\end{align}
\end{subequations}
to parallel~\eqref{eq:BMS-2}. In both cases $u,v$ and $k$ are
integrated over according to an ``integration convention'' that uses
$\frac{d\Omega_{p}}{4\pi}$ as its measure for any of the momenta.
Retracing the steps leading from the definition of the JIMWLK equation
to~\eqref{eq:pre-BK-S} by eye should make it obvious that indeed, the
separation of terms in Eq.~\eqref{eq:H_ng} is such that the first line
generates the Sudakov radiator and the second the nonlinear evolution
inside the jet cones.

A comparison of the result with BMS allows one to determine only the
leading $N_c$ part of the $f^{(i)}$. Already the $1/N_c$ corrections
are not controlled by the matching. To this accuracy the $f$ are $N_c$
independent:
\begin{subequations}
  \label{eq:tilde-cond}
\begin{align}
   f^{(2)}(k) = & u(k)\\
  f^{(1)}(k) = & 1 \\
  \Tilde f^{(1)}(k) = & (1-u(k))  
  \ .
\end{align}
\end{subequations}
Eqns.~\eqref{eq:H_ng-all},~\eqref{eq:tilde-cond}
together with a Fokker-Planck equation of the form
\begin{equation}
  \label{eq:FP-ng}
  E\partial_E \Hat Z_E[U] = -H_{\text{jet}} \Hat Z_E[U]
\end{equation}
define the finite $N_c$ generalization of the BMS equation given
in\cite{Weigert:2003mm}. One might argue that this has to be the
correct result on naturalness reasons --any $N_c$ dependence in these
equations would make it virtually impossible that they should be able
to incorporate {\em all} $N_c$ corrections in a consistent manner--
but an explicit construction of this equation from the amplitude level
is much more instructive and will be reproduced below.

The most important feature of this equation is its infrared finiteness
which directly relates to the real virtual cancellations being
correctly encoded in Eq.~\eqref{eq:H_ng}. Considering the Sudakov term
in $H_{\text{jet}}$ one finds that just as in the BMS limit, the phase
space integration over $k$ is restricted to the outside region where
no hard particles are to be found. Therefore there is no divergence
from the poles in $w_{u v}(k)$. The second term may be recast in
analogy to Eq.~\eqref{eq:JIMWLK-Hamiltonian} and the same finiteness
argument used there clearly carries over to this situation. By looking
at the BMS limit one sees that this same cancellation of terms
corresponds to a real-virtual cancellation in the conventional sense.
On the functional side, real and virtual contributions arise from
different terms.  Contributions with an additional factor $\Tilde U_k$
correspond to real emission, the rest to virtual corrections.

The idea to interpret jet transition probabilities as correlators of
path-ordered exponentials in singlet projections properly takes into
account gauge invariance, just the same way as the formulation of the
JIMWLK equation does.  The equation suggested therefore would appear
to consistently encode the physics it is intended to cover with a set
of very strong constraints. 

%From this perspective it would be rather
%surprising would it not be possible to arrive at this result from
%first principles. 

\subsection{\it An equivalent Langevin description}
\label{sec:an-equiv-lang}

Just as in the small $\xbj$ situation, a translation to a Langevin type
description is possible. The steps are almost identical to the ones
presented in Sec.~\ref{sec:from-fokker-planck}, just slightly
complicated by the added exclusiveness, the distinction into in and
out regions.

Prerequisites are (termwise) positive (semi-) definiteness of the
Fokker-Planck Hamiltonian in question. Only then can one introduce a
bounded noise integral -- if needed separately for each positive term.
If the Hamiltonian can be (termwise) separated into two conjugate
factors in analogy to the structure of
Eq.~\eqref{eq:JIMWLK-Hamiltonian}, any correlation effects can be
included in the Langevin equation. The noise can then be taken to be
(termwise) uncorrelated.

Fortunately all of the prerequisites listed above are also met for
$H_{\text{jet}}$.  In order to achieve termwise factorization, I first
separate off the $f^{(2)}$ terms and further treat the two
contributions proportional to $\Tilde f^{(1)}$ separately.  All of
these are positive due to the definition of $u(k)$:
\begin{equation}
  \label{eq:u-def}
  u(k)=\Theta_{\text{in}}(k)+e^{-\nu\omega} \Theta_{\text{out}}(k)
\end{equation}
ensures $0< u(k)<1$ and the same then for the coefficient functions in
Eq.~\eqref{eq:H_ng}.

To parallel the treatment in Sec.\ref{sec:illustrating-idea}
and~\ref{sec:numer-results-from} one would write the Hamiltonian on
the JIMWLK level as
\begin{align}
  \label{eq:H_ng-2}
  H_{\text{jet}} 
= &\frac{1}{2} i\nabla^a_{u}\chi^{a b}_{u v}i\nabla^b_{v}
\end{align}
where $u,v$ are integrated over with $\frac{d\Omega_u}{4\pi}\frac{
  d\Omega_v}{4\pi}$ and
\begin{align}
  \label{chi_ng}
  \chi^{a b}_{u v} = &
 \   - \int \frac{d\Omega_k}{4\pi} \frac{\alpha_s}{\pi} w_{u v}(k) \Big[
   \Tilde f^{(1)}(k) (1+U_u^\dagger U_v)
   + f^{(2)}(k)(1-U_u^\dagger U_k)(1- U_k^\dagger U_v)\Big]^{a b}
   \ .
\end{align}
Factorization of $\chi$ according to $\chi = {\cal E} {\cal
  E}^\dagger$ is achieved by defining a three component structure
representing the three separately positive terms alluded to before:
\begin{align}
  \label{eq:Edef}
  {\cal E}_{p k}^{a b; \mu} = & \sqrt{\frac{\alpha_s}{\pi}}
\frac{p^\mu}{p.k} \big\{
\sqrt{\Tilde f^{(1)}(k)} \delta^{a b},
\sqrt{\Tilde f^{(1)}(k)}[U^\dagger_p]^{a b},
\sqrt{  f^{(1)}(k) } (1-U_p^\dagger U_k)^{a b}
\big\}
\ .
\end{align}
Correspondingly one has to introduce independent white noise for all
of the components,
\begin{equation}
  \label{eq:noise}
  \Xi_k^{b;\mu} = \big\{(\xi^{(1)})_k^{b;\mu},
  (\xi^{(1')})_k^{b;\mu},
  (\xi^{(2)})_k^{b;\mu}\big\}
\ .
\end{equation}
The only thing left to cope with is the measure and the
$\delta$-functions in the noise correlator, which need to be such
that
\begin{equation}
  \label{eq:noisecond}
\big\langle \big(\int\frac{\Omega_k}{4\pi} 
{\cal E}_{p k}^{a c; \mu}  \Xi_k^{c;\mu} \big) 
\big(\int\frac{\Omega_l}{4\pi} 
{\cal E}_{q l}^{a d; \nu}  \Xi_l^{d;\nu} \big)\big\rangle 
= \chi_{p q}^{a b}
\ .
\end{equation}
Now the measure is $\frac{d\cos\theta d\phi}{4\pi}$ and one 
needs the correlators to read
%\marginpar{check this}
\begin{equation}
  \label{eq:xicorr-1}
  \langle  (\Xi^{i})_p^{a;\mu} (\Xi^{j})_q^{b;\nu} \rangle 
  = 4\pi \delta(\cos\theta_p-\cos\theta_q) \delta(\phi_p-\phi_q) 
  \delta^{i j} 
  g^{\mu\nu}
\ .
\end{equation}
With these preparations and the identification $\y \leftrightarrow
\ln E/E_{\text{out}}$%\marginpar{\color{red}check!} 
one has a Langevin equation that reads
\begin{equation}
  \label{eq:Jet-Langevin}
  \partial_\y\, [U_{\y;p}]_{i j} 
  = [U_{\y;p} i t^a]_{i j} \Big[\int\! \frac{d\Omega_k}{4\pi}\, 
  {\cal E}_{p k}^{a b; \mu}[U_{\y; p}]\ \Xi^{b; \mu}_{\y; k}\Big]
\ .
\end{equation}
In comparison with the JIMWLK case all sigma-terms vanish because
$w_{p q}(k)$ satisfies $w_{p p}(k)=0$.

Note again that any continuum notation is deceptive. The equation to
consider is really a finite difference equation and upon iteration
will allow for an interpretation of subsequent, ordered soft gluon
emission.  To this end it is instructive to rewrite the Langevin
equation in terms of functional derivatives:
\begin{align}
  \label{eq:Jet-Langevin-func}
  \partial_\y\, [U_{\y; p}]_{i j} 
  = & [U_{\y; p} i t^a]_{i j}  \Big[\int\! \frac{d\Omega_k}{4\pi}\, 
  {\cal E}_{p k}^{a b; \mu}[U_{\y; p}]\ \Xi^{b; \mu}_{\y; k}\Big] 
  \nonumber \\ = & 
   \Big\{\int\! \frac{d\Omega_q}{4\pi} \frac{d\Omega_k}{4\pi}\ 
  i {\cal E}_{p q}^{a b; \mu}[U_{\y; p}]\ \Xi^{b; \mu}_{\y; k}%\Big]\Big) 
  i\nabla^a_{U_{\y; q}} \Big\} [U_{\y; p}]_{i j}
\ .
\end{align}
The operator in the last version contains two terms in its third
component, the component responsible for the in-region. Using a
somewhat more compact notation for the $\y$ dependence they read:
\begin{equation}
  \label{eq:exp-real-virt}
  % \Big\{
   \int\! \frac{d\Omega_q}{4\pi} \int\! \frac{d\Omega_k}{4\pi}\, 
  \Big[i({\cal E}^3)_{p q}^{a b; \mu}\ \Xi^{b; \mu}_k%\Big]\Big) 
  i\nabla^a_q\Big]_{\y'} %\Big\}
  =  \int\! \frac{d\Omega_q}{4\pi}\frac{d\Omega_k}{4\pi}\,
  i\sqrt{\frac{\alpha_s}{\pi}} 
  \frac{p^\mu}{p.q} \Big[(i\nabla^a_p+ \Tilde U^{b a}_q i\Bar\nabla^b_p) \ 
  \Xi^{a; \mu}_k\Big]_{\y'} %\Big\}
\ .
\end{equation}
Taking into account the fact that the Langevin description is defined
only in a $\y$-discretized sense, one might be tempted to conclude
that the terms containing factors $\Tilde U_q$ correspond to real
gluon emission, while the others would generate virtual corrections,
not only in this, but also the other components.

A formal solution to~\eqref{eq:Jet-Langevin} is
\begin{align}
   U_{\y, p} =  {\sf P}_\y   \exp \Big\{i \int^\y_{\y_0} d\y'
  \int\! \frac{d\Omega_q}{4\pi}
   \frac{d\Omega_k}{4\pi}\, 
   \Big[{\cal E}_{p q}^{a b; \mu}\ \Xi^{b; \mu}_k 
  i\nabla^a_q\Big]_{\y'} \Big\} U_{\y_0, p}
\ ,
\end{align}
which again is to be interpreted in a discrete sense.  To write this
expression I have also adapted the definition of the functional
derivatives to include a $\delta$-function in $\y$:
$i\nabla^a_{\y,p} U_{\y' q} = -U_{\y,p}t^a
\Tilde\delta(p-q)\delta(\y-\y')$.

The operator
\begin{equation}
  \label{eq:shower-jet-full}
  {\sf P}_\y   \exp \Big\{i \int^\y_{\y_0} d\y'
  \int\! \frac{d\Omega_q}{4\pi}
   \frac{d\Omega_k}{4\pi}\, 
   \Big[{\cal E}_{p q}^{a b; \mu}\ \Xi^{b; \mu}_k 
  i\nabla^a_q\Big]_{\y'} \Big\}
\end{equation}
is the analog of the JIMWLK shower operator~\eqref{eq:small-x-shower}
and is not only relevant to the Langevin formulation itself. It
implements subsequent real emission with virtual corrections correctly
taken into account. This idea will be substantiated below in a three
step process: The first step consists of explicitly constructing the
real emission part of these amplitudes and comparing it to known
results~\cite{Bassetto:1983ik,Fiorani:1988by}. The second step is the
use of this information to recover the evolution
equation~\eqref{eq:FP-ng} --this step reinstates the virtual
corrections-- and the third step is the equivalence of Fokker-Planck
and Langevin treatment which closes the line of argument.

%Instead of jumping to conclusions at this point, I will substantiate
%these ideas with a derivation from first principles that follows the
%strategy of BMS. For this one needs a formulation in terms of
%functionals capable of handling finite $N_c$ corrections. This I will
%turn to next. While doing so, a number of interesting structures, like
%a closed, finite $N_c$ version of Eq.~\eqref{eq:GR-ee}
%will emerge alongside similar expressions for the underlying
%amplitudes.

\subsection{\it Amplitudes in the strongly ordered domain}
\label{sec:ampl-strongly-order}

In order to find a suitable starting point to derive results beyond
the large $N_c$ limit, one has to go rather far back and start with a
general discussion of soft gluon amplitudes along the lines given
already in~\cite{Bassetto:1983ik,Fiorani:1988by}.

To help organize the argument, I will start with the definition of a
generating functional for the (tree level) amplitudes of (real) soft
gluon emission from a $q\Bar q$ pair in the strongly ordered region as
already done in~\cite{Weigert:2003mm}\footnote{Compared to the
  treatment there $\Xi$ has been replaced by $i\Xi$ to make the
  correspondence to the shower operators immediate.}
\begin{equation}
  \label{eq:strong-ordering}
   \omega_{k_n} \ll \omega_{k_{n-1}} \ll \ldots \omega_{k_1} 
  \ll \omega_p < \omega_q 
\end{equation}
written as
\begin{align}
  \label{eq:simple-qqb-gen-func}
  {\sf A}_{p\,q}^{i j}[\Xi] :=\sum\limits_{n=0}^\infty & 
\int \frac{d\Omega_{k_1}}{4\pi}\ldots\frac{d\Omega_{k_n}}{4\pi} 
A(_{q p k_1 \ldots  k_n}^{ i j a_1 \ldots a_n}) 
i\Xi_{k_1}^{a_1} \ldots  i\Xi_{k_n}^{a_n}
\ .
\end{align}
The amplitudes for $n$-gluon emission from a initial $q\Bar q$ pair
are denoted by $A(_{q p k_1 \ldots k_n}^{ i j a_1 \ldots a_n})$, where
the momenta $q, p, k_1 \ldots k_n$ and color indices $i, j, a_1 \ldots
a_n$ are explicitly listed, a corresponding set of Lorentz-indices
$\mu_1\ldots \mu_n$ is suppressed. (For diagrammatic representations,
see Eq.~\eqref{eq:U-qqb-gen-func-graphically}.) They are isolated by
$n$-fold (functional) differentiation w.r.t. $i\Xi$ at $\Xi=0$. At
this stage, $\Xi$ is just an external source, the relationship to the
noise of the Langevin description will become clear shortly.

The $A(_{q p k_1 \ldots k_n}^{ i j a_1 \ldots a_n})$ are known to
satisfy an iterative structure in which the ${\text{color singlet}}\to
q\Bar q g^{n+1}_{\text{soft}}$-amplitude follows by induction from the
corresponding amplitude with $n$ soft gluons. As a result, the
$0$-gluon term in Eq.~\eqref{eq:simple-qqb-gen-func} fully determines
the whole functional, although the explicit construction of the $n$-th
order expression with the tools available to date becomes more and
more cumbersome with growing $n$. The rules of the game have been
explicitly demonstrated in~\cite{Fiorani:1988by} and require no
recourse to the $1/N_c$ limit.

The important observation now is that there
exists~\cite{Weigert:2003mm} a relatively simple, closed operator form
for this generating functional. In fact it consists of an even more
general operator acting on the zeroth order ``hard seed'' term in the
sense that this operator can be applied to any ``hard seed'' to
generate the full soft gluon cloud accompanying it, the full shower
operator already presented in Eq.~\eqref{eq:shower-jet-full}.  This
will be explained in this section using the existing information on
tree level amplitudes and hence the discussion leads to the real
emission part of this operator as given in
Eq.~\eqref{eq:emission-tree}.
%In~\cite{Weigert:2003mm} I have derived a closed operator form of the
%full tower of amplitudes contained in
%Eq.~\eqref{eq:simple-qqb-gen-func} that essentially consists of the
%real emission part of the operator in Eq.~\eqref{eq:shower-jet-full}.
%While explicit generation of large $n$ terms will still be cumbersome,
%the fact that I have a closed mathematical expression instead of a
%prescription for an iterative procedure allows me to use it in
%calculations and eventually in the derivation of the evolution
%equation.

% I {\em choose}
%to implement the strong ordering, by requiring that the $\Xi^a_k$ have
%only support in ranges conforming to
%Eq.~\eqref{eq:strong-ordering}.
%\marginpar{fix the $\omega_k$  integrations here!!!}

For this purpose, a slight generalization of the
definition~\eqref{eq:simple-qqb-gen-func} is operationally somewhat
more useful, in that it allows me to specify an explicit functional
form that casts the iteration step from $q\Bar q g^{n}_{\text{soft}}$
to $q\Bar q g^{n+1}_{\text{soft}}$ of~\cite{Fiorani:1988by} as a
simple functional operation. This operation will involve nothing more
complicated than the functional differentiation rules given in
Eq.~\eqref{Lie-der} and a suitably defined $n=0$ term to start the
process of generating the amplitudes.  The starting point is the
definition
\begin{align}
  \label{eq:U-qqb-gen-func}
  {\sf A}_{p\,q}^{i j}[U,\Xi] :=\sum\limits_{n=0}^\infty & 
\int \frac{d\Omega_{k_1}}{4\pi}\ldots\frac{d\Omega_{k_n}}{4\pi} 
A(_{q p k_1 \ldots  k_n}^{\Tilde i \Tilde j a_1 \ldots a_n}) 
\Tilde U_{k_1}^{a_1 b_1} i\Xi_{k_1}^{b_1}\ldots 
\Tilde U_{k_n}^{a_n b_n} i\Xi_{k_n}^{b_n}\ 
[U^\dagger_p]_{i \Tilde i} [U_q]_{\Tilde j j}
\ .
\end{align}
Here I have simply added a factor of $U$ in the appropriate
representation to each leg of the diagrams contained in the definition
of the ordered amplitudes. Diagrammatically the first few contributions read:
\begin{align}
  \label{eq:U-qqb-gen-func-graphically}
   {\sf A}_{p\,q}^{i j}[U,\Xi] = 
   \begin{minipage}[c]{1.5cm}
   \begin{center}
     \includegraphics[height=1.8cm]{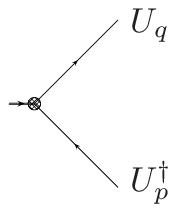}     
   \end{center}
 \end{minipage}
+
   \begin{minipage}[c]{2cm}
   \begin{center}
     \includegraphics[height=1.8cm]{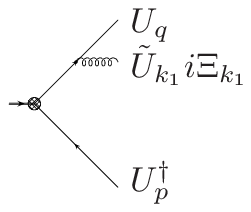}     
   \end{center}
 \end{minipage}
+
   \begin{minipage}[c]{2cm}
   \begin{center}
     \includegraphics[height=1.8cm]{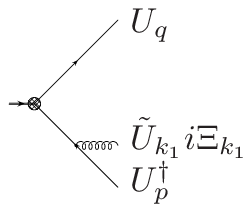}     
   \end{center}
 \end{minipage}
+   
\begin{minipage}[c]{2.1cm}
   \begin{center}
     \includegraphics[height=1.8cm]{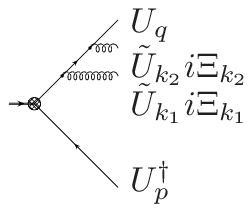}     
   \end{center}
 \end{minipage}
+
  \begin{minipage}[c]{2.1cm}
   \begin{center}
     \includegraphics[height=1.8cm]{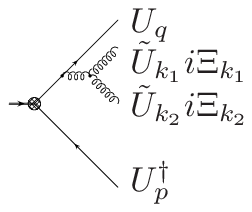}
   \end{center}
 \end{minipage}
+ \ldots
\end{align}
Obviously Eq.~\eqref{eq:simple-qqb-gen-func} is just the special case
at $U=1$: ${\sf A}_{p\,q}^{i j}[\Xi] = {\sf A}_{p\,q}^{i j}[1,\Xi]$.
The rationale behind this definition  lies in the very
nature of soft gluon emission underlying the construction of the
amplitudes. Per definition, soft gluon emission does not change the
direction of the parent (the emitter). In that role as a parent the
entity (be it a quark or a gluon) is therefore written as a path
ordered exponential along the direction of some momentum.  Emission of
a softer gluon, that in later steps will also serve as a parent will
then necessarily add an adjoint factor $\Tilde U_k$ and an eikonal
emission vertex $J_{l k}^\mu t^a$ (with $J^\mu_{p k} :=
\frac{p^\mu}{p.k}$), where the generator is to be taken in the
representation of the emitting object. As all these objects are
encoded as eikonal lines $U$, this will automatically occur if one
writes it via invariant vector fields $i{\Bar\nabla}^a_l$ just as in
the rewrite of the Langevin equation at the end of
Sec.~\ref{sec:an-equiv-lang}.  In functional form, one ends up with
the single emission operator
\begin{equation}
  \label{eq:real-emission-operator}
  i \int\! \frac{d\Omega_l}{4\pi}
   \frac{d\Omega_k}{4\pi}\, 
  g\,J_{l k}^\mu \Tilde U^{a b}_k \ \Xi^{b; \mu}_k%\Big]\Big) 
  i{\Bar\nabla}^a_l
\ .
\end{equation}
Note how closely this resembles the real emission part of the Langevin
equation~\eqref{eq:Jet-Langevin}.  Eq.~\eqref{eq:U-qqb-gen-func}
should then be completely determined by the $n=0$ term in the sum
which describes a $q\Bar q$-pair without additional soft gluons:
higher orders should just follow from repeated application
of~\eqref{eq:real-emission-operator}.  This needs now to be checked
against what is known about the soft gluon amplitudes.

First one needs to give an initial condition for this iteration, the
bare $q\Bar q$ term.  This reads
\begin{equation}
  \label{eq:qqb-iterative-initial}
 [{\sf A}^{(0)}]_{p\,q}^{i j}[U,\Xi] =
 \begin{minipage}[c]{2cm}
   \begin{center}
 \includegraphics[height=1.8cm]{bareqqb-U}     
   \end{center}
 \end{minipage}
:= 
M_2(q,p)  [U^\dagger_p U_q]_{i j} 
\end{equation}
and corresponds to a zero order amplitude of the form $A(_{q p}^{i j})
= M_2(q,p) \delta^{i j}$\footnote{For $M_n$ I have adopted the notation
of~\cite{Bassetto:1983ik,Fiorani:1988by}}.

The above prescription then claims that the $q\Bar q g_{\text{soft}}$
term is given by
\begin{equation}
  \label{eq:qqb-iterative-first}
 [{\sf A}^{(1)}]_{p\,q}^{i j}[U,\Xi]= \Big\{i \int\! \frac{d\Omega_l}{4\pi} 
 \frac{d\Omega_k}{4\pi}\ g\,
  J_{l k}^\mu \Tilde U^{a b}_k \ \Xi^{b; \mu}_k
  i{\Bar\nabla}^a_l \Big\} [{\sf A}^{(0)}]_{p\,q}^{i j}[U,\Xi] 
\ .
\end{equation}
To demonstrate once how to use the machinery, I will go through the
steps explicitly. First insert $[{\sf A}^{(0)}]_{p\,q}^{i j}[U,\Xi] $
from~\eqref{eq:qqb-iterative-initial}. This yields
\begin{equation}
  \label{eq:qqb-iterative-first-1}
 [{\sf A}^{(1)}]_{p\,q}^{i j}[U,\Xi]= \Big\{i \int\! \frac{d\Omega_l}{4\pi} 
 \frac{d\Omega_k}{4\pi}\ g\,
  J_{l k}^\mu \Tilde U^{a b}_k \ \Xi^{b; \mu}_k
  i{\Bar\nabla}^a_l \Big\} 
  M_2(q,p)  [U^\dagger_p U_q]_{i j} 
\ .
\end{equation}
All that is left to do is to use the differentiation
rules~\eqref{Lie-der} (with the adaptation to phase
space~\eqref{eq:nabla-def} taken into account) and differentiate the
two factors $U^\dagger_p$ and $U_q$. The variation eliminates the $l$
integral and one is left with
\begin{align}
  \label{eq:qqb-g-emission}
  [{\sf A}^{(1)}]_{p\,q}^{i j}[U,\Xi]= &
    \begin{minipage}[c]{2.5cm}
   \begin{center}
     \includegraphics[height=1.8cm]{qqb-gu-UXi-i}     
   \end{center}
 \end{minipage}
+
   \begin{minipage}[c]{2.5cm}
   \begin{center}
     \includegraphics[height=1.8cm]{qqb-gd-UXi-i}     
   \end{center}
 \end{minipage}
\nonumber \\ &
=\int\! \frac{d\Omega_k}{4\pi} 
  M_2(p,q) A_{p q}^\mu(k) [U^\dagger_p t^a U_q]_{i j} U^{a b}_k i\Xi^{b; \mu}_k
=
   \begin{minipage}[c]{2.5cm}
   \begin{center}
     \includegraphics[height=1.8cm]{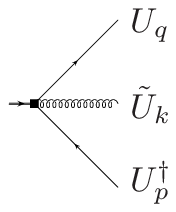}     
   \end{center}
 \end{minipage} i\Xi_k
\end{align}
where
\begin{equation}
  \label{eq:A-def}
   A_{p q}^\mu(k) = J^\mu_{p k}-J^\mu_{q k}
   \ . 
\end{equation}
Now one isolates the amplitude via a variation in $\Xi$ and reads off
\begin{equation}
  \label{eq:qqb-g-amp}
  A(_{p q k}^{i j a}) =  M_2(p,q) A_{p q}^\mu(k) [ t^a ]_{i j} 
  =: M_3(p,k,q) [ t^a ]_{i j}
\end{equation}
with $ M_3(q,k,p)$ again as in the notation
of~\cite{Bassetto:1983ik,Fiorani:1988by}. Eq.~\eqref{eq:qqb-g-amp} in
fact is identical to Eq.~(40) of~\cite{Fiorani:1988by} and an
additional gluon emission, i.e. iteration with
Eq.~\eqref{eq:real-emission-operator}, leads to Eq.~(42)
of~\cite{Fiorani:1988by} for the amplitude with $2$ soft gluons:
\begin{align}
  \label{eq:two-g-emission}
   [{\sf A}^{(2)}]_{p\,q}^{i j}[U,\Xi]= &
\begin{minipage}[c]{2.5cm}
   \begin{center}
     \includegraphics[height=1.8cm]{qqb-g2u-UXi-i}     
   \end{center}
 \end{minipage}
+
  \begin{minipage}[c]{2.5cm}
   \begin{center}
     \includegraphics[height=1.8cm]{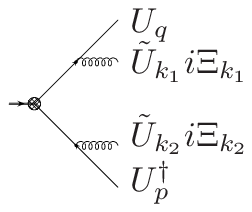}
   \end{center}
 \end{minipage}
+
  \begin{minipage}[c]{2.5cm}
   \begin{center}
     \includegraphics[height=1.8cm]{qqbggsplitu-UXi-i}
   \end{center}
 \end{minipage}
\nonumber \\ &
+
\begin{minipage}[c]{2.5cm}
   \begin{center}
     \includegraphics[height=1.8cm]{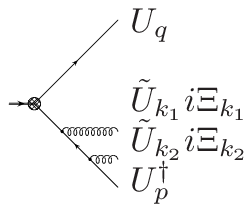}     
   \end{center}
 \end{minipage}
+
  \begin{minipage}[c]{2.5cm}
   \begin{center}
     \includegraphics[height=1.8cm]{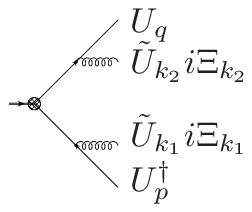}
   \end{center}
 \end{minipage}
+
  \begin{minipage}[c]{2.5cm}
   \begin{center}
     \includegraphics[height=1.8cm]{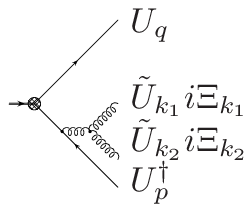}
   \end{center}
 \end{minipage}
\ .
\end{align}
(The color algebra for the explicit comparison is given
in~\cite{Weigert:2003mm}, App. A, case $n=1$.)  The pattern simply
follows the rules of functional differentiation and a comparison of
analytic expressions for the amplitudes as given by the iteration
rules of~\cite{Fiorani:1988by} was performed explicitly using symbolic
algebra tools up to $n=6$, where the task starts to become time
consuming. To compare beyond finite orders, a few additional tools are
required.

In the above I have implemented the ordering ``by hand,'' simply by
requesting the hierarchy~\eqref{eq:strong-ordering} for the
$\omega_{k_i}$. To write the all orders expression advertised
above, I need one more ingredient: the addition of an energy integral
to the emission operator~\eqref{eq:real-emission-operator} and an
appropriate extension of the definition of $i\Bar\nabla^a_p$ to
$i\Bar\nabla^a_{\omega_p,p}$ defined such that
\begin{equation}
  \label{eq:new-nabla}
  i\Bar\nabla^a_{\omega_p,p} U_q =  t^a U_p \Tilde\delta(p-q)
\end{equation}
where $\Tilde\delta(p-q)$ is adapted to the measure used in the
momentum and energy integrations. The natural choice for the latter is
to replace the solid angle integrations
in~\eqref{eq:real-emission-operator} by full fledged phase space
integrations:
\begin{equation}
  \label{eq:now-phase-space}
  \frac{d\Omega_k}{4\pi} \to d\Phi_k:=\frac{d^4k}{(2\pi)^4} 
  \delta(k^2)\theta(\omega_k) 
  = \frac{\omega_k d\omega_k}{(2\pi)^2} 
  \frac{d\Omega_k}{4\pi}\theta(\omega_k)
\ .
\end{equation}
$\Tilde\delta(p-q)$ is then defined as the appropriate $\delta$
function on the forward light cone, i.e. with Jacobian factors such that
\begin{equation}
  \label{eq:delta}
  \int  d\phi_k \Tilde\delta(p-k)f(k) = f(p) 
\ .
\end{equation}
(Note again the similarity of the roles played by $\y$-ordering in
the Langevin description and energy ordering in this case.)  With
these definitions it is straightforward to verify that a replacement
of the solid angle integrations in the above by phase space
integrations leads to no change in the previous results, if one
implements energy ordering for subsequent definitions with ordering
$\theta$-functions, just as one would for the contour parameter in
case of path ordered exponential. To leading logarithmic accuracy this
will not change the result for the evolution equation.\footnote{Also
  the results below for the structure of the multiple ordered soft
  emission can be obtained both ways -- to have a formulation with all
  ingredients (including the ordering) written explicitly is merely a
  convenience from this perspective. What the formalism does, is to
  allow to give a closed expression for the generating functional,
  Eq.~\eqref{eq:exp-amp}.}

By now it should appear to be the natural choice, if I write for the
generating functional (again suppressing the $\omega$-label for compactness)
\begin{align}
  \label{eq:exp-amp}
  {\sf A}_{p\,q}^{i j}[U,\Xi] 
%:=\sum\limits_{n=0}^\infty & 
%\int \frac{d\Omega_{k_1}}{4\pi}\ldots\frac{d\Omega_{k_n}}{4\pi} 
%A(_{q p k_1 \ldots  k_n}^{\Tilde i \Tilde j a_1 \ldots a_n}) 
%\Tilde U_{k_1}^{a_1 b_1}\Xi_{k_1}^{b_1}\ldots 
%\Tilde U_{k_n}^{a_n b_n}\Xi_{k_n}^{b_n}\ 
%[U^\dagger_p]_{i \Tilde i} [U_q]_{\Tilde j j}
%\nonumber \\ = &
= &
 {\sf P}_{\omega_k}  \exp \Big\{i \int\! d\Phi_l
   d\Phi_k\ 
  g\, J_{l k}^\mu \Tilde U^{a b}_k \ \Xi^{b; \mu}_k 
  i{\Bar\nabla}^a_l \Big\} 
[{\sf A}^{(0)}]_{p\,q}^{i j}[U,\Xi]  
%[U^\dagger_p U_q]_{i j} M_2(p, q)
%\nonumber \\ := & 
%\sum\limits_{n=0}^\infty   {\sf P}_{\omega_k} \Big[ 
%\Big\{\int\! \frac{d\Omega_{l_n}}{4\pi}\Big(
%   i\ \Big[\int\! \frac{d\Omega_{k_n}}{4\pi}\, 
%  J_{l_n k_n}^{\mu_n} \Tilde U^{a_n b_n}_{k_n} \ 
%  \Xi^{b_n; \mu_n}_{k_n}\Big]\Big) 
%  i{\Bar\nabla}^{a_n}_{l_n} \Big\} 
%\nonumber \\ & \hspace{1cm}
%\ldots
%\Big\{\int\! \frac{d\Omega_{l_1}}{4\pi}\Big(
%   i\ \Big[\int\! \frac{d\Omega_{k_1}}{4\pi}\, 
%  J_{l_1 k_1}^{\mu_1} \Tilde U^{a_1 b_1}_{k_1} \ \Xi^{b_1; \mu_1}_{k_1}\Big]\Big) 
%  i{\Bar\nabla}^{a_1}_{l_1} \Big\} 
%
%[U^\dagger_p U_q]_{i j} M_2(p, q)
\ ,
\end{align}
where ${\sf P}_{\omega_k}$ implements the strong ordering of
Eq.~\eqref{eq:strong-ordering}. I should actually use a notation that
displays the dependence on the functional form supplied in the zero
emission part and for instance write ${\sf A}_{p\,q}^{i
  j}[U,\Xi,M_2(p,q)]$.

In analogy with path ordered exponentials, this implies that the
exponential series is to be interpreted as
{\small 
\begin{align}
  \label{eq:ordered-exp-series}
%  {\sf A}_{p\,q}^{i j}[U,\Xi] 
%:=\sum\limits_{n=0}^\infty & 
%\int \frac{d\Omega_{k_1}}{4\pi}\ldots\frac{d\Omega_{k_n}}{4\pi} 
%A(_{q p k_1 \ldots  k_n}^{\Tilde i \Tilde j a_1 \ldots a_n}) 
%\Tilde U_{k_1}^{a_1 b_1}\Xi_{k_1}^{b_1}\ldots 
%\Tilde U_{k_n}^{a_n b_n}\Xi_{k_n}^{b_n}\ 
%[U^\dagger_p]_{i \Tilde i} [U_q]_{\Tilde j j}
%\nonumber \\ = &
%= &
  {\sf P}_{\omega_k} &  \exp \Big\{i \int\! d\Phi_k d\Phi_l\ 
  g\,J_{l k}^\mu \Tilde U^{a b}_k \ \Xi^{b; \mu}_k
  i{\Bar\nabla}^a_l \Big\} 
%[{\sf A}^{(0)}]_{p\,q}^{i j}[U,\Xi]  
%[U^\dagger_p U_q]_{i j} M_2(p, q)
\nonumber \\ := & 
\sum\limits_{n=0}^\infty   {\sf P}_{\omega_k} \Big[ 
\Big\{i \int\!d\Phi_{k_n} d\Phi_{l_n} \ 
  g\, J_{l_n k_n}^{\mu_n} \Tilde U^{a_n b_n}_{k_n} \ 
  \Xi^{b_n; \mu_n}_{k_n}%\Big]\Big) 
  i{\Bar\nabla}^{a_n}_{l_n} \Big\} 
%\nonumber \\ & \hspace{1cm}
\ldots
\Big\{i \int\! d\Phi_{k_1} d\Phi_{l_1}\ 
 g\, J_{l_1 k_1}^{\mu_1} 
 \Tilde U^{a_1 b_1}_{k_1} \ \Xi^{b_1; \mu_1}_{k_1}%\Big]\Big) 
  i{\Bar\nabla}^{a_1}_{l_1} \Big\} 
\Big]
%[U^\dagger_p U_q]_{i j} M_2(p, q)
\ ,
\end{align}
}where the $1/n!$ are absent due to the explicit ordering of the
subsequent soft emissions.
%\footnote{ To leading logarithmic accuracy,
%  one may represent the ordering by a product of Heavyside functions,
%  simply writing ${\sf P}_{\omega_k}$ via $\theta(\omega_1-\omega_2) \ldots
%  \theta(\omega_{n-1}-\omega_n)$, i.e.  replace the strong ordering by
%  a mere ordering. The fact that there is ordering at all as it will
%  become clear later, is in direct correspondence to the
%  $\y$-ordering in the Langevin description.}  
For compactness~\cite{Weigert:2003mm} let ${\sf U}[\Xi,U]$ denote the
operator
\begin{equation}
  \label{eq:emission-tree}
  {\sf U}[\Xi,U] :=  {\sf P}_{\omega_k}   
  \exp \Big\{\int\! 
  d\Phi_k d\Phi_l\ 
  g\, J_{l k}^\mu \Tilde U^{a b}_k \ \Xi^{b; \mu}_k%\Big]\Big) 
  i{\Bar\nabla}^a_l \Big\}
\ . 
\end{equation}  The explicit mention of
the $U$-field in the notation will be needed later when different
types of eikonal fields appear.  ${\sf U}[\Xi,U]$ has a meaning
independent of the bare ($n=0$) term it is used to act on. It is
called the (real) shower operator and arises wherever one is forming
that ordered soft gluon cloud around a hard seed.
  
With these tools a direct comparison with the iterative scheme
of~\cite{Fiorani:1988by} at all orders becomes possible. In essence
all that is left to confirm is that the iteration in the shower
operator leads to the same result as the induction step from $n$ to
$n+1$ gluons in~\cite{Fiorani:1988by}. This has been done
in~\cite{Weigert:2003mm}.

It is already apparent that the structure contained in the above is
very similar to the information contained in the Langevin equation.
This has been used to derive the RG equation stated
earlier~\cite{Weigert:2003mm}. The main observation in this regard is
that, as a consequence of the strong ordering, any
$\frac{\delta}{\delta \Xi^{a; \mu}_{k}}$ in which $k$ is at (or near)
the phase space boundary, there will be only a contribution from the
first emitted gluon ($n=1$). As a consequence, for such variations one
gets   (h.b. stands for hard boundary)
\begin{align}
    \label{hard-variation}
    \frac{\delta}{\delta \Xi^{b; \mu}_{k}}\bigg\vert_{\text{h.b.}} 
    {\sf U}[\Xi,U]
={\sf U}[\Xi,U]
\Big\{ i \int\! d\Phi_l\
  %\frac{d\Omega_{l}}{4\pi}\Big(\ \Big[
  J_{l k}^{\mu} \Tilde U^{a b}_{k} %\Big]\Big) 
  i{\Bar\nabla}^a_l \Big\}
\ .
  \end{align}
 Note that
\begin{align}
    \label{eq:qqbg}
     {\sf A}_{p\,q\, k}^{i j b}[U,\Xi,M_3(p,k,q)] := 
 {\sf U}[\Xi,U]
\ 
  \Tilde U^{a b}_{k} [ U^\dagger_p t^a U_q]_{i j}
  M_3(p,k,q)
\end{align}
is the $q\Bar q g$ counterpart to Eq.~\eqref{eq:exp-amp}. In short,
Eq.~\eqref{hard-variation} states, that under variation at the phase
space boundary, one relates the generating functional for a given
tower of amplitudes to that of a tower with an additional hard gluon.
For the above example it relates the amplitudes with a hard $q\Bar q$
to those with hard $q\Bar q g$ content:
\begin{align}
    \label{eq:qqb-qqbg-rel}
     \frac{\delta}{\delta i\Xi^{b; \mu}_{k}}\bigg\vert_{\text{h.b.}} 
     {\sf A}_{p\,q}^{i j}[U,\Xi,M_2(p,q)] 
     =    {\sf A}_{p\,q\, k}^{i j b}[U,\Xi, A_{p\, q}^\mu(k)M_2(p,q)]
\ ,
\end{align}
or diagrammatically
\begin{equation}
  \label{eq:Xivar}
      \frac{\delta}{\delta i\Xi^{b; \mu}_{k}}\bigg\vert_{\text{h.b.}} 
      \> 
    {\sf U}[\Xi,U] \>
   \begin{minipage}[c]{2cm}
   \begin{center}
     \includegraphics[height=1.8cm]{bareqqb-U}     
   \end{center}
 \end{minipage} 
=\  
{\sf U}[\Xi,U]    \begin{minipage}[c]{2.5cm}
   \begin{center}
     \includegraphics[height=1.8cm]{qqbg-U}     
   \end{center}
 \end{minipage} 
\ .
\end{equation}
These two items, the ordered nature leading to~\eqref{hard-variation}
under ``hard'' variations, and the fact that in this situation the
number of hard legs of the soft emission amplitudes is increased by
one, as exemplified in~\eqref{eq:Xivar}, will be the core observations
behind the evolution equation I am aiming at.

\subsection{\it Transition probabilities from $\Xi$ averages}
\label{sec:trans-prob-from}

As a first step to understand how to translate the above into
expressions for transition probabilities, let me take the situation of
a (real) $q\Bar q g^n_{\text{soft}}$ soft emission amplitude in which
one must obtain the standard result
\begin{align}
  \label{eq:qqbgn-prob}
%\int\! d\phi_1\ldots d\phi_n \
A(_{q p k_1 \ldots  k_n }^{i j a_1 \ldots a_n} ) 
[ A(_{q p k_1 \ldots  k_n }^{i j a_1 \ldots a_n} ) ]^\dagger
= &
%\int\! d\phi_1\ldots d\phi_n \
\sum\limits_{\Pi_{n+2}(l)}\sum\limits_{\Pi_{n+2}(l')}
M_n(q,q_{l_1},\ldots, q_{l_n} ,p)
M_n(q,q_{l_1},\ldots, q_{l_n} ,p)^*\nonumber \\ & \hspace{2cm}
\times \tr(t^{a_{l_1}}\ldots\ldots t^{a_{l_n}} t^{a_{l'_n}}\ldots
 \ldots t^{a_{l'_1}})
\ ,
\end{align}
with gluon momenta in the final state labelled by $q_1, \ldots, q_n$.
Indeed, the same expression is obtained, if one considers the $\Xi$
averaged product of functionals
\begin{align}
  \label{eq:prob-from-func-fixed}
  \langle  [{\sf A}^{(n)}]_{p\,q}^{i j}[U,\Xi] 
  [{\sf A}^{(n)}]_{p\,q}^{i j}[U,\Xi]^\dagger \rangle_{n,\Xi}
\end{align}
if the average is defined with a Gaussian distribution for $\Xi$ in
which only strongly ordered modes $q_1, \ldots, q_n$ have support. I
write
\begin{equation}
  \label{eq:xi-average}
  \langle \ldots \rangle_{n\Xi} := 
  \det(M_n)^{\frac{1}{2}} \int\!D[\Xi] \ldots 
  e^{-\frac{1}{2}\int d\phi_p d\phi_q \Xi_p^t M^{-1}_{n,\, p q} \Xi_q}
\end{equation}
where $d\phi_k$ denotes the phase space integral for momentum $k$ and
I have suppressed discrete indices. $M$ defines the $\Xi$ correlator,
which in the fully exclusive case considered above would read
\begin{equation}
  \label{eq:xicorr-2}
  M_{n,\, p q}^{a,\mu\ b,\nu} = 
\langle \Xi_p^{a, \mu} \Xi_q^{b, \nu}\rangle_{n,\Xi} 
= \delta^{a b} g^{\mu\nu} \Tilde \delta(p-q)\ 
  \sum\limits_{i=1}^n \Tilde\delta(q-q_i)
\ .
\end{equation}
In this case, because of the strong ordering of momenta in the
amplitudes, the $\Xi$ integral will strictly {\em pair off} the one
available momentum $k_i$ in the amplitudes falling into the same range
as $q_i$ with the latter.

As such, this is only true if I have ordering in the emission
vertices, either done by hand in the 2-d version, or via ${\sf
  P}_{\omega_k}$ in the 2+1-d formulation. Only then, unwanted cross
terms are excluded.  The $U$ factors just cancel trivially.

To make contact with the non-global observables of the BMS setting,
one needs to depart from the fixed $n$ situation and replace the
$\sum\limits_{i=1}^n \Tilde\delta(q-q_i)$, which selects a given set
of final state momenta in the case where the number of final state
gluons is specified, by a cutoff criterion that restricts real
emission by geometry and energy. By writing
\begin{align}
  \label{eq:M-BMS}
   M_{p q}^{a,\mu\ b,\nu} = 
\langle \Xi_p^{a, \mu} \Xi_q^{b, \nu}\rangle_\Xi  
=
\delta^{a b} g^{\mu\nu} \Tilde \delta(p-q)\ \theta(E-\omega_q) u(q)
\ ,
\end{align}
one can indeed go beyond a final state with a fixed number of gluons
and consider
\begin{align}
  \label{eq:prob-from-func-all}
  \langle  {\sf A}_{p\,q}^{i j}[U,\Xi] 
  {\sf A}_{p\,q}^{i j}[U,\Xi]^\dagger \rangle_\Xi
\ ,
\end{align}
which will provide the finite $N_c$-generalization of $G_{a
  b}^{(\text{real})}$ of BMS.

In fact any degree of inclusiveness  may be imposed by modifying the
restrictions on the allowed modes appearing on the r.h.s. of
Eq.~\eqref{eq:M-BMS}.

%At this point a few comments about the meaning of $\Xi$ as it appears
%in the above are in order. Viewed from a diagrammatic point of view,
%the $\Xi$ stand for nothing else but the final state soft gluons. This
%explains why there is an additional factor of $i$ together with the
%use of the free gluon phase space measure in the averaging, compared
%to structures encountered in Sec.~\ref{sec:an-equiv-lang}. There,
%$\Xi$ should be viewed simply as a perturbative gluon and the same
%ingredients as those encountered here would show up in the formulation
%of $S$-matrix elements via the amputation of external legs.

This realization should make it clear how to include virtual
corrections right from the outset, by a fairly simple modification of
the above. I will refrain from doing so here and instead keep a closer
parallel with the strategy employed
by~\cite{Banfi:2002hw,Bassetto:1983ik,Fiorani:1988by} and use the
expressions for real emission to deduce the structure of the evolution
equation.

\subsection{\it Evolution equations for soft 
  semi-inclusive quantities}
\label{sec:evol-equat-soft}

For semi-inclusive quantities like the non-global observables
of~\cite{Banfi:2002hw}, in which one sums over a given, limited phase
space volume of soft gluons, it is natural to ask for the dependence
on that phase space boundary. Unlike a direct calculation of the
average in Eq.~\eqref{eq:prob-from-func-all} this should lead to a
tractable result that exhibits new structure. In fact, one recovers
the generalization to the BMS equation suggested above.

To arrive at an evolution equation, I will simply take a derivative
w.r.t. the phase space boundary of the emitted gluons in the
semi-inclusive probability~\eqref{eq:prob-from-func-all}. This is in
direct correspondence with the procedure used in BMS. Here it is
essential that one includes the phase space boundary into the
definition of $M$. Then the result is most transparently displayed
using the following simple relationship, based on functional
differentiation and Legendre transformation in the case of a Gaussian
action (``free theory''):
\begin{align}
  \label{eq:average-Legendre}
   \big\langle W[\Xi] \big\rangle_{\Xi}
   = & N \int D[\Xi]\ W[\Xi]\
   e^{-\frac{\Xi^t M^{-1} \Xi}{2}}
   = 
    W[\frac{\delta}{i\, \delta  {\cal J}}] 
    e^{-\frac{{\cal J}^t M {\cal J}}{2}}\Big\vert_{{\cal J}=0}
    = e^{-\frac{\frac{\delta}{i \delta \Xi_0} M 
        \frac{\delta}{i \delta \Xi_0}}{2}} W[\Xi_0]\Big\vert_{\Xi_0=0}
\end{align}
for any functional $W[\Xi]$. Here notation has been condensed even
further, with all momenta, integration signs and measures
suppressed.\footnote{For compactness I will do so throughout this
  section. Where indices and momenta are shown it is with the
  understanding that there will be integration conventions for
  repeated momenta with $d\Phi_k$ as the measure.}  Legendre machinery
is used for the last equality sign. The ``classical field,'' $\Xi_0 =
i M {\cal J}$, vanishes at ${\cal J}=0$.

The canonical example for $W[\Xi]$ here of course is
\begin{equation}
  \label{eq:W-example}
  \big\langle W[\Xi] \big\rangle_{\Xi}\to  
  \big\langle{\sf A}_{p\,q}^{ij}[U,\Xi] 
  {\sf A}_{p\,q}^{i j}[U,\Xi]^\dagger\big\rangle_{\Xi}
\ ,
\end{equation}
which, by construction, is the finite $N_c$ generalization of
$G^{(\text{real})}_{a b}$.

For this example one immediately realizes that the exponential of the
second order differential operator on the right hand side will perform
precisely the ``sewing'' implemented by the Gaussian weight in
Eq.~\eqref{eq:prob-from-func-all}, as long as $M$ is diagonal in
energies as in \eqref{eq:M-BMS}. In this case, the energy ordering
ensures that
\begin{equation}
  \label{eq:sewing-step}
  \frac{1}{2}\frac{\delta}{i \delta \Xi_0} M 
        \frac{\delta}{i \delta \Xi_0}{\sf A}_{p\,q}^{i j}[U,\Xi_0]
 {\sf A}_{p\,q}^{i
  j}[U,\Xi]^\dagger = 
\Big(\frac{\delta}{i \delta \Xi_0}{\sf A}_{p\,q}^{i j}[U,\Xi_0] \Big)
 M 
        \Big(\frac{\delta}{i \delta \Xi_0}
{\sf A}_{p\,q}^{i
  j}[U,\Xi]^\dagger \Big)\ .
\end{equation}

Let me now carry out the derivative with respect to the phase space
boundary in analogy with the derivation of the BMS equation sketched
in Sec.~\ref{sec:analogy}. With $M=M(E)$ a function of the phase space
boundary, one immediately finds a general expression for the
logarithmic $E$-derivative of the above expectation value that reads
\begin{equation}
  \label{eq:Legendre-E-der}
 E\partial_E \ \big\langle W[\Xi] \big\rangle_{\Xi}(E) 
= - \frac{1}{2}e^{-\frac{\frac{\delta}{i \delta \Xi_0} M(E) 
        \frac{\delta}{i \delta \Xi_0}}{2}} \
    \frac{\delta}{i \delta \Xi_0} E\partial_E M(E) 
        \frac{\delta}{i \delta \Xi_0}\ 
        W[\Xi_0]\Big\vert_{\Xi_0=0}
\ .
\end{equation}

The main point about this seemingly trivial exercise is its use in
conjunction with what is already known about functional derivatives of
the amplitudes of interest:
Eqns.~\eqref{hard-variation},~\eqref{eq:qqb-qqbg-rel}. Note that
$\partial_E M(E)$ will force the variations in the factor taken down
from the exponential to be at the phase space boundary. Choosing
$W[\Xi]= {\sf A}_{p\,q}^{i j}[U,\Xi] {\sf A}_{p\,q}^{i
  j}[U,\Xi]^\dagger$, the square of the generating functional of
$q\Bar q\, g^n_{\text{soft}}$ amplitudes, this location at the
boundary will lead to the appearance of the square of the generating
functional of $q\Bar q g\, g^n_{\text{soft}}$ on the right hand side
as follows from~\eqref{eq:qqb-qqbg-rel},~\eqref{eq:Xivar}.

This marks an important difference compared to the BMS case at
infinite $N_c$: the equation does not close. Instead higher and higher
hard correlators will enter: Clearly, an evolution equation for the
$q\Bar q g\, g^n_{\text{soft}}$ amplitudes will then couple in turn to
the corresponding object for $q\Bar q g^2\, g^n_{\text{soft}}$
amplitudes and so forth.  One is faced with an infinite hierarchy. All
quantities encountered in this hierarchy can be defined in complete
analogy to the examples presented in~\eqref{eq:exp-amp} and
\eqref{eq:qqbg}. One simply uses the desired combination of
$U$-factors as the ``bare'' term in the tower of amplitudes one is
interested in.

This implies that firstly, it is not sufficient to consider only the
evolution equation for $q\Bar q g^n_{\text{soft}}$ amplitudes alone.
Instead one has to capture the complete infinite coupled hierarchy of
equations in one. Secondly, it indicates that such a step is possible,
since the objects of interest have already been identified and indeed
can be written down in a general form. The evolution equations then
follow from~\eqref{eq:Legendre-E-der}.

I therefore proceed to define the (tree level) generating functional
based on $m$ hard particles, a general ``antenna pattern,'' by
\begin{equation}
  \label{eq:m-hard-particles}
  {\sf A}_{p_1 \ldots p_m}^{i_1
  \ldots i_n}[U,\Xi] {\sf A}_{p_1\ldots p_m}^{i_1 \ldots i_n}[V,\Xi]^\dagger
  = {\sf U}[\Xi,U]
  \
  (UV^\dagger)_{p_1}^{(\dagger)}\otimes  \ldots  
  \otimes (UV^\dagger)_{p_m}^{(\dagger)}
  \ 
  {\sf U}[\Xi,V]^\dagger \Big\vert_{U=V}
\ .
\end{equation}
The factor $ (UV^\dagger)_{p_1}^{(\dagger)}\otimes \ldots \otimes
(UV^\dagger)_{p_m}^{(\dagger)}$ in this expression generalizes the
expression for the bare particles of the $q\Bar q$ case (the ${\sf
  A}_{p\,q}^{i j}[U,\Xi] {\sf A}_{p\,q}^{i j}[U,\Xi]^\dagger$ from
above) from $2$ to $m$ hard legs, and the ${\sf U}[\Xi,U] \ \ldots \ 
{\sf U}[\Xi,V]^\dagger \Big\vert_{U=V}$ implement the showering.
I have given only $q$ and $\Bar q$ factors as any
gluon can be written in terms of these as $\Tilde{(UV^\dagger)}_k^{a
  b} = 2\tr( t^a (UV^\dagger)_k^\dagger t^b (UV^\dagger)_k)$.

One can now study the evolution of the expectation value of a given
object of the type~\eqref{eq:m-hard-particles} or directly study a
generating functional for {\em all} such objects in one go. The latter
is given by
\begin{align}
  \label{eq:genfunc}
  {\sf W}[j^\dagger,j] := & \Big\langle
  {\sf U}[\Xi,U] \
  e^{i 2\tr j^\dagger U V^\dagger +
  i 2\tr (U V^\dagger)^\dagger j}
  \
  {\sf U}[\Xi,V]^\dagger   
  \Big\rangle_{\Xi}\Big\vert_{U=V}
\end{align}
with integrals in the source exponents understood.  Appropriate
variations with respect to $j$ and $j^\dagger$ at $j=0$ will then
select a given tower of $\text{in}\to\text{hard}\, g^n_{\text{soft}}$
probabilities entering the hierarchy. Since, besides the phase space
constraints contained in $M$, it is the particle content of the bare
terms that define the amplitudes, a notation that emphasizes this is
needed. To this end I will write
\begin{align}
  \label{eq:UV-average}
  \frac{\delta}{i\delta j^{(\dagger)}_{p_1}} \ldots  
  \frac{\delta}{i\delta j^{(\dagger)}_{p_m}}  
  {\sf W}[j^\dagger,j]
  = &\Big\langle
  {\sf U}[\Xi,U] \
  (UV^\dagger)_{p_1}^{(\dagger)}\otimes  \ldots  
  \otimes (UV^\dagger)_{p_m}^{(\dagger)}
  \
  {\sf U}[\Xi,V]^\dagger   
  \Big\rangle_{\Xi}\Big\vert_{U=V}
\nonumber \\ 
  =: 
&
  \Big\langle \ (UV^\dagger)_{p_1}^{(\dagger)}\otimes  \ldots  
  \otimes (UV^\dagger)_{p_m}^{(\dagger)}\
  \Big\rangle_{UV^\dagger}^{\text{real}}
\end{align}
where the $UV^\dagger$ average in the second line is to be taken with
a yet unknown weight that reproduces the l.h.s.. The idea is to
interpret the soft gluon average as an average over $UV^\dagger$
configurations. This is completely legal, if somewhat formal as long
as the only definition of the weight of the average is
through~\eqref{eq:UV-average}.

Nevertheless, it is this reinterpretation that will allow to make
contact with the Fokker-Planck formulation of
Sec.~\ref{sec:from-jimwlk-bk} and that is all that is really needed.

For $m=2$ and an adapted choice for the color structure one encounters
\begin{equation}
  \label{eq:G-at-m=2}
  \langle 
  \tr((UV^\dagger)^\dagger_p (UV^\dagger)_q)/N_c 
  \rangle_{UV^\dagger}^{\text{real}} 
  =: G_{a b}^{(\text{real})}(E)
\end{equation}
and thus provides the precise definition of what was still a bit vague
in~\eqref{eq:interpretation}.  Eq.~\eqref{eq:UV-average} is of course
much more general than that: it defines the real emission part for a
general antenna pattern, as these will all be needed for the evolution
equations.  Inclusion of virtual contributions will be discussed
below.  Note that the eikonal lines appearing in~\eqref{eq:UV-average}
have the interpretation of a product $UV^\dagger$ of contributions
from both the amplitude and the conjugate amplitude.

With~\eqref{eq:average-Legendre} one then has
\begin{align}
  \label{eq:funcform}
   {\sf W}[j^\dagger,j] = &  e^{-\frac{\frac{\delta}{i \delta \Xi_0} M(E) 
        \frac{\delta}{i \delta \Xi_0}}{2}}
{\sf U}[\Xi_0,U]\
  e^{i 2\tr j^\dagger U V^\dagger +
  i 2\tr (U V^\dagger)^\dagger j}
\ 
{\sf U}[\Xi_0,V]^\dagger
\Big\vert_{U=V}
\nonumber \\ = &
%=   
\big\langle   e^{i 2\tr j^\dagger U V^\dagger +
  i 2\tr (U V^\dagger)^\dagger j} \big\rangle_{UV^\dagger}^{\text{real}}
\ .
\end{align}
This demonstrates that $ {\sf W}[j^\dagger,j] $ is in complete analogy
with (the real emission part of) $\Bar Z[J,J^\dagger]$ in the
derivation of the JIMWLK equation of Sec.~\ref{sec:syst-deriv}.

To find the evolution equation for this (meta-) functional and with it
the infinite hierarchy of evolution equations alluded to above, all
that is left to do, is to put together
Eqns.~\eqref{eq:Legendre-E-der},~\eqref{eq:sewing-step},
and~\eqref{hard-variation} to get the dependence on phase space
boundaries for arbitrary $UV$-correlators (bare $m$-jet probabilities).
This yields the r.h.s. for the real emission contribution:
\begin{align}
  \label{eq:real-emission}
    & e^{-\frac{\frac{\delta}{i \delta \Xi_0} M(E) 
        \frac{\delta}{i \delta \Xi_0}}{2}}
{\sf U}[\Xi_0,U]  
\nonumber \\ & 
%\frac{1}{2}
\Big\{%\int\! \frac{d\Omega_{l}}{4\pi}\Big(
   %g\ \Big[
  g\, J_{l k}^{\mu} \Tilde U^{a b}_{k} %\Big]\Big) 
  i{\Bar\nabla}^a_{U_l} \Big\} [E\partial_E M(E)]_{l l'\,\mu\mu'}^{b b'}
  \Big\{%\int\! \frac{d\Omega_{l'}}{4\pi}\Big(
   %g\ \Big[
  g\, J_{l' k'}^{\mu'} \Tilde V^{a' b'}_{k'} %\Big]\Big) 
  i{\Bar\nabla}^a_{V_l'} \Big\} \
  e^{i 2\tr j^\dagger U V^\dagger +
  i 2\tr (U V^\dagger)^\dagger j}
\nonumber \\ & \ {\sf U}[\Xi_0,V]^\dagger \Big\vert_{U=V}
%\nonumber \\ = &
%=   \big\langle   e^{i 2\tr j^\dagger U V^\dagger +
%  i 2\tr (U V^\dagger)^\dagger j} \big\rangle_{UV}
\ .
\end{align}
To arrive at a meaningful answer one still has to supplement virtual
corrections.  They are correctly incorporated as usual, by simply
subtracting the corresponding term proportional to the original
transition probability. This was done in the derivation of the BMS
equation and it also applies to the full hierarchy. For concreteness,
and without loss of generality, I will demonstrate how to achieve this
for the case of $m$ hard particles shown in Eq.~\eqref{eq:UV-average}.

%To facilitate
%calculation, note that ${\sf U}[\Xi_0,U]$ behaves like a functional
%version of a translation operator for which $e^{a.\partial_x} f(x)
%g(x) = (e^{a.\partial_x} f(x)) (e^{a.\partial_x} g(x))$
%\begin{equation}
%  \label{eq:translation-prod}
% ( {\sf U}[\Xi_0,U]\
%{\sf U}[\Xi_0,V]\ (UV)_{p_1}^{(\dagger)})\otimes  \ldots  
%  \otimes ({\sf U}[\Xi_0,U]\
%{\sf U}[\Xi_0,V] (UV)_{p_n}^{(\dagger)} )
%\end{equation}

To understand the constraints on the virtual corrections, let me first
study the real emission contribution in some more detail.  Omitting,
for the moment, the operator that implements the soft gluon shower and
the expectation value in the expression for the real emission part,
\[e^{ -\frac{\frac{\delta}{i \delta \Xi_0} M(E)
    \frac{\delta}{i \delta \Xi_0}}{2}} {\sf U}[\Xi_0,U] \ldots {\sf
  U}[\Xi_0,V]^\dagger\Big\vert_{U=V}\ ,
\]
one has to consider
\begin{align}
  \label{eq:real-emission-1} 
%\frac{1}{2}
\Big\{
  g\,J_{l k}^{\mu} \Tilde U^{a b}_{k} %\Big]\Big) 
  i{\Bar\nabla}^a_{U_l} \Big\}
  [ E\partial_E M(E)]_{l l'}\delta^{b b'}g_{\mu\mu'}
  \Big\{ %\Big(g\ \Big[
  g\, J_{l' k'}^{\mu'} \Tilde V^{a' b'}_{k'} %\Big]\Big) 
  i{\Bar\nabla}^a_{V_l'} \Big\} \
  (UV^\dagger)_{p_1}^{(\dagger)}\otimes  \ldots  
  \otimes (UV^\dagger)_{p_m}^{(\dagger)}
\end{align}
and evaluate the variations. Inserting the BMS choice~\eqref{eq:M-BMS}
for $M$, it is easy to collect all the factors and to make the $k$
integration explicit (I keep the integration convention for $l$ and
$l'$ for readability):
\begin{align}
  \label{eq:get-kernel}
  \eqref{eq:real-emission-1} = &
  %\frac{1}{2} 
  \int \frac{d\omega_k}{\omega_k} E \delta(E-\omega_k) \frac{d\Omega_k}{4\pi}
  \frac{\alpha_s}{\pi} w_{l l'}(k)   (\Tilde{UV}^\dagger)^{a b}_k
   i{\Bar\nabla}^a_{U_l} i{\Bar\nabla}^b_{V_l'} \
  (UV^\dagger)_{p_1}^{(\dagger)}\otimes  \ldots  
  \otimes (UV^\dagger)_{p_m}^{(\dagger)}
\nonumber \\ = &
   %\frac{1}{2} 
\int\!\frac{d\Omega_k}{4\pi}
  \frac{\alpha_s}{\pi} w_{l l'}(k)  (\Tilde{UV}^\dagger)^{a b}_k  
   i{\Bar\nabla}^a_{U_l} i{\Bar\nabla}^b_{V_l'} \
  (UV^\dagger)_{p_1}^{(\dagger)}\otimes  \ldots  
  \otimes (UV^\dagger)_{p_m}^{(\dagger)}\Big\vert_{\omega_k=E}
  \ .
\end{align}
In this expression $\omega_l$ and $\omega_l'$ will be made hard when
the functional derivatives hit the $U$ and $V$. The energy part of the
corresponding phase space integrals will be eliminated at the same
time. In summary one will end up with 3 ``hard'' momenta $k,l,l'$.
% and
%for these one only needs to consider the solid angle integrations as
%the energy ordering is already accounted for.

Carrying out the differentiations, a sum over pairings will emerge,
with different color structures depending on whether one hits $q$ or
$\Bar q$ lines. Concentrating on the individual terms in this sum, it
is obvious that one has to deal with 4 different combinations,
corresponding to the pairings $qq$, $q\Bar q$, $\Bar q q$, and $\Bar
q\Bar q$ completely in parallel with Eqns.~\eqref{eq:barchidef} of the
JIMWLK case.

At this point it is, however, more efficient to directly use the
unifying properties of the left and right invariant vector fields
without exploring these structures explicitly.  This may be done by
noting that Eq.~\eqref{eq:real-emission-1} allows for a very elegant
rewrite:
\begin{align}
  \label{eq:real-rewrite}
  \text{~\eqref{eq:real-emission}} = &
 \frac{\alpha_s}{2\pi} w_{p q}(k)u(k) (\Tilde{UV^\dagger})^{a b}_k 
\big( i \Bar\nabla_{(UV^\dagger)_p}^a  i\nabla^b_{(UV^\dagger)_q} 
+i \Bar\nabla_{(UV^\dagger)_q}^a  i\nabla^b_{(UV^\dagger)_p}\big)\ 
%\Big\{\Big(
%   i\ \Big[
%  J_{l k}^{\mu} \Tilde U^{a b}_{k} \Big]\Big) 
%  i{\Bar\nabla}^a_{U_l} \Big\}[ \partial_E M(E)]_{l l'}^{b b'} 
%  \Big\{\Big(
%   i\ \Big[
%  J_{l' k'}^{\mu'} \Tilde V^{a' b'}_{k'} \Big]\Big) 
%  i{\Bar\nabla}^a_{V_l'} \Big\} \
\nonumber \\ & \hspace{.5\textwidth}
  (UV^\dagger)_{p_1}^{(\dagger)}\otimes  \ldots  
  \otimes (UV^\dagger)_{p_m}^{(\dagger)}
\ .
\end{align}
Eq.~\eqref{eq:real-rewrite} follows from a direct correspondence of
$i\Bar\nabla^a_{(UV^\dagger)_k}$ with $i\Bar \nabla^a_{U_k}$ acting on
the amplitude, and $i\nabla^a_{(UV^\dagger)_k}$ with $i\Bar
\nabla^a_{V_k}$ acting on the complex conjugate factor.  In
Eq.~\eqref{eq:real-rewrite} it is possible to anticipate that the
energies will all be made hard by functional differentiation on the
hard factors $(UV^\dagger)_{p_i}$ and thus to go back to the original
definition of the variations without the energy $\delta$-function.
This allows to use an integration convention that employs the solid
angle integrations only and a symmetric treatment of $k,p,q$ in this
equation. Reinstating the shower-operators and the expectation value,
one arrives at the real emission part of the evolution equation
\begin{align}
  \label{eq:real-evolution}
& \Big\langle 
 \frac{\alpha_s}{2\pi} w_{p q}(k)u(k) (\Tilde{UV^\dagger})^{a b}_k 
\big( i \Bar\nabla_{(UV^\dagger)_p}^a  i\nabla^b_{(UV^\dagger)_q} 
+i \Bar\nabla_{(UV^\dagger)_q}^a  
i\nabla^b_{(UV^\dagger)_p}\big)\
\nonumber \\ & \hspace{.5\textwidth} 
   (UV^\dagger)_{p_1}^{(\dagger)}\otimes  \ldots  
  \otimes (UV^\dagger)_{p_m}^{(\dagger)}
  \Big\rangle_{UV^\dagger}(E) 
\ .
\end{align}

Virtual corrections do not depend on the physical process at hand:
While the real emissions are confined to the {\em inside} regions and
hence carry a factor $u(k)$, virtual corrections appear everywhere and
therefore have no such factor.  Moreover the factor $(\Tilde
UV^\dagger)_k^{a b}$ will be replaced by $\delta^{a b}$ and the vector
fields will have to act twice {\em within} either the amplitude or its
complex conjugate, leading to $i\nabla_{(UV^\dagger)_p}^a
i\nabla^a_{(UV^\dagger)_q}$ and an analogous barred contribution
instead of the mixed ones in Eq.~\eqref{eq:real-evolution}. For
symmetry reasons, one therefore expects the virtual corrections to
read
\begin{align}
  \label{eq:virtual-rewrite}
 \Big\langle \frac{\alpha_s}{2\pi} w_{p q}(k) 
 \big( i \nabla_{(UV^\dagger)_p}^a  i\nabla^a_{(UV^\dagger)_q} 
+i \Bar\nabla_{(UV^\dagger)_q}^a  i\Bar\nabla^a_{(UV^\dagger)_p}\big)\ 
  (UV^\dagger)_{p_1}^{(\dagger)}\otimes  \ldots  
  \otimes (UV^\dagger)_{p_m}^{(\dagger)}
  \Big\rangle_{UV^\dagger}(E)
\ , 
\end{align}
where the overall sign and normalization is fixed by real virtual
cancellation: only with this choice do they add up to an infrared
finite Fokker-Planck Hamiltonian, which happens to coincide with the
expression conjectured in Sec.~\ref{sec:analogy}. Indeed, the operator
appearing in the sum of~\eqref{eq:real-evolution}
and~\eqref{eq:virtual-rewrite} is nothing but $-H_{\text{jet}}$ of
Eq.~\eqref{eq:H_ng-0} with the $f^{(i)}$ completely determined. In
particular, there is no room for any $N_c$ dependence of the
coefficients from this argument.

To summarize, it has been shown that
\begin{equation}
  \label{eq:gen-evol}
  E \partial_E\ \Big\langle   
  e^{i 2\tr j^\dagger U V^\dagger +
  i 2\tr (U V^\dagger)^\dagger j} \Big\rangle_{UV^\dagger}(E)
  = 
- \Big\langle \  H_{\text{jet}} \ 
  e^{i 2\tr j^\dagger U V^\dagger +
  i 2\tr (U V^\dagger)^\dagger j}\ \Big\rangle_{UV^\dagger}(E)
\ .
\end{equation}
The only argument still missing is about how to reverse the step
leading from~\eqref{eq:JIMWLK} to~\eqref{eq:OU-evo}.

This follows from the fact that the identity holds for arbitrary $j$
and $j^\dagger$, i.e.  the fact that the above is completely
independent of the type of correlator considered. The result is, as
advertised in Sec.~\ref{sec:from-jimwlk-bk}, the evolution equation
\begin{equation}
  \label{eq:new-FP}
  E\partial_E\ \Hat Z_E[UV^\dagger] = -H_{\text{jet}}\ \Hat Z_E[UV^\dagger]
\end{equation}
for the weight $Z_E[UV^\dagger]$ used to define the averages $\langle
\ldots \rangle_{UV^\dagger}(E)$. This completes the derivation of
Eq.~\eqref{eq:FP-ng}.

%This procedure just
%retraces the steps used in~\cite{Weigert:2000gi} to derive the JIMWLK
%equation and following them one completes the derivation of the finite
%$N_c$ generalization of the BMS equation.

\subsection{\it The two types of nonlinearities; $N_c$ and 
correlator factorization}
\label{sec:twotypes}

Now that the interpretation of $G_{p q}(E)$ in terms of link operators
suggested in~\eqref{eq:interpretation} is understood, it seems
worthwhile to recapitulate the role of the two types of nonlinearities
discussed repeatedly already in the JIMWLK case: eikonalization of
gluon fields into Wilson lines which enter the definitions
\begin{subequations}
\begin{align}
  \label{eq:S-2-def} 
\Hat G_{p q} := & \tr((UV^\dagger)^\dagger_p (UV^\dagger)_q)/N_c
\\
G_{p q}(E) := &  \langle \Hat G_{p q} \rangle_{UV^\dagger}(E)  =
\frac{1}{N_c}
\frac{\delta}{i \delta (j_p)_{i j} } \frac{\delta}{i \delta (j_p)_{i j} } 
{\sf W}_E[j^\dagger,j]
\ ,
\end{align}  
\end{subequations}
as the first stage and nonlinearities in the Wilson lines themselves
contained in the Fokker-Planck Hamiltonians~\eqref{eq:H_ng} or as
``second eikonalization'' in the shower
operators~\eqref{eq:shower-jet-full} as the second stage. 

Jet observables provide the perfect example that the second stage is
not always visible; in global observables $u(k)$ suppresses the inside
contributions completely and only the outside terms survive. This
completely suppresses the nonlinear terms. How this happens is again
best illustrated with the evolution equation for the two point
function. Starting from \eqref{eq:new-FP}, one immediately finds
\begin{align}
  \label{eq:two-point-to3point-minus-virt-1}
  E & \partial_E   \langle  
  \tr((UV^\dagger)^\dagger_p (UV^\dagger)_q)/N_c\rangle_{UV^\dagger}(E)
\nonumber \\ = & 
 \int\!\frac{d\Omega_k}{4\pi} \frac{\alpha_s}{\pi} w_{p q}(k) \ 
\Big\langle 
u(k)\Tilde {(UV^\dagger)}_k^{a b}\ 2 
\frac{\tr(t^a (UV^\dagger)_p t^b (UV^\dagger)_q)}{N_c} 
- 2 C_{\text{f}} 
\frac{\tr((UV^\dagger)^\dagger_p (UV^\dagger)_q)}{N_c}  
\Big\rangle_{UV^\dagger}(E) 
\ .
\end{align}
The first term on the r.h.s., carrying the factor $u(k)$, originates
from the real contribution, the second from the virtual one.
By~\eqref{eq:Fierz} this is equivalent to
\begin{subequations}
  \label{eq:general-minus-virt}
\begin{align}
   E\partial_E  G_{p q}(E) 
= &   \int\!\frac{d\Omega_k}{4\pi}\frac{\alpha_s}{\pi} w_{p q}(k)
   \Big\langle  u(k)\big(\Hat G_{p k} \Hat G_{k q} N_c 
   - \frac{\Hat G_{p q}}{N_c}\big) 
   -2 C_{\text{f}} \Hat G_{p q} \Big\rangle (E)
 \\
= & \int\!\frac{d\Omega_k}{4\pi}\Bar\alpha_s w_{p q}(k) 
\Big\langle u(k)\big(\Hat G_{p k} \Hat G_{k q} - \Hat G_{p q} \big)
 -\frac{2 C_{\text{f}}}{N_c}(1-u(k))\Hat G_{p q} \Big\rangle (E)
\ .
\end{align}  
\end{subequations}
First and second terms contribute inside and outside the jets, the
former is completely suppressed in global jet observables. In this
case, all traces of the second nonlinearity are gone: since the
equation is linear, nothing couples to higher n-point functions. The
counterpart to the Balitsky hierarchy fully decouples. What is left is
simple Sudakov exponentiation which corresponds to the fact that $G$
still corresponds to a correlator of Wilson lines.

On the other hand, this formulation has a direct link between large
$N_c$ limit and factorization of correlators which leads to the
analogue of the BK equation that triggered these explorations, the BMS
equation. To see this explicitly, use~\eqref{eq:average-Legendre} to
represent the averaging involved in
\begin{equation}
  \label{eq:G-soft-aver}
  G_{p q}(E) =  \langle \Hat G_{p q} \rangle_{UV^\dagger}(E)  
  =\langle  {\sf A}_{p\,q}^{i j}[U,\Xi] 
  {\sf A}_{p\,q}^{i j}[U,\Xi]^\dagger \rangle_\Xi
\end{equation}
and then repeat the argument of\cite{Fiorani:1988by}, Eq.~(50). It is
worth emphasizing that the real-virtual cancellation argument of BMS
restated in Sec.~\ref{sec:analogy} applies already to
Eq.~\eqref{eq:general-minus-virt} itself, without any reference to the
factorization argument. Infrared finiteness in the Sudakov term
follows directly from the argument of BMS, while in the other term it
is the definition of $\Hat G$ that guarantees the necessary
cancellations. This provides an illustration of how the more general
reasoning above will lead to infrared finite results via real-virtual
cancellation in {\em any} evolution equation for color singlet objects
contained in the new evolution equation.

\subsection{Connections with CGC physics}
\label{sec:jets-dense-media}

Here I want to discuss two types of issues concerning the relationship
of the treatments of evolution in the CGC case and that of non-global
jet observables. The first concerns what one can learn from the above
for the interpretation of the JIMWLK equation and the extension of the
underlying methods to describe new, more exclusive observables. The
second type of questions concerns the treatment of jets that are
created in a dense medium such as in heavy ion collisions at RHIC and
the LHC.

Turning to the first of these, let me begin with the observation that
the two types of terms in the JIMWLK Hamiltonian indeed have the
interpretation of virtual correction and real emission terms. Any
restriction on (real) soft gluon emission imposed by a choice of
variables must come along with a modification of the phase space
available on real emission and introduce the analogue of the factor
$u(k)$ appearing in the jet case.  This analogy is important with
respect to a generalization away from purely central collisions in a
way that remains under control perturbatively. The problem is
conceptual and affects both JIMWLK and BK equations, although in the
latter case the problem is easier to state. To this end, return to the
discussion of IR safety of BK (and JIMWLK) evolution as illustrated
with Fig.~\ref{fig:IR-safety}. IR safety was ensured by the presence
of a saturation scale, which in turn emerges from the existence of a
dense gluonic medium. This can confidently be expected for central
collisions, but will certainly fail as soon as one starts to probe the
targets periphery. In this situation the gluon density becomes small,
the evolution equation reduces to the BFKL equation with all its
problems of diffusion into the infrared. The culprit for this in a
practical sense is the long range nature of the emission kernels
${\cal K}$ in both JIMWLK and BK.  As a consequence, if one attempts
to prepare an initial condition to JIMWLK or BK that has large
densities only in a finite spatial range, evolution will not respect
this structure -- the finite size object will not stay finite but grow
exponentially~\cite{Kovner:2001bh,Kovner:2002xa}
%\marginpar{put in all  the others}
. The situation for the saturated region is more
subtle~\cite{Ferreiro:2002kv}, but that does not change the basic
problem of exponential growth. Any calculation that would lead to
finite range of the JIMWLK or BK kernels would necessarily have to
take confinement into account, an idea somewhat at odds with the
perturbative philosophy taken in this review. The same applies to
attempts to phenomenologically treat the impact parameter (or
$\bm{b}$) dependence of real experiments using
$\bm{b}$-profiles~\cite{Kowalski:2003hm, Bondarenko:2003ym,
  Gotsman:2004ra}, although the results are intriguing.

The above treatment suggests a different approach. In the jet case,
space is predominantly empty, in fact the observables (be it in the
non-global or global case) are designed around forbidding radiation
above certain thresholds {\em into} most of space. The radiation veto
then leads to Sudakov logarithms that govern the mostly empty regions,
while the JIMWLK or BK nonlinearities are limited to the regions that
are allowed to become dense by choice of observable.  The only trace
of confinement physics entering this perspective is parton hadron
duality, and often nonperturbative power corrections, which can in
principle be understood. All of these problems appear to be entirely
separate from the evolution equations. A highly attractive alternative
to the DIS-type treatment existent in the literature would thus be the
definition of semi inclusive observables in the small $\xbj$ case of the
type of diffractive and rapidity gap events that treat the empty
regions in a manner similar to the jet case.

While this remains a task for the future, the above developments have
made much of the interpretational lore that dates back to
Gribov~\cite{Gribov:1984tu} to emerge transparently in the actual
technical treatment. While the JIMWLK equation is an equation for the
cross section, the objects entering the formal solutions of the
corresponding Langevin equations, the ``shower-operators''
--Eq.~\eqref{eq:small-x-shower} for JIMWLK and
Eq.~\eqref{eq:shower-jet-full} for the jet case-- give access to the
underlying amplitudes. These amplitudes result from applying the
shower operators to any type of hard ``seed.''  The procedure simply
adds fully developed clouds of soft gluons to the hard objects used as
a ``seed'' as exemplified by DIS at small $\xbj$, where the amplitude
generated by acting on a hard $q\Bar q$ pair may be thought of as the
soft gluon part of the photon wave function.  This, strictly speaking,
is the justification for the description of the structure of the
virtual photon wave function in Sec.~\ref{sec:cgc-dis:-high}. A
derivation of the JIMWLK equation or an evolution equation that
describes diffractive events with rapidity
gaps\footnote{See~\cite{Kovchegov:1999ji} for an early BK based
  attempt.} from this perspective, based on amplitudes and with a full
understanding of the operatorial content would be highly instructive.

\begin{figure}[htb]
  \centering
%\includegraphics[height=4cm]{ng-jet-pic-medium_new}
%\includegraphics[height=4cm]{besen1-neu}
% these are normalized by height
\includegraphics[height=.21\textwidth]{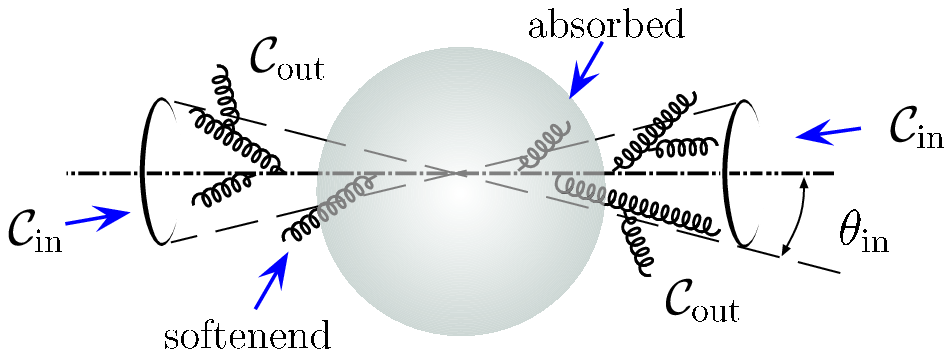}
\hfill
\includegraphics[height=.21\textwidth]{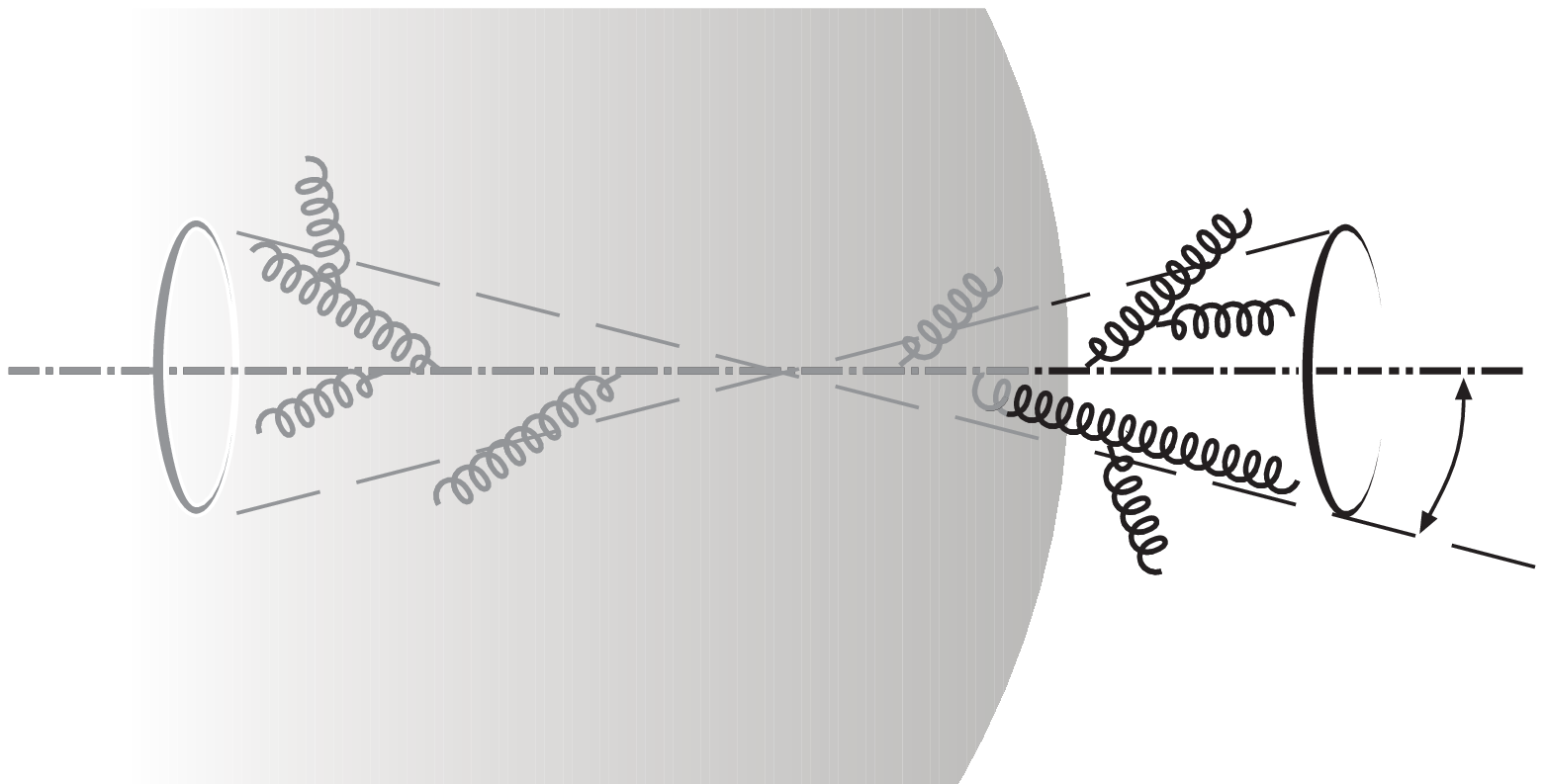}
  \caption{\small%
    In medium analog of the two jet event depicted in
    Fig.~\ref{fig:e+e-jet}. Left: Soft gluons in the inter jet region
    ${\cal C}_{\text{out}}$ now may be totally absorbed by the medium
    or be created with more energy than allowed in the out region but
    then softened until they fall below $E_{\text{out}}$.  Right: If
    the medium is large enough one of the jet cones may even be
    completely be absorbed.}
  \label{fig:e+e-jet-medium}
\end{figure}

On the other hand there are direct applications to jet measurements if
the jets are created in medium. This is the generic case for jet
measurements at the LHC. Let me again consider a measurement of the
inter-jet energy flow~\eqref{eq:Sig-ee} but with the jets created in
medium as depicted in Fig.~\ref{fig:e+e-jet-medium}.
The first obvious modification is that only a subset of the soft
gluons that would be measured without the medium will reach the
detector, the others are absorbed by the medium. Instead
of~\eqref{eq:Sig-ee}, the best one can hope to measure is
\begin{equation}
  \label{eq:Sig-ee-medium}
  \Sigma_{e^+e^-}(Q,E_{\text{out}})=
  \sum_n\int\frac{d\sigma_n(Q)}{\sigma_T(Q)}\cdot
  \Theta\!\left(E_{\text{out}}-E_{\text{loss}}
    -\sum_{h\,\in\,C_{\text{out}}\setminus C_{\text{medium}}}\omega_h\right)
  \ .
\end{equation}
where $E_{\text{loss}}$ is the net energy loss into the medium. Phase
space logarithms now are of the type
$\ln(E/(E_{\text{out}}-E_{\text{loss}}))$ and generically larger than
in the previous case.  Note that one can not lower $E_{\text{out}}$
below $E_{\text{loss}}$ once the medium is present.  Also the
evolution equations would change as new contributions arise. Consider
for example a gluon generated inside the medium with an energy
somewhat above $E_{\text{out}}$ which loses enough energy to fall
below $E_{\text{out}}$ before it reaches the detector. Also this
mechanism enhances the importance of the nonlinear effects.
[Technically this requires a redefinition of the soft emission kernels
and thus of the expressions entering \eqref{eq:Wn}.]

Upon considering to transplant other global jet observables like
thrust or broadening from the $e^+ e^-$ case to the in medium
situation sketched in Fig.~\ref{fig:e+e-jet-medium} one realizes that
global sums like in~\eqref{eq:jet-shape-global} are not experimentally
accessible as part of the contributions will be absorbed in the
medium. This would appear to indicate that also other observables
which are usually considered global will acquire non-global
characteristics. This should lead to nonlinear evolution equations
also in such cases. 

The issues raised here pose interesting and important questions to be
dealt with in future work -- for the remainder of this review I will
return to physics results from the existing evolution equations.

% spellmark 5

\section{Numerical results at fixed coupling: JIMWLK and BK}
\label{sec:numer-results}

\subsection{\it Numerical results from JIMWLK simulations}
\label{sec:numerical-JIMWLK}

\begin{figure}[htbp]
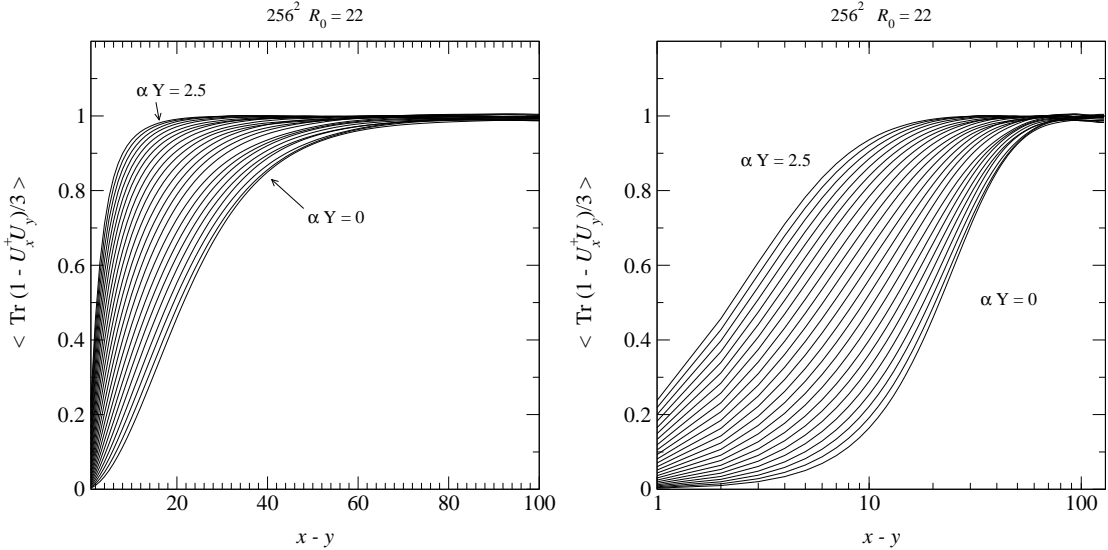

  \centering
  \includegraphics[height=7.3cm]{UUcorr_su3_256b-Y}\hfill
  \includegraphics[height=7.3cm]{UUcorr_su3_256b_log-Y}
  \caption{\small Evolution of the dipole operator from the JIMWLK equation on 
    an $256^2$ lattice, shown on a linear (left) and logarithmic
    (right) distance scale.  The general pattern as predicted from the
    existence of the fixed point and illustrated in
    Fig.~\ref{fig:generic-evol}, shows up clearly. On the right one
    sees a scaling form emerge: a stable shape of the curve that
    merely shifts to the left unaltered. [Plots taken
    from~\cite{Rummukainen:2003ns}]}
  \label{fig:jimwlk-evo}
\end{figure}

While there are many numerical studies of the CGC based on the BK
equation~\cite{Braun:2000wr, Kimber:2001nm,
  Armesto:2001fa,Levin:2001et,Lublinsky:2001bc,Lublinsky:2001yi,Gotsman:2002yy,
  Golec-Biernat:2001if, Albacete:2003iq}
%(see~\cite{Levin:1999mw, Iancu:2002tr,} for analytical treatments and for numerical work)\marginpar{refs}, 
there exists only one set of simulations of the full JIMWLK equation
by Rummukainen and Weigert~\cite{Rummukainen:2003ns}.
Fig.~\ref{fig:jimwlk-evo} shows the evolution of the dipole operator
obtained there, averaged over 8 independent trajectories on a $256^2$
lattice, from interval $\alpha_s \y \equiv \alpha_s \ln(1/\xbj) = 0
\ldots 2.5$.\footnote{In~\cite{Rummukainen:2003ns} one uses volumes
  $30^2$--$512^2$, with 8--20 independent trajectories (with
  independent initial conditions) for each volume.  This amount of
  statistics proved to be sufficient; trajectory-by-trajectory
  fluctuations in the dipole operators are quite small (i.e. already
  one trajectory gives a good estimate of the final result).}  The
main qualitative observation here is that the dipole operator soon
approaches a specific functional scaling form, where the shape is
preserved but the length scale is shortened under the evolution as
discussed in Sec.~\ref{sec:it-geometric-scaling-eqn-idea}. This is
clearly visible on the logarithmic scale plot on the right: the
Gaussian initial condition quickly settles towards the scaling
solution which evolves by moving towards left while approximately
preserving the shape. Such had been seen earlier in numerical studies
of the BK case, non-factorizing contributions do not seem to change
that feature.
With scaling established, one can now measure the evolution
of the saturation length scale $R_s$.  There are several inequivalent
ways to measure $R_s$ from the dipole function; perhaps the most
straightforward and robust method is to measure the distance where the
dipole function reaches some specified value. One defines $R_s(\y)$
to be the distance where the dipole function $N((\bm{x}-\bm{y})^2)=
\tr\langle 1 - U^\dagger_{\bm{x}} U_{\bm{z}}\rangle/N_c $ reaches a
value $c$:
\begin{equation}
  \label{eq:R_s-def-c}
  N(R_s^2):=c
\end{equation}
where $c$ should not be much smaller than 1.  The precise definition
of course is $c$-dependent -- different choices will lead to a
rescaling of the value of $R_s$ -- a common choice is $c=1/2$.
%  Such a
%definition can be used both inside and outside the scaling region and
%will encode some physics information in both.
%Fig.~\ref{fig:rs-tau} shows $R_s(\y)$, for various lattice
%sizes and initial values for $R_s(\y = \y_0)$.  Since the initial
%$\y_0$ is not a physical observable in the sense that one are not in
%a position to match initial conditions to experiment, one is at
%liberty to shift the $\y$-axis for each of the curves in order to
%match them up at large $\y$.  In Fig.~\ref{fig:rs-tau} this
%procedure clearly gives  a section of the $R(\y)$ curve,
%independent of the system size or initial size.
From $R_s(\y)$ one obtains the instantaneous evolution speed
\begin{equation}
\lambda = - \partial_\y \ln R_s(\y)\ .
\end{equation}
Note that contrary to $R_s$ this quantity is unique and independent of
the convention for $c$ in Eq.~\eqref{eq:R_s-def-c} {\em inside} the
scaling regime. Outside this uniqueness is lost.  It was by measuring
$\lambda$ that\cite{Rummukainen:2003ns} discovered a problem with the
continuum limit in both fixed coupling JIMWLK and BK simulations.

\begin{figure}[htbp]
  \centering
%  \begin{minipage}[t]{7.8cm}
\includegraphics[width=8cm]{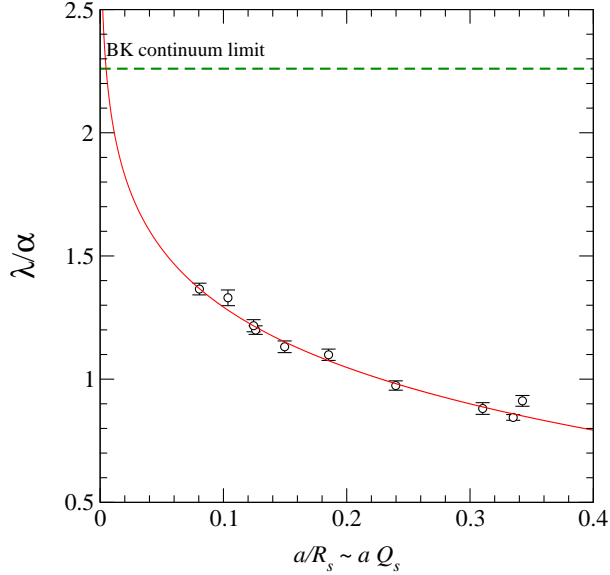}
  \caption{\small Approach to the continuum limit of $\lambda/\alpha$ 
    against $a$  in units of 
    $R_s$. 
    %The values are taken from the maxima in
    %Fig.~\ref{fig:lambda-r}. 
    The dashed line shows the
    continuum value obtained from a simulation of the BK equation. A
    continuum extrapolation from this alone  (continuous line) appears
    unreliable.}% with most of the data away from the steep section.}
  \label{fig:jimwlk-UV-extrap}
%  \end{minipage}
%\hspace{.5cm}
%  \begin{minipage}[t]{7.8cm}
%\includegraphics[width=8cm]{const_R_comp}
%  \caption{\small The evolution of the saturation scale with different IR
%  cutoffs (lattice sizes).  Except for a small initial difference, the
%  differences are well within statistical errors.}
%  \label{fig:jimwlk-IR}
%  \end{minipage}
\end{figure}

While the infinite volume limit constitutes no problem --this is
simply the saturation scale at work to protect against contributions
from the IR-- the continuum extrapolation turns out to be hard. This
is shown in Fig.~\ref{fig:jimwlk-UV-extrap}, which compares the fixed
coupling BK result with the values taken from the JIMWLK simulations.
Any attempt at a continuum extrapolation from these data would appear
to be inherently unstable.
It is simply the necessity of having a very large hierarchy of
relevant scales $a \ll R_s < L$ that interferes with a continuum
extrapolation. This has been found not to be specific to the JIMWLK
equation but to equally affect the BK
equation~\cite{Rummukainen:2003ns}. 
% However, the BK equation is much
%simpler to study numerically than the full JIMWLK equation: firstly,
%there is only a single degree of freedom, a scalar field instead of a
%SU(3) matrix field, and secondly the evolution equation is fully
%deterministic, instead of statistical.  
Comparison with BK has allowed to extend the systematics of the
continuum limit more carefully using a much larger range of volumes and
initial conditions with reduced errors and will be discussed in
Sec.~\ref{sec:bk-jimwlk-compared} after reviewing the size of
violations of correlator factorization~\eqref{eq:factorization} that
distinguishes JIMWLK from BK.

\subsection{\it Size of factorization violations}
\label{sec:size-fact-viol}

A feature that allows a direct comparison of JIMWLK and BK simulations
is the smallness of factorization violations in the simulations
of~\cite{Rummukainen:2003ns}.

\begin{figure}[htbp]
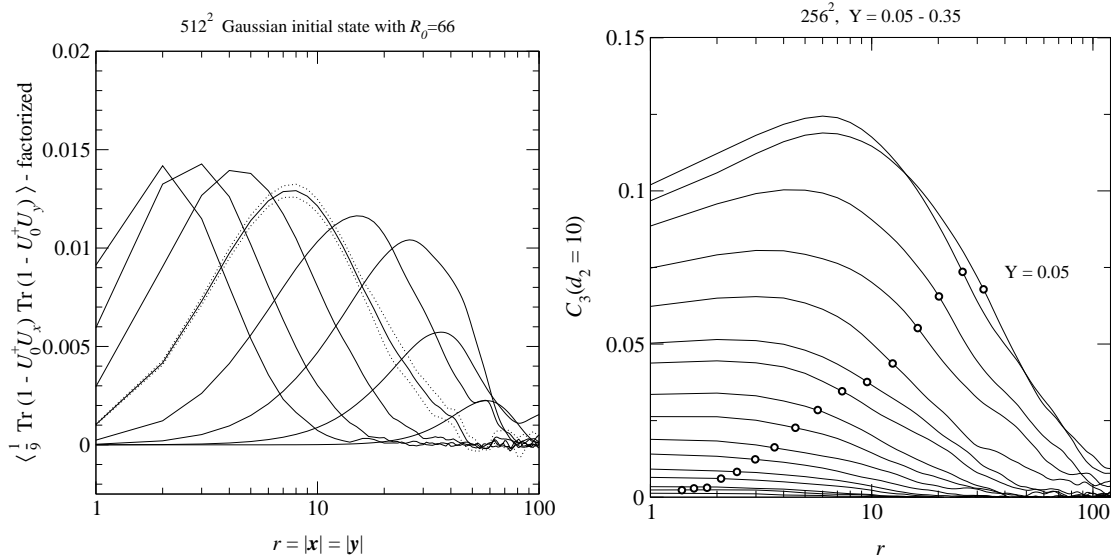

  \centering
  \includegraphics[width=7.3cm]{Lcorr_su3_512_BW}%{UUcorr_su3_fact_80}
  \hfill
 \includegraphics[width=7.3cm]{cL10_su3_256-Y}
  \caption{\small Factorization violations for a $4$-field correlator. 
    The geometry is explained in the text. Unnormalized 
    correlator~\eqref{eq:facviol-unnorm} on the left 
    (from~\cite{Rummukainen:2003ns}). Typical 
    error size is indicated by dashed lines for a curve in the middle.
    Normalized correlator~\eqref{eq:facviol-norm} on the right. The
    dots indicate the value of $R_s(\y)$ for the individual curves.
    Factorization violations are generically small and indicate that
    evolution does not induce strong violations of factorization.
    Consequently, the BK equation will be a good approximation as long
    as the initial condition does not contain strongly nonfactorizing
    contributions.  }
  \label{fig:Nc-violations}
\end{figure}

One might expect that the BK equation  to
be a fairly good approximation of the JIMWLK equation, in
  particular if the initial conditions are uncorrelated and correspond to factorized correlators for which
\begin{subequations}
\label{eq:facviol}
\begin{align}
\label{eq:facviol-unnorm}
& \langle \tr( U_{\bm{x}}^\dagger U_{\bm{z}}) 
        \tr(U_{\bm{z}}^\dagger U_{\bm{z}}) \rangle 
- \langle \tr( U_{\bm{x}}^\dagger U_{\bm{z}})\rangle
        \langle \tr(U_{\bm{z}}^\dagger U_{\bm{z}}) \rangle
\intertext{or}
\label{eq:facviol-norm} 
& \frac{\langle \tr( U_{\bm{x}}^\dagger U_{\bm{z}}) 
        \tr(U_{\bm{z}}^\dagger U_{\bm{z}}) \rangle 
- \langle \tr( U_{\bm{x}}^\dagger U_{\bm{z}})\rangle
        \langle \tr(U_{\bm{z}}^\dagger U_{\bm{z}}) \rangle}{%
        \langle \tr( U_{\bm{x}}^\dagger U_{\bm{z}})\rangle
        \langle \tr(U_{\bm{z}}^\dagger U_{\bm{z}}) \rangle}
\end{align}    
\end{subequations}
vanish. This would appear to be essentially true for generic
separations as shown in Fig.~\ref{fig:Nc-violations}. There, for
concreteness, points are chosen so that $({\bm{x}} - {\bm{z}})
\parallel \bm{e}_1$, $({\bm{z}} - {\bm{z}}) \parallel \bm{e}_2$ and
$|{\bm{x}} -{\bm{z}}| = |{\bm{z}} - {\bm{z}}|$.  The left
shows~\eqref{eq:facviol-unnorm} and one should recall that the natural
magnitude for these correlators is $\sim 1$. The right shows a plot
of~\eqref{eq:facviol-norm} based on the same simulation results with
one leg kept at length 10 in lattice units and the length of the other
leg on the horizontal axis. Evolution would appear to {\em erase}
scaling violations in this plot.  Thus, given the initial conditions
used in\cite{Rummukainen:2003ns}, it would appear that the correlators
cancel with 1--2 \% accuracy, with the violations staying roughly at
the same size throughout the $\y$ interval covered.

Recently, however, Iancu and Mueller~\cite{Iancu:2003zr} have argued
that evolution builds up correlations that violate factorization. They
argue that factorization violations are caused by rare configurations,
build up during evolution and have the effect of reducing the
evolution rate. 
%The correlators they deem most susceptible to
%factorization violations are less symmetric than those studied in
%Fig.~\ref{fig:Nc-violations}, they have, for example $|{\bm{x}} -{\bm{z}}| \gg
%|{\bm{z}} - {\bm{z}}|$ which, at this point we have not studied carefully.
%
%
%On the other hand, 
Although this deserves further study, explicit comparison of shapes for
the dipole correlator from both JIMWLK and BK simulations at same size
lattices also give only very small differences. In particular, as
shown in Fig.~\ref{fig:JIMWL+BK-fixed-extrap}, one does not see strong
deviations between evolution rates of JIMWLK and BK simulations on
comparable lattices.  This seems to indicate that these simulations in
general were not sensitive to the correlations referred to
in~\cite{Iancu:2003zr}.
%While this may by due the limited $\y$
%ranges accessible for a given evolution run --these are limited by the
%fact that the physical scales vary compared to the grid geometry
%during evolution-- a careful search for these effects remains outside
%the scope of this paper. 
For the present purposes it would appear to be safe to conclude that
for the type of initial conditions explored to date, the BK equation
should give a good approximation of the salient features of the
evolution. This is the basis on which BK simulations have been used to
refine the understanding of the UV extrapolation by augmenting
Fig.~\ref{fig:jimwlk-UV-extrap} with results from BK simulations in
which one also uses a lattice representation for transverse space to
parallel the treatment of the JIMWLK equation.

\subsection{\it Continuum limit of BK and JIMWLK at fixed coupling}
\label{sec:bk-jimwlk-compared}

In order to study the cutoff-dependence and to facilitate the
comparison with JIMWLK simulations, one best solves the BK equation
with the same type of discretization in transverse space that is
necessary to implement the former. This is precisely what has been
done in~\cite{Rummukainen:2003ns}. 

Independent results for the continuum have been obtained by another
method based on Fourier transformation presented
in~\cite{Braun:2000wr, Golec-Biernat:2001if}.  In these approaches the
kernel involves only a 1-dimensional integral for which a logarithmic
momentum variable $\Xi \sim \ln k^2$ is used.  Consequently, it becomes
easy to include a very large range of momenta in the calculation;
typically, more than 20 orders of magnitude are used.  Thus, this
allows to obtain a cutoff-independent continuum limit result for the
evolution exponent $\lambda$. The key result obtained with this method
was already shown in Fig.~\ref{fig:jimwlk-UV-extrap} for comparison
with the JIMWLK extrapolation. The large range of momenta needed to
get convergence is already a clear indication of what can be
quantitatively seen in the discretized coordinate space version I will
turn to now.

\begin{figure}[htbp]
  \centering
\begin{minipage}[t]{7.2cm}
  \includegraphics[width=7.3cm]{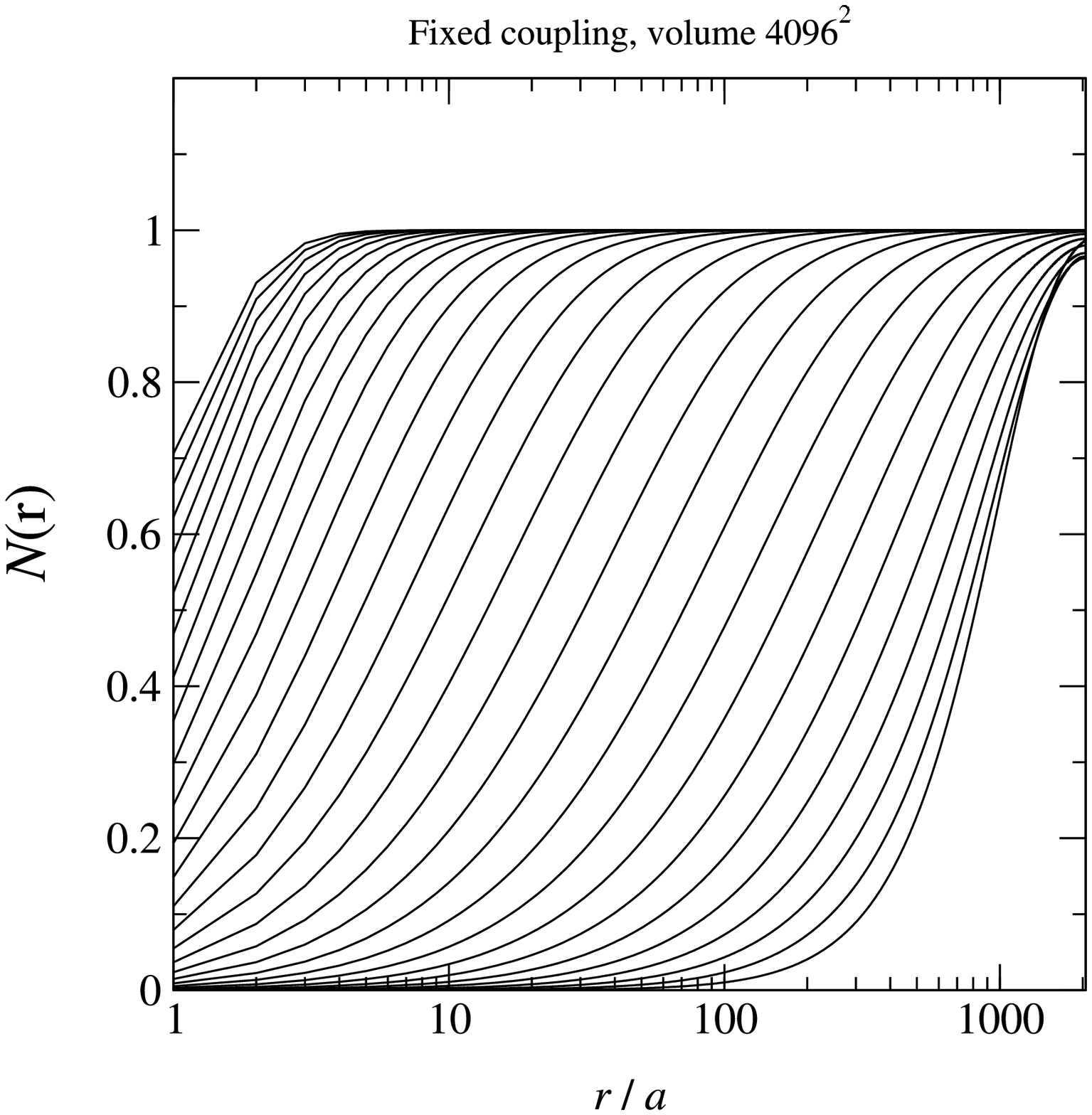}
  \caption{\small (From \cite{Rummukainen:2003ns}) 
    The evolution of $N({\bm{r}})$ on a $4096^2$ lattice, measured
    along on-axis direction.  The curves are plotted with interval
    $\delta\y = 0.25/\alpha$, and the total evolution time is
    $7/\alpha$.}
  \label{fig:N-BK}
\end{minipage}
\hfill
\begin{minipage}[t]{7.2cm}
\includegraphics[width=7.3cm]{lambda_BK_JIMWLK}
  \caption{\small (From \cite{Rummukainen:2003ns}) 
    Continuum extrapolations for $\lambda$ using JIMWLK and 
    BK simulations. The UV cutoff is plotted in units of the physical
    scale $R_s$. The continuous line is 
   $\lambda/\alpha_s = 2.26 -
    2.0(a/R_s)^{0.32}$ }
  \label{fig:JIMWL+BK-fixed-extrap}
\end{minipage}
\end{figure}

Fig.~\ref{fig:N-BK}  shows the
evolution of $N$ measured for a $4096^2$ lattice.  The similarity with
the dipole function obtained with the JIMWLK equation is striking, see
Fig.~\ref{fig:jimwlk-evo}.  Again the initial shape settles towards
the scaling form, which then continues to evolve by shifting towards
smaller $R$ while preserving the shape. Nevertheless, even with a
scaling form reached, the spacing of curves and hence $\lambda$ is not
constant. This again is due to the growing influence of the UV cutoff.

The most sensitive quantity turns out to be $\lambda(\y)$, which in
this case is free of statistical errors.  A function of the form
$\lambda/\alpha_s = c_1 + c_2 (a/R_s)^\nu$ is an excellent fit to the
data and fixes $c_1 = 2.26$, $c_2=2.0$ and $\nu=0.32$.  The
$a\rightarrow 0$ result agrees perfectly with the momentum space
result although this was not included in the fit procedure.  This is
shown in Fig.~\ref{fig:JIMWL+BK-fixed-extrap}.  The agreement of BK
and JIMWLK results at corresponding cutoff scales is remarkable over
the whole range. Most striking is an evaluation of the active phase
space. While on the IR side activity is limited to within one order of
magnitude of the saturation scale $Q_s$, one needs the cutoff on the
UV side to be about 6 orders of magnitude larger than $Q_s$ to get
within 1\% of the continuum value, as is easily extracted from the
power law fit.

This situation is clearly unnatural: large open phase space of this
magnitude usually generates large logarithmic corrections that need to
be resummed to get physically reliable results. This has been used to
argue that the main source for such corrections are due to running
coupling effects and study the effects which, among other new {\em
  qualitative} features, will bring a large reduction of active phase
space in the context of the BK equation that erases the potential for
large logarithmic corrections.

% spellmark 6

\section{Running coupling effects in the BK equation}
\label{sec:runn-coupl-effects}

\subsection{\it Running coupling and UV phase space}
\label{sec:it-running-coupling-phase-space}

If one wants to discuss running coupling effects and how they affect
the UV behavior one has first to state that JIMWLK and thus also BK
have been derived in leading logarithmic approximation and thus,
contrary to state of the art BFKL calculations, one does not know what
sets the scale in these evolution equations from a reliable
calculation. One is therefore left with a discussion of the structure
of the diagrams leading to the driving terms of the evolution
equations to argue a reasonable choice. For
simplicity~\cite{Rummukainen:2003ns} have argued and calculated in the
context of the BK equation. One should always keep in mind that
qualitatively these arguments will have similar implications also for
the full JIMWLK treatment.
  
In the context of both JIMWLK and BK it is common to argue that a good
scale choice is $Q_s(\y)$, since it is there that the r.h.s.  of the
evolution equation peaks.  Since with this choice the scale in the
coupling changes during evolution, this has often been dubbed running
coupling (c.f.\cite{Iancu:2002tr}). In fact, it is quite easy to see,
that by a simple reparametrization of the $\y$ axis such a scale
setting is mapped on a fixed coupling treatment, provided one uses the
same initial condition. For this reason it might be less controversial
to use the term scale setting in this context. To be explicit, fixed
coupling and scale setting at $Q_s$ differ by an additional factor of
$f(\y)=\frac{\alpha_s(Q_s(\y)^2)}{\alpha_s^{\text{fixed}}}$ on the
r.h.s. of the BK equation. If one  uses $\y$ to parametrize evolution
with the $Q_s$ scale setting and $\y'$ in the fixed coupling case,
$f(\y)$ is in fact the Jacobian $d\y'/d\y$ of the
reparametrization
\begin{equation}
  \label{eq:reparametrization}
  \y'-\y'_0 = \int\limits_{\y_0}^\y d\Tilde\y 
  \frac{\alpha_s(Q_s(\y)^2)}{\alpha_s^{\text{fixed}}}
  \ .
\end{equation}
This immediately leads to the conclusion that the phase space
structure of the evolution equation can not differ between the two and
thus, this scale choice will not cure the problems encountered above.
  
The difficulty can be traced to the fact that the scales in a diagram
describing a single emission step during evolution for a fixed size
parent dipole will generically have nothing to do with $Q_s$, if that
parent dipole has a size very different from $1/Q_s$. The situation in
the IR and UV (relative to $Q_s$) will behave very differently: While
the nonlinear effects will generally block evolution in the IR, UV
scales that dominate the diagrams will become visible and enter the
coupling. While on this level one can not calculate (even for BK)
which of the scales (the sizes of parent and daughter dipoles) or
combination thereof truly enters the coupling, a scale of the order of
the parent dipole size appears sensible. Like the authors
of~\cite{Mueller:2002zm, Triantafyllopoulos:2002nz},
also\cite{Rummukainen:2003ns} choose to account for running coupling
effects by having the coupling run at the natural scale present in the
BK-equation, the size of the parent dipole $(\bm{x}-\bm{y})^2$. Such a
treatment is often called ``DGLAP improved'' if the term ``running
coupling'' is already occupied for what is here called $Q_s$ scale
setting.  Explicitly in Eq.~(\ref{eq:BK}), one replaces the constant
$\alpha_s$ by the one loop running coupling at the scale
$1/(\bm{x}-\bm{y})^2$:
\begin{equation}
  \label{eq:d-running}
  \alpha_s\to 
  \alpha_s(1/(\bm{x}-\bm{y})^2) = 
  \frac{4\pi}{\beta_0\ln\frac{1}{(\bm{x}-\bm{y})^2
      \Lambda^2%+\mu^2_0
    }} \hspace{1cm}\text{with}\hspace{1cm} \beta_0=(11 N_c -2 N_f)/3
\ .
\end{equation}
$\Lambda$ in Eq.~\eqref{eq:d-running} is related to
$\Lambda_{\text{QCD}}$ by a simple factor of the order of $2\pi$. The equation then reads
\begin{equation}
  \label{eq:BK-parent-dipole}
  \partial_\y N_{\y; \bm{x y}}  =
  \frac{\alpha_s(1/(\bm{x}-\bm{y})^2) N_c}{2\pi^2}\int\!\!d^2 z
  \ \Tilde{\cal K}_{\bm{x}
    \bm{z} \bm{y}}\,
  %\frac{(\bm{x}-\bm{y})^2}{%
  %  (\bm{x}-\bm{z})^2(\bm{z}-\bm{y})^2} 
  \Big\{  
   N_{\y; \bm{x z}}+N_{\y; \bm{z y}}-
  N_{\y; \bm{x y}} 
  - N_{\y; \bm{x z}} N_{\y; \bm{z y}} 
  \Big\}  
\ .
\end{equation}
Such a simple replacement, of course, is not what we expect to arise
from an exact calculation in which the two loop contributions that
generate the running will emerge underneath the integral over
$\bm{z}$. This comes on top of the fact that at next to leading order
there will be corrections beyond the running coupling effects that are
entirely unaccounted for by this substitution. Nevertheless, it
appears natural to argue that Eq.~(\ref{eq:d-running}), when used in
the BK equation, captures the leading effect associated with the
``extraneous'' phase space uncovered quantitatively in
Sec.~\ref{sec:bk-jimwlk-compared}. This has been the outcome of a long
discussion in the context of the BFKL equation and its next to leading
order corrections~\cite{Fadin:1995xg, Fadin:1996zv, Fadin:1998hr,
  Fadin:1998py, Colferai:1999em, Thorne:1999rb, Thorne:2001nr}. Here
one simply takes this as the starting point with the attitude that a
scale setting technique like the one employed in
Eq.~(\ref{eq:d-running}) is but the easiest way to get even
quantitatively reliable initial results.

%{\color{red} 
With running coupling implemented, one has to face the presence of the
Landau pole in Eq.~\eqref{eq:d-running}. The main point here is, that,
while indeed one does not control what is happening at that scale, one
is justified in providing a prescription for its treatment. With $Q_s
\gg \Lambda_{\text{QCD}}$ the integral on the r.h.s.\ of the BK
equation yields extremely small contributions at scales of the order
of $\Lambda_{\text{QCD}}$, typically of the order of $10^{-7}$ times
what is seen around $Q_s$.  This is the reason why one is allowed to
ignore whatever happens there and to freeze any correlators such as
$N$ at these scales at the values provided by the initial condition.
Note that the same reasoning is already necessary at fixed coupling in
some sense.  If one turns on running coupling, the physical thing to
do is to switch off the influence of the divergence at the Landau
pole, for instance by freezing the coupling below some scale $\mu_0$.
Of course, one has to make sure that the details of where one does that
will not affect the results.

Such an argument about the integrals in the BK equation is valid at
zero impact parameter or infinite transverse target size, where the
nonlinearity is present and sizable everywhere. For large impact
parameter one necessarily enters the region where the nonlinearity is
small and there one would remain sensitive to the value of $\mu_0$.
That is the central reason for not considering such situations.
%\marginpar{where does this really belong?}  
%}

\begin{figure}[htb]
  \centering 
%  \begin{minipage}[t]{7.8cm}
\includegraphics[width=8cm]{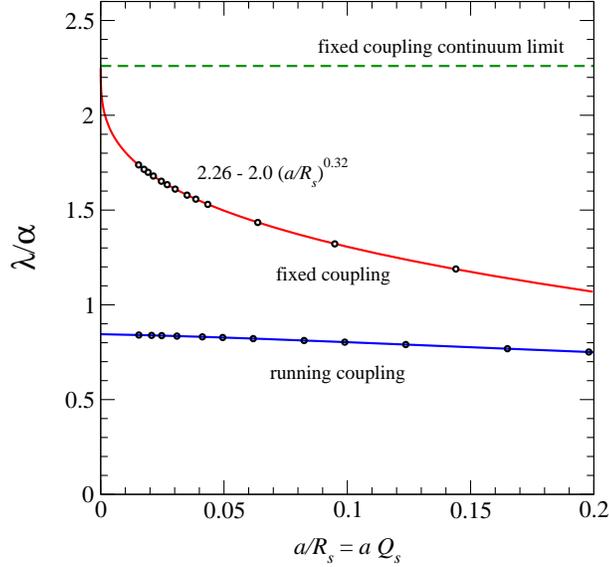}
%  \end{minipage}
%  \hspace{.5cm}
 \caption{\small (From\cite{Rummukainen:2003ns}) 
   example of a continuum extrapolation of 
   $\lambda/\alpha_s(Q_s(\y)^2)$ for a fixed initial condition at a
   $\y$-value where $\alpha_s(Q_s(\y)^2) = 0.2$, compared to the
   fixed coupling case.  }
  \label{fig:runn-extrap}
\end{figure}

\begin{figure}[htb]
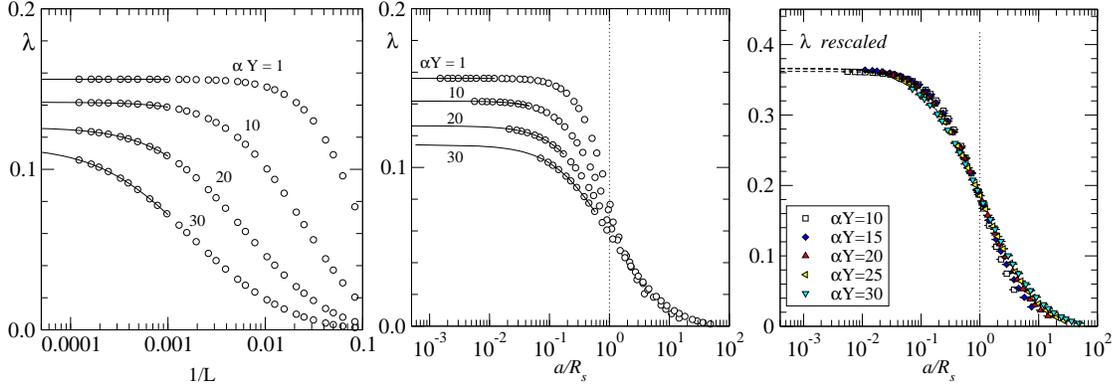

  \centering
%  \begin{minipage}[b]{5cm}       
\includegraphics[height=5.1cm]{bkr_r2b_lsize2-Y}%
\includegraphics[height=5.1cm]{bkr_r2b_aRsize2-Y}%
\includegraphics[height=5cm]{bkr_r2b_aRscaled3-Y}%
\caption{\small (From\cite{Rummukainen:2003ns}) UV dependence of the 
  BK-equation on the asymptotic
  line after the parent dipole scale is taken to determine the
  coupling. Left: $\lambda$ against $a$ in units of lattice size.
  Middle: $\lambda$ against $a$ in units of $R_s$, the natural units.
  Right: the same rescaled with the $\y$ dependence scaled out using
  the results
  of~\cite{Mueller:2002zm,Triantafyllopoulos:2002nz},~\eqref{eq:dionlambda}
  and~\eqref{eq:dionfit} with $\Tilde c = 1.5$ ($\Tilde Y$ just leads to a
  reparametrization). Active phase space is limited in the UV (and IR)
  by $5$ to $10\times Q_s$. }
\label{fig:active-phase-space}
\end{figure}

That the effects of running coupling slow down evolution has been
known from the earliest numerical treatments of BK. The study of the
UV phase space in~\cite{Rummukainen:2003ns} explicitly demonstrates
the reason for this: Fig.~\ref{fig:runn-extrap} shows an example for
the extrapolation of $\lambda/\alpha_s(Q_s(\y)^2)$ to its continuum
value at one value of $\y$ (corresponding to $\alpha_s(Q_s(\y)^2) =
0.2$) and compares it to the fixed coupling case shown earlier. This
feature persists over a large range in $Q_s$ as shown
in~\cite{Rummukainen:2003ns}.
The size of active phase space becomes particularly clear in
Fig.~\ref{fig:active-phase-space}, which shows activity centered
within an order of magnitude around $Q_s$. The rightmost panel shows
remarkable agreement with the analytical results of Mueller and
Triantafyllopoulos if their free parameter is used as a fit parameter.

\begin{figure}[htbp]
  \centering
  \begin{minipage}[t]{.485\textwidth}
 \includegraphics[width=7cm]{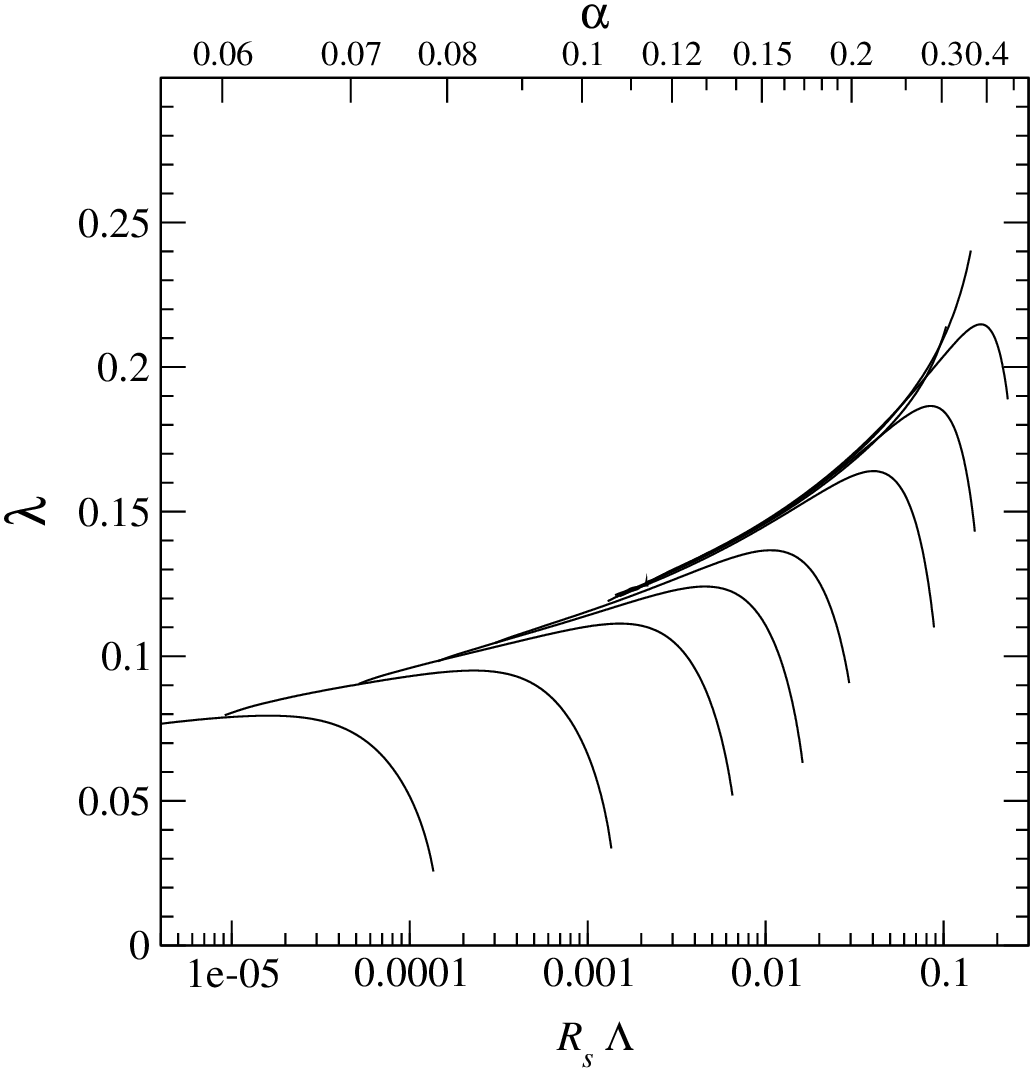}%{bkr2_b_lambda_RL}
  \caption{\small  (From~\cite{Rummukainen:2003ns}) 
    $\lambda=\partial_\y \ln Q_s(\y) $ for different initial
    conditions.  All curves starting from Gaussian initial conditions
    stay below an asymptotic line along which we find (approximately)
    scaling dipole functions $N$.  }
  \label{fig:runn-lambda}
  \end{minipage}
\hfill
  \begin{minipage}[t]{.485\textwidth}
  \includegraphics[width=7cm]{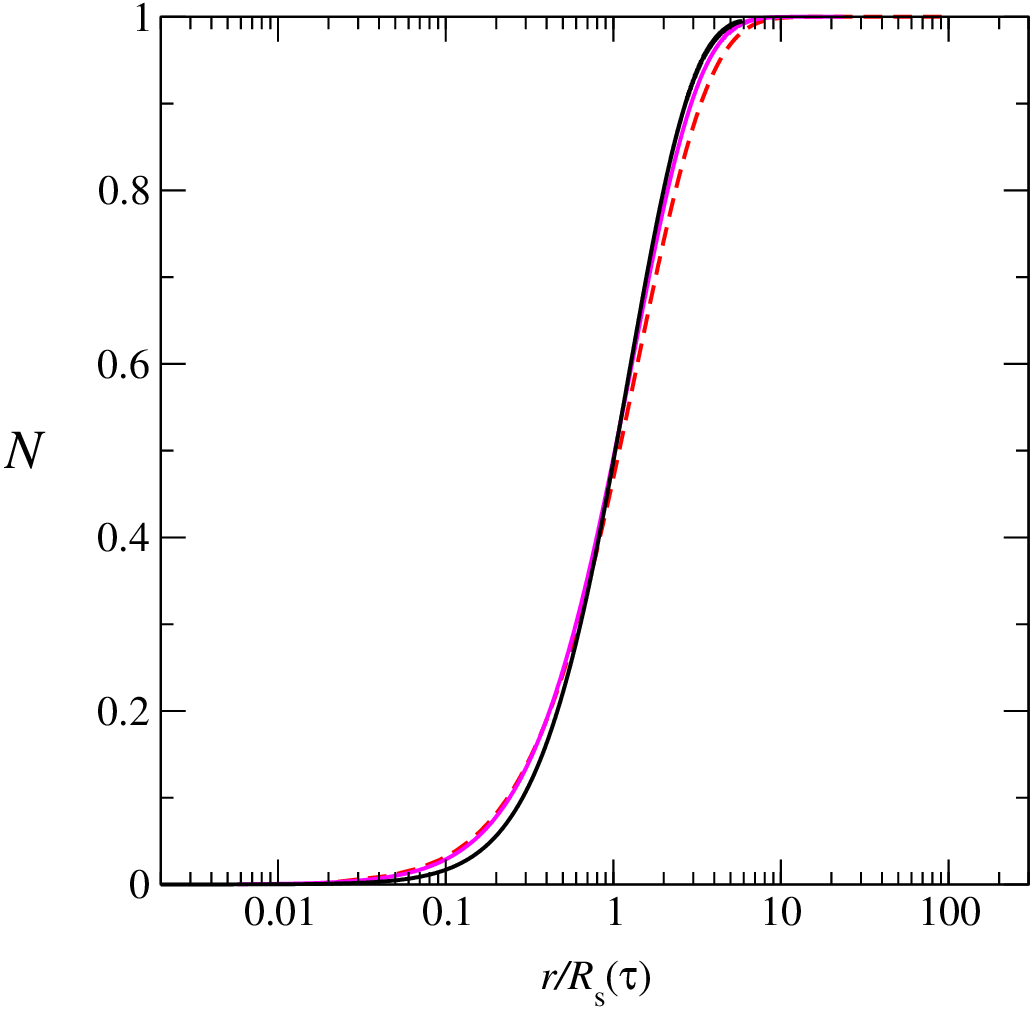}
  \caption{\small (From~\cite{Rummukainen:2003ns}) 
    Approximate scaling behavior of $N_{\y,\bm{x y}}$ on the
    asymptotic line of Fig.~\ref{fig:runn-lambda}.  }
  \label{fig:runn-scaling} 
 \end{minipage}
\end{figure}

\begin{figure}[htbp]
  \centering \includegraphics[width=8cm]{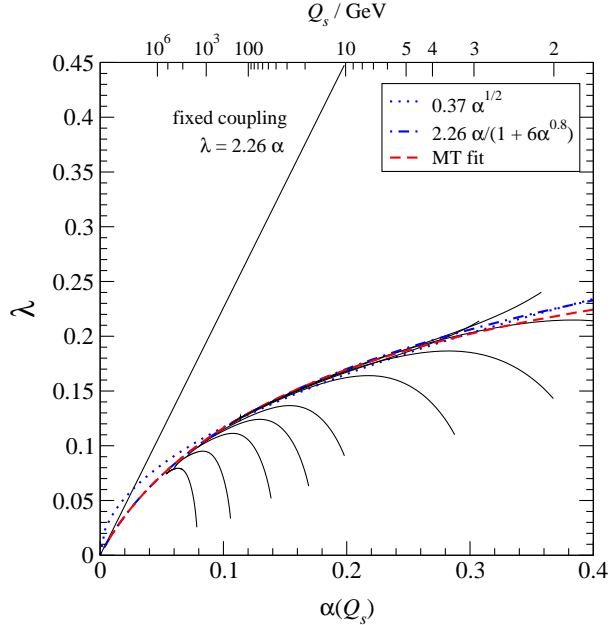}
  \caption{\small (From~\cite{Rummukainen:2003ns}) 
    $\lambda(\y)$ as a function of $\alpha_s(Q_s(\y)^2)$ with running
    coupling.  A perfect overall fit is obtained with the MT result
    Eq.~\eqref{eq:dionfit}. The shape of the curve only depends on
    $\Tilde c$, shown is a fit with $\Tilde c=1.5$. Simple
    parametrizations  emphasize how drastically the
    functional shape is changed away from a simple $c\ 
    \alpha_s(Q_s^2)$.  }
  \label{fig:assesing-sqrt-running}
\end{figure}

Despite the presence of $\Lambda_{\text{QCD}}$ through the running
coupling the solutions show a remarkable degree of scaling, called
near scaling in~\cite{Rummukainen:2003ns}.  Interestingly one may use
the evolution rate $\lambda$ to check if this behavior is reached.
Fig.~\ref{fig:runn-lambda} shows that, starting from G-B+W type
initial conditions of the form
\begin{equation}
  \label{eq:N-GBW}
  N_{\y_0,{\bm{r}}} = 1-e^{-{\bm{r}}^2 Q_s(\y_0)^2/4}\ , 
\end{equation}
with different values for
the characteristic scale the system quickly converges onto an
asymptotic line. On this line the system shows scaling behavior to
very good accuracy as illustrated in Fig.~\ref{fig:runn-scaling}.
In fact, the universal features of this equation have been studied far
into asymptotic region and Fig.~\ref{fig:assesing-sqrt-running}
summarizes many of the results.  Fig.~\ref{fig:assesing-sqrt-running}
extends the range shown in Fig.\ref{fig:runn-lambda} enormously and
adds comparisons with the fixed coupling case and the analytic results
of Mueller and Triantafyllopoulos. Two simple parametrizations are
shown to emphasize how strong the change compared to fixed coupling
is, not only in size but also in functional shape. Let me discuss
these issues in turn.

Fig.~\ref{fig:assesing-sqrt-running} shows $\lambda$ plotted against
$\alpha_s(Q_s^2)$ and the fixed coupling situation will emerge in the
guise of $Q_s$ scale setting. This is represented by the steeply
rising straight line (solid) which simply corresponds to $\lambda(\y) =
2.26\alpha_s(Q_s^2)$, where the coefficient was extracted in a fixed
coupling simulation. Why is this the relevant comparison? Recall the
discussion of Sec.~\ref{sec:it-geometric-scaling-eqn-idea}: in the
fixed coupling case scaling necessarily implies that $Q_s$ is of the
form
\begin{equation}
  \label{eq:Qsfixed}
  Q_s(\y') = e^{\,c\>\alpha_s^{\text{fixed}}(\y'-\y'_0)}\ Q_s(\y'_0)
\end{equation}
with constant $c$. The value of $c$ is uniquely determined by the
shape of the scaling solution in transverse space and turns out to be
$c=2.26$. $Q_s$ scale setting on the other hand amounts to the
reparametrization~\eqref{eq:reparametrization}, and one immediately
reads off how $\lambda$ changes from the fixed coupling case:
\begin{equation}
  \label{eq:lambdaQS}
   \lambda(\y) = \partial_\y \ln Q_s(\y') 
   = c\ \alpha_s^{\text{fixed}}\ \frac{d\y'}{d\y} 
   = c\ \alpha_s(Q_s(\y)^2)
\ .
\end{equation}
$c$ remains the {\em same} constant as in the fixed coupling case and
thus to the fixed coupling curve in
Fig.~\ref{fig:assesing-sqrt-running}.  Eq.~\eqref{eq:lambdaQS} implies
that the form of the leading term with $c=2.26$ simply corresponds to
$Q_s$ scale setting and thus necessarily suffers the same UV problems
as the fixed coupling case.

The Mueller-Triantafyllopoulos result is shown to agree with the
asymptotic line of near scaling throughout the whole range (dashed).
At large $Q_s$ both the fixed coupling and the MT curves match up.
This is due to the fact that there the MT result (which was given in
Eq.\eqref{eq:dionlambda} as a function of $\y$ instead of $Q_s$) at
large $\y$ has an expansion in powers of $\alpha_s(Q_s^2)$ in the form
\begin{equation}
  \label{eq:hypothetical}
  \lambda(\y) = c \alpha_s(Q_s(\y)^2) + \text{small corrections}
\ .
\end{equation}
This can be understood in the following way: One makes the ansatz
\begin{equation}
  \label{eq:lambda-alpha}
 \lambda(\y) := \partial_\y \ln(Q_s(\y)/\Lambda_{\text{QCD}})  
 = \lambda_0\ \big[ \alpha_s(Q_s(\y)^2)\big]^n \hspace{1cm} 
 \alpha_s(Q_s(\y)^2) := 
 \frac{4\pi}{\beta_0 \ln(Q_s(\y)^2/ \Lambda_{\text{QCD}}^2)} 
\end{equation}
with $\beta_0=(11 N_c -2 N_f)/3$, and finds\footnote{To construct the
  solution introduce $f(\y):=\ln(Q_s(\y)^2/\Lambda_{\text{QCD}}^2)$
  and solve by separation of variables in $f$ and $\y$.}
\begin{equation}
  \label{eq:Q_s}
  \frac{Q_s(\y)^2}{\Lambda_{\text{QCD}}^2} = 
  \exp\Big\{ \Big[(n+1) 2 \lambda_0 
  \big(\frac{4\pi}{\beta_0}\big)^n(\y-\y_0)
 +\big[\ln(\frac{Q_s(\y_0)^2}{\Lambda_{\text{QCD}}^2})\big]^{n+1} 
 \Big]^{\frac{1}{n+1}} \Big\} \ .
\end{equation}
Recovering $\lambda$ from this leads to
\begin{equation}
  \label{eq:lambda-as-function-of-tau}
    \lambda(\y) = 
 \frac{ \lambda_0 (\frac{4\pi}{\beta_0})^n}{\Big[
     (n+1) 2 \lambda_0 (\frac{4\pi}{\beta_0})^n (\y-\y_0)
 +\big[\ln(\frac{Q_s(\y_0)^2}{\Lambda_{\text{QCD}}^2})\big]^{n+1} 
 \Big]^{1-\frac{1}{n+1}}} 
 \xrightarrow{n\to 1, \lambda_0\to 2.26} 
 \frac{.88}{\sqrt{\y+\Tilde Y}}
\end{equation}
so that the leading term of Eq.~\eqref{eq:dionlambda} corresponds to
$n=1$ as advertised, with matching numerical coefficients. [I have
followed the spirit of~\cite{Mueller:2002zm,
  Triantafyllopoulos:2002nz} and absorbed the freedom in the initial
condition at $\y_0$ into the corresponding $\Tilde Y$
of~\eqref{eq:dionlambda}.]

This behavior at large $\y$ can also be expected for the numerical
treatment through a simple extension of the scaling argument
of~\cite{Iancu:2002tr}.  Starting from the near scaling observation,
one repeats the steps leading to Eqns.~\eqref{eq:BK-lhs-scaling}
and~\eqref{eq:lambda-integral-def} and ends up with an expression for
$\lambda(\y)$ of the form
\begin{equation}
    \label{eq:running-lambda-integral-def}
    2\pi \lambda(\y) = \frac{N_c}{2\pi^2} \int \frac{d^2r d^2z
    }{\bm{u}^2\bm{v^2}} \alpha_s(1/\bm{r}^2) \big( N(\bm{u}^2Q_s^2) +
    N(\bm{v}^2Q_s^2) - N(\bm{r}^2Q_s^2) - N(\bm{u}^2Q_s^2)
    N(\bm{v}^2Q_s^2) \big)+\ldots
\end{equation}
in which the dots stand for the small scaling violations. The only
source for the difference between the fixed coupling expression
Eq.~\eqref{eq:lambda-integral-def}
and~\eqref{eq:running-lambda-integral-def} or the corresponding lines
in Fig.~\ref{fig:assesing-sqrt-running} lies in the ${\bm{r}}$
dependence of the running coupling that is now inside the integral.
To obtain the strong relative reduction in active phase space observed
in Fig.~\ref{fig:runn-extrap}, $\alpha_s$ must vary considerably over
the range of scales contributing to the integral in the fixed coupling
case.  Wherever this happens, $\lambda(\y)$ will be a nontrivial
function of $\alpha_s(Q_s(\y)^2)$. Conversely, at asymptotically large
$Q_s$, where the $\alpha_s(1/\bm{r}^2)$ probed in
Eq.~\eqref{eq:running-lambda-integral-def} turn essentially constant
over a range comparable to the phase space needed in the fixed
coupling case, one should return to a situation as in
Eq.~\eqref{eq:hypothetical}.

It is essential to remind oneself here that the limit in which the
leading term of~\eqref{eq:hypothetical} is sufficient is not under
control. The price to pay for this behavior at large $\y$ is the
reopening of phase space to the size encountered already at fixed
coupling.

As in~\cite{Rummukainen:2003ns}, this is illustrated by tabulating
(see table~\ref{tab:reopening of phase space}) a measurement of active
phase space in terms of the values of $a/R$ needed to find $\lambda$
within a given percentage of its continuum value at different
$\alpha_s(Q_s(\y)^2)$. This is then compared with the fixed coupling
result.
\begin{table}[htb]
  \centering
  \begin{tabular}[t]{Z|Z|Z|Z|Z|Z}
deviation from & $\alpha_s(Q_s^2)=.2 $ & $\alpha_s(Q_s^2)=.15 $ &
$\alpha_s(Q_s^2)=.12$ & $\alpha_s(Q_s^2)=.1$ & fixed coupling \\
continuum value & $\lambda/\alpha =.845$ & $\lambda/\alpha =.96$ &
$\lambda/\alpha =1.052$ & $\lambda/\alpha =1.14$ & \\ & & & & & \\
10\% & $a/R=0.17$ & $a/R=0.15$ & $a/R=0.12$ & $a/R=0.1$ & $a/R=0.0011$
\\ 5 \% & $a/R=0.1$ & $a/R=0.08$ & $a/R=0.066$ & $a/R=0.05$ &
$a/R=0.00013$ \\ 1 \% & $a/R=0.027$ & $a/R=0.015$ & $a/R=0.007$ &
$a/R=0.0012$ & $a/R=8.2\cdot 10^{-7}$ \\
\end{tabular}
\caption{\small 
Phase space needed to extract continuum values for
    $\lambda$ is reduced for larger values of the coupling. The fixed
    coupling situation reemerges asymptotically for large $Q_s$, i.e.
    small coupling.}
  \label{tab:reopening of phase space}
\end{table}
In the far asymptotic region one is back to the unsatisfactory
situation that important corrections are most likely missed, with the
additional drawback, that now one has no real idea of what those might
be. Fortunately, this would appear to occur only at such extremely
large values of $Q_s$ that this concern is purely academic.

The fact that the asymptotic expansion of Mueller and
Triantafyllopoulos of $\lambda(\y)$ --Eq.~\eqref{eq:dionlambda}--
apparently contains only a single term beyond the questionable $Q_s$
scale setting result, and yet provides a fit over the whole $Q_s$
range with its large deviations from the leading term, may cause
confusion.  The resolution to this simply is hidden in the change of
variables $\y$, used by MT, and $Q_s$, as used here. Any attempt to
translate Eq.~\eqref{eq:dionlambda} from $\y$ to $Q_s$ or
$\alpha_s(Q_s^2)$ leads to a infinite series that will converge only
slowly, so that it is in fact wrong to think of this as a single
correction term from this perspective.

The discussion thus far has been about the behavior of the dipole
within one of two orders of magnitude of $Q_s$. Careful studies have
also been performed at much shorter distances~\cite{Braun:2000wr,
  Albacete:2004gw} in the scaling window, where gluon densities are
still small, but scaling and the presence of $Q_s$ still plays an
important role as it leads to behavior of the
form~\eqref{eq:MT-scaling-running} with
\begin{equation*}
{\cal G}_{\y; r,R_s(\y)} = C
\left(\frac{{\bm{r}}^2}{R_s(\y)^2}\right)^{(1-{\Tilde\nu}_0)}
\left[-\ln\frac{{\bm{r}}^2}{R_s(\y)^2} +
\frac{1}{1-{\Tilde\nu}_0}\right] \ .
\end{equation*}
Both studies attempt to extract the anomalous dimensions of
Eq.~\eqref{eq:MT-scaling-running}, starting from distances typically a
few orders of magnitude smaller than $R_s$.

They find that for this quantity which only makes sense at such short
distances, running coupling effects remain important over an enormous
range. Residual information about the initial conditions used appears
to be more persistent than close to $Q_s$. At the same time running
coupling effects remain particularly important in this range.  Both
these facts can be understood by looking at the BK equation at running
coupling and the integral expression for $\lambda(\y)$.  Writing, for
instance, Eq.\eqref{eq:BK-parent-dipole} in terms of rescaled
variables $\Tilde{\bm{r}}:={\bm{r}}Q_s(\y)$ etc. one obtains
\begin{equation}
  \label{eq:BK-parent-dipole-rescaled}
  \partial_\y N_{ \Tilde{\bm{r}}}  =
  \frac{ N_c}{2\pi^2}\frac{b_0}{\ln(\frac{Q_s^2(\y)}{\Lambda_{\text{QCD}}^2})
    -\ln(\Tilde{\bm{r}}^2)}
  \int\!\!d^2 \Tilde z
  \ \Tilde{\cal K}_{\Tilde{\bm{x}}
    \Tilde{\bm{z}} \Tilde{\bm{y}}}\,
  %\frac{(\bm{x}-\bm{y})^2}{%
  %  (\bm{x}-\bm{z})^2(\bm{z}-\bm{y})^2} 
  \Big\{  
   N_{ \Tilde{\bm{u}}}+N_{\Tilde{\bm{v}}}-N_{\Tilde{\bm{r}}}
  - N_{\Tilde{\bm{u}}}N_{\Tilde{\bm{v}}}
  \Big\}  
\ .
\end{equation}
Clearly, with increasing $Q_s$ the importance of running coupling
effects (in the sense that goes beyond $Q_s$ scale setting) will be
pushed further and further away from the scaling region where
$\Tilde{\bm{r}}={\bm{r}}Q_s(\y)\gtrsim 10^{-2}$, while at shorter
distances, where the anomalous dimensions are measured form a scaling
fit to~\eqref{eq:MT-scaling-running}, $\Tilde{\bm{r}}$ dependence of
the coupling remains important. These features are reflected by the
numerical results of~\cite{Braun:2000wr, Albacete:2004gw} which
emphatically confirm the existence of the scaling window: they cannot
even find any evidence that scaling of the
form~\eqref{eq:MT-scaling-running} would break down at the upper end
of the scaling region.\footnote{At the short distances involved there,
  this might be a problem with short distance corrections
  entirely absent in the BK equation, but this is a question that goes
  beyond the existence and relevance of the scaling window per se.}

To summarize the results of this section, one should state that with
parent dipole running, the contributions to $\lambda$ arise from
within little more than an order of magnitude of the saturation scale,
as Fig.~\ref{fig:active-phase-space} shows impressively. [This
concerns the region in cutoffs over which there is an appreciable
change in the value extracted.] This indicates that the dominant phase
space corrections are indeed taken into account and the basic physics
ideas that had initially motivated the density resummations are
realized.  While the IR side of the resummation was working already at
fixed coupling, the UV side of phase space is only tamed if the
coupling runs at a scale directly visible in the diagrams comprising
the r.h.s.  of the evolution equation such as $1/\bm{r}^2$, the parent
dipole size.  This is equivalent to incorporating subleading terms in
the asymptotic expansions~\eqref{eq:hypothetical} at least to the
order done by Mueller and Triantafyllopoulos in~\eqref{eq:dionlambda}.
Only with both the nonlinearities and a running coupling that goes
beyond $Q_s$ scale setting in the sense explained above are we
justified to claim that the evolution is dominated by distances of the
order of $R_s$ and thus the coupling involved in typical radiation
events during evolution is of order $\alpha_s(Q_s(\y)^2)$.  [Emission
from very small dipoles would simply not be sufficiently suppressed
with $Q_s$ scale setting.] While the result
of~\cite{Mueller:2002zm,Triantafyllopoulos:2002nz} provides a
physically motivated explanation of the curve, a convenient way to
convert this into an explicit function of $Q_s(\y)$ does not exist.
For this purpose the fits shown in
Fig.~\ref{fig:assesing-sqrt-running} provide a useful, quantitative,
rule-of-thumb representation for the numerical result.  The existence
of the scaling window is numerically confirmed: Scaling features
extend into the low density region and persist throughout the scaling
window to distances much shorter than the saturation scale.

\subsection{\it Evolution induces longitudinal 
  correlations: breaking of $A^{\frac{1}{3}}$-scaling in DIS}
\label{sec:evol-a-depend}

Already in the introduction it has been argued and illustrated in
Fig.~\ref{fig:naive-Qs-A-scaling} that for uncorrelated scattering
centers $Q_s$ should scale with larger atomic number like $A^{1/3}$ as
shown in Eq.~\eqref{eq:naive-Qs-A-scaling}. This has recurred in the
context of the MV model and its generalizations
(Sec.\ref{sec:it-mclerr-venug}): generically one expects that going to
larger nuclear targets ought to have an enhancing effect on the
importance of the nonlinearities encountered. To recall the argument:
If one is at small enough $\xbj$ (high enough energies) for the
projectile to punch straight through the target, going to a larger
target means interaction with more scattering centers as indicated in
Fig.~\ref{fig:small-x-dis-geom-rest}. Taking these to be color charges
located in individual nucleons inside the target nucleus the
interaction will occur with more and more uncorrelated color charges
the larger the target. Viewed from up front this amounts to a smaller
correlation length in the transverse direction. In terms of $Q_s$ this
would lead to a variant of relation~\eqref{eq:naive-Qs-A-scaling},
\begin{equation}
  \label{eq:QsptoQsA}
  Q_s^A(\y)^2 = c A^{\frac{1}{3}}  Q_s^p(\y)^2
\end{equation}
where $c$ is some, yet unspecified constant that takes into account
nontrivial geometry and other issues not incorporated in the
McLerran-Venugopalan model (see also the discussion below
Eq.~\eqref{eq:charge-corr} and~\cite{Levin:2003nc}). Note that none of
these arguments would give a clear delineation of the range of $\y$
values for which this would be appropriate. It is clear, however, that
in the context of small $\xbj$ evolution such a statement becomes even
less trivial: the whole idea of deriving a small $\xbj$ evolution
equation is based on the premise that the logarithmic corrections
determining the evolution step are entirely target independent. The
only place where target dependence is allowed to enter is the initial
condition as already indicated in Eq.~\eqref{eq:naive-Qs-A-scaling}. A
careful determination of $c$ can not be done without phenomenological
effort, but a discussion of how evolution changes such a relationship
can be done on the basis of the evolution equations alone. This is
what I focus on here.

In simple cases the two notions --evolution and scaling at all $\y$--
are not conflicting with each other: indeed a rescaling like in
Eq.~\eqref{eq:QsptoQsA} is fully compatible with fixed coupling
evolution on the asymptotic line, where (in the continuum limit)
\begin{equation}
  \label{eq:fixed-A-scaling}
  Q_s(\y)= e^{\lambda(\y-\y_0)} Q_s(\y_0)
\end{equation}
with a universal value for $\lambda$, so that a simple rescaling of
the properties of the initial condition according to
\begin{equation}
  \label{eq:initial-A-scaling}
  Q_s^A(\y_0)^2 = c A^{\frac{1}{3}}  Q_s^p(\y_0)^2 \ 
\end{equation}
automatically yields Eq.~\eqref{eq:QsptoQsA} at all $\y$ declaring
it fully compatible with evolution. 

At running coupling, however, this ceases to be the case. The
straightforward physical argument is that evolution is characterized
by $\alpha_s(Q_s^2)$, which initially is larger for protons than for
nuclei, because of the difference in saturation scales at the initial
condition. In this situation evolution will be faster for smaller
targets and eventually catch up with that of the larger ones.

Technically, with running coupling (and again on the asymptotic line),
one finds a $\y$ dependent $\lambda$ that may be parametrized through
some function $f$ of $\alpha_s(Q_s(\y)^2)$ as in the previous section:
\begin{equation}
  \label{eq:lambda-of-f-alpha}
  \lambda(\y)=\partial_\y \ln Q_s(\y) = f(\alpha_s(Q_s(\y)^2))
\ .
\end{equation}
In this generic case
\begin{equation}
  \label{eq:Qs-lambda-generic}
   Q_s(\y)= e^{\int^\y_{\y_0} d\y'  
     f(\alpha_s(Q_s(\y)^2)} Q_s(\y_0)
\end{equation}
and a rescaling of $Q_s^p(\y_0)^2$ will not simply factor out as in
Eq.~\eqref{eq:fixed-A-scaling}. Therefore a simple rescaling like
Eq.~\eqref{eq:QsptoQsA} will conflict with evolution. All
phenomenological fit functions encountered in Fig.\ref{fig:assesing-sqrt-running}
share this property, which is at the heart of the reasoning used by
Mueller in~\cite{Mueller:2003bz} to argue that evolution will
eventually erase $A$-dependence in $Q_s$.

Instead of a numerical analysis (c.f.~\cite{Albacete:2004gw}) and
purely to illustrate the point, one may trace the memory loss with a
simple parametrization like Eq.~\eqref{eq:lambda-alpha}, equivalent to
a $\y$ dependence of $Q_s$ according to Eq.~\eqref{eq:Q_s}. Recall
that for $n=1/2$ this should be even {\em quantitatively} correct over
the physically interesting domain.

\begin{figure}[htbp]
  \centering
  \includegraphics[width=7cm]{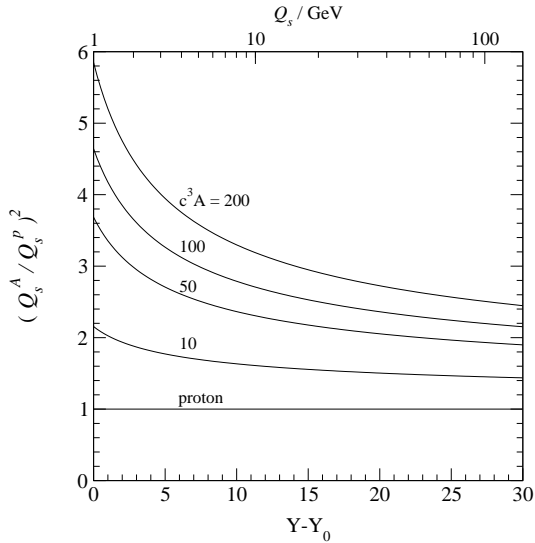}
  \caption{\small (From\cite{Rummukainen:2003ns}) $A$-dependence for 
    proton saturation scales from $1$ to $100$GeV. $c$
    parametrizes possible shape dependence as explained in the text.}
  \label{fig:A-dep}
\end{figure}

Scaling in the $A$-dependence at $\y_0$ according to
$
  Q_s^A(\y_0)^2 = c A^{\frac{1}{3}}  Q_s^p(\y_0)^2 \ 
$,
one uses Eq.~\eqref{eq:Q_s} %~\eqref{eq:Qsasymptotic}
to predict the combined $\y$ and $A$ dependence of the saturation
scale. The result is in qualitative agreement
with~\cite{Mueller:2003bz}: the ratio of saturation scales in the bulk
of the region shown in Fig.~\ref{fig:assesing-sqrt-running} behaves
like
\begin{equation}
  \label{eq:Q_s-scaling}
  \frac{Q^A_s(\y)^2}{Q^p_s(\y)^2} = \frac{
    \exp{ \Big[(n+1) 2\lambda_0 
  \big(\frac{4\pi}{\beta_0}\big)^{n}(\y-\y_0)
 +\big[\ln(\frac{c A^\frac{1}{3} 
   Q_s(\y_0)^2}{\Lambda_{\text{QCD}}^2})\big]^{n+1} 
 \Big]^{\frac{1}{n+1}}
}}{  
    \exp{ \Big[(n+1) 2\lambda_0 
  \big(\frac{4\pi}{\beta_0}\big)^{n}(\y-\y_0)
 +\big[\ln(\frac{ 
   Q_s(\y_0)^2}{\Lambda_{\text{QCD}}^2})\big]^{n+1} 
 \Big]^{\frac{1}{n+1}}
}
} \xrightarrow{\y\to\infty} 1
\ .
\end{equation}
%While the parametrization with $n=1/2$ is not applicable at arbitrary
%large $\y$, this formula nevertheless gives a reasonable estimate of
%the rate of convergence within its window of validity. 
This is plotted in Fig.~\ref{fig:A-dep} over a very large range in
$\y$. Clearly larger nuclei suffer stronger initial ``erasing.'' It
should be kept in mind that realistically one should not expect real
world experiments to cover more than a few orders of magnitude in $\xbj$
and thus a small interval in $\y-\y_0$.

%\marginpar{The fact that Eq.~\eqref{eq:sqrt-lambda} and the ensuing $\y$
%dependence is only valid on the asymptotic line of
%Fig.~\ref{fig:runn-lambda} prevents us from making a direct statement
%about current and future experiments. Such stronger statements require
%a more thorough study of initial conditions in direct comparison with
%existing experiments outside the scope of this paper. If dependence on
%initial conditions can be clarified, $A$ dependence can be studied
%numerically also in the pre-asymptotic region.
%}

% spellmark 7

\section{A brief foray into phenomenology}
\label{sec:pheno}

Phenomenology in the context of the color glass condensate is a very
hot topic, in particular in connection with present activities at
RHIC. As a consequence, many of the topics are not yet settled enough
to be included in a review. Nevertheless, this review would be
incomplete without trying to give a flavor of the ideas and
experiments being discussed. 

The key point in all of them is the appearance of $Q_s$ and associated
scaling in the interpretation of experimental data. This is done with
respect to the $\xbj$ dependence as well as the $A$ dependence where
available, both in the saturation region and the scaling window.

The prime example still is the application of saturation ideas to HERA
data at $\xbj < 10^{-2}$. This will be the first topic discussed below in
Sec.~\ref{sec:geometric-scaling-exp}.

Nucleus nucleus collisions form a much more complicated environment
than lepton nucleus collisions and the step from fully developed
theory to comparison with data is much larger.
%, a generic problem one
%has to face with all phenomenological applications to RHIC data.

There is one approach to heavy ion data which is directly build on the
technology described in this review, and that aims at the initial
condition of heavy ion collisions. This will be briefly touched upon
in Sec.~\ref{sec:initial-exp}.

A different approach based on $({\bm{k}}=)$ $k_T$-factorized formulae
for particle production in terms of unintegrated gluon densities as
defined in Eq.~\eqref{eq:phidef}~\cite{Gribov:1984tu, Laenen:1994gh,
  Kovchegov:1997ke, Gyulassy:1997vt, Kovchegov:1998bi, Braun:2000wr,
  Kovchegov:2001ni, Kovchegov:2001sc}, allows to apply the results of
the evolution equations to $A A'$ experiments:
\begin{equation}
  \label{eq:fact-cross}
  E {d \sigma \over d^3 p} = {4 \pi N_c \over N_c^2 - 1}\ 
  {1 \over {\bm{p}}^2}\  \times
\, \int^{\bm{p}} \, d {\bm{k}}^2 \,
\alpha_s \ \varphi_{A}(\xbj, {\bm{k}}^2)\ 
\varphi_{A'}(\xbj', ({\bm{p}}-{\bm{k}})^2)
\ .
\end{equation}
%This should be well justified if one subscribes to the idea that the
%proton (or a deuteron atom), compared to the large $A$ target, can be
%treated as having a dilute gluon content, much as the leptonic
%projectiles discussed in connection with the evolution equations.
Despite some obvious shortcomings concerning perturbative and
hadronization corrections, a treatment based on this formula is
currently the best one can do. Keeping this in mind, the formula
imprints the consequences of any $Q_s$ scaling found in unintegrated
distributions onto transverse momentum spectra in nucleus nucleus
collisions.

A very illustrative example in this context is the Cronin effect and
its expected behavior under evolution discussed in
Sec.~\ref{sec:cronin-exp}. Particle multiplicities are another
example, but indications are that this topic will come into its full
right only with LHC energies.  These and other ideas relevant to build
a solid understanding on how and exactly in which kinematical domain
the CGC shows up in experiments will briefly reappear in
Sec.~\ref{sec:many-ideas-abound}.

%In all situations
 
%Simple scaling comparisons according to
%Eq.~\eqref{eq:naive-Qs-A-scaling} indicate that $Q_s$ values at HERA
%and RHIC should fall into the same ballpark. The success of scaling
%ideas at HERA should then lead to expect similar features at RHIC.

%Note, however, that all the theoretical methods developed up to nw
%strictly apply to lepton nucleus experiments, so in order to compare
%with RHIC data, a certain amount of modeling is necessary --
%limitations are not yet fully understood. This work is crucial
%nevertheless, since the heavy ion program at the LHC will certainly
%probe $x$ ranges well into the CGC domain and a good idea of what to
%expect should help focus on interesting questions already at the
%beginning of the runs. This is particularly important in view of the
%flood of data expected from these experiments.

%On a somewhat longer timescale, there are experimental plans to
%complement the RHIC facilities with an electron accelerator to be
%called the Electron Ion Collider (EIC) to have a cleaner setting
%directly in line with the theoretical developments described above.

%This can not

%Here, indications would appear to be that one probes the onset
%of CGC physics as one goes from central to forward rapidities.

% can be done on various levels of sophistication,

\subsection{\it Geometric scaling in HERA 
and lepton nucleus collisions}
\label{sec:geometric-scaling-exp}

\begin{figure}[htb]
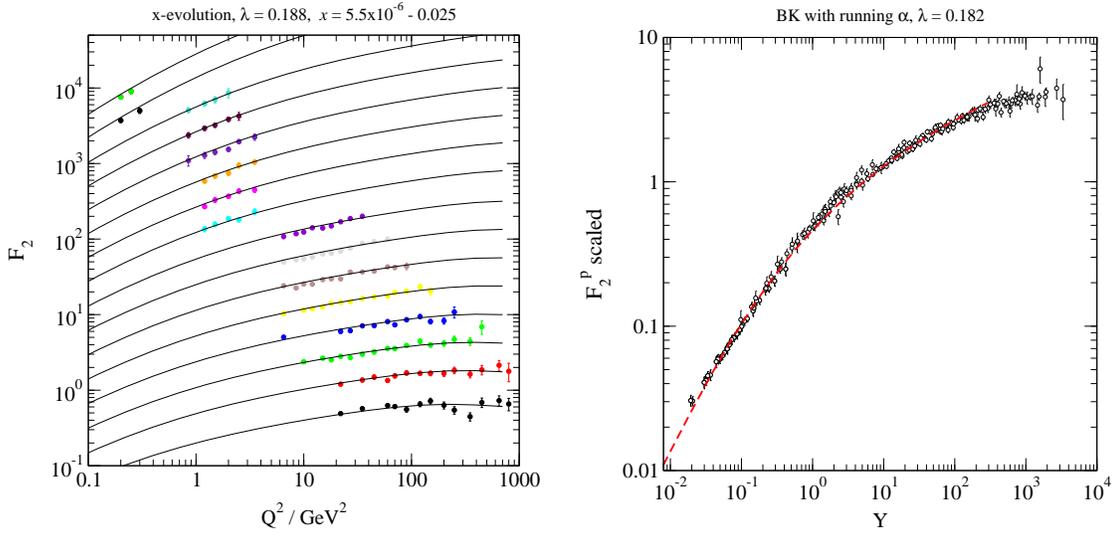

  \centering
  \includegraphics[width=7.1cm]{F2_all_0dot188}  
  \hfill
  \includegraphics[width=7.1cm]{F2_proton_scaled_bkr10_mq014-Y}
  \caption{\small Geometric scaling in HERA data. Shown is  the structure 
    function $F_2$, which at small $\xbj$ is simply related to the
    cross section by a rescaling by $Q^2$: $\sigma(\y,Q^2) \sim
    F_2(\y,Q^2)\cdot Q^2$. }
  \label{fig:geom-scaling-HERA}
\end{figure}

The discovery of geometric scaling in HERA data at $\xbj < 10^{-2}$ as
prompted by the Golec-Biernat and W{\"u}sthoff
fit~\cite{Golec-Biernat:1998js, Stasto:2000er} was the first
phenomenological application of saturation ideas in small $\xbj$ physics,
predating all the theoretical developments around JIMWLK and BK
evolution equations that demonstrated the scaling features of these
evolution equations.  It brought with it the by far most efficient
parametrization of HERA data at $\xbj < 10^{-2}$. Recall the DIS cross
section formula from the introduction (with the momentum fraction
integral restored):
\begin{eqnarray}
\label{eq:stotfact}
\sigma_{\mathrm{tot}}(\y,Q^2)& = & \int d^2{\bm r}\int\limits^1_0\!d\alpha\
|\psi_{\gamma^*}(\alpha,{\bm r}^2,Q^2)|^2
%\nonumber\\& & \hspace{1.5cm}\times
\
\sigma_{\mathrm{dipole}}(\y,{\bm r}^2)
\end{eqnarray}
where $\alpha$ is the longitudinal momentum fraction of the $q$ or
$\Bar q$, and $\bm{z}$ their relative transverse separation.  $Q^2$
and $\xbj$-dependence is clearly separated into wave function and dipole
cross section respectively.  The G-B+W fit parametrizes the latter via
an $\xbj$-dependent saturation scale $Q_s$ according
to~\eqref{eq:GBW-dip} and a normalization factor to carry the
dimensions from the impact parameter integral:
\begin{equation}
  \label{eq:G-B+W-param}
  \sigma_{\text{dipole}}(\y,r^2) = \int d^2b \ 
  \langle\frac{\tr(1-U_{\bm{x}} U^\dagger_{\bm{y}})}{N_c}
  \rangle_{\y} 
  \hspace{1cm}\xrightarrow{\text{ G-B+W } }\hspace{1cm}
 \sigma_0 (1-e^{-{\bm r}^2 Q_s^2(\y)/4})
\ .
\end{equation}
At the same time they preempted the fixed coupling scaling form for
$Q_s$ of Eq.~\eqref{eq:Qspower}, $ Q_s(\y) = e^{\lambda(\y-\y_0)}
Q_s(\y_0)$ $= (\xbj_0/\xbj)^\lambda Q_s(x_0)$. This fit ansatz is just a
particular realization of the scaling behavior later found to result
from JIMWLK and BK, both at fixed and running coupling. Successful
saturation based fits incorporating scaling forms
like~\eqref{eq:MT-scaling-running} have also been performed in the
meantime~\cite{Iancu:2003ge}. The most important phenomenological
result, however, is independent of the exact shape of the dipole cross
section, it is the simple fact that HERA data show geometric scaling.
Once the dipole cross section depends on $\xbj$ only via $Q_s(\y)$ as
$\sigma_{\text{dipole}}(\y,{\bm r}^2) = \sigma_{\text{dipole}}({\bm
  r}^2Q_s^2(\y))$, geometric scaling emerges directly. Rescaling the
$r$ integral in Eq.\eqref{eq:stotfact} and using the fact that the
wave function factor scales like $f({\bm{r}}^2 Q^2)/{\bm{r}}^2$, this
implies
\begin{equation}
  \label{eq:scaling}
 \sigma_{\mathrm{tot}}(\y,Q^2) =
 \sigma_{\mathrm{tot}}\Big(\y_0,\frac{Q^2}{Q_s^2(\y)}Q_s^2(\y_0)\Big)
\end{equation}
and should allow to map all the cross section data onto a single
scaling curve. This is shown in Fig.~\ref{fig:geom-scaling-HERA}
While the explicit parametrizations shown in
Fig.~\ref{fig:geom-scaling-HERA} make use of BK evolution results
from~\cite{Rummukainen:2003ns}, this is not essential, at least for
the scaling plot on the right. In fact, successful fits of HERA data
based on CGC evolution exist~\cite{Gotsman:2002yy,Iancu:2003ge} and
they are of a quality comparable to that of DGLAP fits. It is not the
quality of the fit to the data (leaving aisde comments on negative
gluon distributions often seen in global DGLAP analyses) that
distinguishes the CGC perspective, it is the natural emergence of
$Q_s$ and the scaling features which set it apart.
% This simple argument has been uniformly
%used in~\cite{Freund:2002ux} to relate cross section- (or $F_2$-)
%measurements on protons at HERA with measurements on nuclei at NMC and
%E665. Despite the fact that the nuclear data used are at rather small
%$Q^2$ and large $\xbj$ the agreement found was surprisingly good.
Scaling features have since been looked for also in older
NMC~\cite{Arneodo:1997kd} and E665~\cite{Adams:1995ri,Adams:1995is}
data for the ratios of nuclear to deuterium structure functions
$2F_2^A/AF_2^D$. Despite the fact that both $\xbj$ and $Q^2$ ranges are
such that CGC saturation arguments remain marginal despite the
moderate enhancement via $A$, both~\cite{Freund:2002ux}
and~\cite{Armesto:2004ud} find that scaling in both $\xbj$ and $A$ holds
surprisingly well. The scaling in $A$ was fitted with a fixed power
close to $1/3$ ignoring the refinements of
Sec.~\ref{sec:evol-a-depend}. Due to the limited range in $\xbj$ this can
not be taken as an experimental confirmation of CGC physics -- such
would require a measurement of $\xbj$ dependence and this is clearly a
reason to explore this question more closely in experiments at higher
energies. This is being actively pursued. One of the options is the
Electron Ion Collider (EIC) in planning at
BNL~\cite{EICwhite} as a machine dedicated to such experiments
or via $p A$ collisions at the LHC where higher energies are available
at the price of a more complicated projectile.

%\begin{itemize}
%\item Geometric scaling and Golec-Biernat+W{\"u}sthoff fit.
%\item G-B+W plot (does that contain ghe G-B+W anstatz? I suppose so)
%\item Our plot with the statement that this does not contain the G-B+W
%  anstatz, but only the scaling assumption.
%\item Scaling in nuclear data. Ours~\cite{Freund:2002ux} and
%  theirs~\cite{Armesto:2004ud}, discussion of what is important and surprising.
%\end{itemize}

\subsection{\it Gluon production in the initial 
stages of heavy ion collisions}
\label{sec:initial-exp}

The evolution equations have mainly been discussed in the context of
$e A$ collisions, but it should be intuitively clear that a new
perturbatively large scale $Q_s$, present in the nuclear wave function
at small $\xbj$, would also manifest itself in $A A'$ collisions. The
original idea is laid out in~\cite{Kovner:1995ja}, where the incoming
nuclei are treated as sources of strong color fields in the spirit of
the MV model. While the nonlinear nature of the fields does not allow
to simply add them, it is still possible to add the color currents
that create them and then to solve the resulting Yang Mills equations.
This will only modify the solutions after they are in causal
contact,\footnote{If applied to the gluon field $A$, this is a gauge
  dependent statement!} i.e. in the forward lightcone of the collision
point. At the edge of this forward cone the fields generated by the
individual nuclei serve as an initial condition for what happens after
the collision for a given source configuration. The source
configurations themselves are subject to ensemble averaging, say
with~\eqref{eq:rhodist} (per nucleus), if one stays within the MV
model or with the corresponding field averages expressed alternatively
via $b$, $\alpha$ or $U$, if one goes beyond this model to incorporate
$\xbj$ dependence via BK or JIMWLK equations.

\cite{Kovner:1995ja} provides the generic framework but only attempts
perturbative solutions valid for large times, however, the field
equations for large densities are truly nonlinear in the very initial
stages of the collision of interest. In this region numerical methods
are required. These have been developed and used
by~\cite{Krasnitz:1998ns,Krasnitz:1999wc}. More recent work discusses
the possibility to create ``elliptic
flow''~\cite{Krasnitz:2003nv,Krasnitz:2003jw} and independent checks
have been provided in~\cite{Lappi:2003bi}. The main interest of such
studies is the creation of high density effects in the very initial
stages of heavy ion collisions, which would appear to be necessary to
explain the very short equilibration times often quoted as necessary
to explain the collective phenomena supported by data.

\subsection{\it Evolution erases the Cronin effect}
\label{sec:cronin-exp}

The last experimental topic to be discussed here that has met with
strong interest is the question for the energy dependence of what is
known as the Cronin effect. I should begin this section with the
warning that all input and results in terms of the CGC are formulated
on a partonic level, while the underlying definitions of cross
sections~\eqref{eq:croninratio-gen} and actual measurements are done
on a hadronic level. There are strong indications that there may by
important contributions from hadronization to the measured quantities.
For this reason one might want to talk about what is discussed below
as a Cronin-like effect. Nevertheless, the main message that small
$\xbj$ evolution has a specific and rather strong effect on transverse
momentum spectra remains an exciting result.

The Cronin effect refers to an enhancement of the number of particles
produced in the intermediate transverse momentum range of nucleus
nucleus collisions over those in proton proton collisions which is
commonly attributed to a momentum broadening due to multiple
rescatterings in the final state. At first sight this would appear to
be at odds with the ideas of saturation which would lead to the
expectation that particle production should be suppressed for momenta
below $Q_s$ and even, as argued by~\cite{Kharzeev:2002pc} within the
entire scaling window. This, however, is but less than half the
picture.

The first observation is that the original MV model does contain
Cronin type enhancement\cite{Baier:2003hr, Kharzeev:2003wz,
  Jalilian-Marian:2003mf}. This is based on a sum
rule~\cite{Lam:1999wu} for the gluon distribution in the MV model that
equates the total number of gluons in a (large) nucleus with $A$ times
that of a nucleon. At the same time the momentum spectrum is distorted
by eikonalization: small momenta are indeed depleted, while at very
large momenta a perturbative $1/{\bm{k}}^2$ tail remains undisturbed.
This can be readily recovered from~\eqref{eq:MV-gluedist-mom} together
with~\eqref{eq:Qs-Gnuc} under the assumption of $A^{1/3}$ scaling of
the exponent in either gluon distribution or dipole cross section.
(The remaining $A^{2/3}$ comes from the $d^2b$ integral as a global
prefactor.) By necessity the ``excess'' has to accumulate at moderate
transverse momenta which leads to Cronin enhancement.

The second observation comes from the fact that the sum rule is broken
by evolution. One aspect of this is simply the modification of
$A$-dependence of the saturation scale through evolution reviewed in
Sec.~\ref{sec:evol-a-depend}. The effect on the gluon distribution and
thus the Cronin effect is even much more pronounced: The Cronin peak
is suppressed over a very short range in rapidity.

The underlying object measured is a ratio $R_{A B, C
  D}(\y,{\bm{p}},{\bm{b}})$ that compares numbers of produced
particles per rapidity $\y$ (i.e. $\y$ for present purposes),
transverse momentum $\bm p$ and centrality $\bm b$ in the collision of
two nuclei $A, B$ with the same in the collision of another pair $C,
D$ as a function of transverse momentum, both normalized to the number
of collisions $N_{\text{coll}}(b)$:
\begin{equation}
  \label{eq:croninratio-gen}
    R_{\text{A B;C D}}( \y, {\bm{p}},{\bm{b}})\,:=
\,\frac{\langle N_{\text{coll}}(b) \rangle_{C+D}}{\langle N_{\text{coll}}(b) \rangle_{A+B}}\,
   \frac{dN_{A+B}/ d\y\,d^2{\bm{p}}\,
d^2{\bm{b}}}
{dN_{C+D}/d\y\,d^2{\bm{p}}\,d^2{\bm{b}}
   }\, .
\end{equation}
For practical measurements \eqref{eq:croninratio-gen} is too
differential and so simplified variants are used. On the one hand,
there is the ratio $R_{\text{A B}}$ in which the multiplicities
of~\eqref{eq:croninratio-gen} are integrated over impact parameter
$\bm{b}$, and $C, D$ are both taken to be protons for reference. This
is the standard way to show the Cronin effect. On the other hand,
there is $R_{\text{cp}}$ ($c, p$ stand for central and peripheral
respectively) in which $(A, B)=(C, D)$ and the multiplicities in
numerator and denominator are integrated over central and peripheral
regions respectively. [The same applies to the average number of
collisions used to normalize.]  Both quantities should show similar
trends, loosely speaking, the peripheral region looks like a collision
with a smaller nucleus.

\begin{figure}[htbp]
  \centering
  \includegraphics[width=8cm,angle=-90]{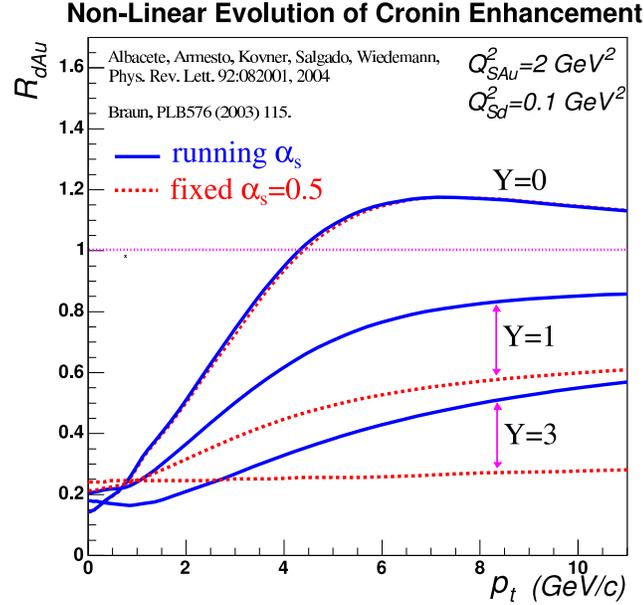}  
  \caption{\small (Courtesy
    of the authors of~\cite{Albacete:2003iq}) Suppression of the Cronin
    peak through BK evolution both at fixed and running coupling.
    Running coupling evolution is slower than the fixed coupling case
    as seen already for generic quantities in
    Sec.~\ref{sec:runn-coupl-effects}. }
  \label{fig:cronin}
\end{figure}

$R_{\text{d Au}}$ had been studied carefully in the context of the BK
equations in~\cite{Albacete:2003iq}. For this theoretical study
initial conditions of the form~\eqref{eq:MV-gluedist-mom} together
with~\eqref{eq:Qs-Gnuc} were used to provide an initial condition that
shows Cronin enhancement. The result is shown in Fig.~\ref{fig:cronin}
and shows the fast disappearing Cronin enhancement. Again
evolution is considerably slower in the running coupling case.

\begin{figure}[htb]
  \centering
\includegraphics[width=\textwidth]{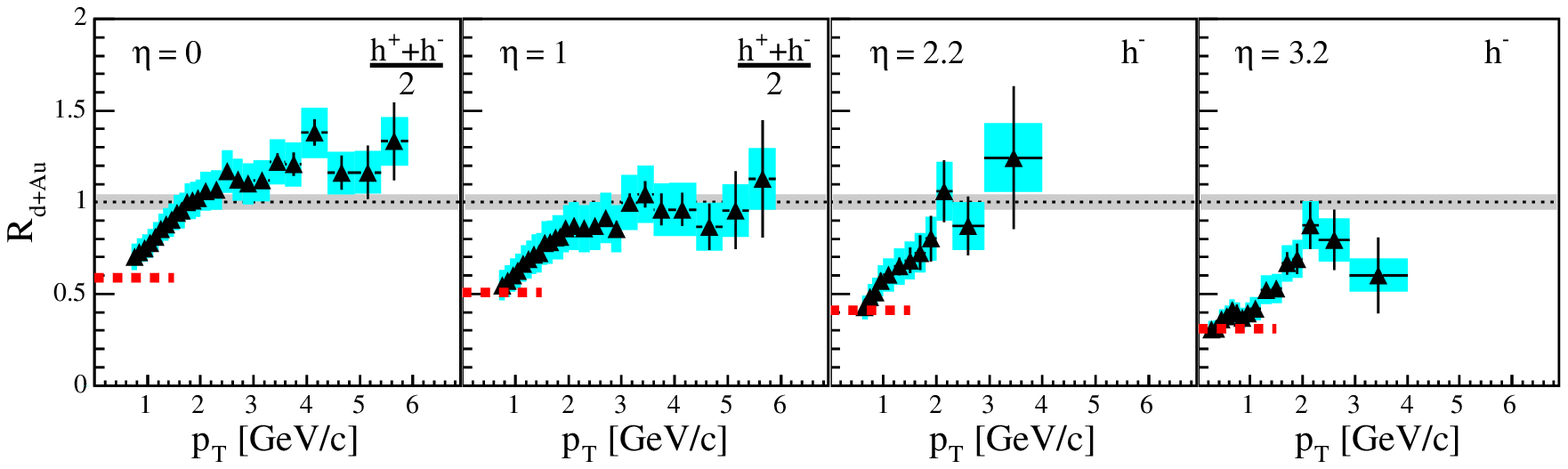}

\includegraphics[width=\textwidth]{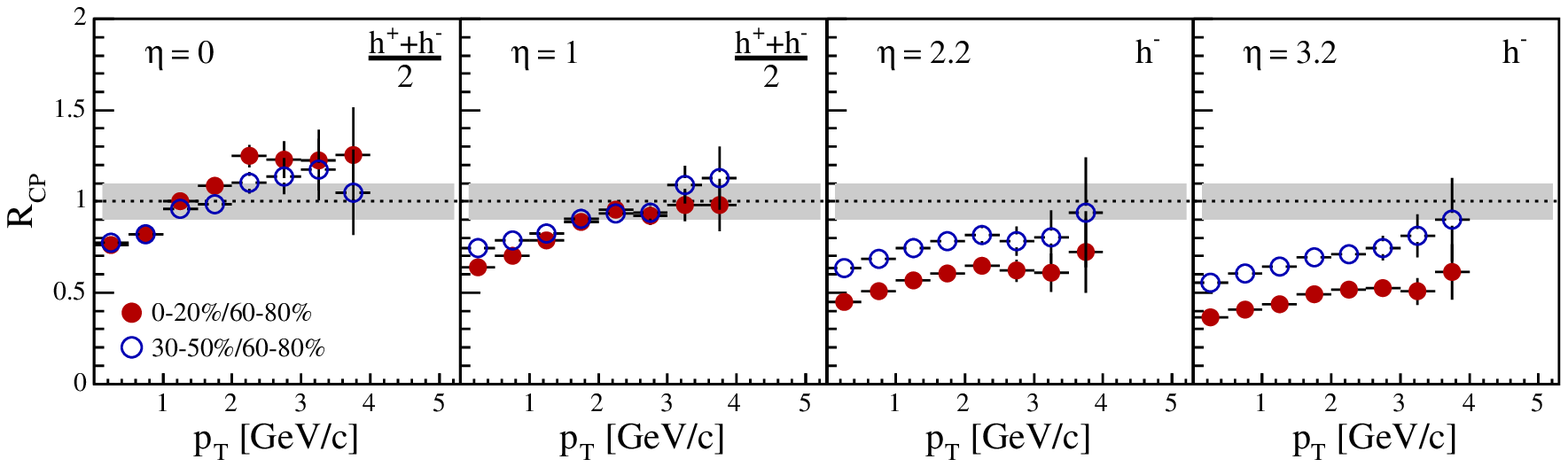}  
  \caption{\small
    Brahms data~\cite{Arsene:2004ux}, Top strip: $R_{\text{d Au}}$
    as a function of ${\bm{p}}=p_T$ for different pseudorapidities $\eta$ (here
    to be identified with $\y$). Bottom strip: The same for
    $R_{\text{cp}}$ in d + Au. }
  \label{fig:brahms}
\end{figure}

Both $R_{\text{d Au}}$ and $R_{\text{cp}}$ have been measured by the
BRAHMS experiment at RHIC~\cite{Arsene:2004ux} and are shown in
Fig.~\ref{fig:brahms}, top and bottom strips respectively.
In both of these measurements it is the depletion at forward
rapidities that is the interesting effect: Other
approaches~\cite{Vitev:2003xu,Accardi:2003jh} that concentrate on
multiple scattering to reshuffle particles in the momentum
distributions typically do not take into account the nonlinearity of
the small $\xbj$ evolution equation which leads to the effect shown in
Fig.~\ref{fig:cronin}. Instead the Cronin enhancement tends to
increase, reflecting the increasing number of scattering centers at
small $\xbj$.

Experimental work and theoretical discussion on this topic is still
ongoing, see for example~\cite{Kharzeev:2004yx} for a recent fit of
available data.

\subsection{\it Many ideas abound}
\label{sec:many-ideas-abound}

The list of ideas given above is by no means exhaustive. The effect of
$Q_s$ on particle multiplicities in $p A$ and $A A'$ collisions has
been discussed as well, initially for RHIC, where they have been
essentially ruled out for the central rapidity region. This is in
keeping with the tentative interpretation of BRAHMS data discussed in
Sec.~\ref{sec:cronin-exp}, but judging from the same data, one would
expect this to be highly relevant at the LHC at central rapidities for
which predictions exist~\cite{Kharzeev:2004if}. The applications
typically combine methods to calculate particle yields with classical
Glauber formulae for multiple scattering situations --used in order to
understand the reaction geometry-- with a factorized ansatz for
inclusive production in terms of the unintegrated gluon densities of
the participants as discussed in Eq.~\eqref{eq:fact-cross}. The latter
are then parametrized using information about the CGC, typically with
some parametrizations for the energy dependence of the saturation
scale.  This should give first ideas about the effects to be expected,
but clearly could profit from a closer integration with the evolution
equations and the concepts of Sec.~\ref{sec:initial-exp}.

Recently two particle and jet azimuthal corrections have been
suggested~\cite{Jalilian-Marian:2004da, Kharzeev:2004bw} to further
study the onset of parton saturation suggested by the BRAHMS data,
ideas to measure the transverse size of the area populated by gluons
have been formulated~\cite{Strikman:2004km} and too many others to
give a complete list of.

It is clear, however, from the added number of theoretical
complications in heavy ion experiments that a lepton nucleus
experiment at high energies would be an important tool to get a clean
baseline for all the applications to these more complicated
experiments. This is the goal of the EIC project currently in the
planning stage at BNL.

% spellmark 8

\section{Conclusions}
\label{sec:conclusions}

The concepts to describe saturation effects in small $\xbj$ have come
a long ways since the early discussions sparked by the
Gribov-Levin-Ryskin (GLR) equation~\cite{Gribov:1981ac}.  Stages on
the way are represented by the Mueller-Qiu\cite{Mueller:1986wy}
equation (aiming at the same double logarithmic limit as the GLR
equation), Mueller's original dipole model\cite{Mueller:1994rr,
  Mueller:1994jq, Mueller:1995gb, Chen:1995pa} designed to give
evolution towards small $\xbj$ and the McLerran-Venugopalan model
meant to describe density effects at small but fixed $\xbj$ in large
nuclei.

JIMWLK and BK evolution equations have provided a new framework,
derived from small coupling QCD, that allows to take into account
gluon induced density effects at small $\xbj$. 

The theoretical treatment contains two types of nonlinearities,
eikonalization of ``soft'' gluon fields into path ordered exponentials
or Wilson lines as a first stage and a second stage that even
exponentiates these objects --inside of shower operators-- to describe
emission of ``hard'' gluons.

The first stage is sufficient to obtain bounded results for cross
sections as evidenced by the dipole formula for the DIS
example~\eqref{eq:sigma-dipole} or Sudakov form factors in jet
observables. Such expressions, since bounded, are always efficiently
parametrized as exponentials as in
Eqns.~\eqref{eq:simple-corr}, without any reference to a specific
averaging procedure for the gluon fields -- interpretation in terms of
a saturation scale as in Eq.~\eqref{eq:MV-gluedist-mom} is most
natural in the small $\xbj$ context.

The second stage exponentiation via the shower
operators~\cite{Weigert:2003mm} implements $\xbj$ dependence of objects
like the photon wavefunction in DIS. Via translation through Langevin
formulations\cite{Weigert:2000gi, Blaizot:2002xy} and JIMWLK
equation~\cite{ Jalilian-Marian:1997xn, Jalilian-Marian:1997jx,
  Jalilian-Marian:1997gr, Jalilian-Marian:1997dw,
  Jalilian-Marian:1998cb, Kovner:2000pt, Weigert:2000gi,
  Iancu:2000hn,Ferreiro:2001qy}, this generates an infinite hierarchy
of coupled equations for cross sections, the Balitsky hierarchy, which
reduces to the BK equation, if correlators factorize into two point
functions.

In whichever guise this is presented, the $\xbj$ dependence is
characterized by the emergence of scaling: universal behavior that
ties the $\xbj$ dependence to dependence on the saturation scale.

The evolution equations impose this behavior quite quickly: Near the
saturation scale details about initial conditions are efficiently
erased, as is evidenced by the quick convergence onto the  near
scaling curve shown by the plot of the evolution rate
$\lambda(\y)=\partial_\y \ln Q_s(\y)$ in
Fig.~\ref{fig:assesing-sqrt-running}. Once near scaling is reached,
the scaling form only changes {\em very slowly} -- scale breaking is
present but weak\cite{Freund:2002ux}.  This feature also manifests
itself in other quantities such as, for example, anomalous dimensions
as measured from BK evolution by~\cite{Braun:2000wr, Albacete:2004gw}.
A subtlety that was manifest --although unappreciated-- already in
numerical simulations at fixed coupling is the importance of running
coupling to tame UV phase space.  Only with running coupling
implemented is the evolution activity centered around the physical
scale $Q_s$.

The generic structures of the phase space diagram
Fig.~\ref{fig:x-Q-plane-density} with saturation, scaling and extended
scaling region have all been seen numerically. The perhaps simplest
physical interpretation is this: Starting from a dipole function that
is saturated at large dipole sizes and perturbatively small at small
dipole sizes, evolution is driven by contributions from small sizes.
This follows the BFKL pattern of exponential growth. Creation of large
objects that would fall into the already saturated domain is
prohibited by the nonlinearities, corresponding contributions are
absorbed by the medium.  Due to the growth at small distances,
densities rise also there and the boundary where the nonlinearities
start to act readjust. This happens selfconsistently in JIMWLK and BK.
In~\cite{Mueller:2002zm,Triantafyllopoulos:2002nz} the picture is
taken literally and an even quantitatively comparable result is
achieved by combining BFKL evolution with an absorptive boundary which
is selfconsistently adjusted to limit growth. This type of analysis
leads to scaling forms for the dipole cross section valid at small
distances, and predicts the size of the scaling window via the
corresponding BFKL diffusion radius.

Scaling features as found in these evolution equations had been seen
earlier in HERA data at small $\xbj$ and called geometric scaling. This
has encouraged phenomenological applications to various observables
like particle multiplicities or the Cronin effect. Besides its $\xbj$
dependence, the $A$ dependence of $Q_s$ as implied by the MV model and
somewhat modified by evolution~\cite{Mueller:2003bz,
  Rummukainen:2003ns, Albacete:2004gw} is an interesting tool to tie
together experimental results in lepton proton, lepton nucleus, and
nucleus nucleus experiments. While qualitative features in existing
experiments --HERA and RHIC featuring most prominently on this list--
appear to be compatible with state of the art CGC phenomenology,
matters are far from settled. Experiments at smaller $\xbj$, like the
LHC, and simpler environments, like the EIC with its electron
projectiles, are needed to fully clarify the situation. Current
experimental activities at RHIC and phenomenological work to match and
assess the enormous amount of new results, for example by suggesting
new measurements and phenomenological consequences, are very
encouraging.

At the same time it would help to close the gap between reliable
theory and phenomenological applications. Our understanding of small
$\xbj$ evolution equations at this point is limited to the central part
of lepton nucleus collisions. The technological tools and insights
gained through comparison with soft emission phenomena in non-global
jet observables should allow to formulate new, IR safe observables
together with the associated evolution equations also in the small $\xbj$
case. This field has matured from its origins, but new developments
are just around the corner.

\paragraph{Acknowledgement:} 
I whish to thank DFG and BMBF for financial support during later
stages of my involvement with this topic. I am grateful to Andreas
Sch\"afer for feedback on the first part of this manuscript, to
Dionysis Triantafyllopoulos and Edmond Iancu for comments on the
sections on the MT model and the scaling region as well as to Urs
Wiedemann for his input on the presentation of the Cronin effect.

\providecommand{\href}[2]{#2}\begingroup\raggedright\endgroup

%\bibliography{newletter,jetstuff,cronin,multiplicities,other-pheno,fact} %<-----------BIBLIOGRAPHIE

\begin{thebibliography}{10%
0}

\bibitem{Bal-Lip}
Y.~Y. Balitsky and L.~N. Lipatov {\em Sov. J. Nucl. Phys.} {\bf 28} (1978) 822.

\bibitem{Lipat}
L.~N. Lipatov {\em Sov. J. Nucl. Phys.} {\bf 23} (1976) 642.

\bibitem{Kuraev:1976ge}
E.~A. Kuraev, L.~N. Lipatov, and V.~S. Fadin {\em Sov. Phys. JETP} {\bf 44}
  (1976) 443--450.

\bibitem{Kuraev:1977fs}
E.~A. Kuraev, L.~N. Lipatov, and V.~S. Fadin {\em Sov. Phys. JETP} {\bf 45}
  (1977) 199--204.

\bibitem{Lipat-2}
L.~N. Lipatov {\em Sov. Phys. JETP} {\bf 63} (1986) 904.

\bibitem{Froissart:1961ux}
M.~Froissart {\em Phys. Rev.} {\bf 123} (1961) 1053--1057.

\bibitem{Jalilian-Marian:1997xn}
J.~Jalilian-Marian, A.~Kovner, L.~D. McLerran, and H.~Weigert {\em Phys. Rev.}
  {\bf D55} (1997) 5414--5428,
  [\href{http://xxx.lanl.gov/abs/hep-ph/9606337}{{\tt hep-ph/9606337}}].

\bibitem{Jalilian-Marian:1997jx}
J.~Jalilian-Marian, A.~Kovner, A.~Leonidov, and H.~Weigert {\em Nucl. Phys.}
  {\bf B504} (1997) 415--431,
  [\href{http://xxx.lanl.gov/abs/hep-ph/9701284}{{\tt hep-ph/9701284}}].

\bibitem{Jalilian-Marian:1997gr}
J.~Jalilian-Marian, A.~Kovner, A.~Leonidov, and H.~Weigert {\em Phys. Rev.}
  {\bf D59} (1999) 014014, [\href{http://xxx.lanl.gov/abs/hep-ph/9706377}{{\tt
  hep-ph/9706377}}].

\bibitem{Jalilian-Marian:1997dw}
J.~Jalilian-Marian, A.~Kovner, and H.~Weigert {\em Phys. Rev.} {\bf D59} (1999)
  014015, [\href{http://xxx.lanl.gov/abs/hep-ph/9709432}{{\tt
  hep-ph/9709432}}].

\bibitem{Jalilian-Marian:1998cb}
J.~Jalilian-Marian, A.~Kovner, A.~Leonidov, and H.~Weigert {\em Phys. Rev.}
  {\bf D59} (1999) 034007, [\href{http://xxx.lanl.gov/abs/hep-ph/9807462}{{\tt
  hep-ph/9807462}}].

\bibitem{Kovner:2000pt}
A.~Kovner, J.~G. Milhano, and H.~Weigert {\em Phys. Rev.} {\bf D62} (2000)
  114005, [\href{http://xxx.lanl.gov/abs/hep-ph/0004014}{{\tt
  hep-ph/0004014}}].

\bibitem{Weigert:2000gi}
H.~Weigert {\em Nucl. Phys.} {\bf A703} (2002) 823--860,
  [\href{http://xxx.lanl.gov/abs/hep-ph/0004044}{{\tt hep-ph/0004044}}].

\bibitem{Iancu:2000hn}
E.~Iancu, A.~Leonidov, and L.~D. McLerran {\em Nucl. Phys.} {\bf A692} (2001)
  583--645, [\href{http://xxx.lanl.gov/abs/hep-ph/0011241}{{\tt
  hep-ph/0011241}}].

\bibitem{Ferreiro:2001qy}
E.~Ferreiro, E.~Iancu, A.~Leonidov, and L.~McLerran {\em Nucl. Phys.} {\bf
  A703} (2002) 489--538, [\href{http://xxx.lanl.gov/abs/hep-ph/0109115}{{\tt
  hep-ph/0109115}}].

\bibitem{Mueller:1986wy}
A.~H. Mueller and J.-w. Qiu {\em Nucl. Phys.} {\bf B268} (1986) 427.

\bibitem{Mueller:1994rr}
A.~H. Mueller {\em Nucl. Phys.} {\bf B415} (1994) 373--385.

\bibitem{Mueller:1994jq}
A.~H. Mueller and B.~Patel {\em Nucl. Phys.} {\bf B425} (1994) 471--488,
  [\href{http://xxx.lanl.gov/abs/hep-ph/9403256}{{\tt hep-ph/9403256}}].

\bibitem{McLerran:1994ni}
L.~D. McLerran and R.~Venugopalan {\em Phys. Rev.} {\bf D49} (1994) 2233--2241,
  [\href{http://xxx.lanl.gov/abs/hep-ph/9309289}{{\tt hep-ph/9309289}}].

\bibitem{McLerran:1994ka}
L.~D. McLerran and R.~Venugopalan {\em Phys. Rev.} {\bf D49} (1994) 3352--3355,
  [\href{http://xxx.lanl.gov/abs/hep-ph/9311205}{{\tt hep-ph/9311205}}].

\bibitem{McLerran:1994vd}
L.~D. McLerran and R.~Venugopalan {\em Phys. Rev.} {\bf D50} (1994) 2225--2233,
  [\href{http://xxx.lanl.gov/abs/hep-ph/9402335}{{\tt hep-ph/9402335}}].

\bibitem{Ayala:1995kg}
A.~Ayala, J.~Jalilian-Marian, L.~D. McLerran, and R.~Venugopalan {\em Phys.
  Rev.} {\bf D52} (1995) 2935--2943,
  [\href{http://xxx.lanl.gov/abs/hep-ph/9501324}{{\tt hep-ph/9501324}}].

\bibitem{Kovner:1995ja}
A.~Kovner, L.~D. McLerran, and H.~Weigert {\em Phys. Rev.} {\bf D52} (1995)
  6231--6237, [\href{http://xxx.lanl.gov/abs/hep-ph/9502289}{{\tt
  hep-ph/9502289}}].

\bibitem{Ayala:1996hx}
A.~Ayala, J.~Jalilian-Marian, L.~D. McLerran, and R.~Venugopalan {\em Phys.
  Rev.} {\bf D53} (1996) 458--475,
  [\href{http://xxx.lanl.gov/abs/hep-ph/9508302}{{\tt hep-ph/9508302}}].

\bibitem{Kovchegov:1996ty}
Y.~V. Kovchegov {\em Phys. Rev.} {\bf D54} (1996) 5463--5469,
  [\href{http://xxx.lanl.gov/abs/hep-ph/9605446}{{\tt hep-ph/9605446}}].

\bibitem{Balitsky:1996ub}
I.~Balitsky {\em Nucl. Phys.} {\bf B463} (1996) 99--160,
  [\href{http://xxx.lanl.gov/abs/hep-ph/9509348}{{\tt hep-ph/9509348}}].

\bibitem{Kovchegov:1997pc}
Y.~V. Kovchegov {\em Phys. Rev.} {\bf D55} (1997) 5445--5455,
  [\href{http://xxx.lanl.gov/abs/hep-ph/9701229}{{\tt hep-ph/9701229}}].

\bibitem{Balitsky:1997mk}
I.~Balitsky \href{http://xxx.lanl.gov/abs/hep-ph/9706411}{{\tt
  hep-ph/9706411}}.

\bibitem{Mueller:1999wm}
A.~H. Mueller {\em Nucl. Phys.} {\bf B558} (1999) 285--303,
  [\href{http://xxx.lanl.gov/abs/hep-ph/9904404}{{\tt hep-ph/9904404}}].

\bibitem{Kovchegov:1999ua}
Y.~V. Kovchegov {\em Phys. Rev.} {\bf D61} (2000) 074018,
  [\href{http://xxx.lanl.gov/abs/hep-ph/9905214}{{\tt hep-ph/9905214}}].

\bibitem{Kovner:1999bj}
A.~Kovner and J.~G. Milhano {\em Phys. Rev.} {\bf D61} (2000) 014012,
  [\href{http://xxx.lanl.gov/abs/hep-ph/9904420}{{\tt hep-ph/9904420}}].

\bibitem{Golec-Biernat:1998js}
K.~Golec-Biernat and M.~{W{\"u}sthoff} {\em Phys. Rev.} {\bf D59} (1999)
  014017, [\href{http://xxx.lanl.gov/abs/hep-ph/9807513}{{\tt
  hep-ph/9807513}}].

\bibitem{Golec-Biernat:1999qd}
K.~Golec-Biernat and M.~{W{\"u}sthoff} {\em Phys. Rev.} {\bf D60} (1999)
  114023, [\href{http://xxx.lanl.gov/abs/hep-ph/9903358}{{\tt
  hep-ph/9903358}}].

\bibitem{Stasto:2000er}
A.~M. Stasto, K.~Golec-Biernat, and J.~Kwiecinski {\em Phys. Rev. Lett.} {\bf
  86} (2001) 596--599, [\href{http://xxx.lanl.gov/abs/hep-ph/0007192}{{\tt
  hep-ph/0007192}}].

\bibitem{Rummukainen:2003ns}
K.~Rummukainen and H.~Weigert
  \href{http://xxx.lanl.gov/abs/hep-ph/0309306}{{\tt hep-ph/0309306}}.

\bibitem{Iancu:2002xk}
E.~Iancu, A.~Leonidov, and L.~McLerran
  \href{http://xxx.lanl.gov/abs/hep-ph/0202270}{{\tt hep-ph/0202270}}.

\bibitem{Iancu:2003xm}
E.~Iancu and R.~Venugopalan \href{http://xxx.lanl.gov/abs/hep-ph/0303204}{{\tt
  hep-ph/0303204}}.

\bibitem{Hebecker:1998kv}
A.~Hebecker and H.~Weigert {\em Phys. Lett.} {\bf B432} (1998) 215--221,
  [\href{http://xxx.lanl.gov/abs/hep-ph/9804217}{{\tt hep-ph/9804217}}].

\bibitem{Buchmuller:1995mr}
W.~Buchmuller and A.~Hebecker {\em Nucl. Phys.} {\bf B476} (1996) 203--224,
  [\href{http://xxx.lanl.gov/abs/hep-ph/9512329}{{\tt hep-ph/9512329}}].

\bibitem{Buchmuller:1996xw}
W.~Buchmuller, M.~F. McDermott, and A.~Hebecker {\em Nucl. Phys.} {\bf B487}
  (1997) 283--310, [\href{http://xxx.lanl.gov/abs/hep-ph/9607290}{{\tt
  hep-ph/9607290}}].

\bibitem{Hebecker:1999ej}
A.~Hebecker {\em Phys. Rept.} {\bf 331} (2000) 1--115,
  [\href{http://xxx.lanl.gov/abs/hep-ph/9905226}{{\tt hep-ph/9905226}}].

\bibitem{Levin:2003nc}
E.~Levin and M.~Lublinsky {\em Nucl. Phys.} {\bf A730} (2004) 191--211,
  [\href{http://xxx.lanl.gov/abs/hep-ph/0308279}{{\tt hep-ph/0308279}}].

\bibitem{Lam:1999wu}
C.~S. Lam and G.~Mahlon {\em Phys. Rev.} {\bf D61} (2000) 014005,
  [\href{http://xxx.lanl.gov/abs/hep-ph/9907281}{{\tt hep-ph/9907281}}].

\bibitem{'tHooft:1974hx}
G.~'t~Hooft {\em Nucl. Phys.} {\bf B75} (1974) 461.

\bibitem{Beneke:1995pq}
M.~Beneke and V.~M. Braun {\em Nucl. Phys.} {\bf B454} (1995) 253--290,
  [\href{http://xxx.lanl.gov/abs/hep-ph/9506452}{{\tt hep-ph/9506452}}].

\bibitem{Kharzeev:2000ph}
D.~Kharzeev and M.~Nardi {\em Phys. Lett.} {\bf B507} (2001) 121--128,
  [\href{http://xxx.lanl.gov/abs/nucl-th/0012025}{{\tt nucl-th/0012025}}].

\bibitem{Kharzeev:2001yq}
D.~Kharzeev, E.~Levin, and M.~Nardi
  \href{http://xxx.lanl.gov/abs/hep-ph/0111315}{{\tt hep-ph/0111315}}.

\bibitem{Kharzeev:2001gp}
D.~Kharzeev and E.~Levin {\em Phys. Lett.} {\bf B523} (2001) 79--87,
  [\href{http://xxx.lanl.gov/abs/nucl-th/0108006}{{\tt nucl-th/0108006}}].

\bibitem{Baier:2003hr}
R.~Baier, A.~Kovner, and U.~A. Wiedemann {\em Phys. Rev.} {\bf D68} (2003)
  054009, [\href{http://xxx.lanl.gov/abs/hep-ph/0305265}{{\tt
  hep-ph/0305265}}].

\bibitem{Kowalski:2003hm}
H.~Kowalski and D.~Teaney {\em Phys. Rev.} {\bf D68} (2003) 114005,
  [\href{http://xxx.lanl.gov/abs/hep-ph/0304189}{{\tt hep-ph/0304189}}].

\bibitem{Bondarenko:2003ym}
S.~Bondarenko, M.~Kozlov, and E.~Levin {\em Nucl. Phys.} {\bf A727} (2003)
  139--178, [\href{http://xxx.lanl.gov/abs/hep-ph/0305150}{{\tt
  hep-ph/0305150}}].

\bibitem{Gotsman:2004ra}
E.~Gotsman, M.~Kozlov, E.~Levin, U.~Maor, and E.~Naftali {\em Nucl. Phys.} {\bf
  A742} (2004) 55--79, [\href{http://xxx.lanl.gov/abs/hep-ph/0401021}{{\tt
  hep-ph/0401021}}].

\bibitem{Kovchegov:1999yj}
Y.~V. Kovchegov {\em Phys. Rev.} {\bf D60} (1999) 034008,
  [\href{http://xxx.lanl.gov/abs/hep-ph/9901281}{{\tt hep-ph/9901281}}].

\bibitem{Weigert:2003mm}
H.~Weigert {\em Nucl. Phys.} {\bf B685} (2004) 321--350,
  [\href{http://xxx.lanl.gov/abs/hep-ph/0312050}{{\tt hep-ph/0312050}}].

\bibitem{Iancu:2002tr}
E.~Iancu, K.~Itakura, and L.~McLerran {\em Nucl. Phys.} {\bf A708} (2002)
  327--352, [\href{http://xxx.lanl.gov/abs/hep-ph/0203137}{{\tt
  hep-ph/0203137}}].

\bibitem{Munier:2003vc}
S.~Munier and R.~Peschanski {\em Phys. Rev. Lett.} {\bf 91} (2003) 232001,
  [\href{http://xxx.lanl.gov/abs/hep-ph/0309177}{{\tt hep-ph/0309177}}].

\bibitem{Munier:2003sj}
S.~Munier and R.~Peschanski {\em Phys. Rev.} {\bf D69} (2004) 034008,
  [\href{http://xxx.lanl.gov/abs/hep-ph/0310357}{{\tt hep-ph/0310357}}].

\bibitem{Munier:2004xu}
S.~Munier and R.~Peschanski \href{http://xxx.lanl.gov/abs/hep-ph/0401215}{{\tt
  hep-ph/0401215}}.

\bibitem{FISHER}
R.~Fisher {\em Ann. Eugenics} {\bf 7} (1937) 355.

\bibitem{KPP}
A.~Kolmogorov, I.~Petrovsky, and N.~Piscounov {\em Moscou Univ. Bull. Math.}
  {\bf A 1} (1937) 1.

\bibitem{Braun:2000wr}
M.~Braun {\em Eur. Phys. J.} {\bf C16} (2000) 337--347,
  [\href{http://xxx.lanl.gov/abs/hep-ph/0001268}{{\tt hep-ph/0001268}}].

\bibitem{Kimber:2001nm}
M.~A. Kimber, J.~Kwiecinski, and A.~D. Martin {\em Phys. Lett.} {\bf B508}
  (2001) 58--64, [\href{http://xxx.lanl.gov/abs/hep-ph/0101099}{{\tt
  hep-ph/0101099}}].

\bibitem{Armesto:2001fa}
N.~Armesto and M.~A. Braun {\em Eur. Phys. J.} {\bf C20} (2001) 517--522,
  [\href{http://xxx.lanl.gov/abs/hep-ph/0104038}{{\tt hep-ph/0104038}}].

\bibitem{Levin:2001et}
E.~Levin and M.~Lublinsky {\em Nucl. Phys.} {\bf A696} (2001) 833--850,
  [\href{http://xxx.lanl.gov/abs/hep-ph/0104108}{{\tt hep-ph/0104108}}].

\bibitem{Lublinsky:2001bc}
M.~Lublinsky {\em Eur. Phys. J.} {\bf C21} (2001) 513--519,
  [\href{http://xxx.lanl.gov/abs/hep-ph/0106112}{{\tt hep-ph/0106112}}].

\bibitem{Lublinsky:2001yi}
M.~Lublinsky, E.~Gotsman, E.~Levin, and U.~Maor {\em Nucl. Phys.} {\bf A696}
  (2001) 851--869, [\href{http://xxx.lanl.gov/abs/hep-ph/0102321}{{\tt
  hep-ph/0102321}}].

\bibitem{Gotsman:2002yy}
E.~Gotsman, E.~Levin, M.~Lublinsky, and U.~Maor {\em Eur. Phys. J.} {\bf C27}
  (2003) 411--425, [\href{http://xxx.lanl.gov/abs/hep-ph/0209074}{{\tt
  hep-ph/0209074}}].

\bibitem{Golec-Biernat:2001if}
K.~Golec-Biernat, L.~Motyka, and A.~M. Stasto {\em Phys. Rev.} {\bf D65} (2002)
  074037, [\href{http://xxx.lanl.gov/abs/hep-ph/0110325}{{\tt
  hep-ph/0110325}}].

\bibitem{Albacete:2003iq}
J.~L. Albacete, N.~Armesto, A.~Kovner, C.~A. Salgado, and U.~A. Wiedemann {\em
  Phys. Rev. Lett.} {\bf 92} (2004) 082001,
  [\href{http://xxx.lanl.gov/abs/hep-ph/0307179}{{\tt hep-ph/0307179}}].

\bibitem{Bartels:2004ef}
J.~Bartels, L.~N. Lipatov, and G.~P. Vacca
  \href{http://xxx.lanl.gov/abs/hep-ph/0404110}{{\tt hep-ph/0404110}}.

\bibitem{Lotter:1996vk}
H.~Lotter \href{http://xxx.lanl.gov/abs/hep-ph/9705288}{{\tt hep-ph/9705288}}.

\bibitem{Banfi:2002hw}
A.~Banfi, G.~Marchesini, and G.~Smye {\em JHEP} {\bf 08} (2002) 006,
  [\href{http://xxx.lanl.gov/abs/hep-ph/0206076}{{\tt hep-ph/0206076}}].

\bibitem{Mueller:2002zm}
A.~H. Mueller and D.~N. Triantafyllopoulos {\em Nucl. Phys.} {\bf B640} (2002)
  331--350, [\href{http://xxx.lanl.gov/abs/hep-ph/0205167}{{\tt
  hep-ph/0205167}}].

\bibitem{Triantafyllopoulos:2002nz}
D.~N. Triantafyllopoulos {\em Nucl. Phys.} {\bf B648} (2003) 293--316,
  [\href{http://xxx.lanl.gov/abs/hep-ph/0209121}{{\tt hep-ph/0209121}}].

\bibitem{Blaizot:2002xy}
J.-P. Blaizot, E.~Iancu, and H.~Weigert {\em Nucl. Phys.} {\bf A713} (2003)
  441--469, [\href{http://xxx.lanl.gov/abs/hep-ph/0206279}{{\tt
  hep-ph/0206279}}].

\bibitem{Dasgupta:2001sh}
M.~Dasgupta and G.~P. Salam {\em Phys. Lett.} {\bf B512} (2001) 323--330,
  [\href{http://xxx.lanl.gov/abs/hep-ph/0104277}{{\tt hep-ph/0104277}}].

\bibitem{Dasgupta:2002bw}
M.~Dasgupta and G.~P. Salam {\em JHEP} {\bf 03} (2002) 017,
  [\href{http://xxx.lanl.gov/abs/hep-ph/0203009}{{\tt hep-ph/0203009}}].

\bibitem{Catani:1993ua}
S.~Catani, L.~Trentadue, G.~Turnock, and B.~R. Webber {\em Nucl. Phys.} {\bf
  B407} (1993) 3--42.

\bibitem{Marchesini:2003nh}
G.~Marchesini and A.~H. Mueller
  \href{http://xxx.lanl.gov/abs/hep-ph/0308284}{{\tt hep-ph/0308284}}.

\bibitem{Bassetto:1983ik}
A.~Bassetto, M.~Ciafaloni, and G.~Marchesini {\em Phys. Rept.} {\bf 100} (1983)
  201--272.

\bibitem{Fiorani:1988by}
F.~Fiorani, G.~Marchesini, and L.~Reina {\em Nucl. Phys.} {\bf B309} (1988)
  439.

\bibitem{Kovner:2001bh}
A.~Kovner and U.~A. Wiedemann {\em Phys. Rev.} {\bf D66} (2002) 051502,
  [\href{http://xxx.lanl.gov/abs/hep-ph/0112140}{{\tt hep-ph/0112140}}].

\bibitem{Kovner:2002xa}
A.~Kovner and U.~A. Wiedemann {\em Phys. Rev.} {\bf D66} (2002) 034031,
  [\href{http://xxx.lanl.gov/abs/hep-ph/0204277}{{\tt hep-ph/0204277}}].

\bibitem{Ferreiro:2002kv}
E.~Ferreiro, E.~Iancu, K.~Itakura, and L.~McLerran {\em Nucl. Phys.} {\bf A710}
  (2002) 373--414, [\href{http://xxx.lanl.gov/abs/hep-ph/0206241}{{\tt
  hep-ph/0206241}}].

\bibitem{Gribov:1984tu}
L.~V. Gribov, E.~M. Levin, and M.~G. Ryskin {\em Phys. Rept.} {\bf 100} (1983)
  1--150.

\bibitem{Kovchegov:1999ji}
Y.~V. Kovchegov and E.~Levin {\em Nucl. Phys.} {\bf B577} (2000) 221--239,
  [\href{http://xxx.lanl.gov/abs/hep-ph/9911523}{{\tt hep-ph/9911523}}].

\bibitem{Iancu:2003zr}
E.~Iancu and A.~H. Mueller {\em Nucl. Phys.} {\bf A730} (2004) 494--513,
  [\href{http://xxx.lanl.gov/abs/hep-ph/0309276}{{\tt hep-ph/0309276}}].

\bibitem{Fadin:1995xg}
V.~S. Fadin, M.~I. Kotsky, and R.~Fiore {\em Phys. Lett.} {\bf B359} (1995)
  181--188.

\bibitem{Fadin:1996zv}
V.~S. Fadin, M.~I. Kotsky, and L.~N. Lipatov
  \href{http://xxx.lanl.gov/abs/hep-ph/9704267}{{\tt hep-ph/9704267}}.

\bibitem{Fadin:1998hr}
V.~S. Fadin, R.~Fiore, A.~Flachi, and M.~I. Kotsky {\em Phys. Lett.} {\bf B422}
  (1998) 287--293, [\href{http://xxx.lanl.gov/abs/hep-ph/9711427}{{\tt
  hep-ph/9711427}}].

\bibitem{Fadin:1998py}
V.~S. Fadin and L.~N. Lipatov {\em Phys. Lett.} {\bf B429} (1998) 127--134,
  [\href{http://xxx.lanl.gov/abs/hep-ph/9802290}{{\tt hep-ph/9802290}}].

\bibitem{Colferai:1999em}
D.~Colferai \href{http://xxx.lanl.gov/abs/hep-ph/0008309}{{\tt
  hep-ph/0008309}}.

\bibitem{Thorne:1999rb}
R.~S. Thorne {\em Phys. Rev.} {\bf D60} (1999) 054031,
  [\href{http://xxx.lanl.gov/abs/hep-ph/9901331}{{\tt hep-ph/9901331}}].

\bibitem{Thorne:2001nr}
R.~S. Thorne {\em Phys. Rev.} {\bf D64} (2001) 074005,
  [\href{http://xxx.lanl.gov/abs/hep-ph/0103210}{{\tt hep-ph/0103210}}].

\bibitem{Albacete:2004gw}
J.~L. Albacete, N.~Armesto, J.~G. Milhano, C.~A. Salgado, and U.~A. Wiedemann
  \href{http://xxx.lanl.gov/abs/hep-ph/0408216}{{\tt hep-ph/0408216}}.

\bibitem{Mueller:2003bz}
A.~H. Mueller \href{http://xxx.lanl.gov/abs/hep-ph/0301109}{{\tt
  hep-ph/0301109}}.

\bibitem{Laenen:1994gh}
E.~Laenen and E.~Levin {\em Ann. Rev. Nucl. Part. Sci.} {\bf 44} (1994)
  199--246.

\bibitem{Kovchegov:1997ke}
Y.~V. Kovchegov and D.~H. Rischke {\em Phys. Rev.} {\bf C56} (1997) 1084--1094,
  [\href{http://xxx.lanl.gov/abs/hep-ph/9704201}{{\tt hep-ph/9704201}}].

\bibitem{Gyulassy:1997vt}
M.~Gyulassy and L.~D. McLerran {\em Phys. Rev.} {\bf C56} (1997) 2219--2228,
  [\href{http://xxx.lanl.gov/abs/nucl-th/9704034}{{\tt nucl-th/9704034}}].

\bibitem{Kovchegov:1998bi}
Y.~V. Kovchegov and A.~H. Mueller {\em Nucl. Phys.} {\bf B529} (1998) 451--479,
  [\href{http://xxx.lanl.gov/abs/hep-ph/9802440}{{\tt hep-ph/9802440}}].

\bibitem{Kovchegov:2001ni}
Y.~V. Kovchegov {\em Phys. Rev.} {\bf D64} (2001) 114016,
  [\href{http://xxx.lanl.gov/abs/hep-ph/0107256}{{\tt hep-ph/0107256}}].

\bibitem{Kovchegov:2001sc}
Y.~V. Kovchegov and K.~Tuchin {\em Phys. Rev.} {\bf D65} (2002) 074026,
  [\href{http://xxx.lanl.gov/abs/hep-ph/0111362}{{\tt hep-ph/0111362}}].

\bibitem{Iancu:2003ge}
E.~Iancu, K.~Itakura, and S.~Munier {\em Phys. Lett.} {\bf B590} (2004)
  199--208, [\href{http://xxx.lanl.gov/abs/hep-ph/0310338}{{\tt
  hep-ph/0310338}}].

\bibitem{Arneodo:1997kd}
{\bf New Muon} Collaboration, M.~Arneodo {\em et.~al.} {\em Nucl. Phys.} {\bf
  B487} (1997) 3--26, [\href{http://xxx.lanl.gov/abs/hep-ex/9611022}{{\tt
  hep-ex/9611022}}].

\bibitem{Adams:1995ri}
{\bf E665} Collaboration, M.~R. Adams {\em et.~al.} {\em Z. Phys.} {\bf C65}
  (1995) 225--244.

\bibitem{Adams:1995is}
{\bf E665} Collaboration, M.~R. Adams {\em et.~al.} {\em Z. Phys.} {\bf C67}
  (1995) 403--410, [\href{http://xxx.lanl.gov/abs/hep-ex/9505006}{{\tt
  hep-ex/9505006}}].

\bibitem{Freund:2002ux}
A.~Freund, K.~Rummukainen, H.~Weigert, and A.~{Sch{\"a}fer} {\em Phys. Rev.
  Lett.} {\bf 90} (2003) 222002,
  [\href{http://xxx.lanl.gov/abs/hep-ph/0210139}{{\tt hep-ph/0210139}}].

\bibitem{Armesto:2004ud}
N.~Armesto, C.~A. Salgado, and U.~A. Wiedemann
  \href{http://xxx.lanl.gov/abs/hep-ph/0407018}{{\tt hep-ph/0407018}}.

\bibitem{EICwhite}
{{\em The Electron Ion Collider: A white paper}, BNL Report
  BNL-68933-02/07-REV, Eds. A. Deshpande, R. Milner and R. Venugopalan. }.

\bibitem{Krasnitz:1998ns}
A.~Krasnitz and R.~Venugopalan {\em Nucl. Phys.} {\bf B557} (1999) 237,
  [\href{http://xxx.lanl.gov/abs/hep-ph/9809433}{{\tt hep-ph/9809433}}].

\bibitem{Krasnitz:1999wc}
A.~Krasnitz and R.~Venugopalan {\em Phys. Rev. Lett.} {\bf 84} (2000)
  4309--4312, [\href{http://xxx.lanl.gov/abs/hep-ph/9909203}{{\tt
  hep-ph/9909203}}].

\bibitem{Krasnitz:2003nv}
A.~Krasnitz, Y.~Nara, and R.~Venugopalan {\em Braz. J. Phys.} {\bf 33} (2003)
  223--230.

\bibitem{Krasnitz:2003jw}
A.~Krasnitz, Y.~Nara, and R.~Venugopalan {\em Nucl. Phys.} {\bf A727} (2003)
  427--436, [\href{http://xxx.lanl.gov/abs/hep-ph/0305112}{{\tt
  hep-ph/0305112}}].

\bibitem{Lappi:2003bi}
T.~Lappi {\em Phys. Rev.} {\bf C67} (2003) 054903,
  [\href{http://xxx.lanl.gov/abs/hep-ph/0303076}{{\tt hep-ph/0303076}}].

\bibitem{Kharzeev:2002pc}
D.~Kharzeev, E.~Levin, and L.~McLerran {\em Phys. Lett.} {\bf B561} (2003)
  93--101, [\href{http://xxx.lanl.gov/abs/hep-ph/0210332}{{\tt
  hep-ph/0210332}}].

\bibitem{Kharzeev:2003wz}
D.~Kharzeev, Y.~V. Kovchegov, and K.~Tuchin {\em Phys. Rev.} {\bf D68} (2003)
  094013, [\href{http://xxx.lanl.gov/abs/hep-ph/0307037}{{\tt
  hep-ph/0307037}}].

\bibitem{Jalilian-Marian:2003mf}
J.~Jalilian-Marian, Y.~Nara, and R.~Venugopalan {\em Phys. Lett.} {\bf B577}
  (2003) 54--60, [\href{http://xxx.lanl.gov/abs/nucl-th/0307022}{{\tt
  nucl-th/0307022}}].

\bibitem{Arsene:2004ux}
{\bf BRAHMS} Collaboration, I.~Arsene {\em et.~al.}
  \href{http://xxx.lanl.gov/abs/nucl-ex/0403005}{{\tt nucl-ex/0403005}}.

\bibitem{Vitev:2003xu}
I.~Vitev {\em Phys. Lett.} {\bf B562} (2003) 36--44,
  [\href{http://xxx.lanl.gov/abs/nucl-th/0302002}{{\tt nucl-th/0302002}}].

\bibitem{Accardi:2003jh}
A.~Accardi and M.~Gyulassy {\em Phys. Lett.} {\bf B586} (2004) 244--253,
  [\href{http://xxx.lanl.gov/abs/nucl-th/0308029}{{\tt nucl-th/0308029}}].

\bibitem{Kharzeev:2004yx}
D.~Kharzeev, Y.~V. Kovchegov, and K.~Tuchin
  \href{http://xxx.lanl.gov/abs/hep-ph/0405045}{{\tt hep-ph/0405045}}.

\bibitem{Kharzeev:2004if}
D.~Kharzeev, E.~Levin, and M.~Nardi
  \href{http://xxx.lanl.gov/abs/hep-ph/0408050}{{\tt hep-ph/0408050}}.

\bibitem{Jalilian-Marian:2004da}
J.~Jalilian-Marian and Y.~V. Kovchegov
  \href{http://xxx.lanl.gov/abs/hep-ph/0405266}{{\tt hep-ph/0405266}}.

\bibitem{Kharzeev:2004bw}
D.~Kharzeev, E.~Levin, and L.~McLerran
  \href{http://xxx.lanl.gov/abs/hep-ph/0403271}{{\tt hep-ph/0403271}}.

\bibitem{Strikman:2004km}
M.~Strikman and C.~Weiss \href{http://xxx.lanl.gov/abs/hep-ph/0408345}{{\tt
  hep-ph/0408345}}.

\bibitem{Gribov:1981ac}
L.~V. Gribov, E.~M. Levin, and M.~G. Ryskin {\em Nucl. Phys.} {\bf B188} (1981)
  555--576.

\bibitem{Mueller:1995gb}
A.~H. Mueller {\em Nucl. Phys.} {\bf B437} (1995) 107--126,
  [\href{http://xxx.lanl.gov/abs/hep-ph/9408245}{{\tt hep-ph/9408245}}].

\bibitem{Chen:1995pa}
Z.~Chen and A.~H. Mueller {\em Nucl. Phys.} {\bf B451} (1995) 579--604.

\end{thebibliography}
%\bibliographystyle{JHEP-short} %<-----------BIBSTYLE 

\end{document}